\documentclass[binding=0.6cm]{sapthesis}
\usepackage{microtype}
\usepackage[english]{babel}
\usepackage[utf8]{inputenc}
\usepackage{hyperref}

\usepackage{slashed}
\usepackage{physics}
\usepackage{ amssymb }
\usepackage{breqn}
\usepackage{halloweenmath}
\usepackage{amsthm}
\newtheorem{theorem}{Theorem}[chapter]
\newtheorem{proposition}[theorem]{Proposition}
\newtheorem{conjecture}[theorem]{Conjecture}
\theoremstyle{remark}
\newtheorem{remark}[theorem]{Remark}
\hypersetup{pdftitle={Two Recursions for the Seiberg-Witten Prepotential: Zamolodchikov's Recursion, the AGT Correspondence and Topological Recursion on the Seiberg-Witten Curve},pdfauthor={Bastam Tajik}}
\title{Two Recursions for the Seiberg-Witten Prepotential}
\subtitle{Zamolodchikov's Recursion, the AGT \mbox{Correspondence} and \mbox{Topological} Recursion on the \mbox{Seiberg-Witten} Curve}
\author{Bastam Tajik}
\IDnumber{1822751}
\course{Theoretical Physics}
\courseorganizer{Facoltà di Fisica}
\AcademicYear{2024/2025}
\advisor{Prof. Francesco Fucito }
\coadvisor{Dr. Fabio Riccioni}
\authoremail{tajik.1822751@studenti.uniroma1.it}
\copyyear{2023}
\thesistype{Master thesis}
\begin{document}
\frontmatter
\maketitle
\dedication{Dedicated to \\ My parents and Aunt}
\begin{abstract}
The low-energy dynamics of $\mathcal{N}=2$ supersymmetric Yang-Mills theory is encoded in a single holomorphic function, the prepotential, which Seiberg and Witten determined exactly from the periods of a family of complex curves \cite{SW2}. This thesis follows the prepotential through the different constructions that have been found for it since, and ends by confronting two of them which are recursive in nature.

The first part is devoted to the microscopic side. The ADHM construction is presented as a hyperKähler quotient, the supersymmetric collective-coordinate measure on the moduli space of instantons is built, and, after a review of the $\mathcal{N}=1$ dualities and of Seiberg-Witten theory with and without matter, the one- and two-instanton contributions to the prepotential are computed, first directly and then through the non-commutative resolution of the moduli space and localization \cite{Algorithm},\cite{All orders}.

The second part turns to Nekrasov's partition function and to the AGT correspondence \cite{AGT}, which identifies it with a Virasoro conformal block of two-dimensional conformal field theory. Zamolodchikov's recursion for the conformal block \cite{ZM} then becomes a recursion for the instanton partition function \cite{Recursion},\cite{Recursion2}. For the pure $SU(2)$ theory this recursion is run explicitly up to five instantons, together with its first corrections in the $\Omega$-background.

In the last part the same quantities are obtained from the topological recursion of Eynard and Orantin \cite{EO} on the Seiberg-Witten curve, which yields the Taylor expansion of the prepotential around an arbitrary point of the moduli space \cite{Chaimanowong}. For pure $SU(2)$ the Bergman kernel and the first steps of the recursion are found in closed form. They reproduce the instanton series generated by Zamolodchikov's recursion, remain valid at strong coupling, where they detect the massless monopole, and at genus one give the first correction of the partition function on the self-dual $\Omega$-background. The two recursions thus appear as complementary expansions of one function: one exact in the vacuum expectation value and perturbative in the instanton number, the other exact in the instanton parameter and perturbative around a fixed vacuum. A conjecture extending the correspondence to all genera, and a list of open problems, close the thesis.

\end{abstract}
\tableofcontents
\mainmatter

\makeatletter
\newcommand*{\rawslashed}[1]{{\mathpalette{\sla@/00}{#1}}}
\makeatother
\setcounter{chapter}{-1}
\chapter{Conventions, Notations and Prerequisites}
\label{Conventions,-Notations-and-Prerequisites}
In this thesis work, the notation is a synthesis of different incompatible references. Therefore, for clarity and ease of reading, one reference \cite{Instanton Calculus} is chosen in particular, and the same notation is adopted to express the content of the other sources.  

Throughout this thesis, the partition function is always assumed in the following format:

\begin{dmath}
    Z[J]= \int[d\Phi] \ \text{exp} \left( -\frac{1}{g^2} S[\Phi] + \int d^4x \  J \Phi \right)
\end{dmath}

where $\Phi$ stands for any type of fields (not superfields, unless the integral is explicitly carried out over the entire superspace) with a source $J$ and \textbf{Euclidean action} $S[\Phi]$ which is real and bounded from below and $g^2$ stands for the gauge theory coupling that can be always managed to appear in front of the action by proper rescaling of the fields.

The \textbf{Super-Poincaré algebra} consists of the generators that satisfy the following relations among themselves:

\begin{dmath}
    [P^{\mu}, P^{\nu}]=0
\end{dmath}
\begin{dmath}
    [M^{\mu \nu} , P^{\sigma}]= i (P^{\mu} {\eta}^{\nu \sigma} -P^{\nu} {\eta}^{\mu \sigma})
\end{dmath}
\begin{dmath}
    [M^{\mu \nu} , M^{\rho \sigma}] = i({\eta}^{\nu \rho} M^{\mu \sigma} - {\eta}^{\nu \sigma} M^{\mu \rho} + {\eta}^{\mu \sigma} M^{\nu \rho} - {\eta}^{\mu \rho} M^{\nu \sigma})
\end{dmath}
\begin{dmath}
    \{Q_{\alpha} , \bar{Q}_{\dot{\alpha}} \} = 2 \sigma^{\mu}_{\alpha \dot{\alpha}} P_{\mu}
\end{dmath}
\begin{dmath}
    [M^{\mu \nu} , Q_{\alpha}] = (\sigma^{\mu \nu})_{\alpha}^{\beta} Q_{\beta}
\end{dmath}
\begin{dmath}
    [M^{\mu \nu} , \bar{Q}^{\dot{\alpha}}]= (\bar{\sigma}^{\mu \nu})_{\dot{\beta}}^{\dot{\alpha}} \bar{Q}^{\dot{\beta}}
\end{dmath}

\begin{dmath}
{[Q_{\alpha} , P^{\mu}]=\{ Q_{\alpha} , Q_{\beta} \} = 0}
\end{dmath}

One can inspect the existence of a $U(1)$ external automorphism on the algebra that acts on the generators by $Q_{\alpha} \rightarrow e^{-i \lambda} Q_{\alpha}$ and $\bar{Q}_{\dot{\alpha}} \rightarrow e^{i \lambda} \bar{Q}_{\dot{\alpha}}$. By denoting the generator of such action by $R$ one can find the commutation relations:
$$[R, Q_{\alpha}] = - Q_{\alpha} \ \ \ \ \ \ \ \ \ [R, \bar{Q}_{\dot{\alpha}}] = + \bar{Q}_{\dot{\alpha}}$$

The representations of the SuperPoincaré algebra are classified by the two Casimir operators:

\begin{dmath}
    C_1 = P_{\mu} P^{\mu}  \ \ \ \ \   {C_2 = X_{\mu \nu} X^{\mu \nu}}
\end{dmath}

where:

\begin{dmath}
    X_{\mu \nu} = Y_{\mu} P_{\nu} - Y_{\nu} P_{\mu} \ \ \ \ \ \ \ \ {Y_{\mu} = W_{\mu} - \frac{1}{4} \bar{Q}_{\dot{\alpha}} \bar{\sigma}^{\dot{\alpha} \beta}_{\mu} Q_{\beta}}
\end{dmath}

in which $W_{\mu}$ is the Pauli-Lubanski vector.
Spacetime and Superspace can be defined as:

\begin{dmath}
{\mathbb{R}^{1,3}= \text{Spacetime} = \frac{\text{Poincaré} \ \text{Group}}{\text{Lorentz} \ \text{Group}} = ISO(1,3)/SO(1,3)}
\end{dmath}

\begin{dmath}
    {\mathbb{R}^{{1,3}|4} = \text{Superspace} = \frac{\text{SuperPoincaré} \ \text{Group}}{\text{Lorentz} \ \text{Group}} = ISO(1,3|1)/SO(1,3)}
\end{dmath}

To include spinorial representations of the Poincaré group, one should elevate the Lorentz group $SO(1,3)$ to its double cover $SL(2, \mathbb{C})$, bearing in mind that the \underline{vectorial representations remain intact}. Said differently: $SL(2, \mathbb{C}) / {\mathbb{Z}_2} = SO(1,3)$.

\

In the \textbf{Extended Super-Poincaré algebra} there are more than one pair of Grassmann generators and their Hermitian conjugate. A new index is introduced: $I$ that counts for such new generators: $Q_{\alpha}^I , {{\bar{Q}}_{\dot{\beta}}}^J$.

\subsection{Massless representation}
Since the spectrum is massless $P_{\mu} = E(1,0,0,1)$ and so ${\sigma}^{\mu} P_{\mu}=\begin{pmatrix}
    0 & 0 \\
    0 & 2E
\end{pmatrix}$
so that $\{Q_{\alpha}^I , {\bar{Q}}_{\dot{\beta}}^J \}=\begin{pmatrix}
    0 & 0 \\
    0 & 4E
\end{pmatrix}_{\alpha \dot{\beta}} {\delta}^{IJ}$ and in particular $\{Q_{1}^I , {\bar{Q}}_{\dot{1}}^J \} = 0 \ \ \ \ \forall I,J$ which shows simply that $Q^I_1 = {\bar{Q}_{\dot{1}}}^J = 0$ by Hilbert space positive definiteness. So there remain only half of the $2N$ generators that can be written as:
\begin{dmath}
    {a_I = \frac{1}{\sqrt{4E}} Q^I_2} \ \ , \ \ {a^{\dagger}_I = \frac{1}{\sqrt{4E}} {\bar{Q}_{{\dot{2}}}}^I}
\end{dmath}
which is subject to simple anticommutation relations:
\begin{dmath}
    {\{a_I,{a^{\dagger}_J}\}= {\delta}_{IJ}} \ \ \ \ {\{a_I,a_J\} = \{{a^{\dagger}_I},{a^{\dagger}_J}\} = 0} 
\end{dmath}

As before, the Clifford vacuum is the state annihilated by all the lowering ladder operators. Clearly giving rise to a spectrum:
$$ |\lambda_0 > \ \ a^{\dagger}_I |\lambda_0>=|\lambda_0 + \frac{1}{2}> \ \ a^{\dagger}_I a^{\dagger}_J |\lambda_0> = |\lambda_0 + 1> \ \ ... \ \ a^{\dagger}_1 a^{\dagger}_2 ... a^{\dagger}_N |\lambda_0> = |\lambda_0 + \frac{N}{2}>$$ and their CPT conjugate, which in general should be added(which are states with a different choice of $\lambda_0$).

\subsection{Massive representation}

For massive representation, without any loss of generality, one can choose the rest frame where $P=(m,0,0,0)$ where the SUSY algebra becomes:

\begin{dmath}
    \{Q^I_{\alpha} , (Q^J_{\beta})^{\dagger} \} = 2m {\delta}_{\alpha \beta} {\delta}^{IJ}
\end{dmath}
\begin{dmath}
    \{Q^I_{\alpha} , Q^J_{\beta} \} = \epsilon_{\alpha \beta} Z^{IJ}
\end{dmath}

\begin{dmath}
    \{(Q^I_{\alpha})^{\dagger} , (Q^J_{\beta})^{\dagger} \} = \epsilon_{\alpha \beta} (Z^{IJ})^{*}
\end{dmath}
where $Z^{IJ}$ is the central charge and can be written in the block-diagonal form by a proper $U(\mathcal{N})$ rotation:

\begin{dmath}
    Z=\begin{pmatrix}
    0 & q_1 & 0 & 0 & . & . & . & \\
    -q_1 & 0 & 0 & 0 \\
    0 & 0 & 0 & q_2 \\
    0 & 0 & -q_2 & 0 \\
. \\
. \\
. 
\end{pmatrix}
\end{dmath}
One should assume $Z$ is even-dimensional otherwise there's an extra zero eigenvalue that can be ignored. One can introduce the new operators:

\begin{dmath}
    a^1_{\alpha} = \frac{1}{\sqrt{2}} (Q^1_{\alpha} + \epsilon_{\alpha \beta} (Q^2_{\beta})^{\dagger})
\end{dmath}

\begin{dmath}
    b^1_{\alpha} = \frac{1}{\sqrt{2}} (Q^1_{\alpha} - \epsilon_{\alpha \beta} (Q^2_{\beta})^{\dagger})
\end{dmath}
\begin{dmath}
   a^2_{\alpha} = \frac{1}{\sqrt{2}} (Q^3_{\alpha} + \epsilon_{\alpha \beta} (Q^4_{\beta})^{\dagger})
\end{dmath}
\begin{dmath}
    b^2_{\alpha} = \frac{1}{\sqrt{2}} (Q^3_{\alpha} - \epsilon_{\alpha \beta}(Q^4_{\beta})^{\dagger})
\end{dmath}
$$...$$
and these new operators satisfy the anti-commutation relations:
\begin{dmath}
    {\{a^r_{\alpha} , (a^s_{\beta})^{\dagger}\} = (2m+q_r) \delta_{rs} \delta_{\alpha \beta}}
\end{dmath}
\begin{dmath}
    \{b^r_{\alpha} , (b^s_{\beta})^{\dagger} \} = (2m - q_r) \delta_{rs} \delta_{\alpha \beta}
\end{dmath}
\begin{dmath}
   {\{ a^r_{\alpha} , (b^s_{\beta}) \} = \{ a^r_{\alpha} , a^s_{\beta} \} = ... =0}
\end{dmath}

Again, positivity of the Hilbert space results in the nontrivial result: $2m \geq |q_n|$ 
Obviously, for the massless case where $m=0$, one finds $q_n = 0$, which means there's no central charge for the massless part of the spectrum.

\chapter{ADHM construction as hyperKähler quotient, Super-Geometry of the Moduli Space}
\section{Poisson Manifolds and Momentum Map}
\label{sec:Poisson-Manifolds-and-Momentum-Map}
A \textbf{Poisson structure/bracket} $\{.\}$ on a smooth manifold $P$ is a bilinear operation on $\mathcal{F}(P)=C^{\infty}(P)$ such that:

\begin{itemize}
    \item $(\mathcal{F}(P),\{.\})$ is an associative algebra closed under $\{.\} : \mathcal{F}(P) \times \mathcal{F}(P) \rightarrow \mathcal{F}(P)$
    \item $\{.\}$ is antisymmetric: $\{F,G\} = -\{G,F\} $
    
    \item $\{.\}$ follows Leibniz rule, that is: $\{ FG,H \}= \{ F,H \}G + F \{ G,H \}$
    \item $\{.\}$ follows Jacobi identity: $\{F,\{G,H\}\}+\{H,\{F,G\}\}+\{G,\{H,F\}\}=0$
\end{itemize}

A \textbf{Poisson manifold} is a smooth manifold $P$ endowed with a Poisson structure on $\mathcal{F}(P)$.

Any \textbf{Symplectic manifold}(Section \ref{sec:Symplectic,-Kählar-and-HyperKähler-Quotients}) is a Poisson manifold.

If $H \in \mathcal{F}(P)$, then there exists a unique vector field $X_H$ on $P$ such that $X_H [G] = \{G,H\}$ for all $G \in \mathcal{F}(P)$ where $X_H [G]:=\mathcal{L}_{X_H} G$. $X_H$ is called the \textbf{Hamiltonian vector field} of $H$. 

Note that in the context of symplectic manifolds(Section \ref{sec:Symplectic,-Kählar-and-HyperKähler-Quotients}), a vector field is called \textbf{Hamiltonian} if there exists a function $H: P \rightarrow \mathbb{R}$ such that ${i_X}\omega=dH$ where $i_X$ is the interior product.

The map $H\mapsto{X_H}$ of $\mathcal{F}(P)$ to $\mathfrak{X}(P)$(Lie algebra of all vector fields on the manifold $P$) is a Lie algebra antihomomorphism. In other words:
\begin{dmath}
    [X_H,X_K]=-X_{\{H,K\}}
\end{dmath}
Let $H \in \mathcal{F}(P)$ and $\phi_t$ be the flow of $X_H$. Then:

\begin{itemize}
    \item $ \frac{\mathrm{d}}{\mathrm{d}t}(F \circ \phi_t)={\{F,H\} \circ \phi_t}={\{F\circ \phi_t,H\}}$ or neatly $\dot{F}=\{F,H\} $
    \item  $H\circ{\phi_t}=H$ (\textit{a.k.a.}. \textbf{conservation of energy})
\end{itemize}

Let $G,H \in \mathcal{F}(P)$. Then $G$ is constant along the integral curves of $X_H$ if and only if $\{G,H\}=0$, if and only if $H$ is constant along the integral curves of $X_G$.

Those elements of $\mathcal{F}(P)$ such as $C$, that $\{C,F\}=0$ for all $F \in \mathcal{F}(P)$ are called \textbf{Casimir functions} of the Poisson structure.

Suppose the Lie group $G$ acts on the Poisson manifold $P$ from left via the map $\Phi: G \times P\rightarrow{P}$. The action is called \textbf{canonical} if and only if it preserves the Poisson structure. In other words: $\Phi^*_g \{F_1,F_2\}=\{\Phi^*_g F_1, \Phi^*_g F_2 \}$ for all $g\in G$ and $F_1 , F_2 \in \mathcal{F}(P)$, where $\Phi_g: P\rightarrow{P}$ is the restriction of $\Phi$ to a fixed $g$.

Consequently, in the case of symplectic manifolds(Section \ref{sec:Symplectic,-Kählar-and-HyperKähler-Quotients}) the canonical actions can be pronounced as: $\Phi^*_g \Omega =\Omega$ where $\Omega$ is the \textbf{symplectic form}(Section \ref{sec:Symplectic,-Kählar-and-HyperKähler-Quotients}).

An \textbf{infinitesimal generator} of the Lie algebra action corresponding to an element $\xi\in \mathfrak{g}$ is the vector field $\xi_P$ on a manifold $P$ given by: 
\begin{dmath}
    {{\xi_P}(p)=\frac{\mathrm{d}}{\mathrm{d}t}[\exp{(t\xi)}.p]|_{t=0} \ \ \ \forall p \in P}
\end{dmath}
Where $\exp: \mathfrak{g} \rightarrow G$ is the \textbf{exponential map} defined as:
\begin{dmath}
    \exp{(\xi)}={\gamma_{\xi}}(1)
\end{dmath}
in which $\gamma_\xi$ is the \textbf{one-parameter subgroup} of $G$:
the unique integral curve $\gamma_\xi: \mathbb{R} \rightarrow G$  of the left-invariant vector field $X_\xi$ for $\xi \in \mathfrak{g}$ that satisfies: $\gamma_{\xi}(0)=e$ and ${\gamma_\xi}'(t)=X_{\xi}(\gamma_\xi (t))$; which together result in: $\gamma_{\xi}(s+t)={ \gamma_{\xi}}(s)\gamma_{\xi}(t)$

One can prove $[\xi_P,\eta_P]=-[\xi,\eta]_P$.

Taking the derivative of the relation $\Phi^*_g \{F_1,F_2\}=\{\Phi^*_g F_1, \Phi^*_g F_2 \}$ with respect to $g$ in the $\xi$ direction, one finds:

\begin{dmath}[label={eq:infinitesimal-poisson-automorphism}]
    \xi_P[\{F_1,F_2\}]=\{\xi_P[F_1],F_2\}+\{F_1,\xi_P[F_2]\}
\end{dmath}
Any vector field on $P$ that satisfies this relation is called an \textbf{infinitesimal Poisson automorphism}.
In case $P$ is a symplectic manifold(Section \ref{sec:Symplectic,-Kählar-and-HyperKähler-Quotients}), it further simplifies to:
\begin{dmath}
    {\mathcal{L}_{\xi_P}\omega=0  \leftrightarrow d(i_{\xi_P} \omega) = 0}
\end{dmath}

${i_{\xi_P}}  \omega \ \text{being closed since} \ d \omega= 0$ where $\mathcal{L}_{\xi_P}$ is the Lie derivative. Using Cartan's magic formula, one finds $\mathcal{L}_{\xi_P}\omega=d({i_{\xi_P} \omega})$ which according to the Poincarè lemma means that $i_{\xi_P}\omega$ is locally exact, i.e. it is \textbf{locally Hamiltonian}. This gives us an idea of the physical meaning of infinitesimal Poisson automorphisms in the general case where $P$  is a Poisson manifold\footnote{Remark: However, one must bear in mind thatin the more general case of Poisson geometry, \eqref{eq:infinitesimal-poisson-automorphism} doesn't necessarily mean locally Hamiltonian!}

We are interested in the case where $\xi_P$ is globally Hamiltonian. That is, there exists a global Hamiltonian $\mu(\xi)\in\mathcal{F}(P)$ such that: $X_{\mu(\xi)}=\xi_P$. However, this equation does not fix the global Hamiltonian $\mu(\xi)$ uniquely but only up to a Casimir element of $\mathcal{C}(P)$. It turns out, if $P$ is symplectic(Section \ref{sec:Symplectic,-Kählar-and-HyperKähler-Quotients}) and connected, then $\mu(\xi)$ \underline{is determined up to a constant}.

For reasons to be clarified later, we shift our attention from the left group action of $G$ on $P$ to the canonical left Lie algebra action $\xi\in\mathfrak{g}\mapsto \xi_P \in \mathfrak{X}(P)$. This canonical action, as alluded to previously, is an antihomomorphism of Lie algebras.

The existence of such a globally Hamiltonian function $\mu: \mathfrak{g}\mapsto \mathcal{F}(P)$ can be equivalently pronounced as the existence of the map $\boldsymbol{\mu}: P \rightarrow \mathfrak{g^*}$ defined by: $\langle \boldsymbol{\mu}(z), \xi \rangle = \mu(\xi)(z)$. $\boldsymbol{\mu}$ is called the \textbf{momentum map}.

Since $\xi\mapsto \xi_P$ and $F\mapsto X_F$ are both Lie algebra antihomomorphisms, for $\xi,\eta \in \mathfrak{g}$ one concludes: $X_{\mu([\xi,\eta])}=[\xi,\eta]_P =-[\xi_P,\eta_P]=-[X_{\mu(\xi)},X_{\mu(\eta)}]=X_{\{\mu(\xi),\mu(\eta)\}}$.

We end this section with an algebraic definition of the momentum map that helps us formulate some sufficient criteria for the case of symplectic manifolds(Section \ref{sec:Symplectic,-Kählar-and-HyperKähler-Quotients}) for the existence of the momentum map.

Let's take the following exact sequence:

\begin{dmath}
    {0 \rightarrow \mathcal{C}(P) \xrightarrow{i} \mathcal{F}(P) \xrightarrow{\mathcal{H}} \mathcal{P}(P) \xrightarrow{\pi} \mathcal{P}(P)/\mathcal{H}(P) \rightarrow 0}
\end{dmath}

where: 

$\mathcal{H}(P)=\{X_F \in \mathfrak{X}(P) | F \in \mathcal{F}(P) \}$ is the Lie algebra of globally Hamiltonian vector fields on $P$,

$\mathcal{P}(P)=\{X \in \mathfrak{X}(P) | X[\{F_1,F_2\}]=\{X[F_1],F_2\}+\{F_1,X[F_2]\} \}$ the Lie algebra of infinitesimal Poisson automprphisms of $P$, and $\mathcal{H}(F) := X_F$.

If the left Lie algebra action is named: $\rho: \mathfrak{g} \rightarrow \mathcal{P}(P)$, one can prove easily that \underline{the existence of the linear momentum map $\mu$(so that $\mathcal{H} \circ \mu=\rho$)} is equivalent to $\pi \circ \rho=0$.

In case $P$ is a symplectic manifold(Section \ref{sec:Symplectic,-Kählar-and-HyperKähler-Quotients}): $\mathcal{P}(P)/\mathcal{H}(P)$ is isomorphic to the first cohomology group $H^1(P,\mathbb{R})$ which is an abelian group. Therefore, $\pi \circ \rho=0$ if and only if the induced mapping $\rho':\mathfrak{g} /[\mathfrak{g},\mathfrak{g}] = H^1(\mathfrak{g},\mathbb{R}) \rightarrow H^1(P,\mathbb{R})$ vanishes.

If $P$ is a symplectic(Section \ref{sec:Symplectic,-Kählar-and-HyperKähler-Quotients}) manifold and $\mathfrak{g}$ is semisimple, $\rho'$ vanishes due to \textbf{Whitehead's first lemma}. \cite{Whitehead's Lemma}

\section{Equivariant Momentum Map, Globally Hamiltonian Group Actions} 

\label{sec:Equivariant-Momentum Map,-Globally-Hamiltonian-Group-Actions}

Given one of the few critical conclusions of the previous section: 
\begin{dmath}[label={eq:momentum-map-bracket}]
    X_{\mu([\xi,\eta])}=X_{\{\mu(\xi),\mu(\eta)\}}
\end{dmath}
one can pose the question: \textit{since $\rho$ and $\mathcal{H}$ are Lie algebra antihomomorphisms, is $\mu:\mathfrak{g} \rightarrow \mathcal{F}(P)$ supposed to be a Lie algebra homomorphism?} 

To answer this question one should notice that from \eqref{eq:momentum-map-bracket} it follows that: 

\begin{dmath}
    {\mu([\xi,\eta])}-{\{\mu(\xi),\mu(\eta)\}} = \sigma(\xi, \eta)
\end{dmath}

is a Casimir element, $\sigma(\xi,\eta) \in \mathcal{C}(P)$.
Clearly $\sigma$ is antisymmetric, bilinear, and fulfills:

\begin{dmath}
    \sigma(\xi, [\eta, \zeta])+\sigma(\zeta, [\xi, \eta])+\sigma(\eta, [\zeta, \xi])=0
\end{dmath}

which is the $\mathcal{C}(P)$-valued 2-cocycle of $\mathfrak{g}$, in other words: $[\sigma] \in H^2 (\mathfrak{g}, \mathcal{C}(P))$ i.e. a 2nd \textbf{Lie algebra cohomology} of $P$. Clearly, if $[\sigma]=0$ the momentum map can be modified so that it becomes a homomorphism of the Lie algebras. Particularly for $P$ a symplectic manifold and $\mathfrak{g}$ a semisimple Lie algebra $[\sigma]=0$. Indeed, $\mathcal{C}(P) = \mathbb{R}$ for the symplectic $P$ and thanks to \textbf{Whitehead's second lemma} $H^2 (\mathfrak{g},\mathbb{R})=0$ \cite{Marsden}, \cite{Whitehead's Lemma}, \cite{WH}.

However, in general, one can at least prove that $[\sigma]=0$ if and only if: $T_z \boldsymbol{\mu} . \eta_P =-ad^*_\eta \boldsymbol{\mu}(z)$ for all $\eta \in \mathfrak{g}$ where $ad^*_\eta$ is the dual linear map of $ad_\eta: \xi \in \mathfrak{g} \mapsto [\eta, \xi] \in \mathfrak{g}$ and evidently $T_z \boldsymbol{\mu}$ is the tangent map of the momentum map $\boldsymbol{\mu}: P \rightarrow \mathfrak{g^*}$ at point $z \in P$. Under this condition, the momentum map $\boldsymbol{\mu}$ is called \textbf{infinitesimally equivariant} and the Lie algebra action is called \textbf{Hamiltonian}.

A momentum map is called \textbf{equivariant} if: 
\begin{dmath}
    {Ad^*_{g^{-1}}} \circ \boldsymbol{\mu}= \boldsymbol{\mu} \circ \Phi_g
\end{dmath}

or equivalently:

\begin{dmath}
    \mu(Ad_g \xi)(g.z)=\mu(\xi)(z)
\end{dmath}
where $Ad^*_g$ is the dual map of $Ad_g: T_e G = \mathfrak{g} \rightarrow T_e G =\mathfrak{g}$ which is the derivative of the conjugation map $\psi_g : G \rightarrow G$ defined by $\psi_g(h) = ghg^{-1}$ around the identity $h=e$. It is also worthy of mention that the derivative at the identity of the map $Ad : G \rightarrow \text{Aut}(\mathfrak{g})$\footnote{$\text{Aut}(\mathfrak{g})$ is a Lie group by a simple application of the closed subgroup theorem, since it's a closed subgroup of the Lie group $GL(\mathfrak{g})$} is the map $ad : \mathfrak{g} \rightarrow \text{Der}(\mathfrak{g})$ where $ad_{\eta} (\xi) = [\eta,\xi]$ which was encountered previously.\footnote{$\text{Der}(\mathfrak{g})$ is the lie algebra of all of the derivatives on $\mathfrak{g}$ where a derivative is an endomorphism of the Lie algebra $\delta : \mathfrak{g} \rightarrow \mathfrak{g}$ that satisfies the Leibniz rule: $\delta [\eta,\xi] = [\delta \eta , \xi] + [\eta , \delta \xi]$}

\
\
A Lie group action is called \textbf{globally Hamiltonian} if it has an equivariant momentum map. Evidently, equivariance implies infinitesimal equivariance.

\

It can be proven that, in case $P,G$ are both connected spaces, global and local equivariance are equivalent \cite{Marsden}.
\
We sum up the conclusion in this theorem:

\textit{A canonical left Lie algebra(group) action is Hamiltonian(globally Hamiltonian) if and only if there is a Lie algebra homomorphism $\mu: \mathfrak{g} \rightarrow \mathcal{F}(P)$ such that $X_{\mu(\xi)}= \xi_P$ for all $\xi \in \mathfrak{g}$. } \cite{Marsden}

\section{Riemannian Manifolds and Holonomy Group}

A Riemannian manifold is a smooth manifold $M$ endowed with a metric tensor $g\in\Gamma((TM \otimes TM)^*)$ (where $TM$ is the tangent bundle and $\Gamma(E)$ stands for the sections of the vector bundle $E$) which is fibrewise non-degenerate and positive-definite. We denote such a pair as $(M,g)$.

A special type of $\mathbb{R}$-linear connection $\nabla: \Gamma(TM)\rightarrow \Gamma(T^* M \otimes TM)$ that is $C^\infty (M)$-Leibniz(for any $f \in C^\infty(M)$ one has $\nabla(fs)=df \otimes s + f\nabla s$) can be defined on Riemannian manifolds called: Levi-Civita connection that is completely fixed by the metric(and its derivatives), at the price of removing torsion so that: $[X,Y]=\nabla_X Y - \nabla_Y X$ for any $X,Y \in \Gamma(TM)$ and with the help of the constraint: $\nabla g=0$.

A coordinate-free expression for the Levi-Civita $\nabla$ is:

\begin{dmath}
    2\langle Z, \nabla_X Y \rangle =X \langle Y,Z \rangle + Y \langle Z,X \rangle - Z \langle X, Y \rangle - \langle [X,Z],Y \rangle - \langle [Y,Z],X \rangle +\langle[X,Y],Z \rangle 
\end{dmath}

A vector field $X$ is called \textbf{parallel along a curve} $\gamma:I \rightarrow M$ if and only if ${\nabla_{\dot{\gamma}}}X=0$. A curve is called \textbf{geodesic} if and only if:
\begin{dmath}[label={eq:geodesic}]
    \nabla_{\dot{\gamma}} \dot{\gamma} =0
\end{dmath}

Conversely, one can define a parallel lift of a curve to a vector bundle which can be proven to be unique. Let's take a closed piecewise differentiable loop based at $p$. For any given vector in the tangent space of $p=\gamma(0)$, one can assign a unique new vector at $p=\gamma(1)$ by means of the unique parallel lift the loop to the vector bundle. This gives a linear map $\mathbb{P}_\gamma:T_{\gamma(0)} M \rightarrow T_{\gamma(1)} M$. Each curve gives rise to one such linear map and this forms a group under linear map composition. One calls such a group the \textbf{holonomy group} of a connection $\nabla$. In other words:

\begin{dmath}
    H(p)=\{ \mathbb{P}_\gamma | \gamma \ any \ loop \ based \ at \ p \}
\end{dmath}

The \textbf{restricted holonomy group} is:

\begin{dmath}
    H^0(p)=\{ \mathbb{P}_\gamma | \gamma \ any \ contractible \ loop \ based \ at \ p \}
\end{dmath}

A manifold is called \textbf{irreducible} with respect to a linear connection if and only if its tangent space is an irreducible representation of the holonomy group.

The holonomy group of an $m$-dimensional Riemannian manifold with respect to the Levi-Civita connection can be shown easily to lie in $SO(m)$ \cite{EM duality}

An important theorem by Ambrose and Singer shows that the generators of the holonomy group Lie algebra are the Riemann curvature tensors \cite{EM duality}:

\begin{dmath}
    {R_{ab} : T_p M \rightarrow T_p M} \ \ \ \ \ \ \ \ \ \ \ \ \ \ \ \ \ \ \ \ \ \ \ \ \
\end{dmath}
\begin{dmath}
 \ \ \ \ \ \ \ \ \ \ \ \ \ \ \ \ \ \ \ \ \ \ \   {R_{ab}: \partial_c \mapsto {R_{abc}}^d \partial_d \ \ \ \ \text{local expression}}
\end{dmath}

Where ${R_{abc}}^d$ are the components of the Riemann curvature tensor with respect to a local coordinate system.

\section{Almost Complex Structure, Complex Manifolds, Integrability}

A smooth tensor field $I \in {TM} \otimes TM^{*}$ such that $I^2 =I \circ I=-1$ is called an \textbf{almost complex structure}.
A manifold [in case it is possible\footnote{Counter example: $S^{4n} \ \text{for} \ n \geq 1$ and $S^{2n}\  \text{for} \ n\geq4$ do not admit any almost complex structure, which means only $S^2$ and $S^6$ survive.}] endowed with such a structure is called an \textbf{almost complex manifold}.

Simply det$(I)^2=$det$(I^2)=$det$(-1)=(-1)^n$ and since $M$ is a real manifold $n$ is an even number.
It is not hard to show that such a manifold must also be orientable.
Every complex manifold has a natural complex structure.

However, the opposite is not always true. Not every almost complex structure is induced by a complex structure. In other words, not every almost complex structure is integrable.
A very interesting theorem by Newlander-Nirenberg proves that an almost complex manifold is \textbf{integrable} if and only if the following tensor, called the Nijenhuis tensor, vanishes($\forall X,Y\in \Gamma(TM) )$ \cite{EM duality}:

\begin{dmath}
    N_I (X,Y)=I[IX,IY]+[X,IY]+[IX,Y]-I[X,Y]
\end{dmath}

A connection $\nabla$ is called complex if and only if the complex structure is parallel relative to it: $\nabla I=0$

One can show the holonomy group of an $2n$-dimensional integrable almost complex manifold: $H\subset U(n)$ \cite{EM duality}

\section{Kähler and HyperKähler manifolds}

There are a few equivalent definitions of the Kähler manifold. In all of these definitions here, the connection $\nabla$ is assumed to be the Levi-Civita connection, which is uniquely determined in terms of the metric $g$ and preserves the metric, and the Holonomy groups are also defined with the same connection.

An almost complex structure is said to be \textbf{compatible} with a metric $g$ if and only if $g(IX,IY)=g(X,Y)$. Another way to say the same is that the metric is \textbf{hermitian} with respect to $I$.

A $2n$-dimensional Riemannian manifold $M$, whose holonomy group lies in $U(n)$, is called \textbf{Kähler}. Another equivalent definition would be: a Riemannian manifold that admits a complex structure $(M,g, I)$ which is compatible with the metric $g$ and $\nabla I = 0$. Equivalently, a Kähler manifold is a symplectic manifold equipped with an integrable complex structure $(M, \omega, I)$ such that the bilinear form $g(X,Y):= \omega(X,IY)$ defines a Riemannian metric over it. 

One can show that $(M,g, I)$ is Kähler if and only if either of the following is fulfilled \cite{EM duality}: 
\begin{itemize}
\item the 2-form $\omega(X,Y):= g(IX,Y)$ is closed i.e. $d \omega=0$
\item $\nabla \omega = 0$
\item $\nabla I = 0$
\end{itemize}
It can be shown that a \textbf{Ricci flat} Kähler manifold has a holonomy group: $H\subset SU(n)$.
And conversely, any simply connected Kähler manifold whose holonomy group lies in $SU(n)$ is Ricci flat \cite{EM duality}.

A celebrated theorem due to Calabi-Yau shows that for any Kähler manifold whose first Chern class vanishes, there exists a unique Ricci flat Kähler manifold in the same \textbf{Kähler class} \cite{EM duality}. Two  Kähler forms(and therefore Kähler manifolds) in a compact manifold $M$ are said to be in the same Kähler class if they both belong to the same second cohomology class $[\omega] \in H^2 (M)$.  

Finally, a HyperKähler manifold is one that is endowed with(in case of possibility) three compatible almost complex structures: $I,J,K \in {TM}^* \otimes TM$ such that $I^2=J^2=K^2 =-1 $, $IJ=K=-JI$, $JK=I=-KJ$, $KI=J=-IK$

Any HyperKähler manifold is $4n$-dimensional \cite{EM duality}.
Any HyperKähler manifold is integrable \cite{EM duality}.
The holonomy group of a HyperKähler manifold lies in $Sp(n)$ \cite{EM duality}.
\underline{All HyperKähler manifolds are Ricci flat.} \cite{EM duality}

We end this section by summarizing the ultimate chain we came up with for the holonomy group $H$ of a $4n$-dimensional Riemannian manifold:

\begin{dmath}
   {H \subset Sp(n) \subset SU(2n) \subset U(2n) \subset SO(4n) \subset O(4n)}
\end{dmath}
\section{Symplectic, Kählar and HyperKähler Quotients}
\label{sec:Symplectic,-Kählar-and-HyperKähler-Quotients}

A \textbf{symplectic form} on a smooth manifold $M$ is a closed, fibrewise nondegenerate, differential 2-form $\omega \in \Gamma({\bigwedge\nolimits^2 T^{*}M})$.

Nondegeneracy means that if $\omega_p (X_p , Y_p)=0$ for all $Y_p \in T_p M$ implies $X_p=0$
It is not hard to see that the nondegeneracy of $\omega$ forces $T_p M$ to be even-dimensional and therefore $M$.

A \textbf{symplectic manifold} is a manifold given a symplectic structure(in case of possibility!).

Suppose that a Lie group $G$ acts on the symplectic manifold $(M,\omega)$ such that it preserves the symplectic structure. The group action can define a vector field on the manifold $M$: 
\begin{dmath}
    {X_a}(p)=\frac{\mathrm{d}}{\mathrm{d}t}[\exp{(te_a)}.p]|_{t=0}
\end{dmath}

for each $p \in M$ and $e_a \in \mathfrak{g}$ where $e_a$ is a basis for $\mathfrak{g}$ and index $a \in 1,...,$dim${\mathfrak{g}}$.

The preservation of the symplectic structure can be articulated in terms of the Lie derivative concept that $\mathcal{L}_{X_a} \omega = 0$ for all $X_a$ in $M$.

This can be pushed further forward using Cartan's formula, $d \omega=0$ and $\mathcal{L}_{X_a} \omega = 0$ to the conclusion that $d{i_{X_a}\omega}=0$. By Poincaré lemma, $i_{X_a} \omega$ is locally exact.

As we showed in Section \ref{sec:Poisson-Manifolds-and-Momentum-Map}, there are particular conditions under which $i_{X_a} \omega$ is globally exact. One of such conditions is $M$ being a symplectic manifold and $\mathfrak{g}$ the Lie Algebra of a semi-simple Lie group $G$. For the applications in this thesis, one is specifically interested in such conditions.

We name such global momentum map $\boldsymbol{\mu}: M \rightarrow \mathfrak{g}^* $. Defining a dual basis for $\mathfrak{g}^*$, $e^a \in \mathfrak{g}^*$  and index $a \in 1,...,$dim${\mathfrak{g}^*}$ one can write $\mu(p)= {\mu_a}(p)e^a$ for each $p \in M$ or equivalently: ${\mu_a}: M \rightarrow \mathbb{R}$, ${\mu_a}(p)= \mu(p)(e_a)$.

$\mu_a$ can be chosen modulo some modest assumptions(such as $\mathfrak{g}$ being the Lie algebra of a semisimple Lie group), to be equivariant.

In that case following section \ref{sec:Equivariant-Momentum Map,-Globally-Hamiltonian-Group-Actions} one can write: 

\begin{dmath}
    \{\mu_a , \mu_b \}= {f_{ab}}^c \mu_c
\end{dmath}

where ${f_{ab}}^c$ are the structure coefficients of the Lie algebra. 

$\mu_a$ are called \textbf{ first-class constraints} whenever the vector fields $X_a$ are tangent to the $\mu^{-1} (0)$.

By an important theorem due to Marsden, Weinstein, and Meyer, one can show that $\mu^{-1} (0)/G$ is a symplectic manifold. We define $\pi: \mu^{-1}(0) \rightarrow \mu^{-1}(0)/G$ as the natural projection, where according to the theorem there exists $\bar{\omega}$ on $\mu^{-1}(0)/G$ where its pullback via projection $\pi^{*} \bar{\omega}$ agrees with the restriction of $\omega$ to $\mu^{-1}(0)$.

For the case of Kähler and HyperKähler quotients, we assume that the group action also preserves the metric; in other words, it also acts by isometries. Consequently, there exists also a metric on $\mu^{-1}(0)/G$.

Let $(M,g,I)$ denote a Kähler manifold. The tangent space at every point like $p$ of the manifold $M$ can be decomposed into: $T_p M=T_p \mu^{-1} (0) \oplus N_p \mu^{-1} (0) $.
One can show that the normal space $N_p \mu^{-1} (0)$ is spanned by the gradients : $grad_p \mu_a$.
 In addition, the tangent space can be decomposed by $T_p \mu^{-1} (0)=H_p \oplus V_p$ where $V_p$ are the vectors tangent to the $G$-orbit. Evidently, this space is spanned by the vector fields $X_a$, and the horizontal vector space can be identified with the orthogonal complement within $T_p \mu^{-1} (0)$ to the vertical space. Given that $\pi: \mu^{-1}(0) \rightarrow \mu^{-1}(0)/G$ is a principle bundle and in such bundle for every curve in the base space there exists a unique horizontal lift, one finds a unique horizontal lift $\tilde{X}$ on $\mu^{-1} (0)$ that projects down to $X=\pi_* \tilde{X}$ for any given vector field $X$ on $\mu^{-1} (0)/G$.

Consequently, the tangent space to $\mu^{-1} (0)/G$ can be uniquely horizontally lifted to the tangent space of the $\mu^{-1} (0)$.

Suppose $Y \in \mathfrak{X}(M)$, then: $g(grad \mu_a , Y)=d \mu_a (Y)= \omega(X_a, Y)=g(IX_a, Y)$ hence $grad \mu_a = IX_a$. As a result for the tangent space over $\mu^{-1}(0)$ in $M$:

\begin{dmath}
    TM=H \oplus V \oplus N \mu^{-1}(0)
\end{dmath}

one can choose a basis $\{X_a \}$ for $V$ and $\{IX_a\}$ for $N \mu^{-1} (0)$ so that the complex structure becomes: $I=\bar{I} \oplus$ 
$\begin{pmatrix}
0 & 1\\
-1 & 0
\end{pmatrix}$
where $\bar{I} \in H \otimes H^{*}$. Note that $\pi_{*} I $ is a complex structure on the quotient and is identified with $\bar{I}$.
Alternatively speaking, $H$ is a complex subspace, or the complex structure commutes with the horizontal projection, and even more explicitly: If $Y \in \mathfrak{X} (\mu^{-1} (0)/G)$ and $\tilde{Y}$ its unique horizontal lift to $\mu^{-1} (0)$ then $I \tilde{Y} = \widetilde{\bar{I} Y}$.
By the Newlander-Nirenberg theorem, $\bar{I}$ is integrable.
Using O'Neil's formula $\nabla_{\tilde{X}} \tilde{Y} = \widetilde{{\bar{\nabla}_X} Y}+1/2[\tilde{X},\tilde{Y}]^v$ (where $\bar{\nabla}$ is the Levi-Civita connection on $\mu^{-1} (0)/G$ and $^v$ stands for the vertical projection on the vertical subspace), one can prove that $\nabla I =0$ on $M$ implies $\bar{\nabla} \bar{I}=0$ on $\mu^{-1} (0)/G$ and so $\mu^{-1} (0)/G$ is a Kähler manifold.

One last point would be that it is more beautiful to think of the tangent space as: 
\begin{dmath}
    TM \cong T(\mu^{-1} (0)/G) \oplus \mathfrak{g}^{\mathbb{C}}
\end{dmath}

Although not precisely, it can be interpreted as if $\mu^{-1} (0)/G$ is the quotient of $M$ by the action of $G^{\mathbb{C}}$!

At the end, to generalize to the HyperKähler case, we can write the momentum map as $\mu: M \rightarrow \mathfrak{g}^* \otimes \mathbb{R}^3$

We can fix one of the structures, say $I$, and define:

\begin{dmath}
    {\nu = \mu^{(J)} + i \mu^{(K)} : M \rightarrow \mathfrak{g}^{*} \otimes \mathbb{C}}
\end{dmath}

For each [killing]vector field $X_a$ and $Y$ both in $\mathfrak{X}(M)$ one can write:

\begin{dmath}[label={eq:dnu-Y}]
    {d\nu (Y)= \omega^{(J)} (X_a , Y) + i \omega^{(K)}(X_a , Y) = g(JX_a , Y) + ig(KX_a , Y)}
\end{dmath}
\begin{dmath}[label={eq:dnu-IY}]
    {d \nu (IY)= g(JX_a , IY) +i g(KX_a, IY)=-g(KX_a, Y) + i g(JX_a, Y)}
\end{dmath}

therefore, by \eqref{eq:dnu-Y} and \eqref{eq:dnu-IY}, $d \nu (IY)=i d \nu (Y)$ so that $\nu$ can be called holomorphic with respect to the complex structure $I$ or $\bar{\partial} \nu = 0$ Consequently, $\nu^{-1} (0)$ is a complex submanifold of a Kähler manifold and so its induced metric is Kähler. $G$ preserves the Kähler structure, while $\nu^{-1} (0)$ is $G$-invariant, and the moment mapping is the restriction of $\mu^{(I)}$ to $\nu^{-1} (0)$.
The resulting manifold is Kähler : $(\nu^{-1} (0) \cap (\mu^{(I)})^{-1} (0) ) / G$ since the induced metric on $(\mu^{(I)})^{-1} (0) / G$ is also Kähler relative to $I$.
The same can be done for $J$ and $K$ So finally $\mu^{-1} (0)/ G$ is HyperKähler.

\section{Projective Vector Fields and Their Lie Subalgebras}
\label{Projective-Vector-Fields-and-Their-Lie-Subalgebras}

A vector field $X$ on $(M,g)$ that preserves the geodesics (the solutions to Equation \eqref{eq:geodesic}) is called a \textbf{Projective Vector Field}. It can be checked that the solutions to \eqref{eq:geodesic} admit a symmetry which is an affine transformation of the parameter (a.k.a., the equation admits such an affine symmetry $t \rightarrow at + b \ \ \forall a,b \in \mathbb{R}, t \in I \subset \mathbb{R}$). The flow of a projective vector field does not necessarily leave the geodesic parameter invariant.

Based on this observation, one can introduce the subalgebras of the projective vector fields $P(M)$.
The first type is those projective vector fields that preserve the affine parameterization of the curve under their flow. Therefore, they are called \textbf{Affine Vector Fields} and represented by $A(M)$. 
Indeed, it turns out quite insightful to decompose the covariant derivative of a vector field in the following manner:

\begin{dmath}
    X_{\mu; \nu} = \frac{1}{2} h_{\mu \nu} + F_{\mu \nu}
\end{dmath}

where $h_{\mu  \nu} = (\mathcal{L}_X g)_{\mu \nu} = X_{\mu ; \nu} + X_{\nu ; \mu}$ and $F_{\mu \nu} = \frac{1}{2} (X_{\mu; \nu} - X_{\nu ; \mu})$ where all indices are evidently covariant. By such definitions, one can prove that an affine vector field can be equivalently formulated as a vector field for which $\nabla h = 0$.
The next subalgebra is that of those projective vector fields that satisfy $h = \mathcal{L}_X g = 2c g$ for some real constant $c$. These preserve the metric up to a constant and are represented by $H(M)$ and are called \textbf{Homothetic Vector Fields}. Finally, the subalgebra of the projective vector fields for which $c =0$ is called the subalgebra of \textbf{Killing Vector Fields} and is represented by $K(M)$.

It is straightforward to verify that the subalgebras can be placed in the following order:

\begin{dmath}
    {K(M) \subset H(M) \subset A(M) \subset P(M)}
\end{dmath}

while $\text{dim}(P(M)) \leq n(n+2)$ and the following upper bounds hold for the dimension of such subalgebras \cite{Projective Vector Field}:

\begin{dmath}
    {\text{dim}(A(M)) \leq n(n+1) ,} \ \ {\text{dim}(H(M)) \leq \frac{1}{2} n (n+1) + 1} , \ \ {\text{dim}(K(M)) \leq \frac{1}{2} n (n+1)}
\end{dmath}

where $n:= \text{dim}(M)$.

\section{Cones, Sasakian Manifolds and 3-Sasakian Manifolds, Infinitesimal Dilation Invariance}\label{sec:Cones-Sasakian-Manifolds}

Indeed, the odd-dimensional counterpoint of Kähler manifolds is the Sasakian manifolds. Kähler manifolds are the intersection of complex, symplectic, and Riemannian manifolds. In the same way, Sasakian manifolds can be viewed as the intersection of CR, contact, and Riemannian geometry \cite{Sasaki}. 

The \textbf{Cone} of the manifold $M$ is $C(M) = M \times \mathbb{R}^{>0}$. The Riemannian cone of the Riemannian manifold $(M,g)$ is $(C(M),\bar{g})$ where $\bar{g} = t^2g + dt^2$.
$(M,g)$ is Sasakian if and only if its Riemannian cone is Kähler. $M$ is necessarily odd-dimensional. Similarly, $(M,g)$ is $3$-Sasakian if and only if its Riemannian cone is a hyper-Kähler manifold.

For $(M,g)$ as a Riemannian manifold, to be invariant under infinitesimal dilation, it is necessary to admit a homothetic vector field; in other words:

\begin{dmath}[label={eq:homothetic-vector-field}]
    {X^{\mu}}_{;\nu} = {\delta^{\mu}}_{\nu}
\end{dmath}

It can be proven that \underline{if $X$ is hypersurface orthogonal} (there exists a function $f: M \rightarrow \mathbb{R}$ such that $X = \nabla f$), the condition \eqref{eq:homothetic-vector-field} alone imposes a conical geometry over $M$ \footnote{However, one must be careful that this is independent of any complex or hyper-complex structure on $M$ and holds in general}. \cite{Cones} That is, $M$ is a Riemannian cone $C(B)$ over the Riemannian base manifold $B$. 

Now, if $M$ is a hyperKähler manifold, the base $B=\mathfrak{F}$ is, by definition, a 3-Sasakian manifold. In this case, the three complex structures are preserved by homothety $X$; in other words:

\begin{dmath}
    \mathcal{L}_X I^c = 0
\end{dmath}

This vector field is called \textbf{ tri-holomorphic}.

The Reeb vector fields tangent to the 3-Sasakian base $\mathfrak{F}$ are:

\begin{dmath}
    ({K^a})^{\mu} = {({I^a})^{\mu}}_{\nu} X^{\nu} 
\end{dmath}

These vector fields are called Reeb vector fields and are, indeed, the killing vector fields, due to an SU(2) isometry action on $M$ since:

\begin{dmath}
    [K^a , K^b] = -2 \epsilon_{abc} K^c
\end{dmath}

Each of the Killing vectors $K^a$(generators of $SU(2)$ isometry) is holomorphic with respect to its complex structure $I^a$ and its momentum map is the function $f$ \footnote{mentioned a few lines above}, respectively, although \underline{the $SU(2)$ action is not tri-holomorphic} as can be checked explicitly:

\begin{dmath}
    \mathcal{L}_{K^a} I^b = -2 \epsilon_{abc} I^{c}
\end{dmath}

Indeed, all four-dimensional hyper-Kähler spaces that admit an $SU(2)$ isometry are classified \cite{Quaternions and Instantons}:
\begin{itemize}
    \item Ricci flat
    \item Taub-Nut space
    \item Atiyah-Hitchin space
\end{itemize}

However, only the Ricci flat space is the viable choice, since among these three spaces, only the Ricci flat space admits a homothetic vector field, and therefore the 3-Sasakian base $\mathfrak{F}$ is an Einstein manifold. An Einstein manifold by definition satisfies $R_{\mu \nu}=\lambda g_{\mu \nu}$ for some $\lambda \in \mathbb{R}$. It turns out that for a Sasaki-Einstein base $\mathfrak{F}$\cite{Sasaki}:

\begin{dmath}
    \lambda_{\mathfrak{F}}= n-2
\end{dmath}

where $n=\text{dim}(C(\mathfrak{F}))$ and the same must hold for a 3-Sasakian-Einstein base $\mathfrak{F}$.

\section{BPST Instanton, Collective Coordinates, Index Theorem}
\label{sec:Moduli-Space-of-Instantons,-Collective-Coordinates-Index-Theorem}

Instantons are known as the finite-action solutions of the Euclidean equations of motion.
In this thesis, we are only concerned with Yang-Mills instantons, both usual and supersymmetric.
As can be read from the following equations, they might come in two types: \textbf{Self-Dual} $F = F^*$ and \textbf{Anti-Self-Dual} $F=-F^*$ connections. However, it is crucial to note that these are not the only possible solutions to the classical Euclidean Yang-Mills solutions over $S^4$. Indeed, non-self-dual solutions exist that minimize the YM action, which are the saddle points of the functional space. However, they might turn out to be unstable, such as instanton-anti-instanton solutions, etc. 

Indeed, it becomes clear that the space of all connections over $S^4$ can be broken into countably many connected components labeled by the \textbf{2nd Chern class} \cite{NSD YM Instantons}:

$$c_2 = \frac{1}{8 \pi^2} \int_{S^4} \text{Tr}_N \{F^* F\} $$

where the self-dual and anti-self-dual solutions turn out to be the absolute minimum of the action for each particular connected component, though globally speaking, they remain the local minima of the action.

Bogomol'nyi–Prasad–Sommerfield Bound can be achieved via \cite{BPST}, \cite{BPST2}:
\begin{dmath}
S=-\frac{1}{2g^2} \int d^4 {x} \Tr_N \{F^2\} = -\frac{1}{4g^2} \int d^4 {x} \Tr_N \{(F \mp F^*)^2\} \mp \frac{1}{2g^2} \int d^4 {x} \Tr_N \{F{F^*}\} \geq \mp \frac{1}{2g^2} \int d^4 {x} \Tr{F F^*} = \frac{8 {\pi}^2}{g^2} (\pm k)
\end{dmath}
where $k$ is the topological charge/sector of the instanton.
The general $SU(N)$ YM one instanton solution can be written as follows:

\begin{dmath}
    {A^{ \ SU(N)}_m =U\begin{pmatrix}
0 & 0\\
0 & A^{ \ SU(2)}_m
\end{pmatrix}U^{\dagger} \ \ \ \  U \in \frac{SU(N)}{SU(N-2) \times U(1)}}
\end{dmath}

where 

\begin{dmath}
  {  A^{a \  SU(2)}_m=-{\bar{\eta}^a}_{mn} \partial_n \ln{\{1+\frac{{\rho}^2}{(x-x_0)^2}\}} }
\end{dmath}

\textbf{in the singular gauge} where 't Hooft symbols are indeed defined as:

\begin{dmath}
    {\sigma_{mn} = \frac{i}{2} \eta^c_{mn} \tau^c \ \ \ \bar{\sigma}_{mn}= \frac{i}{2} \bar{\eta}^c_{mn} \tau^c \ \ \ \ \ \ \text{where} \ \tau^c \ \text{are the Pauli matrices}}
\end{dmath}

and $\rho$ is the scale parameter and $U$ is the \underline{global internal} phase of the instanton solution. These parameters are called \textbf{collective coordinates} and are the natural choice to parameterize the moduli space of the instanton, the space of all solutions with the same action and topological charge, denoted by $\mathcal{M}_k$.

Instantons and anti-instantons definitely interact; therefore, any multi-instanton solution can be achieved only by physically justifiable simplifications (due to the non-linearities of the differential equations). 

The multi-instanton solution presented here is the simplest approximation in which the instantons are widely separated, $\rho_i \rho_j \ll (x_i - x_j)^2$:

\begin{dmath}
    A^{ \ SU(2)}_m = \bar{\sigma}_{mn} \partial_{n} \ln{\phi}
\end{dmath}
\begin{dmath}
    \phi=1+ \sum_{i=1}^{k} \frac{{\rho_i}^2}{(x-{x_0}_i)^2}
\end{dmath}

\textbf{in the singular gauge}.

Now we can find the dimension of $\mathcal{M}_k$ for a generic value of $k$ using index theory. Firstly, all the pure \textbf{local gauge} zero modes should be excluded.
So by fluctuating the gauge field around a self-dual solution such that it remains self-dual:

\begin{dmath}
    {\mathcal{D}_m \delta A_n - \mathcal{D}_n \delta A_m = (\mathcal{D}_m \delta A_n - \mathcal{D}_n \delta A_m)^* = \epsilon_{mnkl} \mathcal{D}_k \delta A_l}
\end{dmath}

one finds: $0={\bar{\eta}^a}_{mn} \mathcal{D}_m \delta A_n$

But to exclude pure local gauge transformations $\delta A_m = \mathcal{D}_m \Lambda$, given that $\Lambda (\infty)=0$ for such modes, imposing orthogonality on these modes gives rise to:

\begin{dmath}
    {0=\int d^4 x \mathcal{D}_m \Lambda \delta A_m= - \int d^4 x \Lambda \mathcal{D}_m \delta A_m}
\end{dmath}

so $\mathcal{D}_m \delta A_m = 0$.

If one wishes to summarize both conditions into one equation, one should introduce the following notation:

\begin{dmath}
    {\sigma_m = \left\{
\begin{array}{ll}
      i {\tau}_m & m=1,2,3 \\
      I & m=4\\
\end{array} 
\right \}     \ \ \ \ \ \ \ \ \ \bar{\sigma}_m = \left\{
\begin{array}{ll}
      -i {\tau}_m & m=1,2,3 \\
      I & m=4\\
\end{array} 
\right \}}\end{dmath} 

in the Euclidean spacetime\footnote{in the Minkowski spacetime one has $\sigma^n = (-1, \vec{\tau})$ and $\bar{\sigma}^n = (-1, -\vec{\tau})$} where $\tau_m$ are Pauli matrices where, by the way, one can rewrite the formerly used $\sigma_{mn}$ as:

\begin{dmath}
    {\sigma_{mn} = \frac{1}{4} (\sigma_m \bar{\sigma}_n - \sigma_n \bar{\sigma}_m )  \ \ \ \ \ \bar{\sigma}_{mn} = \frac{1}{4} (\bar{\sigma}_m \sigma_n - \bar{\sigma}_n \sigma_m )}
\end{dmath}

Using these new symbols, the restrictions can be summarized as follows:

\begin{dmath}
    0=\sigma^{\dagger}_m \sigma_n \mathcal{D}_m \delta A_n
\end{dmath}

and further simplified by introduction of $\Psi= -i \sigma_m \delta A_m $ therefore:

\begin{dmath}[label={eq:zero-mode-Weyl-equation}]
    0=- \sigma^{\dagger}_m \mathcal{D}_m \Psi
\end{dmath}

Therefore, the problem of the number of independent zero modes is reduced to finding the $\dim({ \ker{(\rawslashed{\mathcal{D}})}})$ where $\rawslashed{\mathcal{D}}=\sigma^{\dagger}_m \mathcal{D}_m$

The \textbf{index of an operator} $\rawslashed{\mathcal{D}}$ is defined as:

\begin{dmath}
    {\mathcal{I}(\rawslashed{\mathcal{D}})  =\dim({ \ker{(\rawslashed{\mathcal{D}})}}) - \dim({ \ker{({\rawslashed{\mathcal{D}}}^{\dagger})}}) \leq \dim({ \ker{(\rawslashed{\mathcal{D}})}})}
\end{dmath} 

It is not difficult to prove that $\dim({\ker{(\rawslashed{\mathcal{D}}^{\dagger})}})=0$.
We only quote the result for the $SU(2)$ that, employing \textbf{Index Theory} arguments one can prove:

\begin{dmath}
    \mathcal{I}(\rawslashed{\mathcal{D}})= 2 T(R) k
\end{dmath}

in which $T(R)$ is defined by $\Tr{T^a T^b}= \delta_{ab} T(R)$ that depends on the representation $R$ of the gauge group and $k$ is the topological charge of the instanton.

\section{ADHM Construction as HyperKähler Quotient}
\label{ADHM-Construction-as-HyperKähler-Quotient}

The reward for the first abstract mathematical section is the beautiful simplification of the Atiyah-Drinfeld-Hitchin-Manin or ADHM construction.

We start with the moduli space metric $\mathcal{M}_k$ of the solutions to the Yang-Mills equations with charge $k$ \cite{Instantons}, \cite{Instanton Calculus}:

\begin{dmath}
    g_{\mu \nu} (X)=-2g^2 \int d^{4} {x} \tr_N \{\delta_{\mu} A_{n} (x;X) \delta_{\nu} A_n (x;X) \} 
\end{dmath}

$\mathbb{R}^4$ is naturally a HyperKähler manifold and can be given complex structures ${I^c}_{mn} = - {{\bar{\eta}}^c}_{mn}$ where ${{\bar{\eta}}^c}_{mn}$ are the already introduced 't Hooft symbols (Section \ref{sec:Moduli-Space-of-Instantons,-Collective-Coordinates-Index-Theorem}). The complex structures on $\mathcal{M}_k$ are induced from those on $\mathbb{R}^{4}$ in the following way.
As in \eqref{eq:zero-mode-Weyl-equation}, it was shown that the zero mode equation can be effectively written as $\rawslashed{\mathcal{D}}^{\dot{\alpha} \alpha } \delta A_{\alpha \dot{\beta}} = 0$.  Since every such variation can be decomposed as a sum of linearly independent zero modes $\delta A_{\alpha \dot{\beta}} = (\delta_{\mu} A_{\alpha \dot{\beta}}) (\delta X^{\mu})$, the equation becomes: $\rawslashed{\mathcal{D}}^{\dot{\alpha} \alpha} \delta_{\mu} A_{\alpha \dot{\beta}} =0$. If $\delta_{\mu} A_{\alpha \dot{\beta}}$ is a zero mode, then also $\delta_{\mu} A_{\alpha \dot{\beta}} B^{\dot{\beta}}_{\dot{\alpha}} $ is a zero mode for any constant matrix $B$. Now, one can define three complex structures that coincide with those of $\mathbb{R}^4$ componentwise: 

\begin{dmath}[label={eq:complex-structures-on-zero-modes}]
    (\Vec{\textbf{I}} . \delta_{\mu} A)_{\alpha \dot{\alpha}} = i \delta_{\mu} A_{\alpha \dot{\beta}} \Vec{\tau}^{\dot{\beta}}_{\dot{\alpha}} 
\end{dmath}

But since the right-hand side of \eqref{eq:complex-structures-on-zero-modes} is itself a zero mode, it can be written as a linear combination of zero modes with coefficients ${(I^a)^{\nu}}_{\mu}$ such that

\begin{dmath}
    (\mathbf{I}^a . {\delta_{\mu}}A)_{\alpha \dot{\alpha}} = {\delta}_{\nu} A_{\alpha \dot{\alpha}} ({{I^a})^{\nu}}_{\mu}
\end{dmath}

and therefore, $ {(I^a)^{\nu}}_{\mu}$ are the $\mu \nu$ components of the three complex structures $a=1,2,3$.

It is not hard to prove that $\mathcal{M}_k$ acquires a Kähler potential that is shared by all three complex structures (and so the structures are integrable and $\mathcal{M}_k$ is HyperKähler) \cite{Instantons}:

\begin{dmath}[label={eq:hyperkahler-potential}]
    \chi = -\frac{g^2}{4} \int d^{4} {x} \ x^2 \tr_N \{F^2_{mn}\}
\end{dmath}

and that for any given local holomorphic coordinate system on $\mathcal{M}_k$ say, $(Z^i, \bar{Z}^i)$:

\begin{dmath}
    g(X)=\pdv{\chi}{\bar{Z}^j}{Z^i} dZ^i d \bar{Z}^j
\end{dmath}

The main object in ADHM construction is the $(N+2k) \times 2k $ matrix $\Delta_{{\lambda}i \dot{\alpha}}$ which is linear in quaternionic spacetime coordinates:

\begin{dmath}
    \Delta_{\lambda i \dot{\alpha}}= a_{\lambda i \dot{\alpha}} + b^{\alpha}_{\lambda i} x_{\alpha \dot{\alpha}}
\end{dmath}
\begin{dmath}
    {\bar{\Delta}_i}^{\dot{\alpha} \lambda} = {\bar{a}_i}^{\dot{\alpha} \lambda} + \bar{x}^{\dot{\alpha} \alpha} {\bar{b}_{i\alpha}}^{\lambda}
\end{dmath}

where ${\bar{\Delta}_i}^{\dot{\alpha} \lambda}=(\Delta_{\lambda i \dot{\alpha}} )^*$ and $i,j,..=1,...,k$ are instanton indices and $\mu , \nu,...=1,...,N+2k$ are ADHM indices.

The null space of ${\bar{\Delta}_i}^{\dot{\alpha} \lambda}$ can be written as an $(N+2k) \times N$ dimensional complex matrix $U_{\lambda u}$, $u=1,...,N$. In other words ${\bar{\Delta}_i}^{\dot{\alpha} \lambda} U_{\lambda u}= 0$ where $\bar{U}_u^{\lambda} U_{\lambda v}= \delta_{uv}$.

According to the ADHM construction, the vacuum gauge field configuration can be written as:

\begin{dmath}
    (A_n)_{uv}= g^{-1} \bar{U}^\lambda_u \partial_n U_{\lambda v}
\end{dmath}

The beauty of construction reveals itself when the construction works in all topological sectors! In other words, for any integer $k$. For it to work, one only has to impose the condition ${\bar{\Delta}_i}^{\dot{\alpha} \lambda} \Delta_{\lambda j \dot{\beta}}= {\delta}^{\dot{\alpha}}_{\dot{\beta}} (f^{-1})_{ij}$.

By completeness, one finds the useful identity:

\begin{dmath}[label={eq:ADHM-projector}]
    {\mathcal{P}^{\mu}_{\lambda} = U_{\lambda u} {\bar{U}}^{\mu}_u = \delta^{\mu}_{\lambda} - \Delta_{\lambda i \dot{\alpha}} f_{ij} \bar{\Delta}^{\dot{\alpha} \mu}_j}
\end{dmath}

The gauge field strength is $F_{mn} = 4g^{-1} \bar{U} b {\sigma}_{mn} f \bar{b} U$.
And it can be shown:

\begin{dmath}
    {k= -\frac{g^2}{16 {\pi}^2} \int d^4 x \text{tr}_N \{F^2_{mn}\} =\frac{1}{16 {\pi}^2} \int d^4 x 
\ {\Box}^2 \text{tr} \ {\ln{f}}}
\end{dmath}

There are lots of redundancies in the ADHM construction.
By the following transformations, one can reduce the parameters to $4kN$:

\begin{dmath}[label={eq:ADHM-redundancy}]
    {\Delta \rightarrow \Lambda \Delta {\Gamma}^{-1} 
\ \ \ \ U \rightarrow \Lambda U \ \ \ \ \ f \rightarrow \Gamma f \Gamma^{\dagger}}
\end{dmath}

where $\Lambda \in U(N+2k) $ and $\Gamma \in Gl(k, \mathbb{C})$.

One can put $b$, $\bar{b}$, $a$ and $\bar{a}$ in the proper shape:

\begin{dmath}[label={eq:canonical-b}]
    {b^{\beta}_{\lambda j}= b^{\beta}_{(u+ i \alpha)j}= \begin{pmatrix}
0 \\
\delta^{\beta}_{\alpha} \delta_{ij}
\end{pmatrix} \ \ \ \ \ \ {\bar{b}^{\lambda}}_{\beta j}= {\bar{b}^{(u + i \alpha)}}_{\beta j} = (0 , {{\delta}^{\alpha}_{\beta}} \delta_{ji})}
\end{dmath}

\begin{dmath}
    {a_{\lambda j \dot{\alpha}}=a_{(u+ i \alpha) j \dot{\alpha}}= \begin{pmatrix}
w_{u j \dot{\alpha}} \\
{(a'_{\alpha \dot{\alpha}})}_{ij}
\end{pmatrix} \ \ \ \ \ \ \  {{\bar{a}}^{ \dot{\alpha} \lambda}}_j = {{\bar{a}}^{\dot{\alpha} (u + i \alpha)}}_j = ({{\bar{w}}^{\dot{\alpha}}}_{j u}, ({{\bar{a}}}^{'{\dot{\alpha}} \alpha})_{j i})}
\end{dmath}

The imposed ADHM condition or the \textbf{ADHM constraints} ${\bar{\Delta}_i}^{\dot{\alpha} \lambda} \Delta_{\lambda j \dot{\beta}}= {{\delta}^{\dot{\alpha}}}_{\dot{\beta}} (f^{-1})_{ij}$ translates into:

$${({\tau^a})^{\dot{\alpha}}}_{\dot{\beta}} ({{\bar{a}}^{\dot{\beta}}} a_{\dot{\alpha}}) = 0 \ \ a=1,2,3$$
\begin{dmath}
    (a'_n)^{\dagger} = a'_n
\end{dmath}

Given such simplifications of the ADHM condition, we can write down $f$ in terms of the ADHM variables:

\begin{dmath}
    f=\frac{2}{{\bar{w}}^{\dot{\alpha}} {w}_{\dot{\alpha}} + (a'_n + x_{n} 1_{[k] \times [k]})^2}
\end{dmath}

The residual symmetry group can be characterized as:

\begin{dmath}
    {\Lambda = \begin{pmatrix}
  1_{[N] \times [N]} & 0 \\
  0 & \Xi 1_{[2] \times [2]}  
\end{pmatrix} \  \ \ \ \ \ \Xi \in U(k)}
\end{dmath}

under which the ADHM parameters transform as:

\begin{dmath}
    {w_{u i \dot{\alpha}} \rightarrow w_{\dot{\alpha}} \Xi  \ \ \ \ \ \ \ a'_n \rightarrow {\Xi}^{\dagger} a'_n {\Xi}}
\end{dmath}

So, the relevant moduli space of the $k$-instanton that one is concerned with is: $\mu^{-1}_k (0) / U(k)$.
For this reason, one calls $\mathbb{R}^{4k(N+k)}$ the \textbf{mother space} and $\mu^{-1}_k(0)$ \textbf{the level set}(where the momentum map vanishes) and the $\mathcal{M}_k = \mu^{-1}_k(0) / U(k)$ the \textbf{HyperKähler quotient}.

A natural decomposition of the tangent space of a HyperKähler manifold of dimension $4n$ admits a distinguished $Sp(n) \times SU(2)$ basis of tangent vectors, where the metric can be written as:
\begin{dmath}
    \tilde{g} = {\tilde{\Omega}}_{\tilde{i} \tilde{j}} \epsilon_{\dot{\alpha} \dot{\beta}} {dz}^{\tilde{i} \dot{\alpha}} {dz}^{\tilde{j} \dot{\beta}}
\end{dmath}  where $\tilde{\Omega}$ is the symplectic matrix:

\begin{dmath}
    {z^{\tilde{i} \dot{\alpha}} = \begin{pmatrix}
    {\bar{w}^{\dot{\alpha}}}_{i u} \\
    {(\bar{a}^{' \dot{\alpha} 1})_{ij}} \\
    {\epsilon}^{\dot{\alpha} \dot{\beta}} w_{u i \dot{\beta}} \\
    {\epsilon}^{\dot{\alpha} \dot{\beta}} ({a'}_{1 \dot{\beta}})_{ij}
\end{pmatrix}}\end{dmath}

\begin{dmath}{{\tilde{\Omega}}_{\tilde{i} \tilde{j}} = 4 {\pi}^2 \begin{pmatrix}
    0 & 0 & 1_{[kN] \times [kN]} & 0 \\
    0 & 0 & 0 & 1_{[k^2][k^2]} \\
    -1_{[kN] \times [kN]} & 0 & 0 & 0 \\
    0 & -1_{[k^2][k^2]} & 0 & 0
\end{pmatrix}}
\end{dmath}

where $\tilde{i}$ runs over the composite indices $\{iu,ij,ui,ij \}$, which are all together the $kN + k^2 + Nk + k^2 = 2k(N+k)$ indices.

The $U(k)$ action on the mother space results in the following tri-holomorphic killing vector fields:

\begin{dmath}
    X_{r} = i {T^{r}}_{ij} {{\bar{w}}^{\dot{\alpha}}}_{j u} \frac{\partial}{\partial {\bar{w}^{{\dot{\alpha}}}}_{i u}} - i {T^{r}}_{ji} {w}_{u j \dot{\alpha}} \frac{\partial}{\partial w_{u i \dot{\alpha}}} + i [{T^r} , {a'}_n]_{ij} \frac{\partial}{\partial (a'_n)_{ij}}
\end{dmath}

The momentum maps related to each vector field are:

\begin{dmath}[label={eq:momentum-maps-Uk}]
    i \Vec{{\mu}}_{X_r} = 4 {\pi}^2 {{\Vec{\tau}}^{\dot{\beta}}}_{\dot{\alpha}} \Tr{T^r \bar{a}^{\dot{\alpha}} a_{\dot{\beta}}} - \Vec{\zeta}^r
\end{dmath}

where $\Vec{\zeta}^r$ takes values in the Lie algebra of the $U(1)$ factor of the $U(k)$ gauge group, which turns out to be of critical use in chapter \ref{Localisation}. Indeed, the $\Vec{\zeta}^r \neq 0$ case corresponds to the ADHM construction on a non-commutative spacetime. The set ${\Vec{\mu}}^{-1}_k (0) \subset \mathbb{R}^{4k(N+k)}$ represents the level set which is the subset that satisfies the ADHM constraints and that the HyperKähler quotient ${\Vec{\mu}}^{-1}_k (0) /U(k)$ represents the inequivalent (physically distinct) points of ${\Vec{\mu}}^{-1}_k (0)$. 

REMARK: The " $\vec{}$ " symbol in $\Vec{\mu}_{X_r}$ stands for the vectors in the three dimensional $\mathfrak{su}(2)$ Lie algebra.  

The bosonic zero modes can also be written in terms of the ADHM variables. Given that the zero modes are the solutions of the Weyl equation $\rawslashed{\bar{\mathcal{{D}}}}^{\dot{\alpha} \alpha} \delta A_{\alpha \dot{\beta}} = 0$ one can prove easily that all linear functions:
\begin{dmath}[label={eq:Lambda-of-C}]
    \Lambda_{\alpha} (C) = \bar{U}Cf{\bar{b}_{\alpha}}U - \bar{U}{b_{\alpha}}f \bar{C} U
\end{dmath}

with $C$ a constant $(N+2k) \times k$ matrix $C_{\lambda i}$ is a solution to the Weyl equation as long as it satisfies:

\begin{dmath}
    {\bar{C}_{i \lambda} a_{\lambda j \dot{\alpha}} = - \bar{a}_{i \dot{\alpha} \lambda} C_{\lambda j}}
\end{dmath}
\begin{dmath}
    {\bar{C}_{i \lambda} {b^{\alpha}}_{\lambda j} = {\bar{b}^{\alpha}}_{i \lambda} C_{\lambda j}}
\end{dmath}

Clearly, by taking $C_{\dot{\alpha}} = a_{\dot{\alpha}}$ and further computations one finds:

\begin{dmath}[label={eq:bosonic-zero-modes-ADHM}]
    \delta_{\mu} A_{\alpha \dot{\alpha}}=  2 g^{-1} \Lambda_{\alpha} (\frac{\partial a_{\dot{\alpha}}}{\partial X^{\mu}}) = 2 g^{-1} \bar{U} (\frac{\partial a_{\dot{\alpha}}}{\partial X^{\mu}} f {\bar{b}}_{\alpha} - b_{\alpha} f \frac{\partial \bar{a}_{\dot{\alpha}}}{\partial X^{\mu}})U
\end{dmath}

From the hyperKähler quotient perspective, the zero modes are orthogonal to the $3k^2$ vectors $I^{(c)} X_r$ and $k^2$, $X_r$ killing vector fields, which means the zero modes belong to ${\Vec{\mu}}^{-1} (0) / U(k) = \mathcal{H}$ the horizontal tangent space to ${\Vec{\mu}}^{-1} (0)$.

As for the metric:

\begin{dmath}
    {g_{\mu \nu}(X)= 2 \pi^2 \text{tr}{\frac{\partial {\bar{a}^{\dot{\alpha}}}}{\partial X^{\mu}}(\mathcal{P}_{\infty} +1) \frac{\partial a_{\dot{\alpha}}}{\partial X^{\nu}} + \frac{\partial {\bar{a}^{\dot{\alpha}}}}{\partial X^{\nu}}(\mathcal{P}_{\infty} +1) \frac{\partial a_{\dot{\alpha}}}{\partial{X^{\mu}}}}}
\end{dmath}

where:

\begin{dmath}
    {\mathcal{P}_{\infty}= \lim_{x\to\infty} \mathcal{P}(x) = 1- b \bar{b} = \begin{pmatrix}
    1_{[N] \times [N]} & 0_{[N] \times [2k]} \\
0_{[2k] \times [N]} & 0_{[2k] \times [2k]}
\end{pmatrix}}
\end{dmath}

$g$ can be lifted horizontally uniquely to all of $T {\mu}^{-1} (0)$ and $\mathcal{M}_k$ from $\mathcal{H}$. The projection operator $\mathcal{P}$ is introduced in \eqref{eq:ADHM-projector}.

Finally, we can conclude that the bosonic zero modes in the singular gauge behave as:
$$\delta_{\mu} A_m \propto \mathcal{O} (x^{-3})$$
at large $x$.

\section{Conformal Group Action on ADHM Variables}

Now, the conformal group action on $x_{\alpha \dot{\alpha}}$ coordinates, can be formulated as:
\begin{dmath}
    {x \rightarrow x' = (Ax+B)(Cx+D)^{-1} \ \ \ \text{where} \ \ \  \text{det} \begin{pmatrix}
        A & B \\
        C & D
    \end{pmatrix} = 1}
\end{dmath}
and $A,...$ are indeed quaternions which entail $15$ variables. Consequently, the ADHM matrix transforms as:

\begin{dmath}
    \Delta(x',a ,b) = \Delta(x, aD+bB , aC+bA) (Cx+ D)^{-1}
\end{dmath}

Since the gauge fields depend on $U,\bar{U}$, the $(Cx+ D)^{-1}$ factor is redundant and therefore the ADHM variables transform as:

\begin{dmath}
    {a \rightarrow aD + bB \ \ \ \ \ \ b \rightarrow aC + bA}
\end{dmath}

and by transformations \eqref{eq:ADHM-redundancy}, one can reset $b$ in its canonical form \eqref{eq:canonical-b}. This transformation can be done by the proper choice of $\Lambda, \Gamma$ depending on $a$ and $A$ and $C$: $b \rightarrow \Lambda (aC +bA) \Gamma^{-1}$ while $a \rightarrow \Lambda(aD+bB)\Gamma^{-1}$.

\section{Multi-Instanton Configuration, Clustering}

A realistic multi-instanton solution is surely not the combination of single instantons. Nonetheless, in certain asymptotic regions of the moduli space $\mathcal{M}_k / U(k)$, the multi-instanton can be thought of as a composition of single instantons. It is called the \textbf{completely clustered limit}.

Identifying the center of each of the $k$ single instantons with the diagonal elements of $(a'_n)_{ii}$, one has $X^i_n = -(a'_n)_{ii}$.
Employing a gauge transformation one can set the off diagonal elements of $a'_n$ to zero resulting in $(\tilde{a}'_n)_{ij} (X^i - X^j)_n = 0$ after the gauge transformation, leaving the diagonal symmetry $U(1)^k$ which is the auxiliary symmetry of each of the $k$ single instantons.
In the complete clustering limit $(\tilde{a}^{' \dot{\alpha} \alpha})_{ik}  (\tilde{a}'_{\alpha \dot{\beta}})_{kj} , k \neq i,j$  terms can be ignored in the ADHM constraint. The off-diagonal elements $(\tilde{a}'_{\alpha \dot{\alpha}})_{ij}, i \neq j$ are subject to $(\bar{X}^i - \bar{X}^j)^{\dot{\alpha} \alpha} (\tilde{a}'_{\alpha \dot{\beta}})_{ij} + \bar{w}^{\dot{\alpha}}_{iu} w_{uj \dot{\beta}} \propto \delta^{\dot{\alpha}}_{\dot{\beta}}$ where the diagonals are $\bar{w}^{\dot{\alpha}}_{iu} w_{ui \dot{\beta}} = \rho^2_i \delta^{\dot{\alpha}}_{\dot{\beta}} $ for arbitrary $\rho_i$. Each instanton can be defined by embedding a copy of $SU(2)$ into $SU(N)$ generators as $(T^{c}_i)_{uv} = \rho^{-2}_i w_{ui \dot{\alpha}} \tau^{c \dot{\alpha}}_{\dot{\beta}} \bar{w}^{\dot{\beta}}_{iv}$. Consequently, one can formally write $(X^i - X^j)^2 \gg \rho_i \rho_j \Tr_N (T^c_i T^c_j)$.

The group of gauge transformations acting on the ADHM variable that leave the instanton fixed is called the \textbf{Stability Group}. The stability group is trivial for $N \leq 2k$ and the the analysis for $N > 2k$ is as follows.
As for the multi-instanton(higher charge) solution of the YM one can decompose the ADHM variable $w_{ui\dot{\alpha}}$:

\begin{dmath}
    {w= \mathcal{U} \ . \begin{pmatrix}
    \xi_{11} & \xi_{12} & \cdots & \xi_{1 , 2k} \\
     0 & \xi_{22} & \cdots & \xi_{2, 2k} \\
     \vdots & \vdots & \ddots & \vdots \\
     0 & 0 & 0 & \xi_{2k , 2k} \\
     \vdots & \vdots & \vdots & \vdots \\
      0 & 0 & 0 & 0
\end{pmatrix} = \mathcal{U}\Xi}
\end{dmath}

where $\mathcal{U} \in \frac{SU(N)}{SU(N-2k)}$ and $\Xi$ is an $N \times 2k$ upper-triangular complex matrix whose diagonal elements $\xi_{aa}$ are real. Consequently, the gauge field can be written as:

\begin{dmath}
    {A_n = \mathcal{U}^{\dagger} \begin{pmatrix}
        (A^{\text{inst}}_n)_{[2k]\times[2k]} & 0_{[2k]\times[N-2k]} \\
        0_{[N-2k]\times[2k]} & 0_{[N-2k] \times[N-2k]}
    \end{pmatrix} \mathcal{U}}
\end{dmath}

where $A^{\text{inst}}_n$ is the $k$-instanton solution lying in $SU(2k) \subset SU(N)$.  Therefore, $\mathcal{U} \in \frac{SU(N)}{SU(N-2k)}$ without any further consideration. However, the stability group consists of elements $\text{exp}^{i \lambda \theta} g$ where $g \in SU(N-2k)$ and 

\begin{dmath}
    \lambda = \begin{pmatrix}
        1_{[2k]\times[2k]} & 0_{[2k]\times[N-2k]} \\
        0_{[N-2k]\times[2k]} & -\frac{2k}{N-2k} 1_{[N-2k]\times[N-2k]}
    \end{pmatrix}
\end{dmath}

This results in the stability group: $S(U(N-2k) \times U(1))$. Therefore:

\begin{dmath}
    \mathcal{U} \in \frac{SU(N)}{S(U(N-2k) \times U(1))}
\end{dmath}

\section{The Orbifold Nature of $\mathcal{M}_k$ and Singularities}

The \textbf{Orbifold} nature of the moduli space $\mathcal{M}_k = \mu^{-1} (0) /U(k)$ reveals itself around its singularities, where the action of the gauge group does not appear free. These points can be characterized as the null eigenvectors of a matrix defined on $\mu^{-1} (0)$:

\begin{dmath}
    {L_{rs} = g(X_r , X_s) := 8 \pi^2 \tr{T^r \mathbf{L} T^s}}
\end{dmath}

\begin{dmath}
    {\mathbf{L}. \Omega = \frac{1}{2} \{{\bar{w}}^{\dot{\alpha}} w_{\dot{\alpha}} , \Omega \} + [{a'_n}, [a'_n , \Omega]]}
\end{dmath}

where $\Omega$ is a $k \times k$ Hermitian matrix and $T^r$ are the fundamental representation generators of $U(k)$.

The Classical YM equations have conformal symmetry, and more specifically are invariant under the dilation $D=-ix^{\mu}\partial_{\mu}$ generator of the conformal algebra. Each particular instanton solution spontaneously breaks this symmetry; therefore, the moduli space must contain all these physically distinct solutions. In other words, it must be invariant under the dilation transformation.

It turns out that for $(\mathcal{M}_k,g)$ to admit an infinitesimal conformal invariance, it must necessarily admit a homothetic vector field $X$(Section \ref{Projective-Vector-Fields-and-Their-Lie-Subalgebras}); Such a homothety does indeed exist for $\mathcal{M}_k$ if $\vec{\zeta}^r =0$. To prove its existence, it is better to use another formulation for the hyper-Kähler potential than that of the Section \ref{ADHM-Construction-as-HyperKähler-Quotient}. However, it is rather straightforward to prove that it is equivalent to \eqref{eq:hyperkahler-potential}. It is possible to define a hyper-Kähler potential over the mother space $\mathbb{R}^{4k(N+k)}$:

\begin{dmath}
    \tilde{\chi} = 8 \pi^2 \text{tr}_k (\bar{w}^{\dot{\alpha}} w_{\dot{\alpha}} + a'_n a'_n)
\end{dmath}
One can define a homothety on the mother space with the help of this potential:
\begin{dmath}
    \tilde{X}^{\tilde{i} \dot{\alpha}} = \partial^{\tilde{i} \dot{\alpha}} \tilde{\chi}
\end{dmath}

By definition, this vector field is a hypersurface orthogonal vector field. Since fulfilling the ADHM constrainsts stands for the choice of the level set $\mu^{-1}(0) \subset \mathbb{R}^{4k(N+k)}$, there must exist a parametrization of this hypersurface in terms of the collective coordinates $X^\mu$ such that each $z^{\tilde{i}\dot{\alpha}} = z^{\tilde{i} \dot{\alpha}}(X^{\mu})$ and so the hyper-Kähler potential on the level set $\mu^{-1} (0)$ is simply $\chi_{\mu^{-1} (0)} = z^{*}\tilde{\chi} = \tilde{\chi} \circ z$. Finally, since $\mathcal{M}_k$ is a $U(k)$-principal bundle, the projection $\pi : \mu^{-1} (0) \rightarrow \mu^{-1} (0) / U(k) = \mathcal{M}_k$ uniquely defines the hyper-Kähler potential $\chi = \pi z^{*} \tilde{\chi}$ over $\mathcal{M}_k$. It is helpful to centralize the moduli space $\mathcal{M}_k = \mathbb{R}^4 \times \widehat{\mathcal{M}}_k$. The potential over the centralized moduli space $\widehat{\mathcal{M}}_k$ turns out to be:

\begin{dmath}
    {\widehat{\chi}:=\chi_{\widehat{\mathcal{M}}_k} = 8 \pi^2 \text{tr}_k (\bar{w}^{\dot{\alpha}}(X) w_{\dot{\alpha}}(X))}
\end{dmath}

It is important to bear in mind that $\vec{\zeta}^r$ must vanish in \eqref{eq:momentum-maps-Uk}, otherwise such a homothety cannot be inherited from the mother space to the moduli space.
This is of utmost importance in chapter \ref{Localisation}. The same arguments apply to $\widehat{\mathcal{M}}_k$. The analysis of Section \ref{sec:Cones-Sasakian-Manifolds} applies \textit{mutatis mutandis} to $\widehat{\mathcal{M}}_k$. Mainly that, $(\widehat{\mathcal{M}}_k ,\widehat{g} )$ is a hyper-Kähler Riemannian cone over the Einstein-3-Sasakian base $(\mathfrak{F}, \widehat{g}_{\mathfrak{F}})$ where:

\begin{dmath}
    \widehat{g} = \frac{(d{\widehat{\chi}})^2}{4\widehat{\chi}} + \widehat{\chi} \widehat{g}_{\mathfrak{F}}
\end{dmath}

where $g_{\mathcal{M}_k} = \widehat{g} \oplus g_{\mathbb{R}^4}$ and $\widehat{g}_{\mathfrak{F}}$ is the metric over $\mathfrak{F}$.  The $SU(2)$ isometric group action Killing vector fields in this context are:

\begin{dmath}
    {\tilde{n}^{(c) \tilde{i} \dot{\alpha}} = {\tilde{\textbf{I}}^{(c) \dot{\alpha}}}_{\ \ \ \ \ \dot{\beta}} . z^{\tilde{i} \dot{\beta}} = i \tau^{c \dot{\alpha}}_{\ \ \ \dot{\beta}} z^{\tilde{i} \dot{\beta}}}
\end{dmath}

which are inherited by the moduli space $\mathcal{M}_k$ or $\widehat{\mathcal{M}}_{k}$ if $\vec{\zeta}^r=0$ for all $r$ in the momentum map.

It can be shown that the centralized moduli space base $\mathfrak{F}$ is a non-trivial $Sp(1)$ fibration over a quaternionic Kähler space $\mathfrak{Q}_k$: $Sp(1) \rightarrow \mathfrak{F} \rightarrow \mathfrak{Q}_k$ as a result of admitting an $SU(2)$ isometry group and the dilation symmetry over $\widehat{\mathcal{M}}_k$.

In short:

\begin{dmath}
    \widehat{\mathcal{M}}_k = \mathbb{R}^{>0} \times [Sp(1) \times_f \mathfrak{Q}_k]
\end{dmath}

where $f$ index represents the non-trivial fibration and $k$ represents the instanton charge.
For instance, for the case of $k=1$, the decomposition is as follows:

\begin{dmath}
    \widehat{\mathcal{M}}_1 = \mathbb{R}^{>0} \times \frac{SU(N)}{S(U(N-2) \times U(1))}
\end{dmath}

where $\mathfrak{F} = \frac{SU(N)}{S(U(N-2) \times U(1))} = Sp(1) \times_f \mathfrak{Q}_1$ in an Einstein-3-Sasakian space and $\mathfrak{Q}_1$ is known as the \textbf{Wolf spaces} \cite{Quaternions and Instantons} which are the only compact homogeneous quaternionic Kähler manifolds and the radial direction of the cone is parameterized by the hyper-Kähler potential $\widehat{\chi} \propto \rho^2$. The centralized moduli space is singular at the cone's apex, but this is the only singularity in such a case.

As for the higher charge moduli spaces, the first type of singularities corresponds to the points with $w_{\dot{\alpha}} = 0$ and vanishing non-trace parts of $a'_n =0$. All of $U(k)$ is fixed in this case, meaning all the instantons have shrunk to zero size. 

The next type of singularities are those of the Einstein-3-Sasakian space $\mathfrak{F}$ since it is not necessarily a homogeneous space in general. More explicitly, whenever $w_{i u \dot{\alpha}} \rightarrow 0$ the ADHM construction implies that $(a'_n)_{il} \ , (a'_n)_{li} \rightarrow 0$ when $i \neq l$. In such a limit, the $U(1) \subset U(k)$ subgroup, which acts as follows:
\begin{dmath}
    {w_{ui \dot{\alpha}} \rightarrow e^{i \delta_{il} \phi} {w_{ui \dot{\alpha}}} \ \ \ , \ \ \ (a'_n)_{ij} \rightarrow e^{i (\delta_{jl} - \delta_{il}) \phi} (a'_n)_{ij}} 
\end{dmath}

does not act freely. Indeed, in this region, $\mathcal{M}_k \rightarrow \mathcal{M}_{k-1} \times \mathbb{R}^{4}$, which is the Cartesian product of the $(k-1)$-instanton moduli space with that of the zero size $1$-instanton. 

The decomposition can be generalized to that of an $(k-r)$-instanton and $r$ zero-size instantons by choosing the proper subgroup $U(1)^r \subset U(k)$ that does not act freely:

\begin{dmath}
    \mathcal{M}_k \rightarrow \mathcal{M}_{k-r} \times \text{Sym}^r \mathbb{R}^4
\end{dmath}

where $\text{Sym}^r \mathbb{R}^4$ is the $r$-fold symmetric product $(\mathbb{R}^4)^r / S_r$ of $\mathbb{R}^4$, with $\dim({\text{Sym}^r (\mathbb{R}^4)}) = 4r$. %
Intuitively, it is the subgroup of $U(k)$ that permutes the position of $r$ zero-size instantons, and therefore, when two of these instantons coincide, it becomes a fixed point of the subgroup action, and the quotient by this subgroup results in orbifold singularities on the moduli space. \footnote{This group will appear more than once in the following chapters and is called the Weyl subgroup of $U(k)$.}

And finally, there might exist a non-Abelian subgroup of $U(k)$ that does not act freely. This turns out to correspond to the case of zero-size instantons coming together at the same spacetime point.

\section{SUSic Moduli space, Grassmann Collective Coordinates, SUSic Instantons}
\label{SUSic-Moduli-space,-Grassmann-Collective-Coordinates,-SUSic-Instantons}
To extend the moduli space to that of a supersymmetric one, we should note that there is a problem regarding SUSY in Euclidean spacetime. In four-dimensional Euclidean space, Weyl spinors are in the pseudo-real representation of $SO(4)$, with the reality condition of $\bar{\lambda}^{\dot{\alpha}} = \lambda^{\dagger}_{\alpha}$ for $\alpha = \dot{\alpha}$. Instead, one knows that $\mathfrak{so}(4) = \mathfrak{su}(2) \times \mathfrak{su}(2)$, meaning that the Weyl spinors $\bar{\lambda}^{\dot{\alpha}}, \lambda_{\alpha}$ live in different representations and do not mix with complex conjugation at all. This means there is no Majorana spinor in the Wick rotated Euclidean theory and consequently, no real action. The solution to this problem is not unique. The safest solution would be to ignore the construction of $\mathcal{N}=1$ Euclidean field theory and construct extended SUSic ($\mathcal{N}=2,4$) field theory that can be articulated in terms of Dirac spinors by a combination of Weyl spinors, since Dirac spinors have Euclidean counterparts.

As for fermionic zero modes one should deal with the equations of motions for the Weyl fermion in the charge $k$ instanton background $\rawslashed{\bar{\mathcal{D}}} \lambda^A =0, \slashed{\mathcal{D}} \bar{\lambda}_A = 0$ where the $\slashed{\mathcal{D}}$ is in the adjoint representation.

One can write down the general solution in terms of the $\Lambda(C)$ function \eqref{eq:Lambda-of-C} we found in section \ref{ADHM-Construction-as-HyperKähler-Quotient} as:
\begin{dmath}
    \lambda_{\alpha} = \sum^{2kN}_{i=1} \psi^i \Lambda_{\alpha}(C_i)
\end{dmath}
which would look nicer if one introduced the Grassmann ADHM variable $\mathcal{M}= \psi^i C_i$ where $\psi^i$ are called Grassmann collective coordinates:

\begin{dmath}
    {\lambda_{\alpha} = \Lambda_{\alpha}(\mathcal{M})}
\end{dmath}

$\mathcal{M}$ constitutes the ADHM variables that can be written as:

\begin{dmath}
   { {\mathcal{M}_{\lambda j} = \mathcal{M}_{(u+i \alpha) j} = \begin{pmatrix}
    \mu_{u j} \\
    (\mathcal{M}'_{\alpha})_{ij}
\end{pmatrix}} , \ \ \ \ \bar{\mathcal{M}}^{\lambda}_j = \bar{\mathcal{M}}_{j(u+i \alpha)} = \left( \bar{\mu}_{ju} , (\bar{\mathcal{M}}^{' \alpha})_{ji} \right) }
\end{dmath}
where ADHM constraints require ${\bar{\mathcal{M}}'_{\alpha}} = \mathcal{M}'_{\alpha}$ and as for the other constraint $\bar{\mu} w_{\dot{\alpha}} + \bar{w}_{\dot{\alpha}} \mu + [\mathcal{M}^{' \alpha} , a'_{\alpha \dot{\alpha}}] = 0$.

Geometrically speaking, the Grassmann collective coordinates $\psi^i$ are the Grassmann-valued \textbf{symplectice tangent vectors} of the hyperKähler quotient $\mu^{-1}(0) /U(k)$.
In analogy with the bosonic collective coordinates $z^{\tilde{i} \dot{\alpha}}$ one can defind the Grassmann collective coordinate:
\begin{dmath}
    \mathcal{M}^{\tilde{i}} = \begin{pmatrix}
    \bar{\mu}_{iu} \\
    (\mathcal{M}^{'1})_{ij} \\
     \mu_{ui} \\
     (\mathcal{M}'_1)_{ij}
\end{pmatrix}
\end{dmath}
As a symplectic tangent vector, it should (by definition) satisfy the following: 
\begin{dmath}
    {\mathcal{M}^{\tilde{i}} \tilde{\Omega}_{\tilde{i} \tilde{j}} X^{\tilde{j} \dot{\alpha}}_r = -4 i \pi^2 \text{tr}_k T^r (\bar{\mu} w^{\dot{\alpha}} + \bar{w}^{\dot{\alpha}} \mu + [\bar{a}^{'{\dot{\alpha}}\alpha} , \mathcal{M}'_{\alpha}])=0}
\end{dmath}

which are precisely the Grassmann ADHM constraints ($X_r$ being the vector fields generated by the $U(k)$ action on the moduli space); and $\tilde{\Omega}_{\tilde{i} \tilde{j}}$ is the symplectic form over the moduli space defined by the inner product of the Grassmann zero modes:

\begin{dmath}[label={eq:zero-mode-inner-product}]
    {-\frac{1}{4} \tilde{\Omega} (\mathcal{M}, \mathcal{N}) = \int d^4 x \text{tr}_N \Lambda(\mathcal{M}) \Lambda(\mathcal{N}) = - \frac{\pi^2}{2} \text{tr}_k [\bar{\mathcal{M}} (\mathcal{P}_{\infty} + 1) \mathcal{N} + \bar{\mathcal{N}}(\mathcal{P}_{\infty} + 1) \mathcal{M}]}
\end{dmath}

inducing a symplectic form on the hyperKähler quotient: 

$\Omega_{ij} (X) \psi^i \theta^j = \tilde{\Omega}(\mathcal{M}(\psi , X) , \mathcal{M}(\theta, X)) , i$ running over $i=1,..., 2kN$.

Since in this thesis one is mainly concerned with the $\mathcal{N}=2$ SQCD, we concentrate on the Euclidean version of such a theory, starting with $\mathcal{N}=2$ SYM and adding matter(hypermultiplets) later. The $\mathcal{N}=2$ Euclidean SYM can be written as:
\begin{dmath}
S^{E} = \int d^4 x \tr_N \left\{-\frac{1}{2} F^2_{mn} + \frac{i \theta g^2}{16 \pi^2} F_{mn} \tilde{F}_{mn} -2 \mathcal{D}_n \bar{\lambda} \bar{\sigma}_n \lambda -2 \mathcal{D}_n \bar{\psi} \bar{\sigma}_n \psi + \mathcal{D}_n \phi^{\dagger} \mathcal{D}_n \phi +2ig \bar{\psi} [\phi, \bar{\lambda}] + 2ig [\phi^{\dagger} , \lambda] \psi + \frac{1}{4}g^2 [\phi, \phi^{\dagger}]^2 \right\}
\end{dmath}

where $\phi = \phi_1 + i\phi_2$ and $\phi^{\dagger}= \phi_1 - i \phi_2$ with $\phi_a \ \ a =1,2$  being real scalars. Pay attention that in the Euclidean action $\phi$ and $\phi^{\dagger}$ are independent fields and the reality condition fails in general, meaning $(\phi^{\dagger})^{\dagger} \neq \phi$. 

resulting in the equations of motion:

\begin{dmath}
    \mathcal{D}_m F_{nm} = 2g[\phi_a , \mathcal{D}_n \phi_a] + 2g \bar{\sigma}_n \{\lambda^A , \bar{\lambda}_A \}
\end{dmath}
\begin{dmath}
    \rawslashed{\bar{\mathcal{D}}} \lambda^A = g \Sigma^{AB}_a [\phi_a , \bar{\lambda}_B]
\end{dmath}
\begin{dmath}
    \slashed{\mathcal{D}} \bar{\lambda}_A = g \bar{\Sigma}_{aAB} [\phi_a , \lambda^{B}]
\end{dmath}
\begin{dmath}[label={eq:eom-scalar}]
    {\mathcal{D}^2 \phi_a = g^2 [\phi_b ,[\phi_b , \phi_a]] + g \bar{\Sigma}_{aAB} \lambda^A \lambda^B + g \Sigma^{AB}_{a} \bar{\lambda}_A \bar{\lambda}_B}
\end{dmath}
where $\Sigma$-matrices are related to the $\mathcal{N}=2$ $SU(2)$ R-symmetry:
\begin{dmath}
{\Sigma^{AB}_a = \epsilon^{AB} (i,1) \ \ \ , \ \ \ \bar{\Sigma}_{aAB} = \epsilon_{AB} (-i,1)}
\end{dmath}
$A,B$ being the spinorial indices of $SU(2)_R \subset U(2)_R$.

The on-shell variations of the fields are in order:
\begin{dmath}
    \delta A_n = i \xi^A \sigma_n \bar{\lambda}_A + i \bar{\xi}_A \bar{\sigma}_n \lambda^A
\end{dmath}
\begin{dmath}[label={eq:susy-variation-lambda}]
    \delta \lambda^A = i \sigma_{mn} \xi^{A} F_{mn} - ig \Sigma_{abB}^A \xi^B [\phi_a , \phi_b] -i\Sigma^{AB}_a \slashed{\mathcal{D}} \phi_a \bar{\xi}_B
\end{dmath}
\begin{dmath}[label={eq:susy-variation-lambda-bar}]
    \delta \bar{\lambda}_A = i \bar{\sigma}_{mn} \bar{\xi}_A F_{mn} -ig \bar{\Sigma}_{abA}^{B} \bar{\xi}_B [\phi_a , \phi_b] - i \bar{\Sigma}_{aAB} \bar{\slashed{\mathcal{D}}} \phi_a \xi^B
\end{dmath}
\begin{dmath}
    \delta \phi_a = i \xi^A \bar{\Sigma}_{aAB} \lambda^B + i\bar{\xi}_A {\Sigma}^{AB}_a \bar{\lambda}_B
\end{dmath}

A general expansion in terms of $g$ for all fields looks like:
\begin{dmath}
    A_m = g^{-1} A^{(0)}_m + g A^{(1)}_m + g^3 A^{(2)}_m + ...
\end{dmath}
\begin{dmath}
    \lambda^{A} = g^{-1/2} \lambda^{(0)A} + g^{3/2} \lambda^{(1)A} + g^{7/2} \lambda^{(2)A} +...
\end{dmath}
\begin{dmath}
    \bar{\lambda}_A= g^{1/2} \bar{\lambda}^{(0)}_A + g^{5/2} \bar{\lambda}^{(1)}_A + g^{9/2} \bar{\lambda}^{(2)}_A + ...
\end{dmath}
\begin{dmath}
    \phi_a = g^0 \phi^{(0)}_a + g^2 \phi^{(1)}_a + g^4 \phi^{(2)}_a + ...
\end{dmath}

where one should immediately identify $A^{(0)}_m = \bar{U} \partial_m U$ , $\lambda^{(0)A} = \Lambda (\mathcal{M}^A)$ leading to $A_m = g^{-1} A^{(0)}_m , \lambda^A = g^{-1/2} \lambda^{(0)A} , \bar{\lambda}_A = 0 , \phi_a = g^0 \phi^{(0)}_a$ keeping in mind that for the $\mathcal{N}=1$ case there is no scalar field $\phi_a$. In the absence of VEVs in the $\mathcal{N}=2$, the leading order is indeed, \underline{an exact solution of the full equations of motion}.

\
\

The general solution of $\rawslashed{\bar{\mathcal{D}}} \lambda^A = 0$ in terms of the ADHM variables is:

\begin{dmath}[label={eq:gaugino-zero-mode-ADHM}]
    {\lambda^A_{\alpha} = g^{-1/2} \Lambda_{\alpha} (\mathcal{M^A}) = g^{-1/2} \left(\bar{U} \mathcal{M^A} f \bar{b}_{\alpha} U - \bar{U}b_{\alpha} f \bar{\mathcal{M^A}} U \right)}
\end{dmath}

To the leading order in $g$ as can be seen from the field variations \eqref{eq:susy-variation-lambda} and \eqref{eq:susy-variation-lambda-bar}, that the SUSY algebra turns on chiral zero-modes by acting on the bosonic instanton $\delta \lambda^A = i \sigma_{mn} \xi^A F_{mn}$ while the anti-chiral fermions remain unchanged $\delta \bar{\lambda}_A = i \bar{\sigma}_{mn} \bar{\xi}_A F_{mn} = 0$ under anti-chiral transformations $\bar{\xi}_A$ due to self-duality of $F_{mn}$. In other words half of the SUSY algebra leaves the bosonic solution invariant. Explicitly:
\begin{dmath}
    {\delta\lambda^A_{\alpha} = 4i g^{-1/2} (\sigma_{mn} \xi^A)_{\alpha} \bar{U}b \sigma_{mn} \bar{b} f U } = -4i g^{-1/2} \bar{U} (b \xi^A f \bar{b}_{\alpha} - b_{\alpha} f \xi^{A} \bar{b}) U = g^{-1/2} \Lambda_{\alpha}(-4ib \xi^A)
\end{dmath}
where we re-scaled $\xi^A \rightarrow g^{1/2} \xi^A$ so that one can write for the Grassmann ADHM variables:
\begin{dmath}[label={eq:susy-zero-modes-ADHM}]
  {\delta \mathcal{M}^A_{\lambda i} = -4i \xi^A_{\alpha} b^{\alpha}_{\lambda i} \ \ \ , \ \ \ \delta \bar{\mathcal{M}}^{\lambda A}_i} = -4i \xi^{\alpha A} \bar{b}^{\lambda}_{\alpha i}
\end{dmath}
resulting in $2 \mathcal{N}$ \textbf{supersymmetric zero modes}.

\underline{The key point to take here for the next section is that the SUSY transformations} \underline{are traded for SUSY transformation of the collective coordinates!}

One can find even further zero modes by localising the Grassmann variations $\xi^A \rightarrow \xi^A (x)$. By doing so and Taylor-expanding in Grassmann variables one can write: $\xi^A_{\alpha} (x) = \xi^A_{\alpha} - x_{\alpha \dot{\alpha}} \bar{\eta}^{\dot{\alpha} A} , \bar{\xi}^{\dot{\alpha}}_A (x) = \bar{\xi}^{\dot{\alpha}}_A + \bar{x}^{\dot{\alpha} \alpha} \eta_{\alpha A}$. Obviously this is inclusive of the supersymmetric zero modes generated by $\{\xi_A, \bar{\xi}^A \}$ apart from new directions $\{ \eta^A, \bar{\eta}_A \}$. Analogously, the self-dual bosonic instanton is invariant under super-conformal transformations in the $\eta_A$ directions, while the other half of such super-conformal generators result in chiral fermion zero modes of the sort:
\begin{dmath}
    {\delta\lambda^A_{\alpha} = -4 i g^{-1/2} (\sigma_{mn} x \bar{\eta}^A)_{\alpha} \bar{U} b \sigma_{mn} \bar{b} f U} = {4 i g^{-1/2} \bar{U} \left(bx\bar{\eta}^A f \bar{b}_{\alpha} - b_{\alpha}f \bar{\eta}^A \bar{x} \bar{b} \right) U} =-4i g^{-1/2} \bar{U} \left(a \bar{\eta}^A f \bar{b}_{\alpha} - b_{\alpha} f \bar{\eta}^{A} \bar{a} \right)U = g^{-1/2} \Lambda_{\alpha} (-4ai \bar{\eta}^A)
\end{dmath}
where the transition from $b$ to $a$ is carried out thanks to bosonic ADHM constraints $\bar{U}bx= -\bar{U}a , \bar{x} \bar{b} U = -\bar{a} U$ resulting in the ADHM variables:
\begin{dmath}
    {\delta \mathcal{M}^A_{\lambda i} = -4 i a_{\lambda i \dot{\alpha}} \bar{\eta}^{\dot{\alpha} A} \ \ \ , \ \ \ \delta\bar{\mathcal{M}}^{\lambda A}_{i} = -4i \bar{\eta}^A_{\dot{\alpha}} \bar{a}^{\dot{\alpha}\lambda}_i}
\end{dmath}
leading to $2\mathcal{N}$ \textbf{super-conformal zero modes}.

Expanding the Euclidean action around the $\mathcal{N}=2$ SUSic instanton solution one finds:
\begin{dmath}
    {S=\frac{8 \pi^2 k}{g^2} - i k \theta + \tilde{S} g^0 + \mathcal{O}(g^2)}
\end{dmath}

where:

\begin{dmath}
    \tilde{S} = \int d^4 x \text{tr}_N \{ \mathcal{D}_n \phi^{(0)}_a \mathcal{D}_n \phi^{(0)}_a -\lambda^{(0)A} \bar{\Sigma}_{aAB} [\phi^{(0)}_a , \lambda^{(0)B}] \}
\end{dmath}

To order $g^0$ the equation of motion \eqref{eq:eom-scalar} for the scalar field $\phi^{(0)}_a$ is 

$\mathcal{D}^2 \phi^{(0)}_a = \bar{\Sigma}_{aAB} \lambda^{(0)A} \lambda^{(0)B}$. For vanishing VEV, the solution is that of:

\begin{dmath}[label={eq:scalar-solution-zero-vev}]
    \phi_a = -\frac{1}{4} \bar{\Sigma}_{aAB} \bar{U} \mathcal{M}^A f \bar{\mathcal{M}}^B U + \bar{U} \begin{pmatrix}
    0_{[N] \times [N]} & 0_{[N] \times [2k]} \\
    0_{[2k] \times [N]} & \varphi_a 1_{[2] \times [2]}
\end{pmatrix}U
\end{dmath}
$\varphi_a$ being a $k \times k$ matrix:

\begin{dmath}
    \varphi_a = \frac{1}{4} \bar{\Sigma}_{aAB} \mathbf{L}^{-1} (\bar{\mathcal{M}}^A \mathcal{M}^B)
\end{dmath}

For vanishing VEV, one can verify that $\tilde{S}=0$ along with all the other higher-order corrections in $g$, and the SUSic instanton is degenerate with the bosonic instanton bearing the same action as $-2 \pi i k \tau$.

One should bear in mind that in SUSic instanton calculus $\phi_a$ does not remain the hermitian conjugation of $\phi^{\dagger}_a$ due to the quadratic Grassmann quantum corrections, and this fact is conceivably related to the lack of a reality condition on the Weyl fermions that we alluded to at the beginning of this section; so while $\phi_a$ is nontrivial, $\phi^{\dagger}_a$ remains zero!
\

Moreover, we leave it as a result that: the existence of the $U(1)_R$ symmetry in the $\mathcal{N}=2$ case guarantees that there is no lifting of the $4kN$ Grassmann collective coordinates due to interactions and nonlinearities.

\
\

Now let's see what happens in the case of non-vanishing VEVs.

\
\

\section{Constrained Instantons and the Solution for non-vanishing VEVs}
\label{Constrained-Instantons-and-the-Solution-for-non-vanishing-VEVs}
When the VEVs are turned on, the instantons cease to be the minima of the action! Indeed \underline{a potential develops in the moduli space} where instantons have non-trivial action and entropy. 

However, instantons are substituted with \textbf{constrained instantons}. 

It would be better to take a toy model and leave the general case of fundamental Higgsed $SU(N)$ YM, $\mathcal{N}=2$ SYM, and SQCD to the relevant references.

The easiest model (yet quite technical!) would be the four-dimensional Euclidean theory $\phi^4$ \underline{ with the wrong interaction sign}:

\begin{dmath}
    \mathcal{L} = \frac{1}{g} \left[\frac{1}{2} \left( \partial_{\mu} \phi \right)^2 + \frac{1}{2} \mu^2 \phi^2 -\frac{\phi^4}{4!}\right]
\end{dmath}

where it is easy to check that no instanton(with $\rho \neq 0$) solution exists in the massive theory \\ ($\mu \neq 0$) by a simple application of Derrick's lemma. The massless ($\mu = 0$) theory instanton solution is $\phi_0 (x) = \frac{4 \sqrt{3} \rho}{\rho^2 + x^2} = \frac{1}{\rho} \frac{4 \sqrt{3}}{1+ x^2 / \rho^2}$. The vacuum of this theory $\phi=0$ in infinite volume is meta-stable, and instanton solutions can act as tunneling between this false vacuum and $\phi \rightarrow -\infty$. The tunneling is suppressed by the height of the barrier at $\phi = \sqrt{6} \mu$ (both for the massive and the massless cases).

Here, one argues that a perturbative approach can be applied in case the dimensionless quantity $\rho \mu \ll 1$(It turns out that at the quantum level one reaches a more nuanced bound of $\rho \lesssim \sqrt{g} \mu^{-1}$). The main obstacle to this approach is the \textbf{non-normalizable} [formerly]zero modes $\frac{\partial \phi_0}{\partial \rho}$. Indeed, this mode is lifted due to the mass term and becomes what one calls a \textbf{quasi zero-mode}.  

More explicitly given the leading order perturbative equation of motion \\ $\left( \partial^2 + \frac{1}{2} \phi^2_0 \right) \delta \phi = \mu^2 \phi_0$, by multiplying the zero mode $\frac{\partial \phi_0}{\partial \rho}$ on both sides and integrating over the Euclidean space one finds:
\begin{dmath}
    \int_{|x| = \infty} d S_{\mu} \frac{\partial \phi_0}{\partial \rho} \overset\leftrightarrow{\partial} \delta \phi = \mu^2 \int d^4 x \frac{\partial \phi_0}{\partial \rho} \phi_0
\end{dmath}

so, if $\delta \phi$ is not singular anywhere one cannot expect the $\delta \phi$ to vanish at infinity since $\delta \phi(x) \rightarrow \frac{\mu^2}{16 \sqrt{3} \pi^2} \int d^4 x \frac{\partial \phi_0}{\partial \rho} \phi_0$ when $\mu x \rightarrow \infty$ unless the projection of $\phi_0$ on the zero mode $\frac{\partial \phi_0}{\partial \rho}$ vanishes.

Since in case of $\mu \neq 0$ the only finite action solution is the one with $\rho=0$, one should introduce a new constraint that fixes $\rho$ at a non-zero value, giving rise to a finite action solution of the equation of motion with $\mu \neq 0$ by minimizing the action on each [$\rho$-dependent]level-set of the constraint($F[\phi] = f(\rho, \mu)$).

Assuming analyticity of the the constrained $\phi(x)$ in $0 \leq \rho \mu \ll 1$ \footnote{this premise can be theoretically duobted as $\phi(x)$ might be non-analytic at $\rho \mu = 0$}, a general finite action solution then, can be decomposed in the following manner:

\begin{dmath}
    \phi = \phi_0 + \delta \phi
\end{dmath}

\begin{dmath}
\delta \phi =  (\rho \mu)^0 \left[ \frac{1}{\rho} \phi^0_1 (x^{\mu} / \rho) + \mu \phi^0_2 (x^{\mu} \mu) \right] + (\rho \mu)^1 \left[ \frac{1}{\rho} \phi^1_1 (x^{\mu} / \rho) + \mu \phi^1_2 (x^{\mu} \mu) \right] + (\rho \mu)^2 \left[ \frac{1}{\rho}\phi^2_1 (x^{\mu} / \rho) + \mu \phi^2_2 (x^{\mu} \mu) \right]  + O((\rho \mu)^3)
\end{dmath}

The first term $\phi^0_1 (x^{\mu} / \rho)$ not depending on $\mu$ in any way, should vanish as the massless limit $\rho \mu \rightarrow 0$ should converge to $\phi_0$. Also an explicit perturbative abalysis of the equation of motion, sets $\phi^0_2 = 0$ \footnote{one should be very much cautious that in such analysis $\partial^2 = \mu^2 {\bar{\partial}}^2$, $\phi_0 (x) = \mu (\rho \mu) \bar{\phi}_0 (x \mu)$} One should bear in mind that all terms $\frac{1}{\rho} \phi^n_1 (x^{\mu}/ \rho)$ for any $n$ can be rewritten (by analyticity in $\rho \mu$ premise) as $ \frac{\mu}{\rho \mu} \phi^n_1 (x^{\mu} \mu / \rho \mu) = \frac{\mu}{\rho \mu} \psi^{n-1}_1 (x^{\mu} \mu)(1 + O(\rho \mu))$, meaning they can be converted into lower order terms in perturbation series. By explicit perturbative insertion of the expanded function in the equation of motion one can check that all the odd powers of $\rho \mu$ vanish and finally given the symmetry of the centered solution under Poincaré group and violation of only the conformal symmetry, one expects $\phi(x^{\mu} \mu) = \phi(x \mu)$ leading to(after renaming the indices):

\begin{dmath}
    \delta \phi = \mu (\rho \mu) \left[ \phi^1 (x \mu) + (\rho \mu)^2 \phi^2 (x \mu) + O((\rho \mu)^4) \right]
\end{dmath}

\

The problem with such a construct would be the divergence of the action at the leading order, since:
\begin{dmath}[label={eq:phi4-leading-correction}]
    {\delta \phi (x \mu) = 2 \sqrt{3} \mu (\rho \mu) \left( 2 \frac{1}{x \mu} K_1(x \mu) - \left( \ln \frac{( \rho \mu)^2}{4} + 2\gamma + 2  \right) \frac{1}{x \mu} I_1 (x \mu) \right)}
\end{dmath}

as the integral over $I_1 (x \mu)$ diverges when $ x \mu \rightarrow \infty $. To eliminate such a divergence, one should impose a proper constraint on the solutions. Taking a general operator $\mathcal{O}$ with $\dim [\mathcal{O}] = n$ the natural way to implement the constraint is:

\begin{dmath}
    \int d^4 x \mathcal{O} = c \rho^{4-n}
\end{dmath}

resulting in 
\begin{dmath}
    S = S_{u.c.} + \frac{1}{g}\sigma \left( \int d^4 x \mathcal{O} - c \rho^{4-n} \right)
\end{dmath} 

where u.c. stands for unconstrained and $\sigma$ is the Lagrange coefficient, which can be written perturbatively in terms of $\rho \mu$: 

\begin{dmath}
    \sigma \propto \frac{1}{\rho^{4-n}} (\rho \mu)^2 \ln (\rho \mu \sigma_1 ) + ...
\end{dmath}

(the $\rho\mu$ coefficient inside the logarithm is included since the equations are better looking this way).

To leading order in $\rho \mu$, $\delta \phi^1$ satisfies the equation $\left[ \partial^2 + \frac{1}{2} \phi^2_0 \right] \delta \phi_1 = \sigma_1 \frac{\delta \mathcal{O}}{\delta \phi}$ where $\delta \phi_1$ is the deviation of the solution from that of the unconstrained massive solution $\delta \phi$ such that $\phi = \phi_0 + \delta \phi + \delta \phi_1$ while $\sigma_1 = \frac{6}{\pi^2} \left[ \frac{\partial}{\partial \rho} \int d^4 x \mathcal{O} \right]^{-1}$.

The operator $\mathcal{O}$ can be chosen arbitrarily, however when worked out one faces that they introduce singularities in the $x \ll \rho$ while $\delta \phi \rightarrow 0$ when $\mu x \rightarrow \infty$ which means the perturbation when projected on the zero mode, vanishes but at the cost of introducing a singularity. \cite{Constrained Instantons}

Any constraint should keep the large distance behaviour at bay at the cost of modifying the short distance behaviour such that (e.g. to leading order) \\ $ 2\sqrt{3} \mu (\rho \mu) \left( \ln \frac{( \rho \mu)^2}{4} + 2\gamma + 2  \right) \frac{1}{x \mu} I_1 (\mu x) $ in \eqref{eq:phi4-leading-correction} is removed as its action is divergent for large $\mu x$. 

For $\mathcal{O}(x) = (\phi (x))^3 $ one finds:

\begin{dmath}
    \sigma \propto \frac{1}{\rho} \left(\rho \mu \right)^2 \left( \ln \frac{\left(\rho \mu \right)^2}{4} + 2 \gamma + 2 \right)
\end{dmath}

meaning $\sigma_1 = \frac{1}{2} e^{\gamma + 1}$.

\

The action can be computed perturbatively. The only seemingly disappointing point might be the infrared divergence of the mass term at the leading order $\frac{1}{2} \mu^2 \int_{|x| < L} d^4 x \phi^2_0 = 24 \pi^2 (\rho \mu)^2 \ln(L^2/ \rho^2)$ where $L >> {\mu}^{-1}$ is an artificial cut-off, but it turns out to cancel with higher order surface term $\int_{|x| = L} dS_{\mu} \phi^3 \partial_{\mu} \phi_0$.

The imaginary part of the Green function represents its spectrum; therefore, for the massless instanton:

\begin{dmath}
\Im G^{(n)} (x_1 , x_2 , ... , x_n) \sim g^{-(n+5)/2}_R e^{- \frac{16 \pi^2}{g_R}} \int^{\infty}_{0} \frac{d \rho}{{\rho}^5} (\rho M)^3 \times \int d^4 x_0 \phi_0 (\rho, x_1 - x_0) \cdots \phi_0 (\rho , x_n - x_0)
\end{dmath}
where $\Lambda^3 e^{-16 \pi^2 / g} = M^3 e^{-16 \pi^2 / g_R}$ is assumed after an $\overline{MS}$-scheme renormalisation in which $\Lambda$ is a UV cut-off, $M$ the subtraction scale/point, $g_R = g_R (M)$ and $g$ is the bare UV coupling. After a Fourier transformation, one finds:
\begin{dmath}
    \Im{G^{(n)}} (p_1 , ... , p_n) \sim g^{-(n+5)/2}_{R} e^{-16 \pi^2 / g_R} \int^{\infty}_{0} \frac{d \rho}{{\rho}^5} (\rho M)^3 \tilde{\phi}_0 (\rho , p_1) ... \tilde{\phi}_0(\rho , p_n) 
\end{dmath}

where:

$$
    {\tilde{\phi}_0 (\rho , p) = \int d^4 x e^{ipx} \phi_0 (\rho , x) \rightarrow \frac{16 \sqrt{3} \pi^2 \rho}{p^2} \ \ \ \ |p|\rho \rightarrow 0}$$
\begin{dmath}
    {\ \ \ \ \ \ \ \ \ \ \ \ \ \ \ \ \ \ \ \ \ \ \ \ \ \ \ \ \ \ \  \  \ \ \ \ \ \ \ \sim \ \ \  e^{-|p| \rho} \ \ \ \  \  \ \ |p| \rho \rightarrow \infty}
\end{dmath}

so that as can be seen the $\rho-$integral is convergent both for $\rho \rightarrow 0$ and $ \infty$ where in the former case it cuts off at scales $\sim \frac{1}{p}$. The perturbation series are in terms of $g_{\text{eff}} (\rho^{-1})$ where $g_{\text{eff}} (\rho^{-1}) \ll 1$ (after integrating the UV modes at one-loop order out, an effective coupling $g_{\text{eff}}$ appears that depends on $\rho$ that freezes(stops running, screening effect) above this length scale \textit{a.k.a.} IR region).

As for the massive theory, one should introduce the constraint in the path integral that leads to:

\begin{dmath}
\Im G^{(n)} (p_1 , ... , p_n) \sim g^{-(n+5)/2}_R \int \frac{d \rho}{\rho^{5}} (\rho M)^3 \times e^{-(1/g_R) [16 \pi^2 -24 \pi^2 (\rho \mu)^2 \ln (\rho \mu)^2]} \tilde{\phi} (\rho, p_1) \cdots \tilde{\phi} (\rho , p_n)
\end{dmath}

where $\tilde{\phi}(\rho , p)$ is the Fourier transform of the constrained instanton.

Finally, for fixed $p_i \sim O(\mu)$ and $g_R \rightarrow 0$ the $\tilde{\phi} (\rho , p_i)$ should contribute to the $\rho$-integral only in the $\rho p_i \ll 1/\mu$ regime where $\tilde{\phi} (\rho , p_i) = \frac{16 \pi^2 \sqrt{3} \rho}{p^2_i + \mu^2}$ resulting in (after amputation of the external legs propagators $\frac{1}{p^2_i + \mu^2}$):

\begin{dmath}
    \Im \Gamma^{(n)} (p_1,...,p_n) \sim e^{-16\pi^2 / g_R} g_R^{-(n+5)/2} \int^{\infty}_{0} d \rho M^3 \rho^{n-2} e^{(24 \pi^2 / g_R) (\rho \mu)^2 \ln (\rho \mu)^2}
\end{dmath}

where the integrand maximum is at $(\rho \mu)^2 \sim (n-2) g_R / (48 \pi^2 \ln g^{-1}_R) \ll 1$\\ (meaning approximately $\sqrt{\frac{g_R}{\ln g^{-1}_R}} \ll 1 \leftrightarrow g_R \ll 1$) and then a minimum at $\rho \sim \mu^{-1} e^{-1/2}$ and later it increases with no upper bound.
From the classical limit assumption, one should cut off the integral when $\rho \sim O(\mu^{-1})$ or, semi-classically speaking: $g_{\text{eff}} (\rho^{-1}) \ll 1$.

After a change of variables from $\rho$ to $x$ as $x^2 =(\rho \mu)^2 \ln g^{-1}_R / 2g_R$ one finds(after changing the upper bound from $\infty$ to $(\mu \sqrt{e})^{-1}$):

\begin{dmath}
    \Im \Gamma^{(n)} \sim e^{-16 \pi^2 / g_R} g^{-3}_R \left( \ln g^{-1}_R \right)^{(1-n)/2} M^3 \mu^{1-n} \int^{\infty}_{0} dx x^{n-2} e^{-48 \pi^2 x^2}
\end{dmath}
where the modified running coupling can be derived by imposing 
\begin{dmath}
    d (e^{-16 \pi^2 / g_R} g^{-3}_R (\ln g^{-1}_R)^{(1-n)/2} M^3 ) / dM = 0 
\end{dmath}
(bearing in mind $g_R = g_R (M)$ and that the coupling stops running at $M=\rho^{-1}$ due to screening effect!) and finally, the computational limits used here and an alternative limit can be discussed:

\begin{itemize}
    \item \textbf{weak coupling limit} \\
the final result was obtained via this limit and $\rho \sim \mu^{-1} \sqrt{\frac{g_R}{\ln g^{-1}_R}}$ so that when $g_R \rightarrow 0$ the instanton size approaches $0$ where it behaves just like a free propagator, which results in a local $n$-point vertex.

    \item \textbf{classical limit} \\ 
Alternatively a classical limit like $g_{\text{eff}}(p_i)$ held fixed while $\mu / |p_i| \rightarrow 0$, where the the massless results are recovered: $S \rightarrow 16 \pi^2 / g_R$, $\phi \rightarrow \phi_0$.

\end{itemize}

The same techniques used to find the constrained instantons in this model can be used to find the constrained instantons in the case of Higged-$SU(2)$ YM for Higgs in an arbitrary representation with isospin $q$ in \cite{Constrained Instantons}, and also the $\mathcal{N}=2$ $SU(N)$ SYM and SQCD in \cite{SYM Constrained instantons}.

Here we only mention the final results in both cases:
\begin{itemize}
    \item \textbf{Higged-$SU(2)$ YM (with the Higgs field in the fundamental representation)}

The Lagrangian is:
\begin{dmath}
    \mathcal{L} = - \frac{1}{g^2} \left[ \frac{1}{4} G^{a}_{\mu \nu} G^a_{\mu \nu} + \kappa \left( \left( D_{\mu} \phi \right)^{\dagger} D_{\mu} \phi + \frac{1}{4} \left( \phi^{\dagger} \phi - \mu^2 \right)^2 \right) \right]
\end{dmath}
with $\kappa > 0$ and:
\begin{dmath}
    G^{a}_{\mu \nu} = \partial_{\mu} A^{a}_{\nu} - \partial_{\nu} A^a_{\mu} + \epsilon^{abc} A^{b}_{\mu} A^{c}_{\nu} , \ \ \ \ D_{\mu} = \partial_{\mu} - i \frac{\tau^{a}}{2} A^{a}_{\mu}
\end{dmath}
using the ansatz: 
\begin{dmath}
    {A^a_{\mu} = -\bar{\eta}^a_{\mu \nu} \partial_{\nu} \ln \alpha  \ \ \ \ \  \phi = {\begin{pmatrix}
    0 \\
    f
\end{pmatrix}}}
\end{dmath}
resulting in the equations of motion(without the constraint):
\begin{dmath}
    \alpha \partial_{\nu} \partial^2 \alpha -3 \partial_{\nu} \alpha \partial^2 \alpha = \frac{\kappa}{2} f^2 \alpha \partial_{\nu} \alpha
\end{dmath}
\begin{dmath}
    \alpha^2 \partial^2 f - \frac{3}{4} \left(\partial_{\nu} \alpha\right)^2 f + \frac{1}{2} \alpha^2 f \left( \mu^2 - f^2 \right) = 0
\end{dmath}

A suitable choice of constraint is $S_{\text{const}} = \sigma \left( \Sigma_{\text{prov}} [A] - c \right)$ where:
\begin{dmath}    
\Sigma_{\text{prov}} [A] = -12 \int d^4 x A^{a}_{\nu} \bar{\eta}^{a}_{\nu \lambda} \frac{\rho^2 x_{\lambda}}{x^2 (x^2 + \rho^2)^2} = \frac{3}{\rho} \int d^4 x A^a_{\nu} (x) \left( \frac{\partial A^a_{0 \nu}}{\partial \rho} - \frac{2}{\rho} A^a_{0 \nu} (x) \right)
\end{dmath}
where $A^a_{0 \nu} (x) = \frac{2 \rho^2 \bar{\eta}^a_{\nu\mu} x_{\mu}}{x^2 (x^2 + \rho^2)}$.

The reader must beware that this constraint \textbf{almost} works \cite{Constrained Instantons 2}. This is a good motivation for introducing a more systematic method, or throwing out the whole idea of constrained instantons and establishing a manifestly constraint-independent formal point of view called \textbf{Valley Instanton}.

\item \textbf{$SU(N)$ SYM, SQCD }

For a qualitative discussion, one can check the following works \cite{Instantons} \cite{Instanton Calculus} and for an explicit computation \cite{SYM Constrained instantons}.

\end{itemize}

It is insightful to do the same computations via the valley method.

\section{Valley Instanton}

Given that there is no finite action solution with $ \rho \mu \neq 0$, one can think of the instanton with $\rho=0$ as the only finite action solution with old zero modes lifted, leading to quasi-zero modes. These quasizero modes form a valley whose minima are the instanton solution with $\rho=0$. The valley system of equations is as follows:

\begin{dmath}
    {\int d^{4} y \frac{\delta^2 S }{\delta \phi (x) \delta \phi (y)} F (y) = \lambda F (x) \ \ \ \ , \ \ \ F = \frac{\delta S}{\delta \phi}}
\end{dmath}

where $\frac{\delta^2 S }{\delta \phi(x) \delta \phi(y)}$ is the Hessian matrix and the first equation is the eigenvalue equation and $F$ is the auxiliary field that quantifies the deviation of the field configuration from the one that satisfies the equation of motion($0 = \frac{\delta S}{\delta \phi}$).  

As can be observed, this system of equations means that the [functional] gradient $\frac{\delta S}{\delta \phi}$ is an eigenvector of the Hessian matrix. To understand the meaning of this system, by combining the equations, one can conclude:

\begin{dmath}
    \frac{\delta}{\delta \phi(x)} \left( \frac{1}{2} \int d^4 y \left[ {\frac{\delta S}{\delta \phi (y)}}\right]^2 -\lambda S \right) = 0
\end{dmath}

which is the formulation of the problem of finding the stationary point of the gradient norm under the constraint of $S=\text{constant}$. Alternatively, it can be read as the minimization of action under the constraint of $|\frac{\delta S}{\delta \phi}|^2 = \text{constant}$.

This introduces a natural constraint that works for all values of $0\leq \rho \mu$, especially the eigenvalue due to the quasi-zero mode can be removed from the determinant just like the case with vanishing $\mu$.

Expanding the action around the valley:
\begin{dmath}
S(\phi) = S(\phi_{\alpha}) + \int d^4 x \frac{\delta S}{\delta \phi(x)} |_{\phi_{\alpha}} \left( \phi(x) - \phi_{\alpha}(x) \right) + \frac{1}{2} \int d^4 y d^4 x  {\frac{\delta^2 S}{\delta \phi(y) \delta \phi(x)}}|_{\phi_{\alpha}} ( \phi(x) - \phi_{\alpha}(x)  ) ( \phi(y) - \phi_{\alpha} (y) ) + \cdots
\end{dmath}

where $\phi_{\alpha}$ is the field configuration along the valley parameterized by a generalized quasicollective coordinate $\alpha$, bearing in mind that $\phi_\alpha$ does not satisfy the field equation $\frac{\delta S}{\delta \phi} = 0$. It's helpful to call $D(x,y) = \frac{\delta^2 S }{\delta \phi(x) \delta \phi(y)}|_{\phi= \phi_{\alpha}}$.

By a Faddeev-Popov trick:
\begin{dmath}
    \int d \alpha \ \delta \left( \int d^4 x \left( \phi(x) - \phi_{\alpha} (x) \right) R(x) \right) \Delta (\phi_{\alpha}) = 1
\end{dmath}
where $R(x) = \frac{1}{\sqrt{|\delta S / \delta \phi|^2}} \frac{\delta S}{\delta \phi(x)}$ and equivalently:

\begin{dmath}
    \Delta(\phi_{\alpha}) = \bigg | \frac{\partial \phi_{\alpha} (x)}{\partial \alpha} \left[ R(x) - \frac{\delta R(y)}{\delta \phi(x)} (\phi(y) - \phi_{\alpha} (y)) \right] \bigg |
\end{dmath}

Now, the FP integral can be inserted in the partition function:

\begin{dmath}
    Z = \mathcal{N} \int d\alpha \int \mathcal{D} \phi \delta((\phi -\phi_{\alpha})R) \left|\frac{\partial \phi_{\alpha}}{\partial \alpha} R\ \right| e^{-S (\phi)} = \mathcal{N'} \int d \alpha \left| \frac{\partial \phi_{\alpha}}{\partial \alpha} \right| \frac{1}{\sqrt{\text{det}'(D)}} e^{- S(\phi_{\alpha})}
\end{dmath}

where $\text{det}'(D) = \frac{\text{det}(D)}{\lambda}$ so that the eigenvalue of the quasizero mode is factored out. One must be careful that $\lambda$ is the eigenvalue closest to $0$ and plays the role of $\alpha$ after adjusting the units $(\lambda = \mu^2 \alpha)$. The valley instanton is the perturbative solution to the differential equations(after rescaling the parameters and the field content, assuming spherical symmetry):

\begin{dmath}
    {r=\frac{\sqrt{x^2}}{\rho} \ \ \ \ \ \ \lambda = \mu^2 \alpha \ \ \ \ \ \phi(x) = \frac{h(r)}{\rho} \ \ \ \ \ F(x)=\frac{\mu^2}{\rho} f(r)}
\end{dmath}

\begin{dmath}
    -\frac{1}{r^3} \frac{d}{dr}\left(r^3 \frac{dh}{dr}\right) + (\rho \mu)^2 h - \frac{1}{3!} h^3 =(\rho \mu)^2 f
\end{dmath}

\begin{dmath}
    -\frac{1}{r^3} \frac{d}{dr}\left(r^3 \frac{df}{dr}\right) + (\rho \mu)^2 f - \frac{1}{2!} h^2 f =(\rho \mu)^2 \alpha f
\end{dmath}

In the $\rho \mu = 0$ the equation has the following solutions:

\begin{dmath}
    {h_0 (r) = \frac{4\sqrt{3}}{1+r^2}  \ \ \ \ \ \ f_0 (r) = C \left( \frac{4\sqrt{3}}{1+r^2} -\frac{8\sqrt{3}}{(1+r^2)^2} \right)}
\end{dmath}

One can expand the general solution perturbatively as: 

\begin{dmath}
    {h(r) = h_0 (r) + (\rho \mu)^2 \hat{h}(r) \ \ \ \ \ \ \ \ f(r) = f_0 (r) + (\rho \mu)^2 \hat{f}(r)}
\end{dmath}

and the the \textit{core region} is defined to be:

\begin{dmath}
    {h_0 (r) \gg (\rho \mu)^2 \hat{h}(r) \ \ \ \ \ \ \ f_0 (r) \gg (\rho \mu)^2 \hat{f} (r)}
\end{dmath}

resulting in the equations:

\begin{dmath}[label={eq:valley-core-h}]
    -\frac{1}{r^3} \frac{d}{dr}\left(r^3 \frac{d\hat{h}}{dr}\right) - \frac{1}{2!} h^2_0 \hat{h} =f_0 -h_0
\end{dmath}

\begin{dmath}[label={eq:valley-core-f}]
    -\frac{1}{r^3} \frac{d}{dr}\left(r^3 \frac{d\hat{f}}{dr}\right) - \frac{1}{2!} h^2_0 \hat{f} = (\alpha -1)f_0 + f_0 h_0 \hat{h}
\end{dmath}

where the first equation \eqref{eq:valley-core-h} has a zero mode: $\varphi(r) = \left( \frac{4\sqrt{3}}{1+r^2} -\frac{8\sqrt{3}}{(1+r^2)^2} \right)$ that satisfies the following equation:

\begin{dmath}
    -\frac{1}{r^3} \frac{d}{dr}\left(r^3 \frac{d\varphi}{dr} \right) - \frac{1}{2!} h^2_0 \varphi =0
\end{dmath}

using this zero mode and multiplying both sides of both differential equations \eqref{eq:valley-core-h}, \eqref{eq:valley-core-f} by $r^3 \varphi$ and later integrating, one finds:

\begin{dmath}[label={eq:valley-integrated-h}]
    {-r^3 \varphi \frac{d\hat{h}}{dr} + r^3 \frac{d\varphi}{dr} \hat{h} = \int^{r}_0 dr' {r'}^3 \varphi [f_0 - h_0]}
\end{dmath}

\begin{dmath}[label={eq:valley-integrated-f}]
    {-r^3 \varphi \frac{d \hat{f}}{dr} + r^3 \frac{d\varphi}{dr} \hat{f} = \int^r_0 dr' {r'}^3 \varphi f_0 [(\alpha -1) + h_0 \hat{h}]}
\end{dmath}

Both equations \eqref{eq:valley-integrated-h}, \eqref{eq:valley-integrated-f} can be solved for $r \gg 1$, resulting in:

\begin{dmath}
    {\hat{h} = 2\sqrt{3} (1-C) \text{ln}(r) + ... \ \ \ \ \ \ \ \hat{f} = (1-\alpha) 2 \sqrt{3} C \text{ln}(r) + ...}
\end{dmath}

A different limit to consider is: $h \ll \rho \mu$ which results is the following differential equations:

\begin{dmath}
    -\frac{1}{r^3} \frac{d}{dr}\left(r^3 \frac{dh}{dr}\right) + (\rho \mu)^2 h =(\rho \mu)^2 f
\end{dmath}

\begin{dmath}
    -\frac{1}{r^3} \frac{d}{dr}\left(r^3 \frac{df}{dr}\right) + (\rho \mu)^2 f =(\rho \mu)^2 \alpha f
\end{dmath}

whose solution is:

\begin{dmath}
    h(r) = C_1 G_{\rho \mu} (r) + \frac{f(r)}{\alpha}
\end{dmath}
\begin{dmath}
    f(r) = C_2 G_{\rho \mu \sqrt{1-\alpha}} (r)
\end{dmath}

where $G_m (r) = \frac{mK_1 (mr)}{(2\pi)^2 r}$ in which $K_1$ is the modified Bessel function.

Limiting one's attention to the limits $r\ll (\rho \mu)^{-1}$ and $r \ll (\rho \mu \sqrt{1-\alpha})^{-1}$ respectively, one can expand $h$ and $f$: 

\begin{dmath}
h(r) = \frac{C_1}{(2\pi)^2} \left[ \frac{1}{r^2} + \frac{1}{2}(\rho \mu)^2 \ln (\rho \mu r c) + ... \right] + \frac{C_2}{(2\pi)^2 \alpha} \left[ \frac{1}{r^2} + \frac{1}{2}(\rho \mu)^2 (1-\alpha) \ln (\rho \mu \sqrt{1-\alpha} r c) + ...  \right]
\end{dmath}

\begin{dmath}
    f(r)= \frac{C_2}{(2\pi)^2 } \left[ \frac{1}{r^2} + \frac{1}{2}(\rho \mu)^2 (1-\alpha) \ln (\rho \mu \sqrt{1-\alpha} r c) + ...  \right]
\end{dmath}

where $c=e^{\gamma -1/2} /2$ with $\gamma$ being the Euler's constant.

By matching the two solutions, one can fix the coefficients:

\

$C_1 =0 \ \ \ \ \ C_2 = 4 \sqrt{3} (2\pi)^2 \ \ \ \ \ C=1 \ \ \ \ \text{as} \ \rho\mu \rightarrow 0$

Finally, $h$ can be summarized as \cite{Valley Instantons}:

\begin{dmath}
{h(r) = \begin{cases}
\frac{4\sqrt{3}}{1+r^2} \ \ \ \ \ \ \ \ \ \ \ \ \ \ \ \ \ \ \ \ \ \ \ \ \ \ \ r\ll (\rho \mu)^{-1/2}\\
\frac{4\sqrt{3}}{r^2} + \mathcal{O}((\rho \mu)^2) \ \ \ \ \ \  \ \ \ \ \ \ (\rho \mu)^{-1/2} \ll r \ll (\rho \mu)^{-1} \\
4\sqrt{3} (2 \pi)^2 G_{\rho \mu \sqrt{1-\alpha}} (r) \ \ \ \ \ (\rho \mu)^{-1/2} \ll r 
\end{cases}}
\end{dmath}

\section{SUSY and the Collective Coordinates}
\label{SUSY-and-the-Collective-Coordinates}
The (anti-)chiral SUSY transformations of the (anti-)self-dual bosonic instanton background result in (anti-)chiral fermionic zero modes(as was shown in section \ref{SUSic-Moduli-space,-Grassmann-Collective-Coordinates,-SUSic-Instantons}, and these (anti-)chiral zero modes, when turned on, generate (chiral)anti-chiral transformations of the bosonic instanton, which means that the (chiral)anti-chiral SUSY transformations do not leave the superinstanton invariant any longer!

\
\

This section argues that: (for self-dual bosonic instantons) SUSY transformation can be \underline{traded for} SUSY transforming the collective coordinates among themselves!

\
\

This observation can be violated, though perturbatively, in the case of the Coulomb branch $\mathcal{N}=2$ quasi-instanton, where SUSY transformations not only transform the collective coordinates(as we will see in a while) but also turn on the higher order corrections(the non-zero modes) of the fields in the semi-classical expansion. We should stipulate that such [perturbative]violation of the SUSY transformations is a direct symptom of the fact that the superinstanton is not an exact solution in case of the nonvanishing VEVs on the $\mathcal{N}=2$ Coulomb branch.

Explicitly, since the anti-chiral field vanishes to leading order for a super-instanton $\delta A_{\alpha \dot{\alpha}} = 2i g^{-1/2} \bar{\xi}_{\dot{\alpha} A} \lambda^A_{\alpha} = 2 g^{-1} \Lambda_{\alpha}(i \bar{\xi}_{\dot{\alpha} A} \mathcal{M}^A)$ to leading order in $g$ by \eqref{eq:gaugino-zero-mode-ADHM}, where the re-scaling $\bar{\xi}_A \rightarrow g^{-1/2} \bar{\xi}_A$  is taken for granted. On the other hand, by \eqref{eq:Lambda-of-C} and \eqref{eq:bosonic-zero-modes-ADHM}, one can write:

\begin{dmath}
    \delta A_{\alpha \dot{\alpha}} = (\delta_{\mu} A_{\alpha \dot{\alpha}}) \delta X^{\mu} = 2 g^{-1} \bar{U} ((\frac{\partial a_{\dot{\alpha}}}{\partial X^{\mu}} \delta X^{\mu}) f \bar{b}_{\alpha} - b_{\alpha} f (\frac{\partial \bar{a}_{\dot{\alpha}}}{\partial X^{\mu}} \delta X^{\mu}))U 
= 2 g^{-1} \bar{U} (( \delta a_{\dot{\alpha}}) f \bar{b}_{\alpha} - b_{\alpha} f ( \delta \bar{a}_{\dot{\alpha}}))U = 2g^{-1} \Lambda_{\alpha} (\delta a_{\dot{\alpha}}) 
\end{dmath}

which results in: 

\begin{dmath}
    \delta a_{\dot{\alpha}} = i \bar{\xi}_{\dot{\alpha} A} \mathcal{M}^A
\end{dmath}

And as for the fermionic fields, in the previous section \ref{SUSic-Moduli-space,-Grassmann-Collective-Coordinates,-SUSic-Instantons} one observed that the chiral SUSY transformations can be conceived as Grassmann collective transformations that transform the chiral fermion zero mode. Here one can add the observation that the chiral SUSY transformations, thanks to the presence of non-trivial scalar fields $\phi_a$ in the background modify the chiral fermionic fields at the $\mathcal{O}(g^{3/2})$ which casts itself as $\delta \lambda^{(1)A}$ rather than $\delta \lambda^{(0)A}$ which is once more a symptom of non-exactness of the super-instanton solution on the $\mathcal{N}=2$ Coulomb branch. (As can be checked since this variation vanishes in case of vanishing VEVs). One should bear in mind, in doing so, that the chiral transformation Grassmann parameters are re-scaled by $\xi^A \rightarrow g^{1/2} \xi^A$(in contrast to the anti-chiral transformations). The anti-chiral SUSY transformation however, leads to a zero mode shift that again can be shown after a long ADHM calculation (\cite{Instanton Calculus}) to be again a SUSY transformation of the Grassmann collective coordinates: $\delta \mathcal{M}^A = -4i \xi^A_{\alpha} b^{\alpha} + 2i \Sigma^{AB}_a \mathcal{C}_{a \dot{\alpha}} \bar{\xi}^{\dot{\alpha}}_{B} \ \ , \delta \bar{\mathcal{M}}^A = -4 i \xi^{\alpha A} \bar{b}_{\alpha} + 2i \Sigma^{AB}_a \bar{\xi}_{\dot{\alpha}B} \bar{\mathcal{C}}^{\dot{\alpha}}_a$ in which:

\begin{dmath}
    {\mathcal{C}_{a \dot{\alpha}} = \begin{pmatrix}
    \phi^0_a & 0 \\
    0 & \varphi_a
\end{pmatrix}a_{\dot{\alpha}} - a_{\dot{\alpha}} \varphi_a \ \ \ , \ \ \ \bar{\mathcal{C}}^{\dot{\alpha}}_a = \bar{a}^{\dot{\alpha}}\begin{pmatrix}
    \phi^0_a & 0 \\
    0 & \varphi_a
\end{pmatrix} - \varphi_a \bar{a}^{\dot{\alpha}}}
\end{dmath}

The anti-chiral SUSY transformations modify the anti-chiral fermion fields at the leading order $\mathcal{O}(g^{1/2})$(although one has already seen there exist no anti-chiral zero modes for a self-dual bosonic instanton, meaning this is rather a non-zero mode modification) which is again a result of the nonexactness of the super-instanton solution: $\delta \bar{\lambda}_A = -i g^{1/2} \bar{\Sigma}_{abA}^{B} \bar{\xi}_B [\phi_a, \phi_b]$.

One should bear in mind that the non-exactness of the SUSY transformations of the collective coordinates on the $\mathcal{N}=2$ Coulomb branch is not a flaw, \underline{since they're still a symmetry of the leading order super-instanton solution.} (perturbatively speaking)

It turns out \cite{Instanton Calculus}, that for non-vanishing VEVs the \eqref{eq:scalar-solution-zero-vev} is modified to:

$$ \phi_a = -\frac{1}{4} \bar{\Sigma}_{aAB} \bar{U} \mathcal{M}^A f \bar{\mathcal{M}}^B U + \bar{U} \begin{pmatrix}
    \phi^0_a & 0_{[N] \times [2k]} \\
    0_{[2k] \times [N]} & \varphi_a 1_{[2] \times [2]}
\end{pmatrix}U$$

with $\varphi_a$ given by:

$$ \varphi_a = \mathbf{L}^{-1} \left( \frac{1}{4} \bar{\Sigma}_{aAB} \bar{\mathcal{M}}^A \mathcal{M}^B + \bar{w}^{\dot{\alpha}} \phi^0_a w_{\dot{\alpha}} \right) $$

\section{SUSic Path Integral and Collective Coordinate Measure, Instanton Effective Action}

\label{SUSic-Path-Integral-and-Collective-Coordinate-Measure,-Instanton-Effective-Action}

Starting with the fermionic functional integral measure, it's not hard to show:

\begin{dmath}
    \int \prod^{\mathcal{N}}_{A=1} [d \lambda^A] [d \bar{\lambda}_A] = g^{kN \mathcal{N}} \int \prod^{\mathcal{N}}_{A=1} \left( \frac{\prod^{2kN}_{i=1} d \psi^{iA}}{Pf{\frac{1}{2} \Omega(X)}} [d \tilde{\lambda}^A][d \bar{\lambda}_A] \right)
\end{dmath}

where in writing this one has assumed the semi-classical expansion: 
\
\

$\lambda^A (x) = g^{-1/2} \lambda^{(0)A}(x;X,\psi) + \tilde{\lambda}^{A} (x;X, \psi)$.

After the semi-classical expansion of the fields, the action becomes:
\
\

\begin{dmath}
    S[g^{-1} A^{(0)}_m + \tilde{A}_m , g^{-1/2} \lambda^{(0)A} + \tilde{\lambda}^A , \bar{\lambda}_A , \phi_a] = -2 \pi i k \tau + S_{kin} + S_{int} 
\end{dmath}
\
\

in which:

\begin{dmath}
    S_{kin} = \int d^4 x \ \text{tr}_N \left\{ \frac{1}{2} \tilde{\bar{A}}^{\dot{\alpha} \alpha} \Delta^{(+) \beta}_{\alpha} \tilde{A}_{\beta \dot{\alpha}} -2 \mathcal{D}_n \bar{\lambda}_A \bar{\sigma}_n \tilde{\lambda}^A + \mathcal{D}_n \phi_a \mathcal{D}_n \phi_a \right\}
\end{dmath}

and:
\begin{dmath}
S_{int} = \int d^4 x \tr_N \left\{-\lambda^{(0)A} \bar{\Sigma}_{aAB} [\phi_a , \lambda^{(0)B}] - 2g^{1/2} [\tilde{A}_n , \bar{\lambda}_A]\bar{\sigma}_n \lambda^{(0)A} - 2 g^{1/2} \lambda^{(0)A} \bar{\Sigma}_{aAB} [\phi_a , \tilde{\lambda}^B] + ... \right\}
\end{dmath}
Except for the very first term, the other terms are all of higher order in $g^2$ and we drop such terms at the leading order of the effective action. To maintain unitarity, the ghost field appears, and finally, one can write:

\begin{dmath}
    e^{-S_{eff}} = e^{2 \pi i k \tau} \int [d\tilde{A}][db][dc][d \tilde{\lambda}][d \bar{\lambda}] [d \phi] e^{-S_{kin} - S_{int}- S_{gh}}
\end{dmath}

One can immediately check that the effective action vanishes as VEVs disappear in $\mathcal{N}=2$ SYM.

After a shift of the scalar fields in the functional integral(to get rid of the Grassmann collective coordinate terms that result in the mismatch between $\phi$ and $\phi^{\dagger}$ after complex conjugation\footnote{Remember that $\phi=\phi_1+i\phi_2$}) the inetgration of quantum fluctuations around the super-instanton solution for the gauge field, fermioins and scalar fields (for $\mathcal{N}=2$) are respectively:

\begin{dmath}
    {|\frac{\det(-\mathcal{D}^2)}{{\det}' \Delta^{(+)}}|, \ \ \ \  |{\det}' \begin{pmatrix}
    0 & \slashed{\mathcal{D}} \\
    \rawslashed{\mathcal{{\bar{D}}}} & 0 
\end{pmatrix}^2|^{1/2} = |{\det}' \Delta^{(+)} . \det\Delta^{(-)}|^{1/2}, \ \ \ \ |\det (- \mathcal{D}^2)|^{-1}}
\end{dmath}

resulting in $|\frac{{\det}' {\Delta}^{(+)}}{\det \Delta^{(-)}}|^{-\frac{1}{2}}$. This ratio can be computed thanks to instanton determinant techniques introduced first by 't Hooft. To do so, one introduces the Pauli-Villars regulators, large masses $\mu_i$ and the alternating metric $e_i$ such that:

\begin{dmath}
    \sum^{\nu}_{i=1} e_i = -1 , \ \ \ \ \sum^{\nu}_{i=1} e_i \mu^{p}_i = 0 \ \ (p=1,...,\nu -1)  
\end{dmath}
along with:
\begin{dmath}
    \log \mu = - \sum^{\nu}_{i=1} e_i \log \mu_i
\end{dmath}
Given this regularization, the determinants are:
\begin{dmath}
    \log {\det}' \Delta^{(+)} = \text{tr} \left\{ \log(\Delta^{(+)} + \mathcal{P}_0) + \sum^{\nu}_{i=1} e_i \log (\Delta^{(+)} + \mu^2_i) \right\}
\end{dmath}
\begin{dmath}
    \log \det \Delta^{(-)} =\text{tr}  \left\{ \log \Delta^{(-)} + \sum^{\nu}_{i=1} e_i \log (\Delta^{(-)} + \mu^2_i) \right\}
\end{dmath}

in which $\mathcal{P}_0$ is the projector onto the zero mode subspace of $\Delta^{(+)}$. Finally, the ratio of the determinants can be written in terms of the regulators as:

\begin{dmath}
    {\frac{{\det}' {\Delta}^{(+)}}{\det \Delta^{(-)}} = \exp (2k N \sum^{\nu}_{i=1} e_i \text{log} \mu^2_i) = \mu^{-4kN}}
\end{dmath}
which means $|\frac{{\det}' {\Delta}^{(+)}}{\det \Delta^{(-)}}|^{\frac{1}{2} -1} = \mu^{2kN}$. Consequently, the effective action can be simplified as $S_{eff} = -2 \pi i k \tau + \{\tilde{S} - 2kN \log \mu \}g^0 + \mathcal{O}(g^2)  $.
\\
Finally, the functional integral in the weak-coupling limit $ g\rightarrow 0$ in the sector with instanton charge $k$ can be reduced to the following:

\begin{dmath}
    \int [dA][d \lambda][d \phi] [db][dc] e^{-S} |_{charge-k} \xrightarrow{g \rightarrow 0} \left(\frac{\mu}{g}\right)^{2kN} e^{2 \pi i k \tau} \mathcal{Z}^{(2)}_k
\end{dmath}
where $\mathcal{Z}^{(\mathcal{N})}_k = \mathcal{Z}^{(2)}_k$ is called the \textbf{instanton partition function} since its format is the same as that of a zero-dimensional field theory which can be shown for $\mathcal{N}=2$ to be equal to the \textbf{Dimensional Reduction} of a two-dimensional $\sigma$-model partition function whose target space is the moduli space(as a specific case of the general relation between instanton calculus and \textbf{$D$-branes} in string theory):

\begin{dmath}
    \mathcal{Z}^{(2)}_k = \int_{\mu^{-1}(0)/U(k)} \boldsymbol{\omega}^{(2)} e^{-\tilde{S}}
\end{dmath}
in which $\tilde{S}=\tilde{S}(X, \psi)$ and $\boldsymbol{\omega}^{(2)}$  is the SUSic volume form on the moduli space $\mu^{-1} (0) / U(k)$ and can be written explicitly:

\begin{dmath}
    {\boldsymbol{\omega}^{(2)} = \frac{\sqrt{\det g(X)}}{[Pf \frac{1}{2} \Omega(X)]}^{2} \prod^{4kN}_{\mu=1} \frac{d X^{\mu}}{\sqrt{2 \pi}} \prod^{2}_{A=1} \prod^{2kN}_{i=1} d \psi^{iA}}
\end{dmath}

The factor $g$ in the prefactor of the instanton partition function runs with $\mu$ so that the combination remains invariant under the renormalization group, generating a scale in theory:

\begin{dmath}
    \Lambda^{2N}_{\mathcal{N}=2} = \mu^{2N} e^{-8 \pi^2 / {g(\mu)^2} + i \theta}
\end{dmath}

As for the effective action $\tilde{S}$ one can show that in:

\begin{dmath}
    \tilde{S} = \int d^4 x \{ \partial_n \tr_N (\phi_a \mathcal{D}_n \phi_a) -\frac{1}{2}g \tr_N \lambda^A \bar{\Sigma}_{aAB} [\phi_a , \lambda^B] \}
\end{dmath}

the first term is equal to $4 \pi^2 \tr_k [\frac{1}{4} \bar{\Sigma}_{aAB} \bar{\mu}^A \phi^0_a \mu^B + \bar{w}^{\dot{\alpha}} \phi^0_a \phi^0_a w_{\dot{\alpha}} - \bar{w}^{\dot{\alpha}} \phi^0_a w_{\dot{\alpha}} \varphi_a ]$. As for the second term, one should use the identity $\bar{\Sigma}_{aAB} [\phi_a , \Lambda (\mathcal{M}^B)] = \slashed{\mathcal{D}} \bar{\psi}_A + \Lambda(\mathcal{N}_A)$ which implies:
\begin{dmath}[label={eq:Yukawa-term-split}]
-\frac{1}{2} \int d^4 x \tr_N \Lambda(\mathcal{M}^A) (\slashed{\mathcal{D}} \bar{\psi}_A + \Lambda(\mathcal{N}_A)) \\ = -\frac{1}{2} \int d^4 x ( \partial_n \tr_N \Lambda(\mathcal{M}^A) \sigma_n \bar{\psi}_A + \tr_N \Lambda(\mathcal{M}^A) \Lambda(\mathcal{N_A}) ) \end{dmath}

where the first term does not contribute to the surface at infinity and can be dropped, and for the second term, the inner-product relation \eqref{eq:zero-mode-inner-product} shows:

\begin{dmath}[label={eq:Yukawa-term-result}]
    -\frac{1}{2} \int \tr_N \Lambda(\mathcal{M}^A) \Lambda(\mathcal{N}_A) = \frac{\pi^2}{4} \tr_k [{\bar{\mathcal{M}}}^A (\mathcal{P}_{\infty} + 1) \mathcal{N}_A + \bar{\mathcal{N}}_A (\mathcal{P}_{\infty} + 1) \mathcal{M}^A] = {\pi^2} \bar{\Sigma}_{a A B} \tr_k [\bar{\mu}^A \phi^0_a \mu^B - {\bar{\mathcal{M}}}^A \mathcal{M}^B \varphi_a]
\end{dmath}

where in \eqref{eq:Yukawa-term-split} and \eqref{eq:Yukawa-term-result} $\mathcal{N}_A$ stands for 
\begin{dmath}
-\bar{\Sigma}_{aAB} \left\{  \begin{pmatrix}
    \phi^0_a & 0 \\
     0 & \varphi_a
\end{pmatrix} \mathcal{M}^B - \mathcal{M}^B \varphi_a \right\} + 2 \begin{pmatrix}
    0 & 0 \\
    0 & \mathcal{G}^{\dot{\alpha}}_A
\end{pmatrix} a_{\dot{\alpha}} - 2 a_{\dot{\alpha}} \mathcal{G}^{\dot{\alpha}}_{A}
\end{dmath}

in which $\mathcal{G}^{\dot{\alpha}}_A$ ensures that $\mathcal{N}_A$ fulfills the ADHM constraints while not contributing to the final integral at all. Summing up the previous lines, one can write:

\begin{dmath}
\tilde{S} = 4 \pi^2 \tr_k \left\{ \frac{1}{2} \bar{\Sigma}_{aAB} \bar{\mu}^A \phi^{0}_a \mu^B + \bar{w}^{\dot{\alpha}} \phi^0_a \phi^0_a w_{\dot{\alpha}} - \varphi_a \mathbf{L} \varphi_a \right\} \end{dmath}

using the expression for $\varphi_a$ one can write:
\begin{dmath}
\tilde{S} = 4 \pi^2 \tr_k \left\{ \frac{1}{2} \bar{\Sigma}_{aAB} \bar{\mu}^A \phi^0_a \mu^B  + \bar{w}^{\dot{\alpha}} \phi^0_a \phi^0_a w_{\dot{\alpha}} - \left( \frac{1}{4} \bar{\Sigma}_{aAB} \bar{\mathcal{M}}^A \mathcal{M}^B + \bar{w}^{\dot{\alpha}} \phi^0_a w_{\dot{\alpha}} \right) \mathbf{L}^{-1} \left( \frac{1}{4} \bar{\Sigma}_{aCD} \bar{\mathcal{M}}^C \mathcal{M}^D +\bar{w}^{\dot{\beta}} \phi^0_a w_{\dot{\beta}} \right) \right\}
\end{dmath}

The most staggering point of this formalism reveals itself when one rewrites the effective action in terms of geometry:
\begin{dmath}
\tilde{S}=  - \frac{\pi^2}{2} \epsilon_{ABCD} \tr_k \left( \bar{\mathcal{M}}^A \mathcal{M}^B \mathbf{L}^{-1} \bar{\mathcal{M}}^{C} \mathcal{M}^D \right) = \boxed{\frac{1}{96} \epsilon_{ABCD} R(\mathcal{M}^A , \mathcal{M}^B , \mathcal{M}^C , \mathcal{M}^D)} =\frac{1}{96} \epsilon_{ABCD} R_{ijkl} \psi^{iA} \psi^{jB} \psi^{kC} \psi^{lD} 
\end{dmath}
where $R$ is the Grassmann-valued symplectic curvature tensor over the SUSic hyperKähler quotient (\textit{a.k.a.} SUSic moduli space).
One can already see that there exists a general correspondence: 
\
\
\
\

\ \ \ \ \ \ \ \ \ \ \fbox{{\textit{\textbf{Geometry of the Moduli Space}}} $\rightarrow$ \textbf{\textit{the [effective-]Action}}}

\
\

For the rest of the terms in $\tilde{S}$, the geometric picture remains. To write them geometrically, one introduces the action generators of the maximal torus $U(1)^{N-1} \subset SU(N)$ on the mother space $\mathbb{R}^{4k(N+k)}$:

\begin{dmath}
    \tilde{V}_a = i \bar{w}^{\dot{\alpha}}_{iu} (\phi^0_a)_u \frac{\partial}{\partial \bar{w}^{\dot{\alpha}}_{iu}} - i (\phi^0_a)_u w_{ui \dot{\alpha}} \frac{\partial}{\partial w_{ui \dot{\alpha}}}
\end{dmath}
The $SU(N)$ group action is an isometry group action on the hyperKähler quotient space which can be shown by checking $\tilde{V}_a \mu_{X_r} = 0 $; the group action is also tri-holomorphic $\mathcal{L}_{\tilde{V}_a} \tilde{I}^{(c)} = 0$. The Killing vectors on the moduli space $V_a$ can be horizontally lifted to $\mu^{-1}_{X_r} (0)$ and turns out to be the same as the projection of $\tilde{V}_a$ on the horizontal subspace of the tangent space $\mathcal{H} \subset T \mu^{-1}_{X_r} (0)$ of the constrained surface $\mu^{-1}_{X_r} (0)$. Using the Killing vector field $V_a$ one can argue $4\pi^2 \tr_{k} \left[ \bar{w}^{\dot{\alpha}} \phi^0_a \phi^0_a w_{\dot{\alpha}} - \bar{w}^{\dot{\alpha}} \phi^0_a w_{\dot{\alpha}} \mathbf{L}^{-1} \bar{w}^{\dot{\beta}} \phi^0_a w_{\dot{\beta}} \right] = \frac{1}{2} g_{\mu \nu} (X) V^{\mu}_a V^{\nu}_a$. One important point is that the group action should not be mistaken for the former $U(k)$ group action generated by the vector fields $X_r$, since $X_r$ has no horizontal component while $\tilde{V}_a$ although tangent to the constrained set $\mu^{-1} (0)$ is not necessarily purely vertical. More explicitly the vertical component of $\tilde{V}_a$ is $\tilde{V}^v_a =  \sum_{rs} X_r \mathbf{L}^{-1}_{rs} \tilde{g} (X_s , \tilde{V}_a) = 8\pi^2 \sum_{rs} X_r \mathbf{L}^{-1}_{rs} \tr_k (T^s \bar{w}^{\dot{\alpha}} \phi^0_a w_{\dot{\alpha}}) $.

\
\

As for the maximal torus $U(1)^{N-1}$ action on the Grassmann collective coordinates one has:
\begin{dmath}
    {\delta_a \mu^{A}_{ui} = - i (\phi^0_a)_u \mu^{A}_{ui} , \ \ \ \ \delta_a \bar{\mu}^{A}_{iu} = i \bar{\mu}^{A}_{iu} (\phi^0_a)_u , \ \ \ \ \ \delta_a \mathcal{M}'_{ij \alpha} = 0}
\end{dmath}
or more concisely:

\begin{dmath}
    \delta_a \mathcal{M}^{\tilde{i} A} = (\tilde{\nabla}_{\tilde{j} \dot{\alpha}} \tilde{V}^{\tilde{i} \dot{\alpha}}_{a}) \mathcal{M}^{\tilde{j} A}
\end{dmath}
where $\tilde{\nabla}_{\tilde{j} \dot{\alpha}}$ is the flat connection on the mother space $\mathbb{R}^{4k(N+k)}$ that determines the connection ${\nabla}_{\tilde{j} \dot{\alpha}}$ on the moduli space.

With the aid of this connection, one can write the whole effective action as:
\begin{dmath}
\tilde{S} = \frac{1}{2} \left\{  g_{\mu \nu}(X) V^{\mu}_a V^{\nu}_a + \frac{i}{2} \bar{\Sigma}_{aAB} \Omega_{ij} (X) \psi^{iA} ( \nabla_{k \dot{\alpha}} V^{j\dot{\alpha}}_{a} ) \psi^{kB} + \frac{1}{48} \epsilon_{ABCD} R_{ijkl} \psi^{iA} \psi^{jB} \psi^{kC} \psi^{lD}     \right\}
\end{dmath}

which is purely geometric! Such effective action can be further shown to descend from the compactification of an $\mathcal{N}=2$ $2d$  $\sigma$-model with the moduli space $\mu^{-1} (0) / U(k)$ as its target space.

As the penultimate section of this chapter, there remains only the derivation and symmetry analysis of the SUSic collective coordinate measure. The measure should be studied in two different regimes of $N < 2k$ and $N \geq 2k$.
\begin{itemize}
    \item $N < 2k$

For this case, one starts with the measure over the mother space for the Grassmann collective coordinates $\mathcal{M}_k$:

\begin{dmath}\int \frac{\prod^{2n}_{i=1} d \mathcal{M}^{\tilde{i}}}{Pf \tilde{\Omega}}
\end{dmath}

where $n=k(N+k)$. By inserting a $\delta$-function that limits the functional to the moduli space, one gets:

\begin{dmath}\int \frac{\prod^{2n}_{i=1} d \mathcal{M}^{\tilde{i}}}{Pf\tilde{\Omega}} \frac{1}{J_f} \prod^{\dim G}_{r=1} \prod^2_{\dot{\alpha}=1} \delta (\mathcal{M}^{\tilde{i}} \tilde{\Omega}_{\tilde{i} \tilde{j}} X^{\tilde{j}\dot{\alpha}}_r)
\end{dmath}
where $J_f = |\det_{k^2} \mathbf{L}|$. Having $\mathcal{N}=2$, the result, including the bosonic collective coordinate measure, is:

\begin{dmath}[label={eq:susy-volume-form-N-less-2k}]
\int_{\mu^{-1} (0) / U(k)} \boldsymbol{\omega}^{(2)} = \frac{C^{(2)}_k}{\text{VolU} (k)} \int d^{4k(N+k)} a \prod^{2}_{A=1} d^{2k(N+k)} \mathcal{M}^{A} | {\det}_{k^2} \mathbf{L} |^{-1} \times \\ \prod^{k^2}_{r=1} \left\{ \prod^{3}_{c=1} \delta (\frac{1}{2} \tr_k T^r ({\tau^{c \dot{\alpha}}}_{\dot{\beta}} \bar{a}^{\dot{\beta}} a_{\dot{\alpha}}))  \prod^{2}_{A=1}   \prod^{2}_{\dot{\alpha}} \delta (\tr_k T^r (\bar{\mathcal{M}}^A a_{\dot{\alpha}} + \bar{a}_{\dot{\alpha}} \mathcal{M}^A)) \right\}
\end{dmath}
in which: $C^{(2)}_{k} = 2^{-k(k-1)/2} \pi^{-2kN}$ that can be derived in the specific clustering limit while the Grassmann measure can be further broken down to:

\begin{dmath}
    \int d^{2k(N+k)} \mathcal{M}^A  = \int \prod^{k^2}_{r=1} d^2 (\mathcal{M}^{'Ar}) \prod^{k}_{i=1} \prod^{N}_{u=1} d \bar{\mu}^A_{iu} d \mu^A_{ui}
\end{dmath}

Finally one can check that \underline{$\boldsymbol{\omega}^{(2)}$ is SUSic}.
For that matter, one starts with the SUSic variation of the ADHM variables:
\begin{dmath}
\delta \left( \Vec{\tau}^{\dot{\beta}}_{\dot{\alpha}} \bar{a}^{\dot{\alpha}} a_{\dot{\beta}} \right) = \Vec{\tau}^{\dot{\beta}}_{\dot{\alpha}} \left( -i \bar{\xi}^{\dot{\alpha}}_{A} \bar{\mathcal{M}}^{A} a_{\dot{\beta}} + i \bar{\xi}_{\dot{\beta} A} \bar{a}^{\dot{\alpha}} \mathcal{M}^A \right) = -i \Vec{\tau}^{\dot{\beta}}_{\dot{\alpha}} \bar{\xi}^{\dot{\alpha}}_{A} \left( \bar{\mathcal{M}}^A a_{\dot{\beta}} + \bar{a}_{\dot{\beta}} \mathcal{M}^A \right) =0
\end{dmath}
\begin{dmath}
\delta \left( \bar{\mathcal{M}}^A a_{\dot{\alpha}} + \bar{a}_{\dot{\alpha}} \mathcal{M}^A \right) = 4i \xi^A_{\alpha} \left( b^{\alpha} a_{\dot{\alpha}} - \bar{a}_{\dot{\alpha}} b^{\alpha} \right) + i \bar{\xi}_{\dot{\alpha} B} \left( \bar{\mathcal{M}}^A \mathcal{M}^B - \bar{\mathcal{M}}^B \mathcal{M}^A \right) +2i \bar{\xi}_{\dot{\beta} B} \Sigma^{AB}_{a} \left( \bar{\mathcal{C}}^{\dot{\beta}}_a a_{\dot{\alpha}} - \bar{a}_{\dot{\alpha}} \mathcal{C}^{\dot{\beta}}_a \right)
= i \bar{\xi}_{\dot{\alpha} B} \left( \bar{\mathcal{M}}^A \mathcal{M}^B -\bar{\mathcal{M}}^B \mathcal{M}^A -2\Sigma^{AB}_a \bar{w}^{\dot{\beta}} \phi^0_a w_{\dot{\beta}} + 2 \Sigma^{AB}_a \mathbf{L} \varphi_a \right) = 0
\end{dmath}

we used the SUSY transformations of the collective coordinates in Section \ref{SUSY-and-the-Collective-Coordinates}. This suffices to prove the SUSic invariance of $\boldsymbol{\omega}^{(1)}$ since the super-Jacobian of the $\{ a_{\dot{\alpha}} , \mathcal{M} \}$ at the linear order in $\xi$. But this reasoning does not extend to $\boldsymbol{\omega}^{(2)}$ since, as can be checked in Section \ref{SUSY-and-the-Collective-Coordinates} the variation of Grassmann collective coordinates depends on the Grassmann coordinates themselves though through the dependence of $\mathcal{C}^{\dot{\alpha}}_{a} , \bar{\mathcal{C}}^{\dot{\alpha}}_{a}$ on $\varphi$ which itself depends on $\mathcal{M}^A$. Though this nontrivial super-Jacobian contribution cancels out with the variation of $|\det \mathbf{L}|^{-1}$. For that matter one better remove the dependence of $\delta\mathcal{M}^A$ on the Grassmann collective coordinates by assuming $\varphi$ as an independent auxiliary collective coordinate, meaning the the collective coordinates are $\{ a_{\dot{\alpha}} , \mathcal{M}^A , \varphi \}$. To find the transformation of $\varphi$ one starts with:
\begin{dmath}
    \mathbf{L} . \varphi_a = \frac{1}{4} \bar{\Sigma}_{aAB} \bar{\mathcal{M}}^A \mathcal{M}^B + \bar{w}^{\dot{\alpha}} \phi^0_a w_{\dot{\alpha}}
\end{dmath}
using Leibniz's rule, one finds:

\begin{dmath}
    \mathbf{L} . \delta \left( \Sigma^{AB}_a \varphi_a \right) = i \bar{\xi}^{\dot{\alpha}}_{C} \left( - \Sigma^{BC}_a \mathcal{F}^A_{a \dot{\alpha}} + \Sigma^{AC}_a \mathcal{F}^B_{a \dot{\alpha}} - \Sigma^{AB}_a \mathcal{F}^C_{a \dot{\alpha}} \right)
\end{dmath}

in which $\mathcal{F}^B_{b \dot{\alpha}} = \left( \bar{\mathcal{C}}_{b \dot{\alpha}} \mathcal{M}^B - \bar{\mathcal{M}}^B \mathcal{C}_{b \dot{\alpha}} \right)$. One should bear in mind that $\delta \mathcal{M}^A$ depends on $\Sigma^{AB}_a \varphi_a$ rather than $\varphi$. In our case of $\mathcal{N}=2$, there's only one such non-zero contribution which is $\varphi \equiv \Sigma^{12}_a \varphi_a = i \varphi_1 + \varphi_2$ while $\mathbf{L}. \delta (\varphi) = 0$ as can be checked resulting in $\delta \varphi = 0$.

Finally, the ADHM constraint can be rewritten as: \\ $\mathbf{L}. \varphi = - \frac{1}{2} \left( \bar{\mathcal{M}}^1 \mathcal{M}^2 - \bar{\mathcal{M}}^2 \mathcal{M}^1 \right) + \bar{w}^{\dot{\alpha}} \phi^0 w_{\dot{\alpha}}$ where $\phi^0 \equiv \Sigma^{12}_a \phi^0_a $, permitting the following result:

\begin{dmath}[label={eq:detL-as-integral}]
    |{\det}_{k^2} \mathbf{L}|^{-1} = \int d^{k^2} \varphi \prod^{k^2}_{r=1} \delta \left( \text{tr}_k T^r \left( 
\mathbf{L}. \varphi + \frac{1}{2} \bar{\mathcal{M}}^1 \mathcal{M}^2 -\frac{1}{2} \bar{\mathcal{M}}^2 \mathcal{M}^1 - \bar{w}^{\dot{\alpha}} \phi^0 w_{\dot{\alpha}} \right) \right)
\end{dmath}
As can be directly observed, after substituting the identity \eqref{eq:detL-as-integral} in the volume form \eqref{eq:susy-volume-form-N-less-2k}, since the variation of all the $\delta$-functions vanish to linear order and since the super-Jcobian for $\{a_{\dot{\alpha}}, \mathcal{M}^A, \varphi \}$ variations vanishes or is off-diagonal to linear order in $\xi^A$, $\boldsymbol{\omega}^{(2)}$ is SUSic.

\item $N \geq 2k$

In this case, changing the variables is quite helpful in simplifying the ADHM construction.
\begin{dmath}
    (W^{\dot{\alpha}}_{\dot{\beta}})_{ij} = \bar{w}^{\dot{\alpha}}_{iu} w_{uj \dot{\beta}}
\end{dmath}
which has $4k^2$ variables. Alternatively one can choose: $W^0 = \text{tr}_2 W$, $W^c = \tr_{2} (\tau^{c} W)$ that constitutes four $k \times k$ matrices. The ADHM constraint can be written as:
\begin{dmath}
    W^c = - a'_m a'_n \text{tr}_2 (\tau^c \bar{\sigma}_m \sigma_n)
\end{dmath}
Given that the matrix $w_{u i \dot{\alpha}}$ can be thought of as an $N \times 2k$ matrix $w_{ua}$ where the indices $i \dot{\alpha}$ are composite and are collectively indexed by $a \in \{1,...,2k\}$. 

Generally, one can decompose this matrix into:

\begin{dmath}
    {w= \mathcal{U} \ . \begin{pmatrix}
    \xi_{11} & \xi_{12} & \cdots & \xi_{1 , 2k} \\
     0 & \xi_{22} & \cdots & \xi_{2, 2k} \\
     \vdots & \vdots & \ddots & \vdots \\
     0 & 0 & 0 & \xi_{2k , 2k} \\
     \vdots & \vdots & \vdots & \vdots \\
      0 & 0 & 0 & 0
\end{pmatrix} = \mathcal{U}\Xi}
\end{dmath}
where $\mathcal{U} \in \frac{SU(N)}{SU(N-2k)}$ and $\Xi$ is an $N \times 2k$ upper-triangular complex matrix whose diagonal elements $\xi_{aa}$ are real. One can cut and extend the aforementioned $\Xi$ to a $2k \times 2k$ matrix such that the $0$ components can be removed while instead $w^{\dagger} = \Xi^{\dagger} \mathcal{U}^{\dagger}$ where $w^{\dagger}_{uj \dot{\beta}} =\bar{w}_{uj \dot{\beta}} $.

Consequently, $W= \Xi^{\dagger} \Xi$, keeping in mind the real degrees of freedom in $W$ is the same as that of $\Xi$.
The most important step in rewriting the collective coordinate measure would be finding the Jacobian of the change of variables from the old to the new ones. To find the relevant Jacobian one better use a rather more indirect approach to change of variables which schematically is $\{w_{ui \dot{\alpha}}\} \rightarrow \{\xi_{ab}, u^a\} \rightarrow \{W^0_{ij}, W^c_{ij}\}$ where $\{u^a\}$ is defined as the $2k$ $N$-vectors $u^a$ which is the $a$-th column of the generic matrix $\mathcal{U} \in \frac{SU(N)}{SU(N-2k)}$ such that $(u^{\dagger})^a. \ u^b = \delta^{ab}$.
Then since:

\begin{dmath}
    \frac{SU(N)}{SU(N-2k)} = S^{2N -1} \times S^{2N-3} \times \cdots \times S^{2N-4k +1}
\end{dmath}

homeomorphically, the unit vectors $u^a$ parameterise respectively the unit complex $N$-sphere$(S^{2N-1})$, $N-1$-sphere$(S^{2N-3})$, ..., $N-2k$-sphere$(S^{2N-4k+1})$.

By induction over $k$ one can show:
\begin{dmath}[label={eq:measure-W-Xi}]
    \int d^{4k^2} W = 2^{2k} \int d^{4k^2} \Xi \prod^{2k}_{a=1} \xi^{4k-2a +1}_{aa}
\end{dmath}

Bearing in mind that $w_{ua} = \sum^{a}_{b=1} \xi_{ba} u^b_u$ one can show inductively:

\begin{dmath}[label={eq:measure-w-xi-angles}]
    \int \prod^{N}_{u=1} d w_{ua} d w^{*}_{ua} = 2^{N-a-1} \int \left\{ \prod^{a-1}_{b=1} d \xi_{ba} d \xi^{*}_{ba} \xi^{2N-2a+1}_{aa} d \xi_{aa} d^{2N-2a+1} \hat{\Omega}_a \right\}
\end{dmath}

where $\hat{\Omega}_a$ is the solid angle of the $S^{2N-2a+1}$ sphere.

Putting the former two relations \eqref{eq:measure-W-Xi}, \eqref{eq:measure-w-xi-angles} together and doing the same calculation for $w^{\dagger}$, one concludes:

\begin{dmath}
    {\int d^{2kN} w \ d^{2kN} \bar{w} = 
    A_k \int | {\det}_{2k} W |^{N-2k} d^{k^2} W^0 \prod_{c=1,2,3} d^{k^2} W^c d^{4k(N-k)} \mathcal{U}}
\end{dmath}

where $A_k$ is the normalisation constant that can be derived by defining the unit normalised measure:
\begin{dmath}
    \int d^{4k(N-k)} \mathcal{U} = \frac{1}{\prod^{2k}_{a=1} \text{Vol} S^{2(N-a)+1}} \int \prod^{2k}_{a=1} d^{2N -2a +1} \hat{\Omega}_a
\end{dmath}
so that $A_k = 2^{2kN-4k^2 -k} \prod^{2k}_{a=1} \text{Vol} S^{2(N-a)+1} = \frac{2^{2kN-4k^2 + k} \pi^{2kN-2k^2 +k}}{\prod^{2k}_{a=1} (N-a)!}$

Knowing that the volume form on the moduli space is:

\begin{dmath}
    \int_{{\mu^{-1}(0)}/{U(k)}} \boldsymbol{\omega} = \frac{C_k}{\text{VolU}(k)} \int d^{4k(N+k)} a |{\det}_{k^2} \mathbf{L}| \prod^{k^2}_{r=1} \prod^{3}_{c=1} \delta \left(\frac{1}{2} \tr_k T^r \left({\tau^{c\dot{\alpha}}}_{\dot{\beta}} \bar{a}^{\dot{\beta}} a_{\dot{\alpha}}\right)\right)
\end{dmath}

one can ultimately write down the bosonic collective coordinate:

\begin{dmath}
    \int_{\mu^{-1} (0) / U(k)} \boldsymbol{\omega} = \frac{2^{3k^2} A_k C_k}{\text{VolU}(k)} \int d^{4k^2} a' d^{k^2} W^0 d^{4k(N-k)} \mathcal{U} |{\det}_{k^2} \mathbf{L}| |{\det}_{2k} W|^{N-2k}
\end{dmath}

There remains only the Grassmann collective coordinates contribution to the measure. The trick is to keep in mind that $W= \bar{w} w$ is invariant under global gauge rotations $\mathcal{U} \rightarrow \mathcal{V U}$ where as before $w= \mathcal{U}\Xi$. These transformations are a subset of all infinitesimal transformations that leave $W$ invariant:

\begin{dmath}[label={eq:W-invariance}]
    \bar{w}^{\dot{\alpha}}_{iu} \delta w_{uj \dot{\beta}} + \delta \bar{w}^{\dot{\alpha}}_{iu} w_{uj \dot{\beta}} =0
\end{dmath}
Assuming the chiral/anti-chiral SUSY transformations of the SUSic collective coordinates (Section \ref{SUSY-and-the-Collective-Coordinates}) as a specific case $\delta w_{ui \dot{\alpha}} = i \bar{\xi}_{\dot{\alpha}A}\mu^{A}_{ui}, \ \delta \bar{w}^{\dot{\alpha}}_{iu} = -i\bar{\mu}^{A}_{iu} \bar{\xi}^{\dot{\alpha}}_{A}$, by substitution in \eqref{eq:W-invariance} results in $\bar{w}^{\dot{\alpha}}_{iu} \mu^{A}_{uj} = 0, \ \bar{\mu}^A_{iu} w_{uj \dot{\alpha}} = 0$ which leads us to the conclusion that generally:

\begin{dmath}
    {\mu^{A}_{ui} = w_{uj \dot{\alpha}} (\zeta^{\dot{\alpha}A})_{ji} + \nu^{A}_{ui} \ \ \ \ \bar{\mu}^{A}_{iu} = (\bar{\zeta}^A_{\dot{\alpha}})_{ij} \bar{w}^{\dot{\alpha}}_{ju} + \bar{\nu}^A_{iu}}
\end{dmath}
where $\nu^A, \bar{\nu}^A$ are orthogonal to $ \bar{w}, w$ respectively, and for the same reason, they don't appear in the ADHM construction. This means clearly that the Grassmann superpartners to the $w \rightarrow \mathcal{U}w$ symmetry transformation of $W$ are those of $\{\nu^A , \bar{\nu}^A\}$. 
The Jacobian of the transformation $\{ \mu^A , \bar{\mu}^A \} \rightarrow \{\zeta^A , \bar{\zeta}^A , \nu^A , \bar{\nu}^A \}$ would be:

\begin{dmath}
    \frac{\partial \left(\{\mu^A , \bar{\mu}^A\}\right)}{\partial \left(\{\zeta^A , \bar{\zeta}^A , \nu^A , \bar{\nu}^A\} \right)} = |{\det}_{2k} W|^{-k}
\end{dmath}

Meanwhile the Grassmann ADHM constraint can be spelled as: \\ $\bar{\zeta}^{A}_{\dot{\beta}} W^{\dot{\beta}}_{\dot{\alpha}} + W_{\dot{\alpha} \dot{\beta}} \zeta^{\dot{\beta}A} + [\mathcal{M}^{'\alpha A} , a'_{\alpha \dot{\alpha}}] = 0$ \\
Inserting the $\delta$-function in the Grassmann sector of the mother space integral, one can integrate out the $\bar{\zeta}^A$:

\begin{dmath}
{\int d^{2k^2} \bar{\zeta}^A \prod^{k^2}_{r=1} \prod^{2}_{\dot{\alpha}=1} \delta \left( \tr_k T^r (\bar{\zeta}^A_{\dot{\beta}} {W^{\dot{\beta}}}_{\dot{\alpha}} + W_{\dot{\alpha} \dot{\beta}} \zeta^{\dot{\beta} A} + [\mathcal{M}^{'\alpha A} , a'_{\alpha \dot{\alpha}}]) \right)} \\
= |{\det}_{2k} W|^{k}
\end{dmath}

Finally, the SUSic volume form on the moduli space can be written as:

\begin{dmath}
    \int_{\mu^{-1}(0) / U(k)} \boldsymbol{\omega}^{(2)} 
    = \frac{2^{3k^2} C^{(2)}_{k} A_k}{ \text{VolU}(k)} \int d^{4k^2} a' d^{k^2} W^0 d^{4k(N-k)} \mathcal{U} \\ \prod^{2}_{A=1} \left\{ d^{k(N-2k)} \nu^{A} d^{k(N-2k)} \bar{\nu}^{A} d^{2k^2} \zeta^A d^{2k^2} \mathcal{M}^{'A} \right\} \times |{\det}_{k^2} \mathbf{L}|^{-1} \times |{\det}_{2k} W|^{N-2k}
\end{dmath}

\end{itemize}

\section{Matter Fields in ADHM construction}
\label{Matter Fields-in-ADHM-construction}
We study two cases of interest: $\mathcal{N}=1$ Higgs branch and $\mathcal{N}=2$ Coulomb branch.

\begin{itemize}
    \item $\mathcal{N}=1$ Higgs branch 

The $SU(N)$ gauge field coupled to Dirac fermions(hypermultiplets) or equivalently to $2N_F$ chiral superfields $Q_f$ and $\tilde{Q}_f$ where $1 \leq f \leq N_F$ where $Q_f, \tilde{Q}_f$ transform in the $\mathbf{N}$ and $\mathbf{\bar{N}}$ representations respectively.

The Euclidean action would be:

\begin{dmath}
    S^{E}_{\text{matter}} = \int d^4 x \left\{ \mathcal{D}_n q^{\dagger} \mathcal{D}_n q + \mathcal{D}_n \tilde{q} \mathcal{D}_n \tilde{q}^{\dagger} - \mathcal{D}_n \bar{\chi} \bar{\sigma}_n \chi + \tilde{\chi} \sigma_n \mathcal{D}_n \bar{\tilde{\chi}} - \sqrt{2} ig \bar{\chi}\bar{\lambda}q +i \sqrt{2} g q^{\dagger} \lambda \chi + \sqrt{2} ig \tilde{q} \bar{\lambda} \bar{\tilde{\chi}} - \sqrt{2} ig \tilde{\chi} \lambda \tilde{q}^{\dagger} +\frac{1}{4} g^2 \left(q^{\dagger} q - \tilde{q} \tilde{q}^{\dagger}\right)^2  \right\}
\end{dmath}

where the VEV $N \times N_F$ and $N_F \times N$ matrices after diagonalization are:

\begin{dmath}
    {q^0_{uf} = \begin{pmatrix}
    v_1 & 0 & \cdots & 0 \\
     0  & v_2 & \cdots & 0 \\
     \vdots & \vdots & \ddots & \vdots \\
      0 & 0 & \cdots & v_f \\
      \vdots & \vdots & \ddots & \vdots \\
      0 & 0 & \cdots & 0 
\end{pmatrix} \tilde{q}^{0}_{fu} = \begin{pmatrix}
    \tilde{v}_1 & 0 & \cdots & 0 & \cdots & 0 \\
    0   & \tilde{v}_2 & \cdots & 0 &  \cdots & 0 \\
    \vdots & \vdots & \ddots & 0 & \cdots & 0 \\
     0 & 0 & \cdots & \tilde{v}_f & \cdots & 0 
\end{pmatrix}}
\end{dmath} 

The equation of motion for the chiral fermions(bearng in mind the antichiral fermions are vanishing to leading order, $\bar{\lambda} = \bar{\chi}=0 $) and the scalar fields to leading order are:
\begin{dmath}
    {\rawslashed{\mathcal{\bar{D}}} \chi_{\alpha f} = 0 \ \ \ \ \rawslashed{\mathcal{\bar{D}}} \tilde{\chi}_{f \alpha} = 0 \ \ \ \ \ \ \mathcal{D}^2 q = \sqrt{2}ig \lambda \chi}
\end{dmath}
whose solutions are:

\begin{dmath}
    {\chi_{\alpha f} = g^{-1/2} \bar{U} b_{\alpha} f \mathcal{K}_f \ \ \ \ \ \tilde{\chi}_{f \alpha} = g^{-1/2} \tilde{\mathcal{K}}_f f \bar{b}_{\alpha} U}
\end{dmath}

and:

\begin{dmath}
    {q_f = \bar{U} \begin{pmatrix}
    q^0_f \\
     0
\end{pmatrix} - \frac{i}{2 \sqrt{2}} \bar{U} \mathcal{M} f \mathcal{K}_f }
\end{dmath}

\begin{dmath}
{\tilde{q}_f = \left( \tilde{q}^0_f \ \ 0 \right) U + \frac{i}{2\sqrt{2}} \tilde{\mathcal{K}}_f f \bar{\mathcal{M}} U}
\end{dmath}

The instanton effective action can be written:

\begin{dmath}
    \tilde{S} = \pi^2 \sum^{N_F}_{f=1} \left\{ q^{0 \dagger}_f w_{\dot{\alpha}} \bar{w}^{\dot{\alpha}} q^0_f + \frac{i}{\sqrt{2}} q^{0 \dagger}_f \mu \mathcal{K}_f + \tilde{q}^0_{f} w_{\dot{\alpha}} \bar{w}^{\dot{\alpha}} \tilde{q}^{0 \dagger}_f -\frac{i}{\sqrt{2}} \tilde{\mathcal{K}}_f \bar{\mu} \tilde{q}^{0 \dagger}_f \right\}
\end{dmath}
where only the $\mu, \bar{\mu}$ of $\mathcal{M}, \bar{\mathcal{M}}$ contribute regarding the gaugino modes; which leaves $\mathcal{O}(k)$ unlifted modes. Moreover, $\tilde{S}$ is the disconnected sum of $k$ single instantons, meaning they do not interact with each other. These are specifically the features that appear only in the $\mathcal{N}=1$ case.
Following Section \ref{SUSY-and-the-Collective-Coordinates}, one can argue that the SUSic transformations of the matter collective coordinates are:
\begin{dmath}
    {\delta \mathcal{K}_f = -2 \sqrt{2} \bar{\xi}_{\dot{\alpha}} \bar{w}^{\dot{\alpha}} q^0_f \ \ \ \ \delta \tilde{\mathcal{K}}_f = -2 \sqrt{2} \tilde{q}^0_f w_{\dot{\alpha}} \bar{\xi}^{\dot{\alpha}}}
\end{dmath}

The contribution of the matter zero modes to the effective action can be simplified by:
\begin{dmath}
    \int d^4 x \sum^{N_F}_{f=1} \tilde{\chi}_f \chi_f = \frac{\pi^2}{g} \sum^{N_F}_{f=1} \tilde{\mathcal{K}}_f \mathcal{K}_f
\end{dmath}
The contribution to the path integral measure is
\begin{dmath}
    \left(\frac{g}{\mu \pi^2}\right)^{kN_F} \int d^{kN_F} \mathcal{K}_f d^{k N_F} \tilde{\mathcal{K}}_f
\end{dmath} 

resulting in the new volume form $\boldsymbol{\omega}^{(1 , N_F)} = \pi^{-2k N_F} \boldsymbol{\omega}^{(1)} d^{kN_F} \mathcal{K} d^{kN_F} \tilde{\mathcal{K}}$

leading to a much more compact notation for the instanton partition function:
\begin{dmath}
    \boxed{\mathcal{Z}^{(1 , N_F)}_k = \int_{{\mu^{-1}(0)}/{U(k)}} \boldsymbol{\omega}^{(1, N_F)} e^{-\tilde{S}}}
\end{dmath}

\item $\mathcal{N}=2$ Coulomb branch

On this branch, the hypermultiplets do not acquire VEVs by definition. The Euclidean action can be written in components as:

\begin{dmath}
S_{\text{matter}} = \int d^4 x \left\{ \mathcal{D}_n q^{\dagger} \mathcal{D}_n q + \mathcal{D}_n \tilde{q} \mathcal{D}_n \tilde{q}^{\dagger} - \mathcal{D}_n \bar{\chi} \bar{\sigma}_n \chi + \tilde{\chi} \sigma_n \mathcal{D}_n \bar{\tilde{\chi}} - \sqrt{2} ig \bar{\chi} \bar{\lambda} q + i \sqrt{2} g q^{\dagger} \lambda \chi + \sqrt{2} i g \tilde{q} \bar{\lambda} \bar{\tilde{\chi}} -\sqrt{2}i g \tilde{\chi} \lambda \tilde{q}^{\dagger} -\sqrt{2} g \tilde{\chi} \psi q -\sqrt{2}g \tilde{q} \psi \chi - g\tilde{\chi} \phi \chi - \sqrt{2} g q^{\dagger} \bar{\psi} \bar{\tilde{\chi}} -\sqrt{2}g \bar{\chi} \bar{\psi} \tilde{q}^{\dagger} -g \bar{\chi} \phi^{\dagger} \bar{\tilde{\chi}} \right\} + S_{scalar}
\end{dmath}

where the solution to the equations of motion to leading order is;

\begin{dmath}
    {\chi_{\alpha f} = g^{-1/2} \bar{U} b_{\alpha} f \mathcal{K}_f \ \ \ \ \ \tilde{\chi}_{f \alpha} = g^{-1/2} \tilde{\mathcal{K}}_f f \bar{b}_{\alpha} U}
\end{dmath}

\begin{dmath}
    {q_f =  - \frac{i}{2 \sqrt{2}} \bar{U} \mathcal{M}^1 f \mathcal{K}_f \ \ \ \tilde{q}_f =  \frac{i}{2\sqrt{2}} \tilde{\mathcal{K}}_f f \bar{\mathcal{M}}^1 U}
\end{dmath}
while the adjoint-fermions being in the usual ADHM forms $\lambda = g^{-1/2} \Lambda(\mathcal{M}^1)$ and $\psi = g^{-1/2} \Lambda(\mathcal{M}^2)$ while the scalar is:
\begin{dmath}
    \phi = \frac{i}{2} \bar{U} \mathcal{M}^A f \bar{\mathcal{M}}_A U + \bar{U} \begin{pmatrix}
    \phi^0 & 0 \\
    0 & \varphi 1_{[2] \times [2]}
\end{pmatrix} U
\end{dmath}
where $\varphi = \mathbf{L}^{-1} \left( -\frac{i}{2} \bar{\mathcal{M}}^A \mathcal{M}_A + \bar{w}^{\dot{\alpha}} \phi^0 w_{\dot{\alpha}} \right)$.
The equation of motion (to leading order) for a complex conjugate scalar field is $\mathcal{D}^2 \phi^{\dagger} = -g \chi \tilde{\chi}$ whose solution is:
\begin{dmath}
    \phi^{\dagger} = \bar{U} \begin{pmatrix}
    \phi^{0 \dagger} & 0 \\
    0 & \varphi^{\dagger} 1_{[2] \times [2]}
\end{pmatrix} U
\end{dmath}
where $\varphi^{\dagger} = \mathbf{L}^{-1} \left(-\frac{1}{4} \sum^{N_F}_{f=1} \mathcal{K}_f \tilde{\mathcal{K}}_f + \bar{w}^{\dot{\alpha}} \phi^{0 \dagger} w_{\dot{\alpha}} \right)$
On the same line, the equations of motion for $q^{\dagger}, \tilde{q}^{\dagger}$ are:
\begin{dmath}
    {\mathcal{D}^2 q^{\dagger} = -\sqrt{2} g \tilde{\chi} \psi \ \ \ \ \ \ \mathcal{D}^2 \tilde{q}^{\dagger} = -\sqrt{2} g \psi \chi}
\end{dmath}
whose solutions are: $q^{\dagger}_f = \frac{1}{2 \sqrt{2}} \tilde{\mathcal{K}}_f f \bar{\mathcal{M}}^2 U , \ \tilde{q}^{\dagger}_f = \frac{1}{2 \sqrt{2}} \bar{U} \mathcal{M}^2 f \mathcal{K}_f$

Then, as for the kinetic part and the Yukawa terms of the effective action, after integration by parts:

\begin{dmath}
        \tilde{S} = \int d^4 x \left\{ \partial_n \left(\phi^{\dagger} \mathcal{D}_n \phi\right) + \partial_n \left(q^{\dagger} \mathcal{D}_n q\right) + \partial_n \left(\tilde{q} \mathcal{D}_n \tilde{q}^{\dagger} \right) - g \tilde{\chi} \left(\sqrt{2} i \lambda \tilde{q}^{\dagger} + \sqrt{2} \psi q + \phi \chi \right) \right\}
\end{dmath}

where only $\phi$ has VEV and so the second and the second and third terms vanish at infinity, resulting in:
\begin{dmath}
\tilde{S} = 4 \pi^2 \tr_k \left\{ -\frac{i}{2} \bar{\mu}^{A} \phi^{0 \dagger} \mu_{A} + \bar{w}^{\dot{\alpha}} |\phi^{0}|^2 w_{\dot{\alpha}} - \varphi \bar{w}^{\dot{\alpha}} \phi^{0 \dagger} w_{\dot{\alpha}} \right\} - \\ \ \ \ \ \ \ \ \ \ \ \  g \int d^4 x \tilde{\chi} \left(\sqrt{2} i \lambda \tilde{q}^{\dagger} + \sqrt{2} \psi q + \phi \chi \right)
\end{dmath}

and for the last integral, one can use the trick:

\begin{dmath}
    \sqrt{2} i \lambda \tilde{q}^{\dagger} + \sqrt{2} \psi q + \phi \chi = g^{-1/2} \left(\slashed{\mathcal{D}} \bar{\Upsilon} + \Theta \right)
\end{dmath}
where $\Theta$ is a zero mode in the fundamental representation $\rawslashed{\bar{\mathcal{D}}} \Theta = 0$. With the ADHM algebra, one can check that there exists a solution for $\bar{\Upsilon}, \Theta$:
\begin{dmath}
    {\bar{\Upsilon}^{\dot{\alpha}}_{f} = - \frac{i}{4} \bar{U} \mathcal{M}^A f \bar{\Delta}^{\dot{\alpha}} \mathcal{M}_A f \mathcal{K}_f + \bar{U} \begin{pmatrix}
    \phi^{0} & 0 \\
    0 & \varphi
\end{pmatrix} \Delta^{\dot{\alpha}} f \mathcal{K}_f \ \ , \ \ \ \Theta_f = \bar{U} b_{\alpha} f \varphi \mathcal{K}_f}
\end{dmath}  

meaning the only contribution of the Yukawa terms are $-g^{1/2} \int d^4 x \tilde{\chi} \Theta$ which by using the identity:
\begin{dmath}
    -\pi^2 \sum^{N_F}_{f=1} \tilde{\mathcal{K}}_f \varphi \mathcal{K}_f = \pi^2 \text{tr}_k \left[ 
\left( \sum^{N_F}_{f=1} \mathcal{K}_f \tilde{\mathcal{K}}_f \right) \mathbf{L}^{-1} \left(-\frac{i}{2} \bar{\mathcal{M}}^A \mathcal{M}_A + \bar{w}^{\dot{\alpha}} \phi^0 w_{\dot{\alpha}}\right) \right]
\end{dmath}
and summing with the kinetic terms contribution results in:
\begin{dmath}
\tilde{S} = 4 \pi^2 \tr_k \left\{ -\frac{i}{2} \bar{\mu}^{A} \phi^{0 \dagger} \mu_{A} + \bar{w}^{\dot{\alpha}} |\phi^{0}|^2 w_{\dot{\alpha}} + \left( \frac{1}{4} \sum^{N_F}_{f=1} \mathcal{K}_f \tilde{\mathcal{K}}_f - \bar{w}^{\dot{\alpha}} \phi^{0 \dagger} w_{\dot{\alpha}} \right) \mathbf{L}^{-1} \left( - \frac{i}{2} \bar{\mathcal{M}}^A \mathcal{M}_A + \bar{w}^{\dot{\alpha}} \phi^{0} w_{\dot{\alpha}} \right) \right\}
\end{dmath}

which is $\mathcal{N}=2$ SUSic in terms of collective coordinates SUSic transformations, while since one is on the Coulomb branch $\delta \mathcal{K}_f= \delta \tilde{\mathcal{K}}_f=0$. Finally, the matter Grassmann collective coordinate contribution to the SUSic moduli space volume form is the same as that of the $\mathcal{N}=1$ Higgs branch that was already discussed. 
\end{itemize}

\section{Massive Hypermultiplets}

Regarding $\mathcal{N}=2$ fermions, one can simply add $\mathcal{N}=1$, preserving massive fermions in the $(\mathbf{N} , \bar{\mathbf{N}})$ representation(Dirac fermion) terms, to the lagrangian, resulting in the (Wick-rotated) action:

\begin{dmath}
    S_{\text{mass}} = \int d^4 x \left\{ m \tilde{\chi} \chi + m^{*} \bar{\chi} \bar{\tilde{\chi}} + 2 |m|^2 q^{\dagger} q + 2 |m|^2 \tilde{q} \tilde{q}^{\dagger} \right\}
\end{dmath} 

One can argue that $m^{*} = 0$ as the effective semi-classical action must preserve the principle of holomorphic dependence on action variables. Though this can be formally proven using an appropriate Ward identity, it is left to the reader to investigate it.

By the same manipulations of the previous section, one can simplify writing its contribution to the instanton effective action:

\begin{dmath}
    \tilde{S}_{\text{mass}} = \frac{\pi^2}{g} \sum^{N_F}_{f=1} m_f \tilde{\mathcal{K}}_f \mathcal{K}_f
\end{dmath}

As for $\mathcal{N}=2$ chiral superfields in the adjoint representation one can write quite generally:

\begin{dmath}
    S_{\text{mass}} = \int d^4 x \ m_{AB} \text{tr}_{N} \left\{ \lambda^A \lambda^{B} \right\}
\end{dmath}

where $m_{AB}$ can be diagonalised $m_{AB} = \text{diag}(m_1 , m_2)$ and, quite obviously, one can see that a vanishing eigenvalue results in an $\mathcal{N}=1$ preserving action.

Finally, the contribution to the effective action turns out to be:

\begin{dmath}
    \tilde{S}_{\text{mass}} = - \frac{m_{AB} \pi^2}{g} \text{tr}_{k} \bar{\mathcal{M}}^A (\mathcal{P}_{\infty} + 1) \mathcal{M}^B = - \frac{m_{AB} \pi^2}{g} \text{tr}_k \left[ 2 \bar{\mu}^A \mu^B + \mathcal{M}^{' \alpha A} \mathcal{M}^{'B}_{\alpha} \right]
\end{dmath}

\section{Linearised Instanton Partition Function}
\label{Linearised-Instanton-Partition-Function}
In section \ref{Matter Fields-in-ADHM-construction} we constructed the $\mathcal{N}=2$ partition function. But it's worth(as it will be seen) to further linearise the partition function with the aid of auxiliary fields. 

By adding $2(\mathcal{N} - 1) = 2$ vectors $\chi_a$ of $k \times k$ matrices, a 3-vector $\Vec{D}$ of $k \times k$ matrices and finally Grassmann $k \times k$ matrices $\bar{\psi}^{\dot{\alpha}}_A$, $A=1 , ... , \mathcal{N}=2$ one can rewrite the partition function:

\begin{dmath}
    \mathcal{Z}^{(2)}_k = \frac{\pi^{-4k^2} C^{(2)}_k}{\text{VolU}(k)} \int d^{4k(N+k)} a \ d^{3k^2} D \ d^{2k^2} \chi \prod^{2}_{A=1} \ d^{2k(N+k)} \mathcal{M}^{A} d^{2k^2} \bar{\psi}_A e^{-\tilde{S}}
\end{dmath}

where 
\begin{dmath}
    \tilde{S}= 4 \pi^2 \text{tr}_k \left\{ \chi_a \mathbf{L} \chi_a + \frac{1}{2} \bar{\Sigma}_{aAB} \bar{\mathcal{M}}^A \mathcal{M}^{B} \chi_a \right\} + \tilde{S}_{\text{L.m.}}
\end{dmath}

in which:
\begin{dmath}
    \tilde{S}_{\text{L.m.}} = -4i \pi^2 \text{tr}_k \left\{ \bar{\psi}^{\dot{\alpha}}_A (\bar{\mathcal{M}}^A a_{\dot{\alpha}} + \bar{a}_{\dot{\alpha}} \mathcal{M}^A) + \Vec{D} \ . \ \vec{\tau}^{ \dot{\alpha}}_{\dot{\beta}} \bar{a}^{\dot{\beta}} a_{\dot{\alpha}}  \right\}
\end{dmath}
To generalise $\tilde{S}$ to the $\mathcal{N}=2$ Coulomb branch one must do the following justifiable transformation:

\begin{dmath}
{w_{\dot{\alpha}} \chi_a \rightarrow w_{\dot{\alpha}} \chi_a + \phi^0_a w_{\dot{\alpha}} , \ \ \ \chi_a \bar{w}^{\dot{\alpha}} \rightarrow \chi_a \bar{w}^{\dot{\alpha}} + \bar{w}^{\dot{\alpha}} \phi^0_a}
\end{dmath}

\begin{dmath}
{\mu^{A} \chi_a \rightarrow \mu^{A} \chi_a + \phi^0_a \mu^A , \ \ \chi_a \bar{\mu}^A \rightarrow \chi_a \bar{\mu}^A + \bar{\mu}^A \phi^0_a}    
\end{dmath}

resulting in:

\begin{dmath}[label={eq:linearised-effective-action}]
\tilde{S} = 4 \pi^2 \tr_k \left\{ \big| w_{\dot{\alpha}} \chi_a + \phi^0_a w_{\dot{\alpha}} \big|^2 - \left[ \chi_a , a'_n \right]^2 + \frac{i}{2} \bar{\mu}^A \left( \mu_A \chi^{\dagger} + \phi^{0\dagger} \mu_A \right) + \frac{i}{2} \mathcal{M}^{'A} \mathcal{M}^'_A \chi^{\dagger} + \frac{1}{4} \sum^{N_F}_{f=1} \mathcal{K}_f \tilde{\mathcal{K}}_f \left( \chi - g^{-1} m_f \right) \right\} + \tilde{S}_{\text{L.m.}}
\end{dmath}

The linearised fields and ADHM variables transformations under the $\mathcal{N}=2$ SUSY are then:

\begin{dmath}
{\delta a'_{\alpha \dot{\alpha}} = i \bar{\xi}^{A}_{\dot{\alpha}} \mathcal{M}^{'}_{A \alpha} , \ \ \delta w_{\dot{\alpha}} = i \bar{\xi}^{A}_{\dot{\alpha}} \mu_A, \ \  
\delta \chi_a = - \Sigma_a \bar{\xi}^{A}_{\dot{\alpha}}{\bar{\psi}^{\dot{\alpha}}_{A}}, \ \ {\delta \Vec{D} = - i \Sigma_a \Vec{\tau}^{\dot{\alpha}}_{\dot{\beta}} \bar{\xi}^{A}_{\dot{\alpha}} [\bar{\psi}^{\dot{\beta}}_A , \chi_a]}}
\end{dmath}

\begin{dmath}
{\delta\mathcal{M}'_{\alpha} = -2i \Sigma_a \bar{\xi}^{\dot{\alpha} A} [a'_{\alpha \dot{\alpha}} , \chi_a], \delta \mu^A = -2i \Sigma_a \bar{\xi}^{\dot{\alpha} A} (w_{\dot{\alpha}} \chi_a + \phi^0_a w_{\dot{\alpha}})}  
\end{dmath}

\begin{dmath}[label={eq:susy-variation-psibar}]
{\delta \bar{\psi}^{\dot{\alpha}}_A = 2 \bar{\Sigma}_{ab A}^B [\chi_a , \chi_b] \bar{\xi}^{\dot{\alpha}}_B - i \Vec{D} \ .  \Vec{\tau}^{\dot{\alpha}}_{\dot{\beta}} \bar{\xi}^{\dot{\beta}}_A}
\end{dmath}

where in \eqref{eq:susy-variation-psibar} $\Sigma_{ab} = \frac{1}{4} \left( \Sigma_a \bar{\Sigma}_b - \Sigma_b \bar{\Sigma}_a \right)$ and $\bar{\Sigma}_{ab} = \frac{1}{4} \left( \bar{\Sigma}_a \Sigma_b - \bar{\Sigma}_b \Sigma_a \right)$.

Therefore, the centred moduli space can be written down after factoring out:
\begin{dmath}
    {X_n = -k^{-1} \text{tr}_k a'_n , \ \ , \xi^{A} = \frac{i}{4} k^{-1} \text{tr}_k \mathcal{M}^{'A}}
\end{dmath}
and the moduli space metric is\footnote{here we denote the moduli space $\mu^{-1}(0) /U(k)$ by $\mathcal{M}_k$}:

\begin{dmath}
    ds^2_{\mathcal{M}_k} = 8 \pi^2 k dX_n dX_n + ds^2_{\widehat{\mathcal{M}}_k}
\end{dmath}

resulting in volume forms to be related by:

\begin{dmath}
    \int_{{\mathcal{M}}_k} \boldsymbol{\omega}^{(2, N_F)} = \int (4 \pi k)^2 d^4 X \prod^{2}_{A=1} (32 \pi^2 k)^{-1} d^2 \xi^A \ . \ \int_{{\widehat{\mathcal{M}}}_k} \boldsymbol{\omega}^{(2, N_F)}
\end{dmath}

and the instanton partition function can be generalised to be:

\begin{dmath}[label={eq:instanton-partition-function}]
    \mathcal{Z}^{(2, N_F)}_k = \int_{{\widehat{\mathcal{M}}}_k} \boldsymbol{\omega}^{(2, N_F)} e^{-\tilde{S}}
\end{dmath}

\section{ADHM construction for $Sp(1) \simeq SU(2)$}
\label{ADHM}

Finally, for our future purposes, it's not bad to briefly discuss the ADHM construction in the case of $Sp(N)$ and consequently $Sp(1)\simeq SU(2)$.

The key step is to notice that $Sp(N) \subset SU(2N)$, and the purpose is to impose the proper reality conditions on $SU(2N)$ ADHM conditions so that the gauge potential lies in the $\mathfrak{sp}(N)$ Lie algebra. To do so, it's better to introduce a generalised notion of transposition which is defined with the aid of the $2N \times 2N$ matrix $J$:

\begin{dmath}
    J= \begin{pmatrix}
    0 & 1_{[N] \times [N]} \\
    -1_{[N] \times [N]} & 0
\end{pmatrix}
\end{dmath}
where the transpose of a vector $v$ is denoted by $v^t = v^T J^T$ where $T$ stands for the ordinary transposition of a matrix.

The reality conditions are:

\begin{dmath}
    {\bar{w}^{\dot{\alpha}} = \epsilon^{\dot{\alpha} \dot{\beta}} \left( w^t_{\dot{\beta}} \right) , \ \ \ \ {\left(a'_{\alpha \dot{\alpha}}\right)}^t = a'_{\alpha \dot{\alpha}}}
\end{dmath}

while their Grassmann counterpart would be:

\begin{dmath}
    {\bar{\mu} = \mu^t , \ \ \ \ \left( \mathcal{M}'_{\alpha} \right)^t = \mathcal{M}'_{\alpha}}
\end{dmath}

while not forgetting that $Sp(n)$ is defined to be $Sp(2n; \mathbb{C}) \cap U(2n) = Sp(2n; \mathbb{C}) \cap SU(2n)$.

The subgroup that maintains the reality conditions $H(k) \subset U(k')$-valued ADHM constraints can be determined and turns out to be $O(k)$ where $k=k'$. This means that the ADHM constraints can be thought of as $H(k)$-valued constraints.

As an example, the reality conditions in the case of $Sp(1)$ are explicitly:
\begin{dmath}
    {w^*_{ui \dot{\alpha}} = \epsilon^{\dot{\alpha} \dot{\beta}} J_{uv} w_{vi \dot{\beta}}, \ \ \ \ \left( a'_n \right)_{ij} = \left( a'_n \right)_{ji}}
\end{dmath}

remembering that $a'_n$ are hermitian,we conclude that they're real symmetric $k \times k$ matrices while since $J_{uv} = \epsilon_{uv}$ the gauge group indices can be converted to that of $u \rightarrow \alpha$ turning $w_{ui\alpha}$ into quaternions $w_{i \alpha \dot{\alpha}} = w_{in} \sigma_{n\alpha \dot{\alpha}}$ and the reality condition is more symmetrically:

\begin{dmath}
    {w^*_{i \alpha \dot{\alpha}} = \epsilon^{\dot{\alpha} \dot{\beta}} \epsilon^{\alpha \beta} w_{i \beta \dot{\beta}} \leftrightarrow w_i :\text{quaternion}}
\end{dmath}
for example, when $k=1$:

\begin{dmath}
    a_{\alpha \dot{\alpha}} = \begin{pmatrix}
    w_{\alpha \dot{\alpha}} \\
a'_{\alpha \dot{\alpha}}
\end{pmatrix}
\end{dmath}

with $a'_{\alpha \dot{\alpha}}$ real numbers and $w_{\alpha \dot{\alpha}}$ as quaternions and so more generally for charge $k$:

\begin{dmath}
    {a_{\alpha \dot{\alpha}} = \begin{pmatrix}
    w_{1 \alpha \dot{\alpha}} & \cdots & w_{k \alpha \dot{\alpha}} \\
    a'_{1\dot{1} \alpha \dot{\alpha}} & \cdots & a'_{1\dot{k} \alpha \dot{\alpha}} \\
    \vdots  & \ddots & \vdots \\
    a'_{k\dot{1} \alpha \dot{\alpha}} & \cdots & a'_{k\dot{k} \alpha \dot{\alpha}}
\end{pmatrix}}
\end{dmath}
For an instanton with charge $k$ there exists $2k(k+3)$ variables $a_{\alpha \dot{\alpha}}$ subject to $3k(k-1)/2$ ADHM constraints and $k(k-1)/2$ symmetries resulting a moduli space $\tilde{\mathcal{M}}_k$ of dimension $8k$. There's no doubt that $Sp(1)$ formalism is easier, as there are no constraints at all in the case of $k=1$.

\chapter{$\mathcal{N}=1$ Seiberg Electromagnetic Dualities in SQCD for arbitrary number of flavors $N_f$}
Remark: This section assumes a basic knowledge of Super-Poincaré algebra.

\section{R-Symmetry, Spurions, Holomorphy, \\ Non-Renormalisation}

As for the $R$-charge of the fields belonging to the same representation, one can assign $R[\theta]=+1$, while due to $[R, Q_{\alpha}]= - Q_{\alpha}, \ [R, \bar{Q}_{\dot{\alpha}}]=+\bar{Q}_{\dot{\alpha}}$, if the scalar component of the chiral superfield is assigned a charge of $R[\phi]=r$ it implies: $R[\psi]=r-1$ and $R[F]=r-2$ where $\Phi = \phi + \sqrt{2} \theta \psi + \theta^2 F$. 

One immediately concludes that $R[d^2 \theta]=-2$ and $R[W]=+2$ where $W$ is the superpotential.
In any quantum field theory, the parameters of the theory can be elevated to dynamical, very heavy scalar fields, vacuum expectation values, so that the quantum fluctuations around their vev are extremely suppressed at low energies, so that these fields can be perceived as background classical fields. Such fields are called conveniently \textbf{spurions}.

An example would be a superpotential of the type:
\begin{dmath}
    W= \frac{1}{2} m {\Phi}^2 + \frac{1}{3} \lambda {\Phi}^3
\end{dmath}

By such an argument, one can safely extend the group of global symmetries of the superpotential so that, apart from $U(1)_R$, one has also $U(1)_F$ such that the charges of the superfields, along with the spurions, are:

$$U(1)_R \ : \ R[\Phi]=1 \ R[m]=0 \ \ \ R[\lambda]=-1$$
$$U(1)_F \ : \ F[\Phi]=1 \ F[m]=-2 \ F[\lambda]=-3$$

The most general superpotential with such symmetries is:

\begin{dmath}
    {W_{\text{eff}} = m {\Phi}^2 f(\frac{\lambda \Phi}{m})}
\end{dmath}

The next piece of the puzzle is \textbf{holomorphy}. Expanding $f(x)$ in terms of $x$ and impose regularity on the superpotential in the $m \rightarrow 0$ one finds:

\begin{dmath}
    {W_{\text{eff}} = \frac{1}{2} m {\Phi}^2 +\frac{1}{3} \lambda {\Phi}^3}
\end{dmath}
which coincides with the superpotential at the tree level.

Note that this does not mean that the parameters do not receive quantum corrections themselves, but only that superpotental does not receive perturbative quantum corrections in $\lambda$ which means the shape of the potential does not change while $\lambda$ and $m$ receive multiplicative quantum corrections while \textbf{nonperturbative quantum corrections} are still possible we will find out in section \ref{Effective Superpotential}.

\section{$\mathcal{N}=1$ SYM, SQCD and Moduli Sapce of Vacua}

The general kinetic action for superfields on the superspace consists of a Kähler potential whose only contributing term in the spatial integral is the quadratic term in $\theta, \bar{\theta}$:

\begin{dmath}
    S= \int d^2 \theta d^2 \bar{\theta} d^4 x K = \int d^4 x K_{last}
\end{dmath}
where
\begin{dmath}
    K(x, \theta, \bar{\theta}) = K_{\text{first}} +...+ {\theta}^2 {\bar{\theta}}^2 K_{\text{last}}
\end{dmath}

$K_{last}$ and terms similar to it, are called $D$-terms. $K$ can be, in general, a function of chiral and anti-chiral superfields. 

The next step is to introduce the spinor chiral superfield $W_{\alpha}$:

\begin{dmath}
    {W_{\alpha} = -\frac{1}{8} {\bar{\mathcal{D}}}^2 (e^{-2V} \mathcal{D}_{\alpha} e^{2V})}
\end{dmath}
where $V$ is the real superfield. The field is evidently chiral since $\bar{\mathcal{D}}_{\dot{\alpha}} W_{\alpha} = 0$ thanks to ${\bar{\mathcal{D}}}^3 = 0$.

Expansion in terms of the Grassmann dimensions gives rise to:
\begin{dmath}
    {W_{\alpha} (y, \theta) = \lambda_{\alpha} (y) + \theta_{\alpha} D(y) + ({\sigma}^{\mu \nu} {\theta})_{\alpha} F_{\mu \nu} (y) -i {\theta}^2 {\sigma}^{\mu}_{\alpha \dot{\alpha}} {\partial}_{\mu} \bar{\lambda}^{\dot{\alpha}}}
\end{dmath}
in the Wess-Zumino gauge (where $ V^2 = \frac{1}{2} {\theta} {\theta} \bar{\theta} \bar{\theta} v_{\mu} v^{\mu}, V^n =0 \ , n \geq 3$) and bearing in mind $y^{\mu}= x^{\mu} + i \bar{\theta} {\sigma}^{\mu} \theta$ and that a gauge transformation works as $e^V \rightarrow e^{-i {\Omega}^{\dagger}} e^V e^{i \Omega}$ where $e^{i \Omega} \in G$ for an adjoint valued chiral superfield $\Omega = {\Omega}^A T^A$.
\

The general Lagrangian for $\mathcal{N}=1$ SQCD in the $\theta$-vacuum can be written as:

\begin{dmath}
    S=S_{\text{SYM}} + S_{\text{Kähler}}
\end{dmath}
where:
\begin{dmath}
    {S_{\text{SYM}} = \int d^4 x d^2 \theta  \ \text{Tr}{ \frac{-i \tau}{8 \pi} W^{\alpha} W_{\alpha} + h.c.} \ \ S_{\text{Kähler}} = \int d^4 x d^4 \theta K(\Phi^{\dagger}_{i} e^{2V}, \Phi_i)}
\end{dmath}

To avoid complications due to coupling Weyl fermions to gauge fields causing anomalies, one can introduce Dirac fermions by inclusion of anti-chiral $\tilde{\Phi}$ fields and consequently adding $K({\tilde{\Phi}}^{\dagger}_{i} e^{-2V}, \tilde{\Phi}_i)$ to the Kähler action $S_{\text{Kähler}}$ where if $\phi$ belongs to representatioin $R$ of the gauge field the anti-chiral field $\tilde{\Phi}$ belongs to the conjugate representation $\bar{R}$.

\section{Moduli Space of the $\mathcal{N}=1$ SQCD}
The scalar potential and the $D$-term for the $S_{\text{SQCD}}$ action can be written as:
\begin{dmath}
    {V(\phi, \tilde{\phi}) = \frac{1}{2} D^A D^A \ \ \ \ \ \ D^A = -g \sum^{N_f}_{i=1} ({\phi}^{\dagger}_{i} T^A \phi_i - \tilde{\phi}_i T^A {\tilde{\phi}}^{\dagger}_i)}
\end{dmath}

For the $\mathcal{N}=1$ SQCD with gauge group $G=SU(N)$ and $N_f$ flavors in the fundamental and $N_f$ in the anti-fundamental representation can be written as:

\begin{dmath}
    {\mathcal{M}= \{ \phi | D(\phi)=0 \}}
\end{dmath}

where $D$ is the momentum map.

By the \textbf{Geometric Invariant Theory}, one can prove that the quantum moduli space(as well as the classical one) can be written as:

\begin{dmath}
    {\mathcal{M}= \{ \text{Gauge invariant holomorphic monomials} \} / \{ \text{Algebraic relations} \}}
\end{dmath}

For the analysis of the moduli space and the effective action, one should keep in mind that such monomials depend on the ratio of the number of flavors and colors $N_f / N$ that parametrizes different phases of the theory.

We end this section by introducing the gauge-invariant holomorphic monomials for different phases of the theory/moduli space(different $N_f / N$ ratios):

\begin{itemize}
    \item $N_f < N$ 
In this case the chiral and anti-chiral superfields ${\Phi}^i_a, \tilde{\Phi}^a_i$, can be used to construct the holomorphic monomials as:
\begin{dmath}
    {M^i_j = \langle \tilde{\Phi}^a_j {\Phi}^i_a \rangle}
\end{dmath}
We call these holomorphic monomials \textbf{meson fields}.
The symmetry group breaks from $SU(N)$ to $SU(N-N_f)$, consequently the number of broken generators is $(N^2 -1) -((N-N_f)^2 -1)$. Given this, one finds:
\begin{dmath}
    {\dim{\mathcal{M}} = N^2_f}
\end{dmath}
At the classical level, using the microscopic theory, one can easily show that the Kähler potential can be written as:

\begin{dmath}
    K = 2 \text{Tr}{\sqrt{M^{\dagger} M}}
\end{dmath}

which is defined over the moduli space. It is well-defined except for singular values of $M^{-1}$ such as the origin. This singularity is the point of enhanced gauge symmetry, where the symmetry group is restored to $SU(N)$ and massless particles are introduced into the spectrum. 
     \item $N_f \geq N$ 
     \begin{dmath}
         {B^{i_1,..,i_N} = \langle\Phi^{i_1}_{a_1} ... \Phi^{i_N}_{a_N} \rangle}\epsilon^{a_1 ... a_N}
     \end{dmath}
     \begin{dmath}
         {{\tilde{B}}_{i_1 ,..., i_N} = \langle\tilde{\Phi}^{a_1}_{i_1} ... \tilde{\Phi}^{a_N}_{i_N}  \rangle}{\epsilon}_{a_1, ..., a_N}
     \end{dmath}
These are respectively called the \textbf{baryon fields}.

One should bear in mind that it is necessary for the identification of the moduli space to mod out the constraints that these holomorphic monomials satisfy.

For instance, if $N_f = N$ one can write:
\begin{dmath}[label={eq:classical-constraint-Nf-equal-N}]
    \tilde{B} B = \det{M}
\end{dmath}

so that 
\begin{dmath}
    \dim{\mathcal{M}} = N^2_f +1
\end{dmath}

Though we alarm the reader that this constraint is modified by non-perturbative quantum corrections.

\end{itemize}

\section{Effective Superpotential}
\label{Effective Superpotential}

Before any discussion about the SQCD EFT regimes, one had better outline the symmetries of the theory:

$$SU(N): rep[ \phi]=\mathbf{N} \ \ rep[ \tilde{\phi}]=\mathbf{\bar{N}} \ \ rep[ \psi]=\mathbf{N} \ \ rep[\tilde{\psi}]=\mathbf{\bar{N}} \ \ rep[\lambda ]= adj $$
$$SU(N_f)_L : rep[ \phi ] = \mathbf{N_f} \ \ rep[\tilde{\phi}]= \mathbf{1} \ \ rep[\psi] = \mathbf{N_f} \ \ rep[\tilde{\psi}] = \mathbf{1} \ \ rep[\lambda ] = \mathbf{1} $$
$$SU(N_f)_R: rep[ \phi ] = \mathbf{1} \ \ rep[\tilde{\phi}]= \mathbf{{\bar{N}}_f} \ \ rep[\psi] = \mathbf{1} \ \ rep[\tilde{\psi}] = \mathbf{{\bar{N}}_f} \ \ rep[\lambda ] = \mathbf{1}$$
$$U(1)_B : B[\phi]=1 \ \ B[\tilde{\phi}]=-1 \ \ B[{\psi}]=1 \ \ B[\tilde{\psi}] = -1 \ \ B[\lambda] = 0$$
$$U(1)_A: A[\phi] = 1 \ \ A[\tilde{\phi}]=1 \ \ A[{\psi}]=1 \ \ A[\tilde{\psi}] = 1 \ \ A[\lambda] = 0$$

One can immediately check that both $U(1)_{R'} , U(1)_A$ are anomalous symmetries:
\begin{dmath}
    {\partial}_{\mu} j^{\mu} = \frac{\mathcal{A}}{32 {\pi}^2} \Tr{{F_{\mu \nu}}^* F^{\mu \nu}}
\end{dmath}
\begin{dmath}
    \mathcal{A}_A =2N_f 
\end{dmath}
\begin{dmath}
    \mathcal{A}_{R'} = 2(N - N_f)
\end{dmath}
Nonetheless, a linear combination of the two symmetries survives in the case of $N_f < N$, which is the true $R$-symmetry of the theory:
\begin{dmath}
    R= R' + \frac{N_f - N}{N_f} A
\end{dmath}
Instead, the anomalous symmetry is a good candidate for the spurious symmetry. 
Consequently, the true quantum symmetries of the theory are $SU(N)$, $SU(N_f)_L$, $SU(N_f)_R$,$ U(1)_B$, $U(1)_R$.

As for the dynamically generated holomorphic scale parameter ${\Lambda}^{b_0}$ where $b_0 = 3N-N_f$:
$$U(1)_A : A[{\Lambda}^{b_0}] =2N_f$$
$$U(1)_R : R[{\Lambda}^{b_0}] =0$$
Any effective superpotential can bear only the effective fields(confined) and dimensionful parameters, including the dynamical scales. In our case, the fields are the meson fields $M^i_j$ and the only dimensionful dynamical parameter is $\Lambda^{b_0}$. To reach an effective superpotential with the right charges and correct symmetries, one should take $\det{M}, {\Lambda}^{3N-N_f}$ since:
$$U(1)_B: B[\det{M}]= 0 \ \ B[{\Lambda}^{b_0}]= 0$$
$$U(1)_A: A[\det{M}]= 2 N_f \ \ A[{\Lambda}^{b_0}]= 2 N_f$$ 
$$U(1)_R: R[\det{M}]=2(N_f - N) \ \ R[{\Lambda}^{b_0}]= 0$$
so that the effective superpotential for $N_f < N$ is:
\begin{dmath}
    {W_{\text{eff}} = C \left(\frac{{\Lambda}^{3N-N_f}}{\det{M}}\right)^{\frac{1}{N-N_f}}}
\end{dmath}
called the Affleck-Dine-Seiberg superpotential or ADS.
By a simple trick, one can also find the coefficient $C$ in terms of $N_f$ and $N$. One adds a mass term for the chiral and antichiral fields and then integrates out all the fields again to find the effective superpotential. In doing so, one should add a mass term to the microscopic theory:

\begin{dmath}
    {W_{\text{mass}} = m^j_i \tilde{Q}_j Q^i}
\end{dmath}
The mass matrix can be diagonalized with a proper $SU(N_f)$ symmetry transformation.
The charges of the mass matrix under the symmetries of the theory are:

$$U(1)_B : B[m]=0$$
$$U(1)_A : A[m]=-2$$
$$U(1)_R : R[m]=\frac{2N}{N_f}$$

The most general effective superpotential compatible with all the symmetries of the theory turns out to be \cite{ADS superpotentials},\cite{SUSY},\cite{SUSY2},\cite{SUSY3}:
\begin{dmath}
    {W_{\text{eff}} = \left(\frac{{\Lambda}^{3N-N_f}}{\det{M}}\right)^{\frac{1}{N-N_f}} f(x)}
\end{dmath}
\begin{dmath}
    {x= \text{Tr} \ {mM}\left(\frac{{\det{M}}}{{\Lambda}^{3N-N_f}}\right)^{\frac{1}{N-N_f}}}
\end{dmath}

In the limit of $m/{\Lambda} \rightarrow 0$, for the limit to be smooth, one should stop at $f(x)=C + x$, resulting in\cite{ADS superpotentials},\cite{SUSY},\cite{SUSY2},\cite{SUSY3}:

\begin{dmath}
    {W_{\text{eff}} = (N-N_f) \left(\frac{{\Lambda}^{3N-N_f}}{\det{M}}\right)^{\frac{1}{N-N_f}}   + \text{Tr} \ {mM}}
\end{dmath}
where the minimum is found by solving for $M^{ij}$ that satisfies $\partial W_{eff} / \partial M^{ij} =0$, using the identity $\delta(\det{M}) = (\det{M}) \text{Tr} (M^{-1} \delta M)$ one finds:
\begin{dmath}
    {\det{M} = \frac{1}{\det{m}} \left(\frac{{\Lambda}^{3N-N_f}}{\det{M}}\right)^{\frac{N_f}{N-N_f}}}
\end{dmath}
resulting in:
\begin{dmath}
    {M_j^i = (m^{-1})_j^i (\det{m} {\Lambda}^{3N-N_f})^{1/N}}
\end{dmath}
which introduces $N$ minima that in the limit $m/{\Lambda} \rightarrow 0$ result in a running vacuum, and in case $m/{\Lambda} \gg 1$ the massive field decouples.

By comparing the massless and the massive superpotentials, one finds the coefficient to be\cite{ADS superpotentials},\cite{SUSY},\cite{SUSY2},\cite{SUSY3}:
\begin{dmath}
    {C(N,N_f) = N - N_f}
\end{dmath}

As for the case of $N_f = N$, the gauge invariant holomorphic monomials are $M, B, \tilde{B}$.
As we clarified previously, such monomials should be modded out by the algebraic relations among them, which in this case, at the classical level, is the relation \eqref{eq:classical-constraint-Nf-equal-N}, $\det{M} =\tilde{B} B$, which is not quantum mechanically correct!

To find the quantum correction to the algebraic relation, one should start with the symmetries in the $N_f = N$ case:

$$SU(N_f)_L : rep[\phi]=\mathbf{N_f} \ \ rep[\tilde{\Phi}]=\mathbf{1} \ \ rep[M]=\mathbf{N_f} \ \ rep[B]=\mathbf{1} \ \ rep[\tilde{B}]= \mathbf{1} \ \ rep[{\Lambda}^{2N}]= \mathbf{1}$$
$$SU(N_f)_R : rep[\Phi]= \mathbf{1} \ \ rep[\tilde{\Phi}]= \bar{N}_f \ \ rep[M]= \bar{N}_f \ \ rep[B]= \mathbf{1} \ \ rep[\tilde{B}]= \mathbf{1} \ \ rep[\Lambda]= \mathbf{1}$$
$$U(1)_B : B[\Phi]=1 \ \ B[\tilde{\Phi}]= -1 \ \ B[M]=0 \ \ B[B]= N \ \ B[\tilde{B}]= -N \ \ B[{\Lambda}^{2N}]=0$$
$$U(1)_A : A[\Phi]=1 \ \ A[\tilde{\Phi}]= 1 \ \ A[M]=2 \ \ A[B]= N \ \ A[\tilde{B}]= N \ \ A[{\Lambda}^{2N}]= 2N$$
$$U(1)_R : R[\Phi]=0 \ \ R[\tilde{\Phi}]=0 \ \ R[M]=0 \ \ R[B]=0 \ \ R[\tilde{B}]=0 \ \ R[{\Lambda}^{2N}] = 0$$

One can turn on the mass term for a flavor and integrate it out. The Resulting effective action should match the case for $N_f = N-1$. In doing so, one first introduces the Lagrange multiplier superfield $X$ and adds the new term:
\begin{dmath}[label={eq:quantum-constraint-Nf-equal-N}]
    {W=X(\det{M}- \tilde{B}B - {\Lambda}^{2N})}
\end{dmath}
then by turning on the mass:
\begin{dmath}
    {W= X(\det{M} - \tilde{B}B - {\Lambda}^{2N}_{old}) + \text{Tr} {mM}}
\end{dmath}
solving the equation of motion for $M$, one finds:
\begin{dmath}
    {mM=-X \det{M} \mathbf{1}_{N_f}}
\end{dmath}
where the mass term is $\text{diag}(0,...,0,m)$ and consequently $M= \begin{pmatrix}
    \tilde{M} & 0 \\
     0  &   Z
\end{pmatrix}$ with $Z= M^{N_f}_{N_f}$ leading to $X= -\frac{m}{\det{\tilde{M}}}$. Bearing in mind $\frac{\partial W}{\partial B} = -X \tilde{B} = 0$ and $\frac{\partial W}{\partial \tilde{B}} = -BX = 0$ that due to $X \neq 0$ results in $B , \tilde{B} = 0$. The equation of motion for $X$ gives the constraint and so $Z \det{\tilde{M}} = {\Lambda}^{2N}_{old}$. Finally $\tr{mM}= m Z$ is the only term contributing and gives\cite{ADS superpotentials},\cite{SUSY},\cite{SUSY2},\cite{SUSY3}:

\begin{dmath}
    {W_{\text{eff}}= \frac{{\Lambda}^{2N}_{\text{old}} m}{\det{\tilde{M}}} = \frac{{\Lambda}^{2N+1}_{\text{new}}}{\det{\tilde{M}}}}
\end{dmath}

Therefore, this ends the proof of the necessity of the quantum deformation of the algebraic relation.

The final case studied in this section is that of $N_f = N+1$. The moduli fields are $B^i , {\tilde{B}}_j , M$ and are algebraic relations are:
\begin{dmath}
    {\det{M} (M^{-1})^i_j  = B^{i} \tilde{B}_j}
\end{dmath}
\begin{dmath}
    {M_j^i B^j= M_j^i \tilde{B}_i =0}
\end{dmath}
However, these relations, in contrast to the $N_f = N$ case, are not deformed by quantum corrections.
To prove this, one first introduces the charges and representations.
$$SU(N_f)_L : rep[\Phi]= \mathbf{N}_f \ \ rep[\tilde{\Phi}]= \mathbf{1} \ \ rep[M]= \mathbf{N_f} \ \ rep[B]= \bar{N}_f \ \ rep[\tilde{B}]= \mathbf{1} \ \ rep[{\Lambda}^{2N-1}]=\mathbf{1}$$
$$SU(N_f)_R : rep[\Phi]=\mathbf{1} \ \ rep[\tilde{\Phi}] = \bar{\mathbf{N}}_f \ \ rep[M]= \bar{\mathbf{N}}_f \ \ rep[B]= \mathbf{1} \ \ rep[\tilde{B}]= \mathbf{N}_f \ \ rep[{\Lambda}^{2N-1}]= \mathbf{1} $$
$$U(1)_B : B[\Phi]=1 \ \ B[\tilde{\Phi}]= -1 \ \ B[M]=0 \ \ B[B]= N \ \ B[\tilde{B}]=-N \ \ B[{\Lambda}^{2N-1}]= 0 $$
$$U(1)_A : A[\Phi]=1 \ \ A[\tilde{\Phi}]= 1 \ \ A[M]=2 \ \ A[B]=N \ \ A[\tilde{B}]=N \ \ A[{\Lambda}^{2N-1}] = 2 N_f $$
$$U(1)_R : R[\Phi]=\frac{1}{N_f} \ \ R[\tilde{\phi}] = \frac{1}{N_f} \ \ R[M]= \frac{2}{N_f} \ \ R[B]= \frac{N}{N_f} \ \ R[\tilde{B}] = \frac{N}{N_f} \ \ R[{\Lambda}^{2N-1}]= 0$$

The proof is as before through the proof of \textbf{the necessity} of [this time] no-quantum deformation. For that matter, in contrast to the previous case, one implements the equation of motion without any Lagrange multiplier:

\begin{dmath}
    W = -\frac{1}{{\Lambda}^{2N-1}} (\det{M} - BM\tilde{B})
\end{dmath}

Turning on the mass term for a flavor field, one finds:

\begin{dmath}
    W= -\frac{1}{{\Lambda}^{2N-1}_{\text{old}}} (\det{M} - BM\tilde{B}) + \text{Tr} \ {mM}
\end{dmath}

with $m=diag(0,...,0,m)$. Solving for the meson field gives rise to:
\begin{dmath}[label={eq:meson-eom-Nf-N-plus-1}]
    \det{M} \ (M^{-1})^i_j - B^i \tilde{B}_j = {\Lambda}^{2N-1}_{\text{old}} m^i_j
\end{dmath}
where the meson and baryon fields turn out to be $M=\begin{pmatrix}
    \tilde{M} & 0 \\
    0 & Z
\end{pmatrix}, \ B^i= \begin{pmatrix}
    0 \\
    B
\end{pmatrix}, \ \tilde{B}_j = \begin{pmatrix}
    0 \\
    \tilde{B}
\end{pmatrix}$, with $Z= M_{N_f}^{N_f}$ being the massive flavor. Consequently, the constraints result in $Z=0$ if $B, \tilde{B} \neq 0$. Substituting all the information in the equation of motion \eqref{eq:meson-eom-Nf-N-plus-1}, one finds:
\begin{dmath}
    {\det{\tilde{M}} - \tilde{B}B = m {\Lambda}^{2N-1}_{\text{old}} = {\Lambda}^{2N}_{\text{new}}}
\end{dmath} which precisely reproduces the quantum deformation of the constraint in the case of $N_f = N$, cf. \eqref{eq:quantum-constraint-Nf-equal-N} \cite{ADS superpotentials},\cite{SUSY},\cite{SUSY2},\cite{SUSY3}.

\section{$\mathcal{N}=1$ Electromagnetic Dualities}
In this final section, there is a significant loan from SCFT techniques.
The goal is to analyze the case of $N_f > N$ and, in particular, $\frac{3N}{2}< N_f <3N$.

One knows that the $\beta(g)$ vanishes for $N_f = 3N$ where $g=g(N_f ,N)$. This is enough in 4 dimensions to guarantee the theory has conformal symmetry. \cite{SUSY}  By a technique called $\epsilon$-expansion, one can probe the theory for the values of $N_f= 3N - \epsilon$. Indeed, $\beta({g^{*}})=0$ for not only $N_f = 3N$ but also for $g^{*} = g^{*} (N_f, N)$ as far as $g^{*} \ll 1$ where the existence of an interacting IR SCFT is guaranteed. But the value of $g^{*}$ increases as $N_f$ decreases. The question is \textbf{"how much can one decrease $N_f$ and still have an SCFT?"}

The unitarity bounds in CFT imply $\Delta [\mathcal{O}] \geq \frac{d-2}{2}$ where $\Delta [\mathcal{O}]$ is the \textbf{scaling dimension} of the operator $\mathcal{O}$ defined by $\mathcal{O}(\lambda x) = {\lambda}^{-\Delta} \mathcal{O}(x)$ \cite{SUSY}. In case of SCFT the unitarity bound becomes $\Delta [\mathcal{O}] \geq \frac{3}{2}|R[\mathcal{O}]|$ \cite{RSCI}. The chiral operators saturate this bound $\Delta [\mathcal{O}] = \frac{3}{2} R[\mathcal{O}]$ as well as anti-chiral operators $\Delta[\bar{\mathcal{O}}]=-\frac{3}{2} R[\bar{\mathcal{O}}]$. This introduces a chiral ring structure over chiral operators since $R$ symmetry is additive for chiral operators \cite{RSCI}.

The process to determine the right $R$-symmetry in the theory is called \textbf{a-maximisation} and it guarantees the $R$-symmetry can be defined uniquely.

Given this $R[\Phi]=R[\tilde{\Phi}]= \frac{N_f - N}{N_f}$ resulting in $R[M]= \frac{2(N_f - N)}{N_f}$ and evidently $\Delta[M]=\frac{3(N_f - N)}{N_f}$.
The conformal window where $\frac{3N}{2} \leq N_f \leq 3N$ gives rise to $1 \leq \Delta[M] \leq 2 $. Bearing in mind that $\Delta=1$ means decoupling of the operator, one can assume the end of the conformal window at $N_f = \frac{3N}{2}$.

Now, the claim of Seiberg is that:

The $SU(N)$ gauge theory coupled to $N_f$ flavours $\Phi$ and $\tilde{\Phi}$ for $N_f \geq N+2$ has the same IR physics of the $SU(\tilde{N})$ gauge theory, with $\tilde{N} = N_f - N$, coupled to $N_f$ flavours $q,\tilde{q}$ and $N_f^2$ singlets $\tilde{M}$ with a superpotential $W= \lambda \tilde{q} \tilde{M} q$. 
One calls the first theory the \textbf{electric SQCD} and the second one the \textbf{magnetic SQCD}.

As for the mSQCD the conformal window is at $\frac{3 \tilde{N}}{2} < N_f < 3\tilde{N}$. An $\epsilon$-expansion at $N_f = 3\tilde{N} - \epsilon \tilde{N}$ shows that there exists a perturbative IR fixed point with: ${\tilde{g}}^{*2} = \frac{8 {\pi}^2}{3} \frac{\tilde{N}}{{\tilde{N}}^2 - 1} \left(1+\frac{N_f}{\tilde{N}}\right) \epsilon$ and ${\lambda}_{*}^2 = \frac{16 {\pi}^2}{3 \tilde{N}} \epsilon$. At this point, the superpotential has $R$-charge 2, which means that it is a marginal operator. 

On the other hand, if $\lambda = 0$ then $\tilde{M}$ has no interactions meaning it's a free field, and so $\Delta[\tilde{M}]=1$. Given that $R[q]=R[\tilde{q}]=\frac{N}{N_f}$ one finds $\Delta[W]=1 + \frac{3N}{N_f}$ which for $\frac{3N}{2} < N_f < 3N $ means that the superpotential is always a relevant operator and the fixed point $\tilde{g}^2 = \frac{8 {\pi}^2}{3} \frac{\tilde{N}}{\tilde{N}^2 - 1} \epsilon$ and ${\lambda}^2=0$ is rather an unstable one where if $\lambda$ is set near to zero it flows to the ${\lambda}_{*}^2 = \frac{16 {\pi}^2}{3 \tilde{N}} \epsilon$.

The phase diagram of the theory can be found in the following:

\begin{figure}
    \centering
    \includegraphics[width=0.5\linewidth]{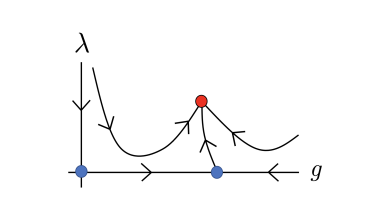}
    \caption{Phase diagram of magnetic SQCD \\ The blue points are the free and the conformal window fixed point, the red point in the middle of the plane is the fixed point induced by the superpotential, and it is conjectured to coincide with the fixed point of SQCD (figure from \cite{SUSY3})}
    \label{fig:phase-diagram}
\end{figure}

The fact that the $R$-charge of $M$ and $\tilde{M}$ are the same for $\frac{3N}{2}< N_f <3N$ is not accidental, as these two fields are the same, meaning the composite mesons of the SQCD become fundamental chiral singlets of the mSQCD. In the $N_f > 3N$ domain, the IR becomes free in SQCD, where the theory the same happens for mSQCD in the $N+1 <N_f <\frac{3N}{2}$ interval. The baryons of the two theories can be matched as:
\begin{dmath}
    {B^{i_1 ... i_N} \propto {\epsilon}^{i_1 ... i_N j_1 ... j_{\tilde{N}}} b_{j_1 ... j_{\tilde{N}}} \ \ \  , \ \ \ \ 
 \tilde{B}_{i_1 ... i_N} \propto {\epsilon}_{i_1 ... i_N j_1 ... j_{\tilde{N}}} \tilde{b}^{j_1 ... j_{\tilde{N}}}}
\end{dmath}

where $B$ transforms in the $\begin{pmatrix}
    N_f \\
    N
\end{pmatrix}$-antisymmetric representation that is equivalent to $\begin{pmatrix}
    N_f \\
    N_f - N
\end{pmatrix}$-antisymmetric representation. Another way to check the duality would be by giving a mass term to the electric/magnetic theory and observing that it results in a Higgs mechanism in the dual magnetic/electric theory.

\chapter{$\mathcal{N}=2$ Infrared Electromagnetic duality and Seiberg-Witten theory}\label{ch:Seiberg-Witten-theory}
In this section, the original approach to Seiberg-Witten curves is briefly discussed.

\section{$\mathcal{N}=2$ Gauge Theory with and without Matter}

\subsection{$\mathcal{N}=2$ SYM}

The Lagrangian for the $\mathcal{N} = 2$ gauge theory can be constructed from the Lagrangian of the $\mathcal{N}=1$ vector multiplet and chiral multiplet.

\begin{dmath}[label={eq:N2-SYM-Lagrangian}]
    \mathcal{L} = \frac{1}{8 \pi} \Im \ \text{Tr} \left\{\tau \left( \int d^2 \theta W^{\alpha} W_{\alpha} + 2 \int d^2 \theta d^2 \bar{\theta} \Phi^{\dagger} e^{-2V} \Phi \right)\right\} = \frac{1}{g^2} \text{Tr} \left\{-\frac{1}{4} F_{\mu \nu} F^{\mu \nu} + g^2 \frac{\theta}{32 \pi^2} F_{\mu \nu} {\tilde{F}}^{\mu \nu}\right\} + \frac{1}{g^2} \text{Tr} \left\{(D_{\mu} A)^{\dagger} D^{\mu} A -\frac{1}{2} [A^{\dagger} , A]^2 \right\} + \frac{1}{g^2} \text{Tr} \left\{-i \lambda {\sigma}^{\mu} D_{\mu} \bar{\lambda} - i \bar{\psi} {\bar{\sigma}}^{\mu} D_{\mu} \psi -i \sqrt{2}[\lambda, \psi] A^{\dagger} -i \sqrt{2} [\bar{\lambda}, \bar{\psi}] A\right\}
\end{dmath}
where $A$ is the scalar field, $\lambda , \psi$ are the Weyl fermionic fields, $A_{\mu}$ is the gauge field all in the $\mathcal{N}=2$ vector multiplet $(A, \psi , \lambda , A_{\mu})$ and $\tau=\frac{\theta}{2\pi} + \frac{4 \pi i}{g^2}$ is the complexified coupling.

But this is not the most general expression for the Lagrangian. To achieve the most general expression, one might move to the superspace formulation of the $\mathcal{N}=2$ gauge theory.

The coordinates on the superspace are: $(x, \theta , \bar{\theta} , \tilde{\theta} , \bar{\tilde{\theta}})$. By generalisation of the definitions of the $\mathcal{N}=1$ chiral superfield one can introduce the $\mathcal{N}=2$ chiral superfield $\Psi$ by the constraints: $\bar{D}_{\dot{\alpha}} \Psi = 0$ and $\bar{\tilde{D}}_{\dot{\alpha}} \Psi = 0$ where $\bar{D}_{\dot{\alpha}}$ and $\bar{\tilde{D}}_{\dot{\alpha}}$ are supercovariant derivatives with respect to $\theta$ and $\tilde{\theta}$.

Given this, any function of such coordinates can be expanded as:

\begin{dmath}
    \Psi (y , \theta , \tilde{\theta}) = \Psi^{(1)}(y, \theta) + \sqrt{2} {\tilde{\theta}}^{\alpha} \Psi^{(2)}_{\alpha} (y, \theta) + \tilde{\theta}^{\alpha} \tilde{\theta}_{\alpha} \Psi^{(3)} (y , \theta)
\end{dmath}
where $y^{\mu} = x^{\mu} + i \theta \sigma^{\mu} \bar{\theta} + i \tilde{\theta} \sigma^{\mu} \bar{\tilde{\theta}}$. By renaming $\Psi^{(1)} = \Phi$ which is an $\mathcal{N}=1$ chiral superfield, and the reality condition one concludes that $\Psi^{(2)}_{\alpha}=W_{\alpha}$ is the $\mathcal{N}=1$ superfield strength, and $\Psi^{(3)} (y , \theta) = \Phi^{\dagger} (y - i \theta \sigma \bar{\theta}, \theta, \bar{\theta}) e^{2g V(y - i \theta \sigma \bar{\theta}, \theta, \bar{\theta})} |_{\bar{\theta} \bar{\theta}}$.

One can now express the previous result \eqref{eq:N2-SYM-Lagrangian} in terms of the more abstract Lagrangian:

$$\mathcal{L} = \frac{1}{4 \pi} \Im{\text{Tr} \left\{\int d^2 \theta d^2 \tilde{\theta} \frac{1}{2} \tau \Psi^2\right\}}$$

However, the most general expression is written in terms of a general holomorphic function of $\Psi$, say $\mathcal{F}(\Psi)$:

\begin{dmath}
    \mathcal{L} =  \frac{1}{4 \pi} \Im{\text{Tr} \ \left\{ \int d^2 \theta d^2 \tilde{\theta} \mathcal{F}(\Psi) \right\}} = \frac{1}{8 \pi} \Im{ \left\{ \int d^2 \theta \mathcal{F}_{ab}(\Phi) W^{a \alpha} W^b_{\alpha} + 2 \int d^2 \theta d^2 \bar{\theta} (\Phi^{\dagger} e^{2gV})^a \mathcal{F}_a (\Phi) \right\}}
\end{dmath}

where $\mathcal{F}_a (\Phi) = \frac{\partial \mathcal{F}}{\partial {\Phi}^a}$ and $\mathcal{F}_{ab} (\Phi) = \frac{\partial^2 \mathcal{F}}{\partial \Phi^a \partial \Phi^b}$ in which $\mathcal{F}$ is called \textbf{ $\mathcal{N}= 2 $ prepotential} and $\Im{\mathcal{F}_{ab} (\Phi)}$ is the Kähler metric on the sapce of fields (It is a special type of metric in that the moduli space admits a special structure that preserves the complex structure covariantly).

To put it briefly, \underline{the purpose of the Seiberg-Witten theory} is \textbf{to determine the prepotential $\mathcal{F}(\Phi)$ in the low energy Wilsonian effective field theory} (after the SuperHiggs mechanism).

\subsection{$\mathcal{N}=2$ SQCD}

To account for the matter field one should introduce $\mathcal{N} = 2$ hypermultiplet that consists of a chiral $Q= (q, \psi_q , F_q)$ and an anti-chiral $\tilde{Q} = (\tilde{q}, \tilde{\psi}_q , F_{\tilde{q}})$ $\mathcal{N}= 1$ chiral multiplets. The general Lagrangian for $N_f$ hypermultiplets in an arbitrary representation of the gauge group $SU(N)$ can be written as:

\begin{dmath}
    \mathcal{L} = \int d^4 \theta ({Q_i}^{\dagger} e^{-2V} Q_i + \tilde{Q}_i e^{2V} {\tilde{Q}}^{\dagger}_i) + \int d^2 \theta ( \sqrt{2} \tilde{Q}_i \Phi Q_i + m_i \tilde{Q}_i Q_i) + h.c.
\end{dmath} 

in which the $i$ index counts the flavours.

The $\sqrt{2} \tilde{Q}_i \Phi Q_i$ term is absent in the case of the $\mathcal{N}=1$ Lagrangian, but to maintain $\mathcal{N} = 2$ supersymmetry, it must be included.
Eliminating the auxiliary fields $F_q , F_{\tilde{q}}$ results in the scalar potential written in terms of the $D$-terms:

\begin{dmath}
    V = \frac{1}{2} g^2 \sum_{a} D_a D^a
\end{dmath}
where 
\begin{dmath}
    D^a = \sum_{i=1}^{N_f} (q^{\dagger}_i T^a q_i - \tilde{q}_i T^a \tilde{q}^{\dagger}_i)
\end{dmath} 
where $T^a$ are gauge group $SU(N)$ generators.
Finally, the $SU(2)_R$ symmetry is not manifest in such a formulation, though it does exist at the Lagrangian level apart from the algebra.

To find the flat directions, one should first note that $q = \tilde{q} = 0$ is the only vacuum possible in case of massive quarks $m_i \neq 0$ and only $A$ can have nonzero vacuum expectation values. If the quarks are massless, then $A$ is zero for a vacuum state, but $q, \tilde{q}$ can acquire any vacuum expectation value.

The minimum of the $D$-term corresponds to $D^a = 0$. $q^{(i)}_{\alpha}$ ($\alpha$ is the color index) are $N$ vectors in $\mathbb{C}^{N_f}$. So we construct a matrix of scalar products:
\begin{dmath}
    {\sum_{i} q^{(i)}_{\alpha} {q^{\dagger (i)}_{\beta}} = q_{\alpha} \ . \ q^{\dagger}_{\beta} = (qq^{\dagger})_{\alpha \beta}}
\end{dmath}
that helps one write:
\begin{dmath}
    {\sum_{i} q^{\dagger}_i T^a q_i = \text{Tr} \left\{q q^{\dagger} T^a \right\}}
\end{dmath}
\begin{dmath}
    {D^a = \text{Tr} \ {(q q^{\dagger} - \tilde{q}^{\dagger} \tilde{q}) T^a} = 0}
\end{dmath}
for $T^a$ is an irreducible representation of the $SU(N)$ one can write:
\begin{dmath}
    {(q q^{\dagger} - (\tilde{q}^{\dagger} \tilde{q})) = c^2 1_N}
\end{dmath}

With the aid of $SU(N) \times SU(N_f) \times U(1)_R$ symmetry the $q q^{\dagger}$ can be put into diagonal form (and so $\tilde{q}^{\dagger} \tilde{q}$).

The solutions for $q$ and $\tilde{q}$ for $N_f < N$ are then:

\begin{dmath}
    {q = \begin{pmatrix}
    v^{(1)}_1 & 0 & ... & 0 \\
    0 & v^{(2)}_2 & ... & 0 \\
    0 & 0 & \ddots & 0 \\
    0 & 0 & ... & v^{(N_f)}_{N_f} \\
    0 & 0 & ... & 0 \\
    \vdots & \vdots & ... & \vdots \\ 
    0 & 0 & ... & 0 \\
\end{pmatrix}
 \ \ \
  \tilde{q} = \begin{pmatrix}
    {\tilde{v}}^{(1)}_1 & 0 & ... & 0 & 0 &...\\
    0 & {\tilde{v}}^{(2)}_2 & ... & 0 & 0 & ... \\
    \vdots &  \vdots &  \ddots  & \vdots & 0 & ... \\
    0 &  0  & ... & {\tilde{v}}^{(N_f)}_{N_f} & 0 & ...
    \end{pmatrix}}
\end{dmath}

the existence of at least $N - N_f $ zero eigenvalues in $q q^{\dagger}$ which results in $c = 0$ and so $v^{(i)}_i = {\tilde{v}}^{(i)}_i$.
As for the $N_f \geq  N$ case, the solutions can be found in the same manner and result in $c \neq 0$ necessarily and so $v^{(i) 2}_i - {\tilde{v}}^{(i) 2}_i = c^2$

\subsection{$\mathcal{N}=2$ Moduli Space with and without Matter, EFT}\label{subsec:N2-Moduli-Space-EFT}

In general, the $\mathcal{N}=2$ moduli space of the theory can always be written locally as the direct product of the vector multiplet moduli and the hypermultiplet moduli $\mathcal{M}^V \otimes \mathcal{M}^H$. The common names for $\mathcal{M}^V$ and $\mathcal{M}^H$ are the Coulomb branch and Higgs branch. A thorough analysis of all branches of the $\mathcal{N}=2$ SQCD moduli space is done in \cite{N=2 Moduli space}.

It's easy to see that SUSY imposes the local direct product structure on the moduli space. Suppose one has the following $\mathcal{N}=1$ chiral multiplets $Q^i$ , $\tilde{Q}^i$ with $i=1,..., \mathcal{H}$ and $\mathcal{N}=1$ chiral multiplets $\Phi^{a}$ with $a=1,...,\mathcal{V}$ while denoting their scalar components as $q^i , \tilde{q}^i$ and ${\phi}^a$. By means of $\mathcal{N}=1$ SUSY one can write the most general Kähler potential $K(Q^i , \tilde{Q}^i , {\Phi}^a , {Q}^{i \dagger} , \tilde{Q}^{i \dagger} , {\Phi}^{a \dagger})$. The kinetic term involves cross terms such as $\partial_i \partial_{\bar{a}} K . \partial_{\mu} q^i \partial^{\mu} \phi^{a \dagger} + c.c.$. However the $\mathcal{N}=2$ hypermultiplet and chiral multiplet dictate a term $\partial_i \partial_{\bar{a}} K . {\bar{\psi}}_{\tilde{q}^i} \slashed{\partial} {\lambda}^a$ in which ${{\psi}_{\tilde{q}^i}}$ belongs to the hypermultiplet and ${\lambda}^a$ is the gaugino. But oppositely, the $\mathcal{N}=1$ variation of this term can not be cancelled by any Lorentz invariant term bearing two derivatives and $A^a_{\mu}$ and scalar fields. This means ${\partial}_i {\partial}_{\bar{a}} K =0$ showing the $\mathcal{N}=2$ Kähler potential decomposes into:
\begin{dmath}
    K= K_H (Q^i , \tilde{Q}^i , {Q}^{i \dagger} , \tilde{Q}^{i \dagger}) + K_V (\Phi^a , \Phi^{a \dagger})
\end{dmath}

As far as the SUSY is maintained quantum mechanically(non-anomalous or non-accidental), such decomposition is maintained at the quantum level, and the moduli space is locally $\mathcal{M}^V \otimes \mathcal{M}^H$ \underline{even after the quantum corrections.}

The moduli coordinates on the Higgs branch can be written in terms of the following mesonic and baryonic operator vacuum expectation values:

\begin{dmath}
    M^i_j = \langle \tilde{Q}^a_j Q^i_a \rangle
\end{dmath}
\begin{dmath}
    B^{i_1 ... i_N} = \langle Q^{i_1}_{a_1} ... Q^{i_N}_{a_N} \rangle \epsilon^{a_1 ... a_N}
\end{dmath}
\begin{dmath}
    \tilde{B}_{i_1 ... i_N} = \langle\tilde{Q}^{a_1}_{i_1} ... \tilde{Q}^{a_N}_{i_N} \rangle \epsilon_{a_1 ... a_N}
\end{dmath}

where the baryon fields $B$ are defined if only $N_f \geq N$. Non-baryonic branch consists of the subspace where $B=\tilde{B}=0$. 
The Higgs branch is a hyperKähler quotient of the squark field space by the gauge group, where the $D$-terms and $F$-terms are the moment maps.

The $D$-terms constraints are:
\begin{dmath}
    {(*B) \tilde{B} = * M^{N}}
\end{dmath}
\begin{dmath}
    {M \ . \ (*B) = M \ . \ (* \tilde{B}) = 0}
\end{dmath}

where the $"."$ represents the contraction of an upper flavour index with a lower one and the $"*"$ represents:

\begin{dmath}
    (*B)_{i_{N+1} ... i_{N_f}}= \epsilon_{i_1 ... i_{N_f}} B^{i_1 ... i_N}
\end{dmath} 
and:

\begin{dmath}
    M^N = M^{[i_1}_{j_1} ... M^{i_N]}_{j_N}
\end{dmath}
in which the square brackets represent the anti-symmetrisation. The two constraints together imply $0=\tilde{B}(M \ . \ (*B))= M \ . \ (*(M^N)) = *(M^{N+1})$ which means that $\text{rank}(M) \leq N$.

The $F$-terms constraints are as follows:

\begin{dmath}
    M' \ . \ B = \tilde{B} \ . \ M' = 0
\end{dmath}
\begin{dmath}
    M \ . \ M' = 0
\end{dmath}

where 
\begin{dmath}
    M' = M - \frac{1}{N}\Tr{M} \mathbf{1}
\end{dmath}

The following results hold at the classical level for the Moduli space:
\begin{itemize}
    \item The Baryonic branch exists for $N_f \geq N$ with $\mathcal{V}=0$ and $\mathcal{H}=N_f N - N^2 +1$
    \item The non-baryonic branches exist for $N_f \geq 2$ with $\mathcal{V} = N - 1 - r$ and $\mathcal{H}=r(N_f - r)$ where $1 \leq r \leq \min{\{[N_f / 2], N-2 \}}$.
    \item The Coulomb branch exists for all $N_f$ with $\mathcal{V} = N-1$ and $\mathcal{H}=0$
\end{itemize}

where $\mathcal{H}$ is the number of masslass neutral hypermultiplets(the quaternionic dimension of $\mathcal{M}^H$) and $\mathcal{V}$ is the number of the massless $U(1)$ vector multiplets. Indeed $(\mathcal{V}, \mathcal{H})$ marks distinct branches of the moduli space. 

Two Higgs branches are distinct if and only if any path that connects them passes through a point of enhanced gauge symmetry. 

The moduli space can be modified in general due to quantum effects, though, as it was shown, the local product structure $\mathcal{M}^V \otimes \mathcal{M}^H$ survives the quantum corrections. Elevating the parameters of the Lagrangian to \textbf{spurions}, they become vector superfields, and using this technique, one can show that the Higgs branch is immune from quantum corrections and only the Coulomb branch is modified quantum mechanically.

To write down the Wilsonian low-energy effective action for the SYM, one starts with the general $\mathcal{N}=2$ lagrangian in terms of an analytic function $\mathcal{F}$ and the $\mathcal{N}=2$ chiral superfield $\Psi$:

\begin{dmath}    
\mathcal{L} =  \frac{1}{4 \pi} \Im \left\{ \text{Tr} \ \int d^2 \theta d^2 \tilde{\theta} \mathcal{F}(\Psi) \right\}
= \frac{1}{8 \pi} \Im \left\{ \int d^2 \theta \mathcal{F}_{ab}(\Phi) W^{a \alpha} W^b_{\alpha} + 2 \int d^2 \theta d^2 \bar{\theta} (\Phi^{\dagger} e^{2gV})^a \mathcal{F}_a (\Phi) \right\} 
\end{dmath}
Consequently, the metric on the Coulomb branch is given by:

\begin{dmath}
    g_{i \bar{j}} = \Im \left\{\frac{\partial^2 \mathcal{F}}{\partial a_i \partial a_j}\right\}
\end{dmath}

in particular for $SU(2)$ one has locally:

\begin{dmath}
    {ds^2 = \Im{\tau} da d\bar{a} \ \ \ \ \ \text{where} \ \ \ \ \ \tau(a)=\frac{\partial^2 \mathcal{F}}{\partial a^2}}
\end{dmath}

To find the one-loop[exact] quantum prepotential, one starts with the anomalous $U(1)_R$ symmetry. The current divergence for such symmetry is:

\begin{dmath}
    \partial^{\mu} J^5_{\mu} = -\frac{N}{8 {\pi}^2} F_{\mu \nu} \tilde{F}^{\mu \nu}
\end{dmath}
while the effective Lagrangian should be changed precisely by the same amount under symmetry transformation, meaning:

\begin{dmath}
    \delta \mathcal{L}_{eff} = - \frac{\alpha N}{8 {\pi}^2} F_{\mu \nu} \tilde{F}^{\mu \nu}
\end{dmath}

The $U(1)_R$ symmetry breaks down to a $\mathbb{Z}_{4N - 2N_f}$ in presence of $N_f$ flavor of matters and particularly to $\mathbb{Z}_{4N}$.
The symmetry transformation at the effective level results in:

\begin{dmath}
    \mathcal{F}''(e^{2i \alpha} A) = \mathcal{F}''(A) - \frac{2 \alpha N}{\pi}
\end{dmath}
and for infinitesimal $\alpha$ one finds:

\begin{dmath}
    \frac{\partial^3 \mathcal{F}}{\partial A^3} = \frac{N}{\pi} \frac{i}{A}
\end{dmath}

which by integration gives rise to:

\begin{dmath}[label={eq:perturbative-prepotential}]
    \mathcal{F}_{\text{pert}}(A) = \frac{i}{2 \pi} A^2 \ln{\frac{A^2}{\Lambda^2}}
\end{dmath}

in which $\Lambda$ is a dynamically generated(dimensional transmutation).
This result is \textbf{one-loop} exact since the $\beta$ function of the $\mathcal{N}=2$ is proven not to receive higher order perturbative quantum corrections. This leaves the door still open for \textbf{non-perturbative} corrections. \cite{Non-perturbative beta function}

Given that the $k$-instanton contribution is proportional to $e^{-8 \pi^2 k / g^2}$ and that the $\beta$-function is $\beta(g) = - g^3 /4 \pi^2 $ one can write:

\begin{dmath}
    e^{-8 \pi^2 k / g^2} = (\frac{\Lambda}{a})^{4k}
\end{dmath}
and by elevating $\Lambda$ to a spurious field with charge 2 under the $U(1)_R$ symmetry, the prepotential should transform under the $U(1)_R$ with charge 4. This gives rise to:

\begin{dmath}
    \mathcal{F}(A) = \frac{i}{2 \pi} A^2 \ln{\frac{A^2}{\Lambda^2}} + \sum^{\infty}_{k=1} \mathcal{F}_k (\frac{\Lambda}{A})^{4k} A^2
\end{dmath}

where $\mathcal{F}_k$ are field-independent coefficients. $\mathcal{F}_1$ can be shown not to vanish.

The crucial observation is that given $\mathcal{F}(A)$, the complexified coupling constant $\tau$ turns out to be a multi-valued function in terms of $a$; however, the metric $\Im\{\tau\}$ is single-valued and more specifically a \textbf{harmonic} function. Since harmonic functions do not have any global minimum so they become negative, which contradicts the fact that the kinetic energy has to be positive-definite. Finally, one concludes that $\Im{\tau}$ can not be defined globally and a change of chart over the moduli space is inevitable \cite{SW theory}, which means the moduli space has a nontrivial topology. The new charts are related to the old ones by \textbf{duality transformations}.

In section \ref{Witten---Seiberg-Theory}, one will find the exact geometry of the quantum moduli space with and without matter.

\subsection{Central Charges in $\mathcal{N}=2$ SYM}

Now we can concentrate on the central piece of this section, \textbf{computation of the central charge in $\mathcal{N} = 2$ SYM and SQCD} \cite{SW theory}. 

The $\mathcal{N} = 2$ SYM supercurrent is:

\begin{dmath}
    g^2 S^{\rho}_{(1)} = - \frac{i}{2} ({\bar{\lambda}}^a_1 \bar{\sigma}^{\rho} \sigma^{\mu \nu} \epsilon + \bar{\epsilon} \bar{\sigma}^{\mu \nu} {\bar{\sigma}}^{\rho} {\lambda}^a_1) F^a_{\mu \nu} - (\bar{\epsilon} \bar{\sigma}^{\rho} \lambda^a_1 + {\bar{\lambda}}^a_1 \bar{\sigma}^{\rho} \epsilon) A^{\dagger} T^a A + \sqrt{2} \epsilon \sigma^{\mu} {\bar{\sigma}}^{\rho} \lambda^a_2 D_{\mu} A^{\dagger a} + \sqrt{2} \bar{\lambda}^a_2 {\bar{\sigma}}^{\rho} \sigma^{\mu} \bar{\epsilon} D_{\mu} A^a 
\end{dmath}

\begin{dmath}
    g^2 S^{\rho}_{(2)} = - \frac{i}{2} ({\bar{\lambda}}^a_2 \bar{\sigma}^{\rho} \sigma^{\mu \nu} \epsilon' + \bar{\epsilon}' \bar{\sigma}^{\mu \nu} {\bar{\sigma}}^{\rho} {\lambda}^a_2) F^a_{\mu \nu} - (\bar{\epsilon}' \bar{\sigma}^{\rho} \lambda^a_2 + {\bar{\lambda}}^a_2 \bar{\sigma}^{\rho} \epsilon') A^{\dagger} T^a A - \sqrt{2} \epsilon' \sigma^{\mu} {\bar{\sigma}}^{\rho} \lambda^a_1 D_{\mu} A^{\dagger a} - \sqrt{2} \bar{\lambda}^a_1 {\bar{\sigma}}^{\rho} \sigma^{\mu} \bar{\epsilon}' D_{\mu} A^a
\end{dmath} 

that can be split further into its Grassmann components by $S^{\rho} = \epsilon^{\alpha} S^{\rho}_{\alpha} + \bar{\epsilon}_{\dot{\alpha}} \bar{S}^{\rho \dot{\alpha}} $. Doing so, the $\mu =0$ component would be:

\begin{dmath}
    g^2 S^0_{(1) \alpha} = -i (\Vec{\sigma} {\sigma}_y {\lambda}^{\dagger a}_1)_{\alpha} \ . \ (i \Vec{F}^a + \Vec{\tilde{F}}^a) + \sqrt{2} \lambda^a_{2 \alpha} D_0 A^{\dagger a} + \sqrt{2} (\Vec{\sigma} . \Vec{D} A^{\dagger a} \lambda^a_2)_{\alpha} + i (\sigma_y \lambda^{\dagger a}_1)_{\alpha} A^{\dagger} T^a A
\end{dmath}

where $\Vec{F}^a = F^{a 0i}$ and $\Vec{\tilde{F}}^a = \tilde{F}^{a 0i}$ and that $\bar{\lambda}^{\dot{\alpha}}_1 = \epsilon^{\dot{\alpha} \dot{\beta}} \bar{\lambda}_{1 \dot{\beta}} = i ({\sigma}_y \lambda^{\dagger}_1)^{\dot{\alpha}}$.

Simply, $S^0_{(2) \alpha}$ is obtained by replacing $\lambda_1 \rightarrow \lambda_2 $ and $\lambda_2 \rightarrow -\lambda_1$.

Given the definition of the Noether charge (which is, in our case, a supercharge) that:
\begin{dmath}
    Q_{(1) \alpha} = \int d^3 x S^0_{(1) \alpha} (\Vec{x} , 0)
\end{dmath}

one can find the supercharge anti-commutators:

\begin{dmath}
    \{Q_{(1) \alpha} , Q_{(2) \beta} \} = - \frac{2 \sqrt{2}}{g^2} \epsilon_{\alpha \beta} \int d^3 x (i F^{a 0i} + \tilde{F}^{a 0i}) D_i A^{\dagger a} = -\frac{2 \sqrt{2}}{g^2} \epsilon_{\alpha \beta} \int d^3 x {\partial}_i [(i F^{a 0i} + \tilde{F}^{a 0i}) A^{\dagger a}]
\end{dmath} 

and so:

\begin{dmath}
    {Q_{\text{electric}} = -\frac{1}{ag} \int d^3 x {\partial}_i (F^{a 0i} A^a) = g n_e}
\end{dmath}
\begin{dmath}
    {Q_{\text{magnetic}} = -\frac{1}{ag} \int d^3 x {\partial}_i (\tilde{F}^{a 0i} A^a) = \frac{4 \pi}{g} n_m}
\end{dmath}

where $a^2$ is the vacuum expectation value of the scalar field $A^2 (\Vec{x})$ at $|\Vec{x}| \gg a^{-1}$.
The results can be extended to the case of $\theta$-vacuum (instantonic vacuum) by the explicit inclusion of a term in the Lagrangian: $\frac{\theta g^2}{32 \pi^2} F^a_{\mu \nu} \tilde{F}^{a \mu \nu}$.

Finally, one can conclude that: 

\begin{dmath}[label={eq:central-charge}]
    {Z = a(n_e + \tau n_m) \ \ \ , \text{where} \ \ \ \tau = \frac{\theta}{2 \pi} + \frac{4 \pi i}{g^2}} 
\end{dmath}

\begin{dmath}[label={eq:BPS-bound}]
    M \geq \sqrt{2} |Z| = \sqrt{2} |a (n_e + \tau n_m)|
\end{dmath}

that is the \textbf{BPS bound}.

One can notice a duality in the electric and magnetic charges (whose survival can be questioned at the quantum level in the microscopic $\mathcal{N}=2$ SYM). Namely: $n_e \leftrightarrow n_m$ ,  $\tau^{-1} \leftrightarrow \tau$, $a \tau \leftrightarrow a$.

\subsection{Witten-Seiberg Theory}
\label{Witten---Seiberg-Theory}
The task of the Seiberg-Witten theory is to investigate the previous section duality at the quantum level, utilising Effective Field Theory techniques in the IR regime. One must bear in mind that the aforementioned electromagnetic duality, known as \textbf{Montonen-Olive} duality, is fully realised only in the case of $\mathcal{N}= 4$ SYM\cite{BPST}.

The $\mathcal{N}=2$ SYM potential is already shown in \eqref{eq:N2-SYM-Lagrangian} to be:
\begin{dmath}
    V(A) = \frac{1}{2g^2} \text{Tr} \ {[A^{\dagger},A]^2}
\end{dmath}

The vacuum solutions of such a potential satisfy $[A, A^{\dagger}] = 0$, meaning $A \in \mathfrak{h}$, where $\mathfrak{h}$ is the Cartan subalgebra of the gauge group $G$.
Let $H$ denote the group generated by the Cartan subalgebra, the so-called \textbf{maximal torus} of $G$.
The maximal torus of the generic group $SU(N)$ is $U(1)^{N-1}$.

The group of $G/H$ does not leave the vacuum invariant, and spans partially(the compact component, so to speak) the vacuum manifold.
The symmetry breaks down from $G$ to $T \times W_G$, where $W_G$ is the Weyl group of $G$ and can be defined as ${N_G (T)}/T$ in which $N_G (T)$ is the normaliser of the maximal torus $T$ in $G$.
For the generic case of $G = SU(N)$, the Weyl group is $S_{N}$, which is the permutation group of $N$ elements.

A general solution to the vacuum is of the form:
\begin{dmath}
    {\langle A \rangle = \begin{pmatrix}
    a^1 & 0 & ... & 0 \\
    0   & a^2 & ... & 0 \\
    \vdots &  \vdots & \ddots & \vdots  \\
    0 & 0 & ... & a^N
\end{pmatrix} \ \ \ \ \text{where} \ \ \ \text{Tr} \ {\langle A \rangle} = \sum_{i=1}^{N} a^i = 0}
\end{dmath}

But due to the residual $W_G$ symmetry, one needs a $T \times W_G$-invariant set of coordinates on the non-compact component of the vacuum manifold; the so-called \textbf{moduli space}.

Such coordinates are mathematically the \textbf{Chern Polynomials} in $\langle A \rangle$ obtained by:

\begin{dmath}
    \det \left( \lambda - \langle A \rangle \right) = \sum_{k=0}^{N} \lambda^{N-k} c_k (\langle A \rangle)
\end{dmath}
where $c_k (\langle A \rangle)$ are the Chern Polynomials in $\langle A \rangle$.
For the $SU(N)$ case, one finds the proper coordinates to be:

\begin{dmath}[label={eq:Chern-polynomials-SUN}]
    {c_2 (\langle A \rangle) = \sum_{i < j} a^i a^j \ \ \ \ \ \ \ c_3 (\langle A \rangle) = - \sum_{i < j < k} a^i a^j a^k \ \ \ \ \ \ \ \ \cdots \ \ \ \ \ \ \ c_N (\langle A \rangle) = (-1)^N a^1 a^2 ... a^N}
\end{dmath}

which are invariant under permutations of $a^i$. So the moduli space is a complex \textbf{Orbifold} $\mathcal{M}^V = \mathbb{C}^{N-1} / S_N$ for the generic case of $SU(N)$. In particular, for $SU(2)$ the moduli space is $\mathcal{M}^V = \mathbb{C}/S_2$.

In Section \ref{subsec:N2-Moduli-Space-EFT}, we ended our discussion with the necessity of duality transformations. Now that we know the topology of the moduli space, we can implement such transformations.

Before that, one introduces the Legendre transformation of the prepotential($\phi$ is another way to represent the same field of $A$ throughout this thesis):
$$\mathcal{F}_D (\phi_D) = \mathcal{F}(\phi) - \phi \phi_D \ \ \ \ \ \text{where} \ \ \ \ \ \phi_D = \mathcal{F}' (\phi) \ \ , \ \ \mathcal{F}^{'}_D (\phi_D) = - \phi$$

By Bianchi identity $\Im\left\{\mathcal{D}W\right\} = 0$. To impose such a constraint, one can introduce a vector superfield $V_D$ as the Lagrange multiplier and write:

\begin{dmath}
    \frac{1}{4\pi} \Im\left\{\int d^4 x d^4 \theta V_D \mathcal{D}W\right\} = \frac{1}{4 \pi} \Re\left\{\int d^4 x d^4 \theta i \mathcal{D}V_D W\right\} = - \frac{1}{4 \pi} \Im\left\{\int d^4 x d^2 \theta W_D W\right\}
\end{dmath}

By adding this to the action and integrating the $W$ superfield from the path integral, the Lagrangian turns into: 
\begin{dmath}
\frac{1}{8\pi} \Im\left\{\int d^2 \theta \left(-\frac{1}{\tau(\phi)}\right) W^2_D\right\}    
\end{dmath}

so that the dual complexified coupling can be written as $\tau_D(\phi_D)= \mathcal{F}''_D(\phi_D) = - \frac{d \phi}{d \phi_D} = - \frac{1}{\mathcal{F}'' (\phi)} = - \frac{1}{\tau(\phi)}$.

Given this, the dual action can be summarised into:
\begin{dmath}
    \frac{1}{4 \pi} \Im \left\{\int d^4 x \left[\frac{1}{2} \int d^2 \theta \mathcal{F}^{''}_D (\phi_D) W^{\alpha}_D W_{D \alpha} + \int d^2 \theta d^2 \bar{\theta} {\phi}^{\dagger}_D {\mathcal{F}}^{'}_D ({\phi}_D) \right]\right\}
\end{dmath}

The dual transformation $\tau \rightarrow - \frac{1}{\tau}$ along with the replacement $\tau \rightarrow \tau +1 $ generate a group $SL(2, \mathbb{Z})$ action on the complex coupling constant $\tau$ such that a generic group action looks like:

\begin{dmath}
    {g \ . \ \tau = \frac{a \tau + b}{c \tau + d} \ \ \ \ \ \text{where} \ \ \ \ \ g= \begin{pmatrix}
    a & b \\
    c & d
\end{pmatrix} \in SL(2, \mathbb{Z})}
\end{dmath}

where $ad-bc=1$ and $a,b,c,d \in \mathbb{Z}$.

These transformations can be represented by: $S=\begin{pmatrix}
    0 & 1 \\
    -1 & 0 
\end{pmatrix} \ , \  T=\begin{pmatrix}
    1 & 1 \\
    0 & 1
\end{pmatrix} $ and are called $S$ and $T$ transformations respectively. For the $SU(N)$ gauge theory the group becomes $Sp(2r, \mathbb{Z})$ where $r=\rank{SU(N)}$

This way, one can elevate the moduli space to a fibre bundle with structure group $SL(2, \mathbb{Z})$. The fibre bundle is the non-twisted\footnote{will be shown in a while} product $\mathcal{M}^V \times W$, where  $W= \mathbb{C}^2$, whose coordinates are denoted by $a, a_D$ and $u= \langle\text{Tr} A^2 \rangle$ is a local chart on $\mathcal{M}^V$. For the more general case of $SU(N)$ gauge theory with rank $r$, the moduli space can be elevated to a vector bundle with fibres $W = \mathbb{C}^{2r}$. Choosing coordinates $a^i, a_{D j} = \partial \mathcal{F} / \partial a^j$ on $W$ and endowing it with a symplectic form $\omega = \frac{i}{2} \sum^{r}_{i=1} (d a^i \land d \bar{a}_{Di} - d a_{D i} \land d \bar{a}^i)$ and holomorphic 2-form $\omega_h = \sum^{r}_{i=1} d a^i \land d a_{Di}$. A local trivialisation on $\mathcal{M}^V$ would be $u^s (s=1,...,r)$. A section in such local trivialisation can be written as $f: U \subset \mathcal{M}^V \rightarrow W$ such that $f^{*} (\omega_h) = 0$, where this condition should be fulfilled to ensure $a_{Dj} = \partial \mathcal{F} / \partial a^j$. Finally, the metric turns out to be 

\begin{dmath}
    g_{r \bar{s}} = \Im\left\{\frac{\partial a_{D i}}{\partial u^r} \frac{\partial \bar{a}^i}{\partial {\bar{u}}^s} \right\}
\end{dmath}

which coincides with the components of the metric associated with $f^{*} (\omega)$. From the BPS relation \eqref{eq:BPS-bound} one knows that $M \geq \sqrt{2}|Z|$ where, by \eqref{eq:central-charge}, $Z=a(n_e + \tau n_m)$. In the effective theory, one finds $Z=a^i n_{e, i} + a_{Dj} n^j_m$ for the gauge group $SU(N)$ with rank $r$. Under $M \in Sp(2r,\mathbb{Z})$, $v=(a_D , a)^{T}$ transforms to $Mv$ consequently to leave $Z$(which is an observable) invariant by $w=(n_m , n_e)$ transforms to $w M^{-1}$. The $\tau_{ij}$ matrix transforms by $M \in Sp(2r, \mathbb{Z})$ \cite{SW2}. This can be elegantly interpreted as the transformation of the period matrix of the genus $r$ Riemann surfaces under the action of the monodromy group on the moduli space of the genus $r$ Riemann surfaces.

\
\

The analysis of the \textbf{quantum moduli space} and its singularities is in order\cite{SUSY2}.

\
\

For that matter, one should make a few points about the $R$-symmetries of the SYM and SQCD vacuum.
First, the $U(1)_R$ part of the $U(2)_R$ symmetry group in the $\mathcal{N}=2$ case does not have the merit of becoming non-anomalous by adding flavors(hypermultiplets).
Consequently, $U(1)_R$ breaks down to $\mathbb{Z}_{4N}$ which along with the nonabelian part amounts to $SU(2) \times \mathbb{Z}_{4N}$. (In the case of SQCD the discrete symmetry group is $\mathbb{Z}_{4N-2N_f}$).
One should keep in mind that the $SU(2)_R$ already contains a $\mathbb{Z}_2$ symmetry that should be modded out. So the reamining symmetry group is $\frac{SU(2)_R \times \mathbb{Z}_{4N}}{\mathbb{Z}_2}$. 
But given that the moduli space coordinate is $u=\frac{1}{2} \text{Tr} {\phi^2}$, such a symmetry group is broken further down. In case of $N=2$ the final surviving $R$-symmetry group is $\frac{SU(2)_R \times \mathbb{Z}_{4}}{\mathbb{Z}_2}$ in which the $\mathbb{Z}_8$ is broken down to $\mathbb{Z}_4$, which means a $\frac{\mathbb{Z}_8}{\mathbb{Z}_4} = \mathbb{Z}_2$ transformation leads to distinct vacuums. This  $\mathbb{Z}_2$ transformation plays a critical role in the singularity analysis of the moduli space.

\
\
 
In case of SQCD the electrically charged matter fields(flavors) and their gauge interaction terms are explicitly added to the Lagrangian $\sqrt{2} n_e A M \tilde{M}$ in a local fashion, where $A$ is the $\mathcal{N}=1$ chiral multiplet in the $\mathcal{N}=2$ vector multiplet $\Psi$ and $M, \tilde{M}$ are $\mathcal{N}=1$ chiral muultiplets in the $\mathcal{N}=2$ hypermultiplet.
\
\

The prepotential computed previously, \eqref{eq:perturbative-prepotential}, results in:
\begin{dmath}
    {a_D (a) = \frac{\partial \mathcal{F}}{\partial a}= \frac{i}{\pi} a (\ln{\frac{a^2}{{\Lambda}^2}}+1)}
\end{dmath}

making a counter at $u= \infty$ one observes that:
\begin{dmath}
    a_D \rightarrow -a_D + 2a
\end{dmath}
\begin{dmath}
    a \rightarrow -a
\end{dmath}

that stands for a monodromy at infinity that acts on the vector $(a_D, a)$:
\begin{dmath}
    {\begin{pmatrix}
    a_D \\
    a
\end{pmatrix} \rightarrow M_{\infty} \begin{pmatrix}
    a_D \\
    a
\end{pmatrix} \ \ \ \text{where} \ \ \ M_{\infty} = \begin{pmatrix}
    -1 & 2 \\
     0 & -1
\end{pmatrix} \in SL(2, \mathbb{Z})}
\end{dmath}
and in terms of the $T$ transformation one can write $M_{\infty} = - T^{-2}$.

Given that the singularities come in pairs due to the $\mathbb{Z}_2$ symmetry, one should expect that the other singularity should coincide with the classical singularity at $u=0$. But this is flawed. If such a unique singular point exists, due to the general relation $M_{\infty}= M_1 M_2 ... M_{n}$, one expects $M_0 = M_{\infty}$. Since $a^2$ is left invariant under $M_{\infty}$, then $u=a^2$ is a good coordinate system on the whole moduli space and not only on the weakly coupled region. Hence $\mathcal{F}(a)$ will be a holomorphic function of $a$ and consequently $\Im\{\tau(a)\}$ is a harmonic function, which is unacceptable since it is against the positive-definiteness of the kinetic term.

The next possibility would be two singularities at $u= \pm u_0$. The question is the nature of these singularities. Suppose these points stand for enhanced gauge symmetry as the $\pm u_0$ points in the classical moduli space where the gauge symmetry is restored. In fact, one is guessing that quantum corrections might shift the classical singularities to two finite points. An interacting non-Abelian Coulomb phase in the IR ends with an SCFT (IR fixed point), and necessitates a conserved $R$-current. At such a singularity, one expects the dimension of $u$ to be zero as one is dealing with an SCFT. Given that in any SCFT, $D[O] \propto R[O]$ for a generic operator $O$, one has $D[u] \propto R[u]$. But since $R[\phi]=2$ and hence $R[u]=4$, $D[u] \neq 0$. Meaning importantly that there's no interacting non-Abelian IR fixed point, and these massless particles, whatever they are, cannot be massless gauge bosons.

\
\

Remembering that one still has dyons in the spectrum, one can think of monopoles and, in general, dyons as the next candidates for such massless particles. Furthermore, these particles \textbf{must belong to a hypermultiplet} and not an $\mathcal{N}=2$ chiral multiplet that contains spin-$1$ particles for the same reason given in the previous paragraph to exclude the interacting non-Abelian IR fixed point possibility.

\
\

A similar analysis rules out the existence of three singularities at $u=0, \pm u_0$.

Now, before trying to find the monodromies around each of the singularities, one should bear in mind that the topological sectors cannot be added to the Lagrangian in a local fashion like the purely electrically charged hypermultiplets as $\sqrt{2} n_e A M \tilde{M}$. But fortunately, thanks to a duality transformation, which stands for transition to another chart, one can pin down the non-perturbative regions. Given the proper chart the dyons transform into purely electrically charged particles and as their mass goes to zero, according to the BPS bound \eqref{eq:BPS-bound}, ${M} \geq \sqrt{2} |Z|= \sqrt{2} |a n_e|$, one finds $a(u) =0$ at a finite $u= u_q$ at which $a(u)$ is not singular so $a(u)$ can be expanded around $u_q$, say $a(u)= c_q (u - u_q)$. (For convenience, one could denote the new variables as $a'(u), a'_D(u)$ to clarify that the $SL(2, \mathbb{Z})$-rotated $a(u)$ is at the focus here)

By means of $\tau(a(u))= - \frac{i}{\pi} \ln{\frac{a(u)}{\Lambda}}$ near the $u_q$ vacuum, whose containing chart (after the $SL(2 ,\mathbb{Z})$ transition) represents purely electric theories, one finds $a_D(u) = -\frac{i}{\pi} a(u) \ln{\frac{a(u)}{\Lambda}} + \frac{i}{\pi} a(u)$. For the monodromy group, the loop $(u- u_q) \rightarrow e^{2 \pi i} (u - u_q)$ results in the following:
\begin{dmath}[label={eq:monodromy-electric-aD}]
    a_D (u) \rightarrow a_D (u) + 2a(u)
\end{dmath}
\begin{dmath}[label={eq:monodromy-electric-a}]
    a(u) \rightarrow a(u)
\end{dmath}

Now under a general $SL(2, \mathbb{Z})$ one can always transform any state of charge $(n_m , n_e)$ into $(n^{'}_m , n^{'}_e)$:

\begin{dmath}
    {\begin{pmatrix}
    a^{'}_D \\
    a'
\end{pmatrix} = \begin{pmatrix}
    \alpha a_D + \beta a \\
    \gamma a_D + \delta a
\end{pmatrix} , \ \ \ \ \ \ \begin{pmatrix}
    n^{'}_m \\
    n^{'}_e 
\end{pmatrix} = \begin{pmatrix}
    n_m \delta - n_e \gamma \\
    -n_m \beta + n_e \alpha 
\end{pmatrix}}
\end{dmath}

\

where $\alpha \delta - \beta \gamma = 1$ helps us to transform all states into purely electrically charged states whose monodromy we already found in \eqref{eq:monodromy-electric-aD} and \eqref{eq:monodromy-electric-a}.

Consequently, the monodromy matrix for a general state of charge $(n_m, n_e)$ is simply:

\begin{dmath}
    \begin{pmatrix}
    1+2n_e n_m & 2 n^2_e \\
    -2n^2_m & 1-2n_e n_m
\end{pmatrix}
\end{dmath}

The monodromy for singularities at $\pm u_0$ can be found easily by taking two massless dyons with charges $(m,n), (m',n')$, since they should satisfy $M_{u_0} M_{-u_0} = M(\infty)$ where $u=-u_{0}$ is chosen as the base point otherwise the monodromy matrix composition would be in a different order. The explicit equations lead to:

\begin{dmath}
    {\begin{pmatrix}
    1+2mn & 2 n^2 \\
    -2 m^2 & 1-2mn
\end{pmatrix} = \begin{pmatrix}
    -1 & 2 \\
    0 &  -1
\end{pmatrix} \begin{pmatrix}
    1-2m' n' & -2n^{'2} \\
    2m^{'2} & 1+2m'n'
\end{pmatrix}}
\end{dmath}

The solutions being: 
\begin{dmath}
    (m,n) : (1, n), (-1, n), (-1, n), (1, n)
\end{dmath}
\begin{dmath}
    (m', n') : (1, n-1), (1,-n-1), (-1, n+1), (-1, -n+1)
\end{dmath}

Finally, $M_{\infty}$ transforms the electric charge by two units, which means that the electric charge is defined modulo 2, which leaves only a finite number of solutions after identifying seemingly distinct solutions.

The critical point of the moduli space singularity analysis is that the moduli space is a modular curve. One such variety is the real torus, accompanied by a necessarily integrable complex structure that is a compact Riemann surface. The monodromy matrices $M_{\pm u_{0}}, M_{\infty}$ form a subgroup of the modular group called the \textbf{Fuchsian group} (or the \textbf{congruence group}) denoted by $\Gamma (2) \subset SL(2, \mathbb{Z})$ \footnote{One could more properly call the projective special linear group $PSL(2,\mathbb{Z})$ the modular group where the group is divided by a $\mathbb{Z}_2$ factor but it does not matter for this thesis so we keep calling the $SL(2, \mathbb{Z})$ the modular group.} with index $[SL(2, \mathbb{Z}), \Gamma(2)]=6$, meaning the quotient $\mathfrak{h}/ {\Gamma(2)}$ is the six-fold covering of the $u$-plane \cite{SW differential form}. It is not hard to check that the quantum moduli space is the quotient of the complex upper half plane by the Fuchsian group, $\mathfrak{h} / \Gamma (2)$. Such space has three cusps that stand for the three singularities and the surface can be compactified by addition of finite number of points(3 in our case) to form a compact surface that is the \textbf{modular curve} $X(2) = \mathfrak{h}^{*} / \Gamma(2)$ where $\mathfrak{h}^{*} = \mathfrak{h} \cup \mathbb{Q} \cup \{\infty \}$ \cite{Modular curves}. 
This in turn is the \textbf{moduli stack of elliptic curves} where each point corresponds to \textit{a framed lattices isomorphism class} \footnote{two lattices $\Lambda, \Lambda'$ are isomorphic iff there exists $w \neq 0 \in \mathbb{C}$ such that $\Lambda'= w \Lambda $} \textit{modded out by $\Gamma(2)$ action on the framing}, that is in bijective relation with \textit{the framed elliptic curves isomorphism classes modded out by the same group action on the framing} \footnote{two elliptic curves are isomorphic if and only if their relevant framed lattices under the Weierstrass functor are isomorphic} \cite{Modular curves}. 
\
\

A typical point of the modular curve $X$ can be algebraically characterized as a specific type of complex projective plane curve in $P (\mathbb{C}^2)$ called an \textbf{elliptic curve}:

\begin{dmath}[label={eq:SU2-elliptic-curve}]
    y^2 (x , u) = (x- u_0)(x+ u_0)(x- u)
\end{dmath}

where $u \in X(2)$, resulting in a family of curves $E_u$ whose period matrix is by construction the gauge coupling $\tau(u)$, so naturally $\Im{\tau(u)} > 0$ as a result of \textbf{Riemann's second relation}. 

As a result of de Rham's theorem, Poincaré duality, and the universal coefficient theorem of homology theory one knows that for the vector space of closed holomorphic $1$-forms up to an exact term, on the Riemann surface $X$, $\dim{H^{10}(X, \mathbb{Z})} = g$ (the same applies to the vector space of closed anti-holomorphic $1$-forms up to an exact term, $\dim{H^{01}(X, \mathbb{Z})} = g$).

In our case, the genus of the moduli space $g=1$ as for a torus. 
One can take such a holomorphic $1$-form and its following period integrals to be \cite{SW differential form}:
\begin{dmath}
    {\omega = -\frac{1}{2 \sqrt{2} \pi} \frac{dx}{y(x,u)}}
\end{dmath}
\begin{dmath}
    {\bar{\omega}_D (u) = \oint_{\beta} \omega \ \ \ \ \ \bar{\omega}(u) = \oint_{\alpha} \omega}
\end{dmath}

where $\alpha$ and $\beta$ are the basis for the vector space of the closed 1-cycles up to a boundary term on the Riemann surface $X$ in $H_{1} (X, \mathbb{Z})$.

By the relation $\tau = \frac{\partial a_D}{\partial a}$ one gets:
\begin{dmath}
    {\bar{\omega}_D (u) = \frac{\partial a_D (u)}{\partial u}  \ \ \ \ \ \ \bar{\omega}(u) = \frac{\partial a(u)}{\partial u}}
\end{dmath}
and finally:
\begin{dmath}
    {a_D (u) = \oint_{\beta} \lambda_{SW} \ \ \ \ \ \ a(u)= \oint_{\alpha} {\lambda}_{SW}}
\end{dmath}
where ${\lambda}_{SW} = \frac{1}{\sqrt{2} \pi} (x-u) \frac{dx}{y(x, u)}$ up to exact forms is a meromorphic 1-form and is called the \textbf{Seiberg-Witten Form}. Therefore:

\begin{dmath}
    {a (u) = \frac{\sqrt{2}}{\pi} \int_{-1}^{1} dx \frac{\sqrt{x-u}}{\sqrt{x^2 - 1}} \ \ \ a_D (u) = \frac{\sqrt{2}}{\pi} \int_{1}^{u} dx \frac{\sqrt{x-u}}{\sqrt{x^2 -1}}}
\end{dmath}

These results can be represented in terms of \textbf{Hypergeometric} functions and in particular, \textbf{Elliptic} integrals:

\begin{dmath}[label={eq:a-hypergeometric}]
    {a (u) = {\sqrt{2(1+u)}} F(-1/2, 1/2, 1; \frac{2}{1+u}) = \frac{4}{\pi} \sqrt{\frac{1+u}{2}} E(\sqrt{\frac{2}{u+1}})}
\end{dmath}

\begin{dmath}[label={eq:aD-hypergeometric}]
    a_D (u) = \frac{i}{2} (u-1) F(1/2, 1/2, 2; \frac{1-u}{2}) = \frac{4}{i \pi} E(\sqrt{\frac{1-u}2}) + \frac{2i}{\pi} (1+u) K(\sqrt{\frac{1-u}{2}})
\end{dmath}

where $E'(k) = E(k') , K'(k) = K(k')$ in which ${k'}^2 = 1 - k^2$ and
\begin{dmath}
    K(k) = \frac{\pi}{2} F(1/2, 1/2, 1; k^2)
\end{dmath}
\begin{dmath}
    E(k) = \frac{\pi}{2} F(-1/2, 1/2, 1; k^2)
\end{dmath}

where 

\begin{dmath}
    F(\alpha, \beta, \gamma; z) = \frac{\Gamma(\gamma)}{\Gamma(\beta) \Gamma(\gamma - \beta)} \int_{0}^{1} dx x^{\beta -1} (1-x)^{\gamma - \beta -1} (1-zx)^{-\alpha} = \frac{\Gamma(\gamma)}{\Gamma(\alpha) \Gamma(\beta)} \sum_{n \geq 0} \frac{\Gamma(\alpha + n) \Gamma(\beta + n)}{\Gamma(\gamma + n)} \frac{z^n}{n!}
\end{dmath}

which is called the hypergeometric function.

Finally, the effective coupling $\tau$ can be written as:

\begin{dmath}[label={eq:tau-elliptic}]
    {\tau (u) = \frac{\partial a_D}{\partial a} = \frac{d a_D / du}{d a / du} = \frac{i K'(\sqrt{\frac{2}{1+u}})}{K(\sqrt{\frac{2}{1+u}})}}
\end{dmath}

This analysis of the moduli space, hitherto, has been limited to SYM without matter being included ab initio(although one found matter content or hypermultiplets as topological defects at the end). The beauty of the Seiberg-Witten framework reveals itself evermore when matter is included initially and observing that in certain limits the distinction between fundamental matter and SYM topological defect disappears \cite{SUSY2} which is the formal incarnation of the \textbf{Montonen-Olive} duality (at the effective and low energy here) where the distinction between elementary and composite, fundamental and emergent, simple and complex reduces to a linguistic matter depending on the language (QFT Lagrangian) chosen to explain the same Physics \cite{BPST}.

\section{Seiberg-Witten Curves for $\mathcal{N}=2$ SQCD with $N_f$ Hypermultiplets}\label{sec:SW-curves-SQCD}

\subsection{General Remarks}\label{subsec:SW-curves-general-remarks}

One can include $N_f$ hypermultiplets into the theory and extend the gauge group to $SU(N)$, resulting in SQCD. Bearing in mind the one-loop exact $\beta$-function of the $\mathcal{N}=2$, $SU(N)$ SQCD:
\begin{dmath}
    \beta(g) = - \frac{2N-N_f}{16{\pi}^2} g^3
\end{dmath}
where $N_f \leq 2N$ since asymptotic freedom is lost for a larger number of flavors and the theory is scale invariant at $N_f = 2N$.

The classical moduli space Coulomb branch coordinates are as we saw in \eqref{eq:Chern-polynomials-SUN} of Section \ref{Witten---Seiberg-Theory}:
\begin{dmath}
    {c_k (a) = (-1)^k \sum_{i_1 < ... < i_k} a_{i_1} ... a_{i_k} \ \ \ k=2,...,N}
\end{dmath}

where the effective theory is explained by $N-1$ multiplets $(\Phi_a , W_a)$ whose symmetry group is $U(1)^{N-1}$.
The general prepotential can be written as:
\begin{dmath}
    \mathcal{F}_{cl} (a) = \frac{\tau}{2} \sum^{N}_{i=1} \left( a_i \right)^2
\end{dmath}

and the one-loop exact contribution to the prepotential is:
\begin{dmath}[label={eq:one-loop-prepotential}]
    \mathcal{F}_1 (a) = i \frac{2N-N_f}{16 \pi} \sum_{i < j} (a_i - a_j)^2 \ln{\frac{(a_i - a_j)^2}{{\Lambda}^2}}
\end{dmath}

There remain only non-perturbative corrections.
The general classical moduli space is the family of genus $g=N-1$ hyper-elliptic curves:

\begin{dmath}
    {y^2 = \mathcal{C}^2_{N} (x) = \prod^{N}_{i=1} (x-a_i)^2 \ \ \text{where} \ \ \mathcal{C}_N (x) = x^N + \sum^{N}_{i=2} c_i x^{N-i}}
\end{dmath}

where $y$ is the double cover of the $x$ plane with $N$ branch points corresponding to the $N$ roots. The singularities of the classical moduli space are those points where the symmetry is enhanced either due to two of the VEVs being equal $a_i = a_j$ for some $i,j \in {1,..., N}$ or a vanishing VEV.

The quantum moduli space, however, can generally be written as:

\begin{dmath}
    y^2 = \prod^{2g +2}_{i=1} (x - e_i)
\end{dmath}

where $e_i$ are the roots of the curve, being functions of $\Lambda, m_i$.

One can write $a^i_D , a^i$ as the periods of a meromorphic one-form $\lambda$:

\begin{dmath}
    {a^i_D = \oint_{\beta_i} \lambda \ \ , \ \ a^i = \oint_{\alpha_i} \lambda}
\end{dmath}

The main assumption to determine $\lambda$ is:

\begin{dmath}
    {\frac{\partial a^i_D}{\partial s_k} \propto \oint_{\beta_i} x^{N-k} \frac{dx}{y} \ \ \ , \ \ \frac{\partial a^i}{\partial s_k} \propto \oint_{\alpha_i} x^{N-k} \frac{dx}{y} \ , \ k= 2,...,N}
\end{dmath} 

where $a^i_D = \frac{\partial \mathcal{F}(a)}{\partial a_i}$ and $x^{N-k} \frac{dx}{y}$ being the basis of the holomorphic one-forms on the curve and $\alpha_i, \beta_i$ are the generators of the Abelian group of all cycles of the hyper-elliptic curve where $s_k := (-1)^k \sum_{i_1 <...<i_k} a_{i_1} ... a_{i_k} ; k=2,...,N$ are used as the classical moduli space definition of coordinates and $ks_k + \sum^k_{i=1} s_{k-i} u_i =0 ; k=\{0,..., N\}$ with $s_0 =1$ and $s_1=u_1 =0$ definition, is used at the quantum level. At the same time, the constant of proportionality can be found by matching with the weak coupling limit relations, which turns out to be $\frac{1}{2 \pi i}$ \cite{Singularity analysis}.

 The residue of $\lambda$ vanishes in the case of vanishing masses and in the case of non-vanishing masses:
 \begin{dmath}
     {\text{res}(\lambda) = \frac{1}{2 \pi i} \sum_{i} n_i m_i \ \ , \ \ n_i \in \frac{1}{2} \mathbb{Z}}
 \end{dmath}

that results in jumps in $a^i_D , a^i$ when crossing the poles of $\lambda$ by deforming the cycles.  

\subsection{On the Singularity Structure, Monodromy and the Global Symmetry}\label{subsec:Singularity-Structure}

The singular submanifold of the $N-1$ dimensional moduli space $\mathcal{M}$ is precisely where the discriminant of the curve vanishes $\Delta[s_k] = \prod_{i<j} (e_i - e_j)^2 = 0$ and we call it $\mathcal{L} \subset \mathcal{M}$.

In the semiclassical limit $\Lambda \rightarrow 0$ it factorises as:
\begin{dmath}
    {\lim_{\Lambda \rightarrow 0} \Delta[\Lambda] = {\Lambda}^{N(2N-N_f)} \Delta^2_N \Delta_{N_f , N}}
\end{dmath}

where $\Delta_N = \prod^{N}_{i<j} (a_i - a_j)^2 $ and $\Delta_{N_f , N} = \prod^{N_f}_{j=1} \left(\sum^{N}_{i=0} s_i (-m_j)^{N-i} \right)$. $\Delta_N$ being where the symmetry group is enhanced from $U(1)^{N-1}$ to $SU(k) \times U(1)^{N-k-1}$ due to $k$ scalar fields acquiring the same VEVs on the classical moduli space and $\Delta_{N_f , N}$ defines a complex co-dimension 1 variety where a quark field becomes massless. 

This factorisation sustains the quantum corrections and is valid for all values of $\Lambda$ except that $s_i$ receives quantum corrections. 

Massless states also appear on singularities in the case of quantum moduli spaces. When two of the branch points coincide, there's no explanation(effective Lagrangian) that includes both massless states or is so-called \textbf{mutually nonlocal} \cite{Singularity analysis}.

The general $R$-charges of the hyper-elliptic curve parameters are as follows, bearing in mind that a generic operator $\mathcal{O}$ transforms as:
\begin{dmath}
    \mathcal{O} \rightarrow \exp{\frac{2 \pi i R(\mathcal{O})}{4(2N-N_f)}} \mathcal{O}
\end{dmath}

\begin{dmath}
    {R[\mathcal{O}]: \ \ \  \ R[y]=2N \ \ R[x]=2 \ \ R[m_i]=2 \ \ R[\Lambda]=2 \ \ R[u_k]=2k \ \ R[s_k]=2k}
\end{dmath}

The general method of constructing the curves is to write all possible terms that contain $y,x,m_i, \Lambda, u_k, s_k$ such that they respect the $R$-symmetries. The instanton corrections are always proportional to $\Lambda^{2N-N_f}$ for $N_f < 2N$. Furthermore, matching with the classical limit, the residues and integrating a massive quark almost completely determine the curve.

Here we only cover the final results:

\begin{itemize}
     \item $N_f < N$

$y^2 = {\mathcal{C}_N (x)}^2 - {\Lambda}^{2N-N_f}_{N_f} \prod^{N_f}_{i=1} (x+ m_i)$
    \item $2 < N_f = N $

$y^2 = \left(\mathcal{C}_N (x) + \frac{{\Lambda}^N}{4} \right)^2 - {\Lambda}^{N} \prod^{N_f}_{i=1} (x + m_i)$
    \item $N_f = N = 2$

$y^2 = \left( x^2 - u + \frac{\Lambda^2_2}{8} \right)^2 - \Lambda^2_2 (x + m_1)(x + m_2)$

    \item $2<N < N_f  < 2N $

$y^2 = \left[ \mathcal{C}_N (x) + \frac{\Lambda^{2N-N_f}}{4} \sum^{N_f - N}_{i=0} x^{N_f -N -i} t_i (m) \right]^2 - \Lambda^{2N-N_f} \prod^{N_f}_{i=1}(x + m_i) $

      where $t_i (m)= \sum_{j_1 < ... < j_i} m_{j_1} \cdots m_{j_i}$.

\end{itemize}

Finally, we mentioned a beautiful result without giving the proof that the order of the discriminant vanishing at a singular point is equal to the dimension of the representation of the global symmetry group at the same point, the massless states belong to.

In the final section of this chapter, we study the results of explicit soft SUSY breaking, such as Dyon/monopole condensation.

\section{Less Supersymmetric SYM}\label{sec:Less-Supersymmetric-SYM}
The $\mathcal{N}=2, SU(2)$ Lagrangian can be perturbed by the addition of new soft SUSY breaking terms:

\begin{dmath}
    \mathcal{L} = \left( \int d^2 \theta \frac{-i}{8 \pi} \tau \text{Tr} \ {W_{\alpha} W^{\alpha}} +c.c. \right) +\frac{\Im{\tau}}{4 \pi} \int d^4 \theta {\Phi}^{\dagger} {\Phi} + \left( \int d^2 \theta \frac{m}{2} \text{Tr} {\Phi^2} + c.c. \right) + \left( \mu \lambda_{\alpha} \lambda^{\alpha} + c.c. \right)
\end{dmath}

When $\mu=m=0$, the theory is exactly $N=2$ SUSic. Setting $\mu=0$ and then letting $m \neq 0$ results in breaking the SUSY from $\mathcal{N}=2$ to $\mathcal{N}=1$, and by sending $m \rightarrow \infty$, one can decouple the chiral multiplet $\Phi$ and end up with the pure $\mathcal{N}=1, SU(2)$ SYM \cite{Breaking SUSY}

One rewrites the new Lagrangian as a perturbation around the new vacuum:

\begin{dmath}
    \int d^2 \theta m \text{Tr} {\Phi^2} \sim \int d^2 \theta m u + \text{fluctuations}
\end{dmath}

The $F$-term equation with respect to $u$ is not fulfilled unless $u$ is a singularity. Expanding around the singular point $u_0 = 2 \Lambda^2$ after a proper monodromy transformation that sets the magnetic charge equal to zero, one knows that the local Lagrangian term for the massless electric multiplets $Q, \tilde{Q}$ can be written as:

\begin{dmath}
    \int d^2 \theta Q a(u) \tilde{Q} = \int d^2 \theta c_q (u - u_0) Q \tilde{Q}
\end{dmath}

then the equation of motion for the term is:

\begin{dmath}
    {m= - c_q Q \tilde{Q} , \ \ \  (u- u_0) \tilde{Q} = 0 , \ \ \  (u - u_0) Q = 0}
\end{dmath}

The solution of which is:

\begin{dmath}
    {u= u_0 , \ \ \ Q \tilde{Q} = - \frac{m}{c_q}}
\end{dmath}

which means the monopoles condensate at $u = u_0$ while the same analysis applies to the case of $u= - u_0$ where the dyons condensate and result in:

\begin{dmath}
    {u = - u_0 , \ \ \ Q' \tilde{Q}' = - \frac{m}{c'_q}}
\end{dmath}

Given that the $R$-symmetry is anomalous, by the method of spurious fields, one compensates for the anomaly by a transformation in $\theta_{UV}$:

\begin{dmath}
    {\Phi_{UV} \rightarrow e^{i \phi} \Phi_{UV} \ \ \ \ \theta_{UV} \rightarrow \theta_{UV} + 4 \phi}
\end{dmath}
resulting in:
\begin{dmath}
    m \langle \text{Tr} {\Phi^2} \rangle = -\frac{i}{2 \pi} \langle \text{Tr} { W^{\alpha} W_{\alpha}} \rangle
\end{dmath}
which means:
\begin{dmath}
    {\langle \lambda_{\alpha} \lambda^{\alpha} \rangle  \propto \pm 2 \pi i m \Lambda^2 = \pm \Lambda^3_{\mathcal{N}=1}}
\end{dmath}
where one can finally take the $m \rightarrow \infty$ limit while fixing $\Lambda_{\mathcal{N}=1}$ where the chiral multiplet $\Phi$ decouples and the theory becomes the pure $\mathcal{N}=1 , SU(2)$ SYM. One should note that the vacuum manifold is $\mathbb{Z}_2$ symmetric.

To further break the $\mathcal{N}=1 , SU(2)$ softly to the $\mathcal{N}=0 , SU(2)$ theory, one should keep in mind that $|\mu| \ll |\Lambda_{\mathcal{N}=1}|$. Bearing in mind this note, one can conclude for the vacuum energy:

\begin{dmath}
    {V \propto \Re{ \left\{ \pm \mu \Lambda^3_{ \mathcal{N}=1} \right\}} \propto \Lambda^4_{\mathcal{N}=0} \Re{\left\{ \pm e^{i \theta_{UV} /2} \right\}}}
\end{dmath}

The two degenerate vacua split ($\mathbb{Z}_2$ symmetry is explicitly broken). When $\mu$ is sufficiently small, there exists a first-order phase transition at $\theta_{UV} = \pi$  when $\theta_{UV}$ changes from $0$ to $2 \pi$.

For finite $\mu \ll \Lambda_{\mathcal{N}=1}$ it is not hard to show that the unique vacuum of the theory confines by magnetic monopole/dyonic condensation, but to generalize the result to $\mathcal{N}=0, SU(2)$ pure gauge theory by decoupling the adjoint fermion fields in the $m \rightarrow \infty$ limit, one cannot say more than conjecturing that such picture holds, though there is already a candidate for violating such conjecture, which is $\mu / \Lambda$ as an order parameter that can parameterizes a phase transitions between $ \mu \ll \Lambda$ and $ \mu \gg \Lambda$ phases! \cite{Breaking SUSY}

\begin{figure}
    \centering
    \includegraphics[width=0.5\linewidth]{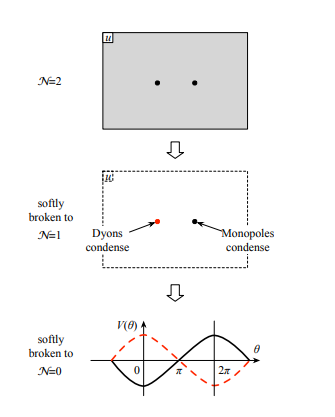}
    \caption{Vacuua in order of $\mathcal{N}=2$ and $\mathcal{N}=1,0$ after soft breaking (figure from \cite{Dulatities})}
    \label{fig:SW-curve-Tachikawa}
\end{figure}

\chapter{Non-Commutative Resolution and Localisation}
\label{Localisation} 
\section{The Way \textit{ Not to Go}! or Why the \textit{Localization Technique}}

The purpose of this digression is to show that, for computational reasons, it seems almost impossible to determine the Seiberg-Witten coefficients, not only in the original way but also utilising the \textbf{instanton calculus method}. Despite attempts to find formal algorithms based on the localization method to determine all such coefficients purely algebro-geometrically \cite{Algorithm} and general theorems about features of coefficients that are determined to all orders \cite{All orders}, there are several questions to pose and answer, regarding the \textbf{efficiency} and \textbf{practicality} of such algorithms in (any) formal sense of the word. 

As we have already observed, the gauge-invariant non-singular charts(excluding the singularity at infinity) are $u_n = 2^{-n/2} g^n \langle \tr_N {\phi}^n \rangle$ where $n= 2,..., N$ and oppositely one writes the former singular charts in terms of this coordinate system: $\phi^0_u = \phi^0_u (u_n)$ and $\phi^0_{Du} = \phi^0_{Du} (u_n)$.

Our approach in this section is matching the four-point function of the chiral fermions of the $\mathcal{N}=2$ chiral multiplet $\Psi$(\textit{a.k.a.} vector multiplet) written in terms of a generalised effective action and its functional derivatives\footnote{or the prepotential $\mathcal{F}(\phi^0_u)$ as a complex line bundle over the moduli space and its derivatives though with respect to the $\phi^0_u$ coordinate system instead of $u_n$, though this turns into major source of problems and difficulty since this coordinate system is singular while one knows that this enhanced symmetry point of singularity on the moduli space is indeed \textbf{resolved} at the quantum level causing a topological modification of the moduli space. We will be back to this argument later with even more care! }, with the same correlation function derived via the instanton partition functions found in Section \ref{Linearised-Instanton-Partition-Function}.

Formally, after a tedious, lengthy calculation, one can write explicitly:

\begin{dmath}[label={eq:four-fermion-correlator-prepotential}]
\langle \bar{\lambda}^{\dot{\alpha}}_{u_1} \left( x^{(1)} \right) \bar{\lambda}^{\dot{\beta}}_{u_2} \left( x^{(2)} \right) \bar{\psi}^{\dot{\gamma}}_{u_3} \left( x^{(3)}\right)\bar{\psi}^{\dot{\delta}}_{u_4} \left( x^{(4)} \right) \rangle = \frac{1}{2 \pi i} \frac{\partial^4 \mathcal{F}}{\partial \phi^0_{u_1} \partial \phi^0_{u_2} \partial \phi^0_{u_3} \partial \phi^0_{u_4}} \int d^4 X \bar{S}^{\dot{\alpha} \alpha} \left( x^{(1)} , X \right) \bar{S}^{\dot{\beta}}_{\alpha} \left( x^{(2)} , X \right) \bar{S}^{\dot{\gamma} \gamma} \left( x^{(3)} , X \right) \bar{S}^{\dot{\delta}}_{\gamma} \left(x^{(4)} , X \right)
\end{dmath}

where:

\begin{dmath}
    \bar{S}(x , X) = \frac{1}{4 \pi^2} \rawslashed{\bar{\mathcal{\partial}}} \frac{1}{(x- X)^2}
\end{dmath}

is the anti-Weyl spinor propagator, and, as it turns out, $X$ will be the coordinate of the instanton centre.

As we understood in Section \ref{Constrained-Instantons-and-the-Solution-for-non-vanishing-VEVs}, the tail of the instantons $SU(N)$ in the weak coupling regime decays exponentially unless they belong to the $U(1)^{N-1}$ subgroup distinguished by the direction of $\phi^0$, which remain massless even after the Higgs mechanism. This results in the massless diagonal elements of the adjoint $ N\times N$ antichiral fermions $\bar{\lambda}_A =(\bar{\lambda}, \bar{\psi})$ dominating the long-distance behaviour of the fermions, which in the semi-classical expansion are precisely equal to their ADHM solutions.

Knowing from the first chapter, \eqref{eq:susy-zero-modes-ADHM}, that the Grassmann collective coordinates are proportional to $\xi^A$ along with the relations(after re-scaling by $g^{1/2}$):

\begin{dmath}
    \bar{\lambda}_A = - g^{1/2} \rawslashed{\bar{\mathcal{D}}} \phi^{\dagger} \xi_A
\end{dmath}
\begin{dmath}
    \phi^{\dagger} = \bar{U} \begin{pmatrix}
    \phi^{0 \dagger} & 0 \\
0 & \varphi^{\dagger} \mathbf{1}_{[2] \times [2]}
\end{pmatrix} U
\end{dmath}
\begin{dmath}
    \varphi^{\dagger} = \mathbf{L}^{-1} \left( -\frac{1}{4} \sum^{N_F}_{f=1} \mathcal{K}_f \tilde{\mathcal{K}}_f + \bar{w}^{\dot{\alpha}} \phi^{0 \dagger} w_{\dot{\alpha}} \right)
\end{dmath}
\begin{dmath}
\tilde{S} = 4 \pi^2 \tr_k \left\{ - \frac{i}{2} \bar{\mu}^A \phi^{0 \dagger} \mu_A + \bar{w}^{\dot{\alpha}} \big| \phi^0 \big|^2 w_{\dot{\alpha}} + \left( \frac{1}{4} \sum^{N_F}_{f=1} \mathcal{K}_f \tilde{\mathcal{K}}_f - \bar{w}^{\dot{\alpha}} \phi^{0 \dagger} w_{\dot{\alpha}} \right) \mathbf{L}^{-1} \left( -\frac{i}{2} \bar{\mathcal{M}}^A \mathcal{M}_A + \bar{w}^{\dot{\alpha}} \phi^0 w_{\dot{\alpha}} \right)  \right\} + S_{\text{mass term}} 
\end{dmath}

\begin{dmath}
    {\Delta_{\dot{\alpha}} \rightarrow b^{\alpha} x_{\alpha \dot{\alpha}} \ \ \ \ f_{ij} \rightarrow \frac{1}{x^2} \delta_{ij} \ \ \ \ U' \rightarrow -\frac{x_{\alpha \dot{\alpha}}}{x^2} \bar{w}^{\dot{\alpha}}} 
\end{dmath}
\begin{dmath}
    {V \rightarrow \mathbf{1}_{[N] \times [N]} \ \ \ \ 
A_n \rightarrow g^{-1} \frac{x_m}{x^4} w_{\dot{\alpha}} \bar{\sigma}^{\dot{\alpha}}_{mn \dot{\beta}} \bar{w}^{\dot{\beta}}}
\end{dmath}

One concludes that:

\begin{dmath}
    \left(\rawslashed{\bar{\mathcal{D}}} \phi^{\dagger}\right)_{uu} = - \rawslashed{\bar{\partial}} \frac{1}{(x- X)^2} w_{u \dot{\alpha}} \left\{ \phi^{0 \dagger}_u \mathbf{1}_{[k]\times[k]} - \mathbf{L}^{-1} \left( \bar{w}^{\dot{\beta}} \phi^{0 \dagger} w_{\dot{\beta}} - \frac{1}{4} \sum^{N_F}_{f=1} \mathcal{K}_f \tilde{\mathcal{K}}_f \right) \right\} \bar{w}^{\dot{\alpha}}_u 
\end{dmath}
far from the instanton core, and finally more compactly:

\begin{dmath}
    \left( \bar{\lambda}^{\dot{\alpha}}_A \right)_{uu} = 2 \sqrt{g} \bar{S}^{\dot{\alpha} \alpha}(x , X) \epsilon_{AB} \xi^B_{\alpha} \frac{\partial \tilde{S}}{\partial \phi^0_u}
\end{dmath}

so finally, one can beautifully write the correlation in the topological charge of the $k$ sector background:

\begin{dmath}[label={eq:four-fermion-correlator-instanton}]
\left\langle \bar{\lambda}^{\dot{\alpha}}_{u_1} \left( x^{(1)} \right) \bar{\lambda}^{\dot{\beta}}_{u_2} \left( x^{(2)} \right) \bar{\psi}^{\dot{\gamma}}_{u_3} \left( x^{(3)}\right) \bar{\psi}^{\dot{\delta}}_{u_4} \left( x^{(4)}\right) \right\rangle_{k} = \left( \frac{\mu}{g} \right)^{k(2N-N_F)} e^{2 \pi i k \tau} \int_{\mathcal{M}_k} \boldsymbol{\omega}^{(2 , N_F)} e^{-\tilde{S}} \bar{\lambda}^{\dot{\alpha}}_{u_1} \left( x^{(1)}\right) \bar{\lambda}^{\dot{\beta}}_{u_2} \left( x^{(2)}\right) \bar{\psi}^{\dot{\gamma}}_{u_3} \left( x^{(3)}\right) \bar{\psi}^{\dot{\delta}}_{u_4} \left( x^{(4)}\right) =
\frac{1}{4\pi^2} g^2 \left( \frac{\mu}{g} \right)^{k(2N-N_F)} e^{2 \pi i k \tau} \frac{\partial^4}{\partial \phi^{0}_{u_1} \partial \phi^{0}_{u_2} \partial \phi^{0}_{u_3} \partial \phi^{0}_{u_4}} \int_{{\widehat{\mathcal{M}}}_{k}} \boldsymbol{\omega}^{(2 , N_F)} e^{-\tilde{S}} \times \int d^4 X \bar{S}^{\dot{\alpha} \alpha} \left(x^{(1)} , X \right) \bar{S}^{\dot{\beta}}_{\alpha} \left( x^{(2)} , X \right) \bar{S}^{\dot{\gamma} \gamma} \left( x^{(3)} , X \right) \bar{S}^{\dot{\delta}}_{\gamma} \left( x^{(4)} , X \right)
\end{dmath}

So, comparing \eqref{eq:four-fermion-correlator-prepotential} with \eqref{eq:four-fermion-correlator-instanton}, one finds:

\begin{dmath}
    \mathcal{F}_k = g^{-k (2N - N_F) + 2} \widehat{\mathcal{Z}}^{(2, N_F)}_k
\end{dmath}

The apparent freedom in $\mathcal{F}_k$ (up to functions whose fourth derivative with respect to VEVs vanishes) can be reduced to second derivative vanishing functions with the aid of other correlators, though since there does not exist any linear function of VEVs and masses to write, $\mathcal{F}_k$ can be determined \underline{up to a constant} that does not matter as what matters in writing the Witten-Seiberg effective action is the derivatives of $\mathcal{F}$.

By re-scaling the fields according to:

\begin{dmath}
    {\phi^0 \rightarrow g^{-1} \phi^0 \ \ \ \ a_{\dot{\alpha}} \rightarrow g a_{\dot{\alpha}} \ \ \ \ \mathcal{M}^A \rightarrow g^{1/2} \mathcal{M}^A \ \ \ \ \mathcal{K} \rightarrow g^{1/2} \mathcal{K} \ \ \ \ \tilde{\mathcal{K}} \rightarrow g^{1/2} \tilde{\mathcal{K}}}
\end{dmath}

The prepotential can be written in a way that does not depend explicitly on $g$:

\begin{dmath}
    \widehat{\mathcal{Z}}^{(2,N_F)}_k \rightarrow g^{k(2N-N_F) - 2} \widehat{\mathcal{Z}}^{(2 , N_F)}_k
\end{dmath}

and consequently:

\begin{dmath}
    \boxed{\mathcal{F}_k = \widehat{\mathcal{Z}}^{(2,N_F)}_k}
\end{dmath}

\begin{itemize}
    \item Determining the $\mathcal{F}_1$ coefficient in $SU(N)$

Starting with the effective action \eqref{eq:linearised-effective-action} for $k=1$:

\begin{dmath}[label={eq:effective-action-k1}]\tilde{S} = 4 \pi^2 \left\{ \big| w_{u \dot{\alpha}} \chi  + \phi^0 w_{u \dot{\alpha}} \big|^2 + \frac{i}{2} \bar{\mu}^A_u \left( \mu_{u A} \chi^* + \phi^{0*} \mu_{u A} \right) + \frac{1}{4} \sum^{N_F}_{f=1} \mathcal{K}_f \tilde{\mathcal{K}}_f \left( \chi - m_f \right) \right\} + \tilde{S}_{\text{L.m.}}\end{dmath}

where

\begin{dmath}
    \tilde{S}_{\text{L.m.}} = -4 i \pi^2 \left\{ \bar{\psi}^{\dot{\alpha}}_A \left( \bar{\mu}^A_u w_{u \dot{\alpha}} + \bar{w}_{u \dot{\alpha}} \mu^A_u \right) + \Vec{D} \ . \ \tau^{\dot{\alpha}}_{\dot{\beta}} \bar{w}^{\dot{\beta}}_u w_{u \dot{\alpha}} \right\}
\end{dmath}

For $k=1$, $\chi$ is a complex variable, and by the linear shifts:

\begin{dmath}
    {\mu^A_u \rightarrow \mu^A_u - \frac{2 w_{u \dot{\alpha}}}{\alpha^{*}_u} \bar{\psi}^{\dot{\alpha} A}  \ \ \ \  {\bar{\mu}}^A_u \rightarrow {\bar{\mu}}^A_u - \frac{2 \bar{w}_{u \dot{\alpha}}}{\alpha^{*}_u} \bar{\psi}^{\dot{\alpha} A}}
\end{dmath}
After integrating over the Grassmann variables, one finds:

\begin{dmath}
    \prod^{N}_{u=1} \left( 2 \pi^2 \alpha^{*}_{u} \right)^2
\end{dmath}
where:
\begin{dmath}
    {\alpha_u = \chi + \phi^0_u \ \ \ \ \alpha^*_u = \chi^* + \phi^{0*}_u}
\end{dmath}
After carrying out all the integrals, one ends up with the expression:

\begin{dmath}[label={eq:Z-k1-chi-integral}]
    \widehat{\mathcal{Z}}^{(2,N_F)}_{k=1} = - \pi^{-1} \int d^2 \chi \frac{\partial^2}{\partial \chi^{*2}} f_1 (\chi , \chi^*) f_2 (\chi)
\end{dmath}
in which:
\begin{dmath}
    {f_1 (\chi, \chi^*) = \sum^{N}_{u=1} \frac{\alpha^*_u}{\alpha_u} \prod^{N}_{\substack{v=1 \\ v \neq u}} \frac{\alpha^{*2}_v}{|\alpha_{v}|^4 -|\alpha_u|^4} \ \ \ \ f_2 (\chi) = \prod^{N_F}_{f=1} \left( m_f - \chi \right)}
\end{dmath}

There are $N$ singularities on the $\chi$-plane which are the points at which $\alpha_u = 0$ or equivalently $\chi = - \phi^0_u$. In the vicinity of the singularity in the polar coordinate system $\alpha_u = r e^{i \theta}$, one can take the boundary of a small circle of radius $r \rightarrow 0$. The integral would be:

\begin{dmath}
    \lim_{r \rightarrow 0} \frac{1}{4 \pi} \int^{2\pi}_0 d \theta e^{2i \theta} f_1 (r , \theta) f_2 (re^{i \theta})
\end{dmath} 

resulting in the expression:
\begin{dmath}
    \prod^{N}_{\substack{v=1 \\ v \neq u}} \frac{1}{(\phi^0_u - \phi^0_v)^2} \prod^{N_F}_{f=1} \left( 
m_f + \phi^0_u \right)
\end{dmath}

The same technique applied to the sphere at infinity leads to a contribution:

\begin{dmath}[label={eq:contribution-at-infinity}]
    - \lim_{r \rightarrow \infty} \frac{1}{4 \pi} \left( 2 + r\frac{\partial}{\partial r} \right) \int^{2 \pi}_{0} d\theta e^{2i \theta} f_1 (r , \theta) f_2 (r e^{i \theta})
\end{dmath}

while the asymptotic for $f_1 (r , \theta)$ and $f_2 (r e^{i \theta})$:

\begin{dmath}[label={eq:f1-f2-asymptotics}]
    {f_1 (r , \theta) \sim \frac{e^{-2iN \theta}}{r^{2(N-1)}} \ \ \ \ \ f_2 (r e^{i \theta}) \sim e^{i N_F \theta} r^{N_F}}
\end{dmath}

which limits $N_F \geq 2 (N-1)$ which means $N_F = 2N-2 , 2N-1, 2N$. Taking into account the integral over $\theta$ the only contributions to the asymptotic expansion for $f_1 (r, \theta)$ are:

\begin{dmath}
    {f_1 (r , \theta) \sim \alpha_1 \frac{e^{-2iN \theta}}{r^{2(N-1)}} + \alpha_2 \frac{e^{-2i(N+1) \theta}}{r^{2N}} \sum^{N}_{u=1} (\phi^0_u)^2}
\end{dmath}

where

\begin{dmath}
    {\alpha_1 = 2^{3- 2N} \begin{pmatrix}
    2N-3 \\
    N-1
\end{pmatrix} \ \ \ \ \ \ \ \ \alpha_2 = 2^{-2N} \begin{pmatrix}
    2N \\
N-1
\end{pmatrix}}
\end{dmath}

All the other terms are sub-leading and $f_2$ is a polynomial in $e^{i\theta}$. All in all, the contribution at infinity amounts to:

\begin{equation*}
\mathcal{S}^{(N_F)}_{k=1}=\begin{cases}
            0 \quad &\text{if} \, N_F < 2N-2 \\
            - \alpha_1  \quad  &\text{if} \, N_F =2N-2 \\
            - \alpha_1 \sum^{N_F}_{f=1} m_f  \quad &\text{if} \, N_F = 2N-1 \\
            - \alpha_1 \sum_{f < f'} m_f m_{f'} - \alpha_2 \sum^{N}_{u=1} (\phi^0_u)^2 \quad  &\text{if} \, N_F = 2N
     \end{cases}
\end{equation*}

leading to the conclusion:

\begin{dmath}[label={eq:F1-commutative}]
    \mathcal{F}_1 = \widehat{\mathcal{Z}}^{(2, N_F)}_{k=1} = \sum^{N}_{u=1} \prod^{N}_{\substack{v=1 \\ v \neq u}} \frac{1}{(\phi^0_v - \phi^0_u)^2} \prod^{N_F}_{f=1} \left( m_f + \phi^0_u \right) + \mathcal{S}^{(N_F)}_1
\end{dmath}

\end{itemize}

To really grasp the difficulty of computations via the instanton calculus method, one might try to compute the $\mathcal{F}_2$ coefficient in the $SU(2)$ gauge group case.

\begin{itemize}
    \item Determining the $\mathcal{F}_2$ for $k=2$: 
    With the help of our digression in section \ref{ADHM}, one can facilitate the computation by converting the $SU(2)$ ADHM construction to the $Sp(1)$ ADHM construction.
Therefore, the ADHM data for $k=2$ is as such:

\begin{dmath}
    {a= \begin{pmatrix}
    w_1 & w_2 \\
-X + a_3 & a_1 \\
a_1 & -X-a_3
\end{pmatrix} \ \ \ \ \ \ \ \ \ \ \mathcal{M}^A = \begin{pmatrix}
    \mu^A_1 & \mu^A_2 \\
-4i\xi^A + \mathcal{M}^{'A}_3 & \mathcal{M}^{'A}_1 \\
\mathcal{M}^{'A}_1 & -4i\xi^A - \mathcal{M}^{'A}_3
\end{pmatrix}}
\end{dmath}
where elements of $\mathcal{M}^A$ is a Weyl spinor and of $a$ is a quaternion. 
There are also the $4N_F$ fundamental zero modes Grassmann numbers $\mathcal{K}_{if}$ and $\tilde{\mathcal{K}}_{fi}$.
By defining:

\begin{dmath}
    {L= \big| w_1 \big|^2 + \big| w_2 \big|^2 \ \ \ 
H= \big| w_1 \big|^2 + \big| w_2 \big|^2 + 4 \big| a_1 \big|^2 + 4 \big| a_3 \big|^2}
\end{dmath}
     
\begin{dmath}
    {\Omega = w_1 \bar{w}_2 - w_2 \bar{w}_1 \ \ \ \omega= \frac{1}{2} \phi^0 \text{tr}_2 \left( \bar{w}_2 \tau^3 w_1 - \bar{w}_1 \tau^3 w_2 \right)}
\end{dmath}
\begin{dmath}
    {Y= -\mu^A_1 \mu_{2A} - 2 \mathcal{M}^{'A}_3 \mathcal{M}^{'}_{1A} \ \ \ Z= \sum^{N_F}_{f=1} \left( \mathcal{K}_{1f} \tilde{\mathcal{K}}_{f2} - \mathcal{K}_{2f} \tilde{\mathcal{K}}_{f1} \right)}
\end{dmath}

one can rewrite everything compactly, especially that $\mathbf{L}$ for $k=2$ is just $H$.

The effective action is then:
\begin{dmath}
\tilde{S} = 8 \pi^2 L \big| \phi^0 \big|^2 - 2\pi^2 i \phi^{0 \dagger} \left( \mu^A_1 \mu_{1A} - \mu^{A}_2 \mu_{2A} \right) - \frac{8 \pi^2}{H} \left( \bar{\omega} - \frac{1}{8} Z \right) \left( \omega - \frac{i}{2}Y \right) + \pi^2 \sum^{N_F}_{f=1} m_f \left( \mathcal{K}_{1f} \tilde{\mathcal{K}}_{f1} + \mathcal{K}_{2f} \tilde{\mathcal{K}}_{f2} \right)
\end{dmath}

Solving for the ADHM variables and integrating over the rest of them, inclusive of the adjoint fermions and fundamental hypermultiplet ones, results in:

\begin{dmath}[label={eq:F2-SU2-commutative}]
    \mathcal{F}_2 \bigg|^{N=2}_{N_F} = \frac{5}{(\phi^0)^6} M^{(N_F)}_{N_F} - \frac{3}{4 (\phi^0)^4} M^{(N_F)}_{N_F - 1} + \frac{1}{16 (\phi^0)^2} M^{(N_F)}_{N_F - 2} - \frac{5}{2^6 3^3} M^{(N_F)}_{N_F -3} + \frac{7 (\phi^0)^2}{2^8 3^5} M^{(N_F)}_{N_F -4}
\end{dmath}

where $\phi^0 := \phi^0_1 - \phi^0_2$ and

\begin{dmath}
    {M^{(N_F)}_l = \sum^{N_F}_{f_1 < f_2 < ... < f_l =1} m^2_{f_1} m^2_{f_2} \cdots m^2_{f_l} \ \ \ \text{with} \ \ \ M^{(N_F)}_0 = 1 , \ \ \ M^{(N_F)}_l =0 \ \ \ l<0}
\end{dmath}

Although we summarized the whole computation, specifically the integrals that must be carried out delicately, it is rather predictable how difficult it might get to find $\mathcal{F}_k$ for higher $k$. 

\end{itemize}

To end this section and motivate further the following one, it is worth noticing that the above computation, especially the final integral over $\chi$ in \eqref{eq:Z-k1-chi-integral}, implies a sort of \textbf{Localisation}.
Indeed, the integral over $\chi$ is carried out by the Stokes' theorem, where the only contributions come from $N$ isolated points and the sphere at infinity.

Clearly, these points are the \textbf{Critical Points} of the instanton effective action corresponding to the solutions to:

\begin{dmath}
    {w_{\dot{\alpha}} \chi + \phi^0 w_{\dot{\alpha}} = 0}
\end{dmath}

that can be indexed by $u$:
\begin{dmath}
    {\chi = - \phi^0_u  \ \ \ \ \ \ \ \ \ w_{v \dot{\alpha}} \propto \delta_{uv}}
\end{dmath}

The contribution at infinity, though, comes from the points where the instanton size shrinks to zero \textit{i.e.} more precisely $w_{\dot{\alpha}} = 0$. Indeed, the equation of motion for $\chi$ results in $\chi \propto \rho^{-1}$ which means $\chi \rightarrow \infty$ as $\rho \rightarrow 0$.

\section{The SW prepotential via the Localization Technique}

\subsection{Non-Commutative Spacetime, Modified ADHM Constraints}
\label{Non-Commutative Spacetime, Modified ADHM Constraints}
A non-commutative spacetime indeed, elevates the spacetime coordinate to operators satisfying:
\begin{dmath}
    [x_m , x_n] = i \theta_{mn}
\end{dmath}
As to whether only the space coordinates anti-commute or spacetime altogether, is a matter of debate, and different conclusions might be based on either of the assumptions.
In this thesis, the assumption is that all of spacetime anti-commutes.

The non-commutativity parameter $\theta_{mn}$, can be decomposed further into:

\begin{dmath}
    {\zeta^{c}_{(+)} = \bar{\eta}^c_{mn} \theta_{mn} \ \ \  \text{anti-self-dual}  \ \ \ \  \zeta^{c}_{(-)} = {\eta}^c_{mn} \theta_{mn} \ \ \ \text{self-dual}}
\end{dmath}

where $\vec{\zeta}_{(+)}$ and $\vec{\zeta}_{(-)}$ are two $3$-vectors that appear in the instanton and anti-instanton ADHM constraints for gauge theories defined on the non-commutative spacetime, respectively.

Obviously, for the Euclidean $4$-dimensional space, $\theta_{mn}$ is real and anti-symmetric. Meaning, the only parameters are $\theta_{12}, \theta_{34}$, which can be even further reduced by straightforward re-scaling of the coordinates such that $\theta_{12} = \theta_{34} = -\frac{\zeta}{4}$ for the self-dual part and $\theta_{12} = -\theta_{34} = -\frac{\zeta}{4}$ for the anti-self-dual.

After the complexification of the coordinate system, the algebra becomes:

\begin{dmath}
    {[z_i , \bar{z}_j] = -\frac{\zeta}{2} \delta_{ij} \ \ \ \ \ \ \ [z_i , z_j] = 0} 
\end{dmath}

where:

\begin{dmath}
    {z_1 = x_2 + ix_1 \ \ \ \ \ z_2 = x_4 + i x_3 \ \ \ \ \ \bar{z}_1 = x_2 - ix_1 \ \ \ \ \ \ \bar{z}_2 = x_4 - i x_3}
\end{dmath}

Consequently:

\begin{dmath}
    \boxed{\mathbb{R}^4_{\text{NC}} = \mathbb{R}^2_{\text{NC}} \times \mathbb{R}^2_{\text{NC}}}
\end{dmath}

The spacetime states are visiblly those of the Fock space of two simple harmonic oscillators, where the coordinates are the creation and annihilation operators:

\begin{dmath}
    {z_1 |n_1 , n_2 \rangle = \sqrt{\frac{\zeta}{2}} \sqrt{n_1 + 1} |n_1 +1 , n_2 \rangle \ \ \ \ z_2 |n_1 , n_2 \rangle = \sqrt{\frac{\zeta}{2}} \sqrt{n_2 + 1} |n_1  , n_2 + 1\rangle}
\end{dmath}

\begin{dmath}
    {\bar{z}_1 |n_1 , n_2 \rangle = \sqrt{\frac{\zeta}{2}} \sqrt{n_1} |n_1 -1 , n_2 \rangle \ \ \ \ 
    \bar{z}_2 |n_1 , n_2 \rangle = \sqrt{\frac{\zeta}{2}} \sqrt{n_2} |n_1 , n_2 -1 \rangle}
\end{dmath}

The integral over $\mathbb{R}^{4}_{\text{NC}}$ stands for the trace operator:

\begin{dmath}
    {\int d^4 x (*) = (2 \pi)^2 \sqrt{\det\theta} \ \text{Tr} (*) = (\frac{\zeta \pi}{2})^2 \text{Tr} (*)} 
\end{dmath}

Now, as for the modified ADHM constraints, one merely cites equation \eqref{eq:momentum-maps-Uk}:
\begin{dmath}[label={eq:modified-ADHM-constraints}]
    {\vec{\tau}^{\dot{\alpha}}_{\dot{\beta}} \bar{a}^{\dot{\beta}} a_{\dot{\alpha}} = \vec{\zeta}_{(+)} 1_{[k] \times [k]} (\text{Instanton}) \ \ \ \ \vec{\tau}^{\dot{\alpha}}_{\dot{\beta}} \bar{a}^{\dot{\beta}} a_{\dot{\alpha}} = \vec{\zeta}_{(-)} 1_{[k] \times [k]} (\text{anti-Instanton})}  
\end{dmath}

Clearly, the ADHM constraints for instantons on the self-dual background appear just like those of instantons on commutative $\mathbb{R}^4$ and vice versa.

Intuitively, the one-instanton modified ADHM constraints indeed result in the resolution of the original moduli space $\mathcal{M}_1$; since, as one knows, the original ADHM constraints are:

\begin{dmath}
    \bar{w}^{\dot{\alpha}}_{u} w_{u \dot{\beta}} = \rho^2 \delta^{\dot{\alpha}}_{\dot{\beta}} + \frac{1}{2} \vec{\tau}^{\dot{\alpha}}_{\ \ \dot{\beta}} . \vec{\zeta}_{(+)}
\end{dmath}
Without loss of generality, one can choose ${\vec{\zeta}} \propto (0,0,1)$ and observe that:

\begin{dmath}
    \rho^2 \geq \frac{1}{2} |\vec{\zeta}_{(+)}|
\end{dmath}

A stunning result. This can be further generalized to asymptotic regions of the multiinstanton moduli space, and it turns out that $\mathcal{M}^{(\zeta)}_k$ is the resolution of the singular moduli space $\mathcal{M}_k$.
For instance, the $U(1)$ non-commutative moduli space $\text{Sym}^k(\mathbb{R}^4)$ is resolved into $\mathcal{M}^{(\zeta)}_k$ which has no singularity any longer.

\subsection{Non-Commutative $U(1)$ Yang-Mills, Infrared Divergence}
\label{Non-Commutative $U(1)$ Yang-Mills, Infrared Divergence}

The non-commutative counterpart of the gauge theory is constructed in this section. To do so, one must first introduce the \textbf{Moyal Product}:

\begin{dmath}
    {(\phi_1 * \phi_2 ) (x) := e^{\frac{i}{2} \theta^{\mu \nu} \partial^{y}_{\mu} \partial^z_{\nu}} \phi_1 (y) \phi_2 (z) |_{y=z=x}}
\end{dmath}

The non-commutative Lagrangian is written by elevating all the products to Moyal products, meaning, that the $U(1)$ Yang-Mills action\footnote{Alarm: In  this particular section, the action style adopted is not precisely that of the conventions of the rest of this thesis outlined in Chapter \ref{Conventions,-Notations-and-Prerequisites}} is:

\begin{dmath}
    {S= -\frac{1}{4} \int d^4 x F^2 \ \ \ \text{where} \ \ \ F_{\mu \nu} = \partial_{\mu} A_{\nu} - \partial_{\nu} A_{\mu} -ig [A_{\mu} , A_{\nu}]_{*}} 
\end{dmath}

in which the commutation relation is defined with the help of the Moyal product:

\begin{dmath}
    [A_\mu , A_{\nu}]_{*} = A_{\mu} * A_{\nu} - A_{\nu} * A_{\mu}
\end{dmath}

The theory turns out to be invariant under a non-commutative gauge transformation:

\begin{dmath}
    \delta_{\lambda} A_{\mu} = \partial_{\mu} \lambda - gi(A_{\mu} * \lambda - \lambda * A_{\mu})
\end{dmath}

One can add matter and scalar fields alike, using the same prescription:

\begin{dmath}
    {S_{\text{matter}} = \int d^4 x \bar{\psi} * \gamma^{\mu} D_{\mu} \psi \ , \ \ S_{\text{scalar}} = \frac{1}{2} \int d^4 x (D_{\mu} \Phi)^2}
\end{dmath}
where the covariant derivative acts as follows:

\begin{dmath}
    D_{\mu}X = \partial_{\mu} X -ig[A_{\mu}, X]_{*}
\end{dmath}

By Taylor expanding the exponential and ignoring irrelevant operators, one might speculate that the theory converges to the same commutative theory in the infrared ($k^2 \ll \theta^{-1}$) as they seemingly belong to the same \textit{universality class}. However, a one-loop computation shows otherwise.

The photon self-energy turns out to be \cite{Non-commutative GT}:

\begin{dmath}
    i \Pi^{\mu \nu} (p) = -4 g^2 N_f \int \frac{d^4 l}{(2\pi)^4} \frac{\text{Tr} [\gamma^{\mu} (\rawslashed{l} - \rawslashed{p}) \gamma^{\nu} \rawslashed{l}]}{(l-p)^2 l^2} \text{sin}^2 (\frac{1}{2}\tilde{p} l)
\end{dmath}

where $\tilde{p}^{\mu}:= \theta^{\mu \nu} p_{\nu}$ and using the elementary trigonometric identity $\text{sin}^{2} (\frac{x}{2}) = \frac{1}{2} (1- \text{cos}(x))$, one can separate the planar and nonplanar shares from each other; indeed, the coefficient of $\text{cos}(\tilde{p}l)$ turns out to be the contribution due to the nonplanar part. The planar loop diagrams turn out to be indeed precisely the same as the commutative theory, while the nonplanar diagrams are exactly those that result in the exceptional IR behaviour of the non-commutative theory. This phenomenon, that the diagrams' UV behavior influences the amplitude's IR behavior, is coined as \textbf{UV/IR Mixing}.

It is easy to verify that the integral is UV-finite. Indeed, one can assume that $l$ is very large and rewrite the integral \footnote{Assuming cut-off $\Lambda$ as an upper bound on the loop momentum $l$}. Finally, one finds that to leading order(the superficially quadratically divergent contribution)
\begin{dmath}
    i\Pi^{\mu \nu} (p) = - 8 i g^2 (2N_f)  \alpha \frac{\tilde{p}^{\mu} \tilde{p}^{\nu}}{(\tilde{p})^4}
\end{dmath}

which is finite and totally independent of the cut-off. Nonetheless, it is evidently IR-divergent as $\tilde{p} \rightarrow 0$ \footnote{The non-commutative momentum $\tilde{p}$ acts as the IR regulator}. Evidently, the divergence is of order $\theta^{-2}$.

The same computation can be done to include the effect of scalars and matter fields, and the final result to leading order in superficial UV-divergence is:

\begin{dmath}
    i \Pi^{\mu \nu} (p) = i c g^2  \alpha \frac{\tilde{p}^{\mu} \tilde{p}^{\nu}}{\tilde{p}^4}
\end{dmath}

The coefficient $c=(N_s + 2 -2N_f)$ turns out to be $c \propto N_b - N_f$ where $N_b$ is the number of bosons. As for the higher-order terms\footnote{Still the one-loop order}(more convergent), the UV-finiteness still holds, though the IR-divergence persists and appears logarithmically.

Fortunately, this can be cured precisely as much as SUSY exists in the theory. Indeed, \underline{$c=0$ in supersymmetric theories.} Even if the SUSY is softly broken, this still holds as the loop contribution is a UV effect.

Although $\mathcal{N}=1$ SUSY can solve the problem at the leading order, there are still logarithmic IR-divergences at higher orders. 

The same arguments hold for the vertex quantum computations, except that the IR-divergence is of order $\theta^{-1}$.

It turns out that in $\mathcal{N}=4$ SUSic non-commutative $U(1)$ Yang-Mills, the problem of IR-divergence can be completely solved.  \cite{WEANCGT}

\subsection{$\mathcal{N}=2$ Non-Commutative $U(N)$ Gauge Theory Prepotential}

In this section, the background field method is used, where the gauge field is decomposed into:

\begin{dmath}
     A_{\mu} = B_{\mu} + N_{\mu} 
\end{dmath}
in which $B$ is the background classical non-commutative field and $N$ is the quantum fluctuations.

Also, the signature of the metric is Euclidean, and the rest of the conventions are as follows:

$$t^0 = (1/ i \sqrt{2N} ) 1_N , \ \ t^A = t^{(0, a)} \ \ \text{where} \ \ a= 1 ,..., N^2 - 1 \ \ (\text{which are the} \ SU(N) \ \text{generators})$$

and $\text{Tr}(t^A t^B) = - \frac{\delta^{AB}}{2}$. The commutation and anti-commutation relations are:

$$[t^A , t^B] = f^{ABC} t^C \ \ \ \ \ \ \ \ \{t^A , t^B\} = -\frac{\delta^{AB}}{N} -i d^{ABC} t^C$$

in which $f^{ABC}$ is totally anti-symmetric and $d^{ABC}$ is totally symmetric, and they are the same as those of $SU(N)$ along with  $f^{0BC} = 0$, $d^{0BC} = \sqrt{\frac{2}{N}} \delta^{BC}$, $d^{00A} = 0$ and $d^{000} = \sqrt{\frac{2}{N}}$.

Based on dimensional analysis, it is not hard to show that the Wilsonian polarization tensor is:

\begin{dmath}
    \Pi^{AB}_{\mu \nu} (k) = \Pi^{AB}_1 (k^2 , \tilde{k}^2) (k^2 \delta_{\mu \nu} - k^{\mu} k^{\nu}) + \Pi^{AB}_2 (k^2 , \tilde{k}^2) \frac{\tilde{k}^{\mu} \tilde{k}^{\nu}}{\tilde{k}^4} 
\end{dmath}

where $\Pi_2$ appears only in the non-commutative theory, indeed, while $\Pi_1$ receives planar and nonplanar contributions, $\Pi_2$ arises only because of nonplanar diagrams. At the end of this section, it becomes clear that $\Pi_1$ receives logarithmic IR divergence due to nonplanar diagrams that maintain their presence even in $\mathcal{N}=2$ SYM.

The general action for a spin $j$ field in representation $\mathbf{r}$ is as follows:

\begin{dmath}
    S[\phi] = -\int d^4 x \phi_{m , a} * (-D^2 (B) \delta_{mn} \delta^{ab} +2i (F^{B}_{\mu \nu})^{ab} \frac{1}{2} J^{\mu \nu}_{mn}) * \phi_{n, b} := - \int d^4 x \phi_{m,a} \left[ \Delta_{j, \mathbf{r}} \right]^{ab}_{mn} * \phi_{n,b}   
\end{dmath}
where $a,b$ indices of representation $\mathbf{r}$ and $J^{\mu \nu}_{mn}$ are the generators of the Euclidean Lorentz group for the relevant spin $j$:

\begin{dmath}
    {J=0 \ \ \ \ \  \text{for spin 0 fields} \ \ \ \ \ J^{\mu \nu}_{\rho \sigma}= i (\delta^{\mu}_{\rho} \delta^{\nu}_{\sigma} - \delta^{\nu}_{\rho} \delta^{\mu}_{\sigma}) \ \ \ \ \ \text{for 4-vectors}}
\end{dmath}

$${[J^{\mu \nu}]_{\alpha}}^{\beta} = i {[\sigma^{\mu \nu}]_{\alpha}}^{\beta} \ \ \ \ \ \text{for Weyl fermions} $$

The effective action \footnote{In this section, the action format convention is as the rest of the thesis} of the theory is:

\begin{dmath}
    S_{\text{eff}} [B]= - \frac{1}{2g^2} \int d^4 x \text{Tr} F^{B}_{\mu \nu}  * F^{B}_{\mu \nu} - \sum_{j ,\textbf{r}} \alpha_j \text{log} \text{det}_{*} \Delta_{j, \mathbf{r}}
\end{dmath}

in which $\alpha_j = +1$ for ghosts, $-1$ for scalars, $+1/2$ for Weyl fermions and $-1/2$ for gauge fields.

\begin{dmath}
    {\text{log} \text{det}_{*} \Delta_{j , \mathbf{r}} := \text{log} \text{det}_{*} (-\partial^2 + \mathcal{K}(B)_{j, \mathbf{r}})} = \text{log} \text{det}_{*} (-\partial^2) + \text{tr}_{*} \text{log} (1+ (-\partial^2)^{-1} \mathcal{K}_{j , \mathbf{r}} (B))
\end{dmath}

Indeed, the action of $\Delta_{j ,\mathbf{r}}$ on any adjoint of any spin is:

\begin{dmath}[label={eq:Delta-on-adjoint}]
    \Delta_{j, \mathbf{G}} * \phi:= - \partial^2 \phi + \mathcal{K}(B)_{j, \mathbf{G}} * \phi
    = - \partial^2 \phi - [\partial_{\mu} B^{\mu} , \phi]_{*} - 2 [B_{\mu} , \partial^{\mu} \phi]_{*} -[B_{\mu} , [B_{\mu} , \phi]_*]_* + 2i \left( \frac{1}{2} J^{\mu \nu} [F^{B}_{\mu \nu} , \phi]_{*} \right)
\end{dmath}

where

\begin{dmath}
    [\phi_1 , \phi_2 ]_{*} = \left( -\frac{i}{2} [\phi^A_1 , \phi^B_2]_{*} d^{ABC} + \frac{1}{2} \left\{ \phi^A_1 , \phi^B_2 \right\}_{*} f^{ABC}  \right) t^C
\end{dmath}

The expansion of \eqref{eq:Delta-on-adjoint} results in the loop diagrams that are of three types:

\begin{dmath}[label={eq:loop-diagram-a}]
    -\frac{1}{2} \int \frac{d^4 k}{(2 \pi)^4} B^{A}_{\mu} (k) B^{B}_{\nu} (-k) \int \frac{d^D p}{ (2 \pi)^D} \text{Tr} {\frac{-(2p + k)_{\mu} (2p + k)_{\nu} M^{AB} (k, p)}{p^2 (p+k)^2}}
\end{dmath}
\begin{dmath}[label={eq:loop-diagram-b}]
    \int \frac{d^4 k}{ (2 \pi)^4} B^{A}_{\mu} (k) B^{B}_{\nu} (-k) \int \frac{d^D p}{(2 \pi)^D} \text{Tr} \frac{\delta_{\mu \nu} M^{AB} (k , p)}{p^2}
\end{dmath}

in which 

\begin{dmath}[label={eq:M-AB}]
    M^{AB} (k,p) = (-d \text{sin}(\frac{k\tilde{p}}{2}) + f \text{cos}(\frac{k\tilde{p}}{2}))^{ALM} (d \text{sin}(\frac{k\tilde{p}}{2}) + f \text{cos}(\frac{k\tilde{p}}{2}))^{BML}
\end{dmath}

and $\text{Tr} \mathbb{I}_j = d(j) $ where $d(j)$ is the number of the spin components of the field $\phi$:

\begin{dmath}
    d(j) =1 \ \ \text{for scalars} \ \ \ 2 \ \ \ \text{for fermions} \ \ \ 4 \ \ \ \text{for vectors}
\end{dmath}

In a supersymmetric theory, one can ensure that $\sum_j \alpha_j d(j) =0$, even if the SUSY is broken softly. Therefore, both contributions \eqref{eq:loop-diagram-a} and \eqref{eq:loop-diagram-b} vanish separately.

Ultimately, the only surviving contribution comes from coupling to the background field:

\begin{dmath}
    -\frac{1}{2} \int \frac{d^4 k}{(2 \pi)^4} B^{A}_{\mu} (k) B^{B}_{\nu} (-k) \int \frac{d^D p}{(2 \pi)^D} \text{Tr} \frac{-4 J^{\mu \rho} J^{\nu \lambda} k_{\lambda} k_{\rho} M^{AB} (k,p)}{p^2 (p+k)^2}
\end{dmath}

where:

\begin{dmath}
    \text{Tr}( J^{\mu \rho} J^{\nu \lambda})_j = C(j) (\delta^{\mu \nu} \delta^{\rho \lambda} - \delta^{\mu \lambda} \delta^{\nu \rho})
\end{dmath}

in which:

\begin{dmath}
    C(j) = 0 \ \ \ \ \text{for scalars}, \ \ \ \frac{1}{2} \ \ \  \text{for Weyl fermions}, \ \ \  2  \ \ \ \text{for vectors}
\end{dmath}

The integrals can be further simplified with the aid of the relations:
\begin{dmath}
    {f^{ALM} f^{BML} = -N c_A \delta^{AB}, \ \ \ \ d^{ALM} d^{BML} = N d_A \delta^{AB}, \ \ \ \ f^{ALM} d^{BML} = 0}
\end{dmath}

where 

\begin{dmath}
    {c_A = 1 - \delta_{0A} \ \ \ \ \ \ d_{A} = 2 - c_A}
\end{dmath}

One can rewrite \eqref{eq:M-AB}:

\begin{dmath}
    {M^{AB} (k , p) = - N \delta^{AB} (1- \delta_{0A} \text{cos}(k \tilde{p}))}
\end{dmath}

which separates the planar from the nonplanar contributions. This is actually excellent news, as is now possible to propose.

\begin{dmath}
    [\Pi^{AB}_{\mu \nu}]^{\text{planar}} [U(N)] = N \delta^{AB} \Pi^{\text{planar}}_{\mu \nu} [U(1)]
\end{dmath}

\begin{dmath}[label={eq:nonplanar-polarization-UN}]
    [\Pi^{AB}_{\mu \nu}]^{\text{nplanar}} [U(N)] = N \delta^{A0} \delta^{B0} \Pi^{\text{nplanar}}_{\mu \nu} [U(1)]
\end{dmath}

where from \eqref{eq:nonplanar-polarization-UN} it can be read that there is no nonplanar contributions to $\Pi_{\mu \nu}$ apart from that of the $U(1) \subset U(1) \times SU(N) = U(N)$ at one-loop order. As it was shown in Section \ref{Non-Commutative $U(1)$ Yang-Mills, Infrared Divergence}, the quadratic IR-divergence in $\Pi^{\text{nplanar}}_{\mu \nu} [U(1)]$ vanishes in any supersymmetric theory even if the SUSY is softly broken. The question is if there might be any logarithmic IR-divergences in $\mathcal{N}=2$ pure Yang-Mills. The answer is yes, as there is a logarithmic contribution in the infrared; \footnote{however, it is worth mentioning that this residual UV/IR mixing disappears in the $\mathcal{N}=4$ SYM \cite{WEANCGT}} 

Indeed \footnote{the normalizatioin of the generator of the $U(1)$ is changed mildly but does not cause any inconsistancy $t^0 = \frac{1}{i}$ } :

\begin{dmath}
    \Pi^{\text{nplanar}}_{\mu \nu} [U(1)] = \Pi^{\text{nplanar}}_{\mu \nu} (k^2 , \tilde{k}^2)= - \frac{4C(\mathbf{G})}{(4\pi)^2} \left( \sum_j C(j) \alpha_j \right) \int^1_0 dx K_0 \left(\sqrt{ \tilde{k}^2 k^2 x(1-x)}\right) 
    (k^2 \delta_{\mu \nu} - k_{\mu} k_{\nu}) 
\end{dmath}

where $\mathbf{G}$ stands for the adjoint representation of non-commutative $U(1)$; in this case $C(\mathbf{G}) = 1$. As can be directly checked, the modified Bessel function of the second kind $\lim_{r\rightarrow 0} K_0 (r) \propto \text{log}(r)$ and so $\Pi^{\text{nplanar}}_{\mu \nu} [U(1)]$ has a logarithmic singularity in the $\tilde{k}^2 \rightarrow 0$.

From \eqref{eq:nonplanar-polarization-UN}, it is clear that the $U(1)$ \textbf{Centre} of $U(N)$ is decoupled from the rest of the group in the effective theory\footnote{also take a look at the initial discussion of \cite{Non-commutative GT2} to check the same phenomenon for the more general case of $\mathcal{N}=0$ $U(N)$ YM}, and the nonplanar $SU(N)$ do not contribute to the one-loop effective action; resulting in:

\begin{dmath}
    {\frac{1}{g^2_{\text{eff}} (k^2)}}_{U(N)} = {{\frac{1}{{g'}^2_{\text{eff}} (k^2)}}}_{U(1)} \oplus \frac{1}{g^2_{\text{eff}} (k^2)}_{SU(N)}
\end{dmath}

It is argued that in the effective low-energy action of $\mathcal{N}=2$ $U(N)$ NCSYM, there are $N$ massless vector multiplets; however, there are only $N-1$ moduli over the Coulomb branch \cite{ERNSGT}, \cite{NCN2}.

\subsection{Instanton Localization}

\subsection{Does the "resolution of the moduli space" idea work?!}

Further investigation ensures that the effective Lagrangian of $\mathcal{N}=2$ NCSYM is \cite{NCN2}:

\begin{dmath}
    \mathcal{L} = \int d^2 \theta \left(  \frac{\partial^2 \mathcal{F} ( A , A_{\{ u\}})}{\partial A^2} W_{\alpha} W^{\alpha} + \sum^{N-1}_{u,v =1} \frac{\partial^{2} \mathcal{F} (A , A_{\{ u \}})}{\partial A_u \partial A_v} W^{u}_{\alpha} W^{\alpha v} \right)
\end{dmath}

where the prepotential with the aid of symmetry arguments can be written as:

\begin{dmath}
    \mathcal{F} = \frac{1}{2} \tau_0 A^2 + f(A_{\{ u \}})
\end{dmath}

which explicitly decouples $A_{\{u\}}$, $u=1,...,N-1$ moduli from the non-commutative one $A$.

Indeed, explicit computations will agree with this, in that the one-instanton and $2$-instanton coefficients in the multi-instanton expansion of the prepotential in $U(N)$ agree with those of $SU(N)$.

As discussed in Section \ref{Non-Commutative Spacetime, Modified ADHM Constraints}, the ADHM constraints are modified on the non-commutative background spacetime. 

Indeed, the relation \eqref{eq:instanton-partition-function} is modified such that the integrals are carried out over the (centred) resolved moduli space $\widehat{\mathcal{M}}^{(\zeta)}_{\pm k}$ instead of $\widehat{\mathcal{M}}_{\pm k}$ to find the instanton partition functions:

\begin{dmath}
    \widehat{\mathcal{Z}}^{(2)}_k = \int_{\widehat{\mathcal{M}}^{(\zeta)}_{\pm k}} \boldsymbol{\omega}^{(2 )} e^{-\tilde{S}}
\end{dmath}

where the SUSic $\boldsymbol{\omega}^{(2,N_F)}$ is modified to:

\begin{dmath}
\int_{{\widehat{\mathcal{M}}^{(\zeta)}}_{\pm k}}  \boldsymbol{\omega}^{(2)} = \frac{C^{(2)}_k}{\text{VolU} (k)} \int d^{4k(N+k)} \hat{a} \prod^{2}_{A=1} d^{2k(N+k)} \widehat{\mathcal{M}}^{A} | {\det}_{k^2} \mathbf{L} |^{-1} \times \\ \prod^{k^2}_{r=1} \left\{ \prod^{3}_{c=1} \delta (\frac{1}{2} \tr_k T^r ({\tau^{c \dot{\alpha}}}_{\dot{\beta}} \bar{a}^{\dot{\beta}} a_{\dot{\alpha}} - \zeta^c_{(+)}))  \prod^{2}_{A=1}   \prod^{2}_{\dot{\alpha}} \delta (\tr_k T^r (\bar{\mathcal{M}}^A a_{\dot{\alpha}} + \bar{a}_{\dot{\alpha}} \mathcal{M}^A)) \right\}
\end{dmath}

in which $C^{(2)}_k = 2^{-k(k-1)/2} \pi^{-2kN}$.

Indeed, the integral over $\vec{D}$ is modified, due to the addition of $4 \pi^2 i \vec{D} \ . \ \vec{\zeta}_{(+)}$ term to the effective action $\tilde{S}$ in \eqref{eq:effective-action-k1}:

\begin{dmath}
    \int \frac{d^3 D}{\vec{D}^2} {e^{-4 \pi^2 i \vec{D} \ . \ \vec{\zeta}_{(+)}}}  {\prod^{N}_{u=1}}  \frac{\alpha^{* 2}_u}{|\alpha_{u}|^4 + \vec{D}^2} 
\end{dmath}

where now, the angular integrals over $\vec{D} = (|\vec{D}| , \theta , \phi)$ are non-trivial:

\begin{dmath}
    \int d(\text{cos} \theta) d \phi \ e^{-4 \pi^2 i \vec{D} \ . \ \vec{\zeta}_{(+)}} = \frac{\text{sin} (4 \pi^2 |\vec{D}| |\vec{\zeta}_{(+)}|)}{\pi |\vec{D}| |\vec{\zeta}_{(+)}|}
\end{dmath}

which after contour integration over $|\vec{D}|$ from $-\infty$ to $+\infty$ gives rise to:

\begin{dmath}
    \frac{1}{2 |\vec{\zeta}_{(+)}|} \left( \prod^{N}_{r=1} \frac{1}{\alpha^2_{r}} - \sum^{N}_{u=1} \frac{1}{\alpha^2_{u}} e^{-4 \pi^2 |\vec{\zeta}_{(+)}| |\alpha_{u}|^2} \prod^{N}_{\substack{s=1 \\ s \neq u}} \frac{{\alpha^{* 2}_s}}{|\alpha_s|^4-|\alpha_u|^4}\right)
\end{dmath}

which agrees with that of the integration of the singular moduli space in the neighborhood of singularities $\chi = -\phi^0_u$(equivalently $\alpha_u = 0$)

\begin{dmath}
    2 \pi^2 \frac{\alpha^{*}_u}{\alpha_u} \prod^{N}_{\substack{s=1 \\ s \neq u}} \frac{{\alpha^{* 2}_s}}{|\alpha_s|^4-|\alpha_u|^4}
\end{dmath}

This is an exciting result! However, one must be careful that the contributions arising from infinity in the $\chi$-space disappear in this case, since $f_1 (r, \theta)$ is modified from \eqref{eq:f1-f2-asymptotics} to:

\begin{dmath}
    f_1 (r, \theta) \sim \frac{e^{-2i N \theta}}{r^{2N}}
\end{dmath}

and therefore \eqref{eq:contribution-at-infinity} falls too fast for $N_F < 2N$. (or due to a neat cancellation in the angular integral when $N_F = 2N$).

Finally, the non-commutative prepotential is found to be:

\begin{dmath}
    \mathcal{F}^{\text{nc}}_1 = \sum^{N}_{u=1} \prod^{N}_{\substack{v=1 \\ v \neq u}} \frac{1}{(\phi^0_v - \phi^0_u)^2} \prod^{N_F}_{f=1} (m_f + \phi^0_u)
\end{dmath}

which lacks the term $\mathcal{S}^{(N_F)}_1$ compared with \eqref{eq:F1-commutative}, which is precisely what one expects, since these terms rise from the vicinity of the singularities and the resolved moduli space does not have any singularity any longer. However, the case of $N_F = 2N$ is rather tricky and a redefinition of the coupling is necessary to include this case. Indeed, the non-commutative and the commutative couplings are no longer the same:

\begin{dmath}[label={eq:tau-nc}]
    \tau^{\text{nc}} = \tau + \sum^{\infty}_{k=1} c_k e^{2 \pi i k \tau} 
\end{dmath}

where, for instance:

\begin{dmath}[label={eq:c1}]
    c_1 = - 2^{-2N}\frac{\begin{pmatrix}
    2N \\
    N-1
\end{pmatrix}}{i \pi}
\end{dmath}

The prepotentials are related:

\begin{dmath}[label={eq:F-nc-versus-F}]
    \mathcal{F}^{\text{nc}}_k (\phi^0_u , \vec{\zeta}_{(+)} ) = \mathcal{F}_k -2 \pi i d_k \sum^N_{u=1} (\phi^0_u)^2
\end{dmath}

up to constants that do not change the result. To prove this relation, the more advanced technique of localization must be introduced first. It turns out that $\mathcal{F}^{\text{nc}} - \mathcal{F}$ does not depend on $\vec{\zeta}_{(+)}$. This, along with the assumption that the difference is due to $\chi$-space sphere at infinity contribution (as seen in the one instanton case ) and that the prepotential has mass dimension two, result in $\mathcal{F}^{\text{nc}} - \mathcal{F}$ being a polynomial in masses $m_f$ and $\phi^0_u$ of degree two, which leaves no other possibility than \eqref{eq:F-nc-versus-F}.

\subsection{Localization}\label{subsec:Localization}

The localization technique is based on the introduction of a new type of field variation that mixes the $R$-symmetry indices and the quaternionic spacetime indices, which is called \textbf{Topological Twisting}:

\begin{dmath}
    \delta = \bar{\xi}_{\dot{\alpha} A} Q^{\dot{\alpha} A} 
\end{dmath}

Under this variational operator, the fields transform as follows:

\begin{dmath}
    {\delta a'_{\alpha \dot{\alpha}} = i \bar{\xi}_{\dot{\alpha} A} \mathcal{M}^{' A}_{\alpha}}
    \\
    {\delta \mathcal{M}^{' A}_{\alpha} = 2 \Sigma^{aAB} \bar{\xi}^{\dot{\alpha}}_B \mathcal{D}_a a^{'}_{\alpha \dot{\alpha}}}
    \\
    {\delta \chi_a = i \Sigma^{AB}_a \bar{\xi}_A \bar{\psi}_B}
    \\
    {\delta \bar{\psi}_A = {{\bar{\Sigma}^{ab \  B}}_{ \ A}} F_{ab} \bar{\xi}_B - i
    \bar{\sigma}_{mn} D_{mn} \bar{\xi}_A}
    \\
    {\delta w_{\dot{\alpha}} = i \bar{\xi}_{\dot{\alpha} A} \mu^A}
    \\
    {\delta \mu^A = 2 \Sigma^{aAB} \bar{\xi}^{\dot{\alpha}}_B \mathcal{D}_a w_{\dot{\alpha}}}
\end{dmath}

Now, the crucial element of this technique is the introduction of the relevant charge for such variations:

\begin{dmath}[label={eq:BRST-charge}]
    \mathcal{Q} = \epsilon_{\dot{\alpha} A} Q^{\dot{\alpha} A} \ \ \ \ \text{BRST Charge}
\end{dmath}

Using this charge, one can rewrite the transformations:

\begin{dmath}
    {\mathcal{Q} w_{\dot{\alpha}} = i \epsilon_{\dot{\alpha} A} \mu^A} 
    \\
    {\mathcal{Q}  a'_{\alpha \dot{\alpha}} = i \epsilon_{\dot{\alpha} A} \mathcal{M}^{' A}_{\alpha}}
    \\
    {\mathcal{Q} \chi = 0}
    \\
    {\mathcal{Q} \bar{\psi}^{\dot{\alpha}}_A = \frac{1}{2} \delta^{\dot{\alpha}}_{ \ A} [\chi^{\dagger} , \chi] - i \vec{D} \ . \vec{\tau}^{\dot{\alpha}}_{\ \dot{\beta}} \delta^{\dot{\beta}}_{A}}
    \\
    {\mathcal{Q} \mathcal{K} = \mathcal{Q} \tilde{\mathcal{K}} = 0}
    \\
    {\mathcal{Q} \mu^{A} = -2 \epsilon^{\dot{\alpha} A} \left( w_{\dot{\alpha}} \chi + \phi^0 w_{\dot{\alpha}} \right)}
    \\
    {\mathcal{Q} \mathcal{M}^{'A}_{\alpha} = -2 \epsilon^{\dot{\alpha} A} [a'_{\alpha \dot{\alpha}} , \chi]}
    \\
    \mathcal{Q} \chi^{\dagger} = 2i \delta^{A}_{\ \ \dot{\alpha}} \bar{\psi}^{\dot{\alpha}}_{A}
    \\
    {\mathcal{Q} \vec{D} = \delta^{ \ A}_{\dot{\alpha}} \vec{\tau}^{\dot{\alpha}}_{\dot{\beta}} [\bar{\psi}^{\dot{\beta}}_A, \chi]}
\end{dmath}

Indeed, the BRST charge turns out to be a nilpotent modulo infinitesimal $U(k) \times SU(N)$ transformation; for instance:

$$\mathcal{Q}^2 w_{\dot{\alpha}} = 2i(w_{\dot{\alpha}} \chi + \phi^0 w_{\dot{\alpha}})$$

The elegance of introducing this charge is revealed when the effective action can be written:

\begin{dmath}
    \tilde{S} = \mathcal{Q} \Xi + \Gamma 
\end{dmath}

where 

\begin{dmath}
    \Xi = 4 \pi^2 \text{tr}_k \left\{ \frac{1}{2} \epsilon_{\dot{\alpha} A} \bar{w}^{\dot{\alpha}} (\mu^A \chi^{\dagger} + \phi^{0 \dagger} \mu^{A}) + \frac{1}{4} \epsilon_{\dot{\alpha} A} \bar{a}^{'\dot{\alpha} \alpha} [\mathcal{M'}^{A}_{\alpha} , \chi^{\dagger}] + \delta^A_{\ \dot{\alpha}} \bar{\psi}^{\dot{\beta}}_A (\bar{a}^{\dot{\alpha}} a_{\dot{\beta}} - \frac{1}{2} \vec{\zeta}_{(+)} \ . \vec{\tau}^{\dot{\alpha}}_{\dot{\beta}}) \right\}
\end{dmath}

and 

\begin{dmath}
    \Gamma = - \pi^2 \sum^{N_F}_{f=1} \text{tr}_k \left( (m_f - \chi) \mathcal{K}_f \tilde{\mathcal{K}}_f \right)
\end{dmath}

where:

\begin{dmath}
    \boxed{\mathcal{Q} \Gamma =0}
\end{dmath}

meaning, that $\mathcal{Q} \tilde{S} = 0$; the effective action is $\mathcal{Q}$-closed.

Now, one can define the generalized instanton partition function:

\begin{dmath}
    \widehat{\mathcal{Z}}^{(2, N_F)}_k (\lambda) = \int_{\widehat{\mathcal{M}}^{(\zeta)}_k} \boldsymbol{\omega}^{(2,N_F)} \text{exp} \left( - \lambda^{-1} \mathcal{Q} \Xi - \Gamma \right)
\end{dmath}

The fascinating observation comes after taking a derivative with respect to $\lambda$ and using the fact that $\omega^{(2,N_F)}$ is SUSY-invariant and that $\mathcal{Q}^2 \Xi = \mathcal{Q} \Gamma = 0$ is that:

\begin{dmath}
    \frac{\partial \widehat{\mathcal{Z}}^{(2, N_F)}_k (\lambda)}{\partial \lambda} = \lambda^{-2} \int_{\widehat{\mathcal{M}}^{(\zeta)}_k} \boldsymbol{\omega}^{(2,N_F)} \mathcal{Q} \left\{ \Xi \text{exp} \left( - \lambda^{-1} \mathcal{Q} \Xi - \Gamma \right) \right\} = 0 \ \ \ !
\end{dmath}

Therefore, one can choose $\lambda$ as one wills! The best choice would be $\lambda \rightarrow 0$, since at this limit the integral is dominated by the critical points of $\mathcal{Q} \Xi$.

Intuitively, one can set $\vec{\zeta}_{(+)} = 0$ and check the critical points of $\mathcal{Q} \Xi$. By doing so one finds:

\begin{dmath}
    |w_{\dot{\alpha}} \chi_a + \phi^0_a w_{\dot{\alpha}}|^2 - [\chi_a , a'_n]^2 \geq 0
\end{dmath}

The critical points are those for which:

\begin{dmath}[label={eq:critical-point-equations}]
    {w_{\dot{\alpha}} \chi_a + \phi^0_a w_{\dot{\alpha}} = [\chi_a , a'_n] = 0}
\end{dmath}

These are precisely the points of singularity on the centred moduli space; namely: $\text{Sym}^k \mathbb{R}^4 \subset \widehat{\mathcal{M}}_k$.

Indeed, it can be shown \cite{Instanton Calculus},\cite{Integral} that the solutions to Equations \eqref{eq:critical-point-equations}, $\chi_a$, which are $k \times k$ matrices, are partitioned into:

\begin{dmath}
    k \rightarrow k_1 + k_2 + ... + k_N 
\end{dmath}

up to $U(k)$ auxiliary transformations, each corresponding to a $U(1)$ instanton of charge $k_u$.

Indeed, tracing the modified ADHM constraints \eqref{eq:modified-ADHM-constraints}, one finds the following:

\begin{dmath}
    \vec{\tau}^{\dot{\alpha}}_{ \ \ \dot{\beta}} \sum^{k_u}_{i= k_{u-1} +1} \bar{w}^{\dot{\beta}}_{iu} w_{ui \dot{\alpha}} = k_u \vec{\zeta}_{(+)}
\end{dmath}

However, the central term guarantees a nontrivial solution for $w_{ui \dot{\alpha}}$; indeed, the critical point set turns out to be the product of noncommutative $U(1)$ instanton moduli spaces:

\begin{dmath}
    \frac{\widehat{\mathcal{M}}^{(\zeta)}_{k_1} |_{N=1} \times \cdots \times \widehat{\mathcal{M}}^{(\zeta)}_{k_N} |_{N=1}}{\mathbb{R}^4}
\end{dmath}

As mentioned in Section \ref{Non-Commutative Spacetime, Modified ADHM Constraints}, each (non-singular) $\widehat{\mathcal{M}}^{(\zeta)}_k |_{N=1}$ is the resolution of the (singular) symmetric product $\text{Sym}^k \mathbb{R}^4$. Now, exact solutions do exist even in the presence of VEVs and are called \textbf{Topicons}. Their important feature is that they have no size modulus and are localized, unlike instantons. $N$ flavours of topicons exist corresponding to each of the $N$ blocks of $\chi_a$. Therefore, the \textbf{Topicons Moduli Space} lies in the greater intanton moduli space:

\begin{dmath}[label={eq:topicon-moduli-space}]
    {\widehat{\mathcal{M}}_k \xrightarrow{\text{resolution}} \widehat{\mathcal{M}}^{(\zeta)}_k \overset{\text{exact}}{\supset} \widehat{\mathcal{M}}^{(\zeta)}_k |_{\text{topicons}} = \bigcup_{\substack{\text{partitions of} \ k \\ \sum^N_{i=1} k_i = k}} \frac{\widehat{\mathcal{M}}^{(\zeta)}_{k_1} |_{N=1} \times \cdots \times \widehat{\mathcal{M}}^{(\zeta)}_{k_N} |_{N=1}}{\mathbb{R}^4}}
\end{dmath}

Now, it is time to put it to the test!
The objective is to find the one and two-instanton contributions to the prepotential of $\mathcal{N}=2$ SYM with $N_F$ hypermultiplets using the localization technique.

\begin{itemize}
    \item One Instanton

The effective action $\tilde{S}$ \eqref{eq:effective-action-k1} has $N$ critical points, labelled by $v \in \{ 1,2,...,N \}$, at which:

\begin{dmath}
    {\chi_a = -(\phi^0_a)_v \ \ \ \ \ \ w_{u \dot{\alpha}} \propto \delta_{u v}}
\end{dmath}

where in one-instanton sector $a'_n =0$. From here onwards, the $v$ index is fixed. One can assume the non-commutativity parameters to be:
\begin{dmath}
    {\zeta^1_{(+)} = \zeta^2_{(+)} = 0 \ , \ \ \ \ \zeta^3_{(+)} := \zeta > 0}
\end{dmath}
The ADHM constraints' solutions turn out to be:
    \begin{dmath}
        w_{u \dot{\alpha}} = \sqrt{\zeta} e^{i \theta} \delta_{uv} \delta_{\dot{\alpha} 1}
    \end{dmath}
    for arbitrary phase $\theta$.
The integrals can be carried out, bearing in mind that the Dirac delta function nullifies the integrals over $w_{v\dot{\alpha}}$:

\begin{dmath}[label={eq:w-integral-k1}]
\int d^2 w_v d^2 \bar{w}_v   \prod^3_{c=1} \delta \left( \frac{1}{2} \tau^{c \dot{\alpha}}_{\dot{\beta}}  \bar{w}^{\dot{\beta}}_{v} w_{v \dot{\alpha}} - \frac{1}{2} \zeta \delta^{c3} \right) = 8 \pi \zeta^{-1} 
\end{dmath}

and correspondingly, the delta functions of the Grassmann ADHM constraints saturate the integrals over $\{ \mu^A_v, \bar{\mu}^A_v \}$:

\begin{dmath}[label={eq:mu-integral-k1}]
    \int d \mu^{A}_v d \bar{\mu}^A_v \prod^2_{\dot{\alpha}} \delta \left( \bar{w}_{v \dot{\alpha}} \mu^A_v + w_{v \dot{\alpha}} \bar{\mu}^A_v \right) = \zeta 
\end{dmath}

for $A=1,2$ separately. The rest of the variables $\{ w_{u \dot{\alpha}}, \mu^A_u, \bar{\mu}^A_u \}_{u \neq v}$ and $\{ \mathcal{K}_{if}, \tilde{\mathcal{K}}_{fi} \}$ are treated as Gaussian fluctuations around the critical points leading to the effective action:

\begin{dmath}
    \tilde{S} = 4 \pi^2 \left\{ \zeta \chi^2_a + \sum^N_{\substack{u = 1 \\ u \neq v}} \left( (\phi^0_a)^2_{vu} |w_{u \dot{\alpha}}|^2 + \frac{i}{2}\phi^{0 \dagger}_{vu} \bar{\mu}^A_u \mu_{u A} \right) -\frac{1}{4} \sum^{N_F}_{f=1} (m_f + \phi^0_v) \mathcal{K}_f \tilde{\mathcal{K}}_f \right\} + \cdots
\end{dmath}

in which $(\phi^0_a)_{uv} = - (\phi^0_a)_{vu} := (\phi^0_a)_u - (\phi^0_a)_v$. The integral over $\chi_a$ leads to the natural cancellation of the $\zeta$ factor in \eqref{eq:w-integral-k1} and \eqref{eq:mu-integral-k1}! Summing over the $N$ critical point sets gives rise to the prepotential:

\begin{dmath}[label={eq:F1-nc-localization}]
    {\mathcal{F}^{\text{nc}}_1 := \widehat{\mathcal{Z}}^{(2,N_F)}_1 = \sum^N_{v=1} \left\{ \prod^N_{\substack{u=1 \\ u \neq v}} \frac{1}{(\phi^0_v - \phi^0_u)^2} \prod^{N_F}_{f=1} (m_f + \phi^0_v) \right\}}
\end{dmath}

This result is manifestly holomorphic in VEVs and independent of $\zeta$ as suggested initially.
This holds for the case of $N_F < 2N$, and as discussed previously (see \eqref{eq:tau-nc}), a redefinition of the coupling for the finite theory $N_F = 2N$ is necessary to extend the result to this case.

   \item Two instantons

   In this case, there are two types of critical points:
   \begin{itemize}
       \item Two topicons of distinct flavours

In this case, the ADHM data can be written as:

\begin{dmath}
    {w_{u i \dot{\alpha}} = \sqrt{\zeta} e^{i \theta_i} \delta_{u u_i} \delta_{\dot{\alpha} 1} , \ \ \ \ \ \  a'_n = \begin{pmatrix}
        Y_n & 0 \\
        0 & -Y_n
    \end{pmatrix}}
\end{dmath}
The phases $\theta_i$, $i=1,2$, are not genuinely moduli parameters, as they can be separately rotated by the $U(2)$ auxiliary group transformation, and $Y_n$ are the coordinates of the relative positions of the topicons from each other. As for the Grassmann ADHM counterparts:

\begin{dmath}
    {\mu^A = \bar{\mu}^A = 0 \ \ \ \ \ \ \ \mathcal{M}^{'A}_{\alpha} = \begin{pmatrix}
        \rho^A_{\alpha} & 0 \\
        0 & -\rho^A_{\alpha}
    \end{pmatrix}}
\end{dmath}
       where $\rho^A_{\alpha}$ are the superpartners of $Y_n$. Since $Y_n \in \mathbb{R}^4$, the moduli space is non-compact. It turns out that the integral over $Y_n$ is, nevertheless, convergent!

       Expanding around the critical points, the fluctuations are:

\begin{dmath}
    {\delta a'_n = \begin{pmatrix}
        0 & Z_n \\
        Z^{*}_n & 0 
    \end{pmatrix} , \ \ \ \ \ \delta \mathcal{M}^{' A}_{\alpha} = \begin{pmatrix}
        0 & \sigma^A_{\alpha} \\
        \varepsilon^A_{\alpha} & 0
    \end{pmatrix}}
\end{dmath}
and making the following shift:

\begin{dmath}
    \chi_a \rightarrow \chi_a - \begin{pmatrix}
        (\phi^0_a)_{u_1} & 0 \\
        0 & (\phi^0_a)_{u_2}
    \end{pmatrix} 
\end{dmath}

so that $\chi_a = 0$ on the critical point set submanifold.

Integrating over the Lagrange multiplier $\vec{D}$ and $\bar{\psi}^{\dot{\alpha}}_{A}$ impose the bosonic and Grassmann ADHM constraints, respectively. The off-diagonal entries $(i \neq j)$ vanish on the critical set; there, they must be expanded to linear order, and the diagonal components are the ADHM constraints of the two single $U(1)$ instantons. For bosonic variables:

\begin{dmath}
    {\sqrt{\zeta} e^{-i \theta_1} (w_{u_1 2})_2 + \sqrt{\zeta} e^{i \theta_2} (w_{u_2 1})^{*}_2 + 4i \bar{\eta}^1_{mn} Y_m Z_n = 0}
    \\
    {-i \sqrt{\zeta} e^{-i \theta_1} (w_{u_1 2})_2 + i \sqrt{\zeta} e^{i \theta_2} (w_{u_2 1})^{*}_2 + 4i \bar{\eta}^{2}_{mn} Y_m Z_n = 0}
    \\
    {\sqrt{\zeta} e^{-i \theta_1} (w_{u_1 2})_1 + \sqrt{\zeta} e^{i \theta_2} (w_{u_2 1})^{*}_1 + 4i \bar{\eta}^3_{mn} Y_m Z_n = 0}
\end{dmath}

where $\bar{\eta}^{c}_{mn} = -\frac{i}{2} \text{tr} (\tau^c \bar{\sigma}_m \sigma_n)$ are the 't Hooft symbols.

As for the Grassmann ADHM constraints:

\begin{dmath}
    {{\sqrt{\zeta} e^{i \theta_2} \bar{\mu}^{A}_{1 u_2} + 2 (\rho^{\alpha A} Z_{\alpha 1} - \sigma^{\alpha A} Y_{\alpha 1})} =0}
    \\
    {\sqrt{\zeta} e^{-i \theta_1} {\mu}^{A}_{u_1 2} + 2 (\rho^{\alpha A} {Z_{\alpha 2} - \sigma^{\alpha A} Y_{\alpha 2})}= 0}
    \\
    {{\sqrt{\zeta} e^{i \theta_1} \bar{\mu}^{A}_{2 u_1} + 2 (\varepsilon^{\alpha A} Y_{\alpha 1} - \rho^{\alpha A} Z^{*}_{\alpha 1})}= 0}
    \\
    {{\sqrt{\zeta} e^{-i \theta_2} {\mu}^{A}_{u_2 1} + 2 (\varepsilon^{\alpha A} Y_{\alpha 2} - \rho^{\alpha A} Z^{*}_{\alpha 2})}= 0}
\end{dmath}

where $Y_{\alpha \dot{\alpha}} = Y_n\sigma_{\alpha \dot{\alpha} n}$.
Before solving the equations, it is better to introduce new formulations for the ADHM data:

\begin{dmath}
    {(w_{u_1 2})_1 = e^{i \theta_1} (\xi + \lambda)} 
    \\
    {(w_{u_2 1})^{*}_1 = e^{-i \theta_2} (-\xi + \lambda)}
\end{dmath}

Now, one can use the $U(2)$ symmetry to rotate the $Z_n$ fluctuations enough to become orthogonal to $Y_n$; namely, $Z_n Y_n =0$ and also set $\theta_i = 0$. The Jacobian for the transformation is:

\begin{dmath}
    \frac{1}{\text{Vol U}(2)} \int d^{12} a' \rightarrow \frac{16}{\pi^2} \int d^4 Y d^3 Z d^3 Z^{*} Y^2
\end{dmath}
The expansion of the bosonic effective action to Gaussian order is then:

\begin{dmath}
    \tilde{S}_b = \tilde{S}^{(1)}_b + \tilde{S}^{(2)}_b + \cdots
\end{dmath}
in which:

\begin{dmath}
    \frac{1}{4 \pi^2} \tilde{S}^{(1)}_b = \zeta ((\chi_a)^2_{11} + (\chi_a)^2_{22}) + 8Y^2 |(\chi_a)_{12}|^2 + 2 |(\phi^0_a)_{u_1 u_2} \xi + \sqrt{\zeta} (\chi_a)_{12}|^2 + 2 (\phi^0_a)^2_{u_1 u_2} (1 + 4 \zeta^{-1} Y^2) |Z|^2
\end{dmath}
and
\begin{dmath}
    \frac{1}{4\pi^2} \tilde{S}^{(2)}_b = \sum^2_{i=1} \sum^{N}_{\substack{u=1 \\ u \neq u_1 , u_2}} (\phi^0_a)^2_{u u_i} |w_{ui \dot{\alpha}}|^2
\end{dmath}

It is helpful to apply the shift:
\begin{dmath}
    \xi \rightarrow \xi - \sqrt{\zeta} \frac{(\phi^0_a)_{u_1 u_2} (\chi_a)_{12}}{(\phi^0_a)^2_{u_1 u_2}}
\end{dmath}

and define:

\begin{dmath}
    {\chi_a = \chi^{\parallel}_a + \chi^{\perp}_a , \ \ \ \ \ \ \chi^{\perp}_a (\phi^0_a)_{u_1 u_2} = 0}
\end{dmath}

and rewrite:

\begin{dmath}
    \frac{1}{4 \pi^2} \tilde{S}^{(1)}_b = \zeta \left( (\chi_a)^2_{11} + (\chi_a)^{2}_{22} \right) + 2 \zeta (1 + 4 \zeta^{-1} Y^2) |(\chi^{\perp})_{12}|^2 + 8 Y^2 |(\chi^{\parallel})_{12}|^2 + 2 (\phi^0_a)^2_{u_1 u_2} \left( |\xi|^2 + (1+ 4 \zeta^{-1} Y^2) |Z|^2 \right)
\end{dmath}

While the Grassmann effective action can also be expanded as:

\begin{dmath}
    \tilde{S}_f = \tilde{S}^{(1)}_f + \tilde{S}^{(2)}_f + \cdots
\end{dmath}

where:
\begin{dmath}
   \frac{1}{4 \pi^2} \tilde{S}^{(1)}_f = - \frac{i}{2} \phi^{0 \dagger}_{u_1 u_2} (1 + 4 \zeta^{-1} Y^2) \sigma^{\alpha A} \varepsilon_{\alpha A} + i \rho^{\alpha A} (2 \zeta^{-1} (\phi^0_{u_1 u_2})^{\dagger} Z_{\alpha \dot{\alpha}} \bar{Y}^{\dot{\alpha} \beta} + \chi^{\dagger}_{12} {\delta_{\alpha}}^{\beta}) \varepsilon_{\beta A} + i \sigma^{\alpha A} (2 \zeta^{-1} (\phi^0_{u_1 u_2})^{\dagger} Y_{\alpha \dot{\alpha}} \bar{Z}^{* \dot{\alpha} \beta} - \chi^{\dagger}_{21} {\delta_{\alpha}}^{\beta}) \rho_{\beta A} - 2i \zeta^{-1} \phi^{0 \dagger}_{u_1 u_2} \rho^{\alpha A} Z_{\alpha \dot{\alpha}} \bar{Z}^{* \dot{\alpha} \beta} \rho_{\beta A}
\end{dmath}

and

\begin{dmath}
    \frac{1}{4 \pi^2} \tilde{S}^{(2)}_f = \frac{i}{2} \sum^2_{i=1} \sum^N_{\substack{u=1 \\ u \neq u_1 , u_2}} \phi^{0 \dagger}_{u u_i} \bar{\mu}^A_{iu} \mu_{uiA} - \frac{1}{4} \sum^2_{i=1} \sum^{N_F}_{f=1} (m_f + \phi^0_{u_i}) \mathcal{K}_{if} \tilde{\mathcal{K}}_{fi}
\end{dmath}
After a shift in $\sigma^A$ and $\varepsilon^A$ as much as $\rho^A$ one can complete the square and write:

\begin{dmath}
    \frac{1}{4 \pi^2} \tilde{S}^{(1)}_f = -\frac{i}{2} (\phi^0_{u_1 u_2})^{\dagger} (1 + 4 \zeta^{-1} Y^2) \sigma^{\alpha A} \varepsilon_{\alpha A} -2 i \zeta^{-1} (1+4 \zeta^{-1}Y^2)^{-1} \rho^{\alpha A} \left( \phi^{0 \dagger}_{u_1 u_2} Z_{\alpha \dot{\alpha}} \bar{Z}^{* \dot{\alpha} \beta} + 2 \chi^{\dagger}_{12} Y_{\alpha \dot{\alpha}} \bar{Z}^{* \dot{\alpha} \beta} + 2 \chi^{\dagger}_{21} Z_{\alpha \dot{\alpha}} \bar{Y}^{\dot{\alpha} \beta} \right) \rho_{\beta A}
\end{dmath}

The contribution to the instanton partition function from the critical point-set submanifold is proportional to:

\begin{dmath}
    \int d^4 Y d \xi d \xi^* d^3 Z d^3 Z^* d^8 \chi_a d^4 \rho d^4 \sigma d^4 \varepsilon \times \prod^2_{i=1} \left\{ \prod^N_{\substack{u=1 \\ u \neq u_1 , u_2}} d^2 w_{ui} d^2 \bar{w}_{iu} d^2 \mu_{ui} d^2 \bar{\mu}_{iu} \prod^{N_F}_{f=1} d \mathcal{K}_{if} d \tilde{\mathcal{K}}_{fi} \right\} Y^2 
    \\
    \times {\text{exp} (-\tilde{S}^{(1)}_b - \tilde{S}^{(2)}_b - \tilde{S}^{(1)}_f - \tilde{S}^{(2)}_f)}
\end{dmath}

The integrals over the Grassmann variables $\{ \sigma^A_{\alpha} , \varepsilon^A_{\alpha} , \rho^A_{\alpha} \}$ give rise to:

\begin{dmath}
    {(\phi^{0 \dagger}_{u_1 u_2})^4 \zeta^{-2} (1+4\zeta^{-1} Y^2)^2}
    \\
    {\left( 4 Y^2 (\chi^{\dagger}_{21} Z - \chi^{\dagger}_{12} Z^{*})^2 + (\phi^{0 \dagger}_{u_1 u_2})^2 (Z^2 Z^{*2} - (Z.Z^{*})^2 ) \right)}
\end{dmath}
and as for the integrals over the Grassmann variables $\{ \mu^A_{ui} , \bar{\mu}^A_{iu} , \mathcal{K}_{if} , \tilde{\mathcal{K}}_{fi} \}_{u \neq u_1 , u_2}$ give rise to:

\begin{dmath}
    \prod^2_{i=1} \prod^{N}_{\substack{u=1 \\ u \neq u_1 , u_2}} (\phi^{0 \dagger}_{u u_i})^2 \prod^{N_F}_{f=1} (m_f + \phi^0_{u_1})(m_f + \phi^0_{u_2})
\end{dmath}

The integrals over $\{ Z, \xi , \chi_a \}$ become:

\begin{dmath}
    \int d \xi d \xi^{*} d^3 Z d^3 Z^{*} d^8 \chi_a 
    \\
    {\left( 4 Y^2 (\chi^{\dagger}_{21} Z - \chi^{\dagger}_{12} Z^{*})^2 + (\phi^{0 \dagger}_{u_1 u_2})^2 (Z^2 Z^{*2} - (Z.Z^{*})^2) \right) e^{-\tilde{S}^{(1)}_b}}
\end{dmath}

which amounts to:

\begin{dmath}
    \frac{(\phi^{0 \dagger}_{u_1 u_2})^2}{\zeta^3 (\phi^0_a)^{12}_{u_1 u_2} Y^2 (1+ 4 \zeta^{-1} Y^2)^6}
\end{dmath}

while those over $\{w_{ui \dot{\alpha}}\}_{u \neq u_1 , u_2}$ lead to a factor of:

\begin{dmath}
\prod^2_{i=1} \prod^N_{\substack{u=1 \\ u \neq u_1 , u_2}} \frac{1}{(\phi^0_a)^4_{u u_i}}    
\end{dmath}

As anticipated earlier, the integral over the relative position turns out to be finite:

\begin{dmath}[label={eq:relative-position-integral}]
    \int d^4 Y \frac{\zeta^2}{(\zeta + 4 Y^2)^4} = \frac{\pi^2}{96}
\end{dmath}

Therefore, gathering all the pieces together, the result is:

\begin{dmath}
    \frac{2}{(\phi^0_{u_1 u_2})^6} \prod^2_{i=1} \prod^N_{\substack{u = 1 \\ u \neq u_1 , u_2}} \frac{1}{(\phi^0_{u u_i})^2}
    \\
    {\prod^{N_F}_{f=1} (m_f + \phi^0_{u_1})(m_f + \phi^0_{u_2})}
\end{dmath}

which is manifestly holomorphic in the VEVs. Summation over the $\frac{1}{2} (N)(N-1)$ critical points for two topicons of distinct flavours leads to:

\begin{dmath}
    \sum^N_{\substack{u,v = 1 \\ u \neq v}} \frac{S_u (\phi^0_u) S_v (\phi^0_v)}{(\phi^0_u - \phi^0_v)^2}
\end{dmath}
in which:

\begin{dmath}[label={eq:S-u-definition}]
    {S_u (x) := \prod^N_{\substack{v=1 \\ v \neq u}} \frac{1}{(x- \phi^0_v)^2} \prod^{N_F}_{f=1} (m_f + x)}
\end{dmath}

       \item Two topicons of the same flavour

There are $N$ critical points of this type; two topicons of the same flavour $u_1 = u_2 = v \in \{ 1,..,N \}$. The ADHM constraints are all satisfied over the critical point-set manifold; two instantons in non-commutative $U(1)$ theory. All the rest of the variables vanish and are treated as fluctuations around the critical point-set submanifold.

Firstly, the shift of $\chi_a$ is necessary again:

\begin{dmath}
    \chi_a \rightarrow \chi_a - (\phi^0_a)_v 1_{[2] \times [2]}
\end{dmath}

Similar computations lead to the following factor for integrating out the fluctuations:

\begin{dmath}[label={eq:same-flavour-fluctuation-factor}]
    {\prod^N_{\substack{u=1 \\ u \neq v}} \frac{1}{(\text{det}_2 (\chi + \phi^0_{uv} 1_{[2] \times [2]}))^2}} 
    \\
    \times {\prod^{N_F}_{f=1} \text{det}_2 ((m_f + \phi^0_v) 1_{[2] \times [2]} - \chi) = S_{v} (\phi^0_v - \lambda_1) S_v (\phi^0_v - \lambda_2)}
\end{dmath}

where $\lambda_i \ ,i=1,2$ are the eigenvalues of the $2 \times 2$ matrix $\chi$ and $S_{u} (x)$ is as defined in \eqref{eq:S-u-definition}.

Now, as for the integral $\int_{{\widehat{\mathcal{M}}^{(\zeta)}_k}|_{N=1}}$, the integral vanishes to linear order in $\chi$. Consequently, one must expand \eqref{eq:same-flavour-fluctuation-factor} in powers of $\lambda_i$, up to quadratic order:

\begin{dmath}
    \frac{1}{2} S_v (\phi^0_v) \frac{\partial^2 S_v (\phi^0_v)}{\partial (\phi^0_v)^2} (\lambda^2_1 + \lambda^2_2) + \frac{\partial S_v (\phi^0_v)}{\partial \phi^0_v} \frac{\partial S_v (\phi^0_v)}{ \partial \phi^0_v} \lambda_1 \lambda_2
\end{dmath}

resulting in a non-vanishing integral above this time:

\begin{dmath}
    \mathcal{J}_1 S_v (\phi^0_v) \frac{\partial^2 S_v (\phi^0_v)}{\partial (\phi^0_v)^2} + \mathcal{J}_2 \frac{\partial S_v (\phi^0_v)}{\partial \phi^0_v} \frac{\partial S_v (\phi^0_v)}{\partial \phi^0_v}
\end{dmath}

in which:

\begin{dmath}
    {\mathcal{J}_1 = \frac{1}{2} \int_{{\widehat{\mathcal{M}}_2}|_{N=1}} \boldsymbol{\omega}^{(2)} (\lambda^2_1 + \lambda^2_2) := \int_{{\widehat{\mathcal{M}}_2}|_{N=1}} \boldsymbol{\omega}^{(2)} \left( \frac{1}{2} ( \text{tr}_2 \chi)^2 - \text{det}_2 \chi \right)}
    \\
    {\mathcal{J}_2 = \int_{{\widehat{\mathcal{M}}_2}|_{N=1}} \boldsymbol{\omega}^{(2)} (\lambda_1 \lambda_2) := \int_{{\widehat{\mathcal{M}}_2}|_{N=1}} \boldsymbol{\omega}^{(2)} \left(  \text{det}_2 \chi \right )}
\end{dmath}

Indeed, the moduli space $\widehat{\mathcal{M}}^{(\zeta)}_2 |_{N=1}$ is the well-known $4$-dimensional hyperKähler Eguchi-Hanson manifold \cite{Eguchi-Hanson}. After integrating over all the Grassmann variables and $\chi_a$, one finds (as is done in the appendix of \cite{Integral}):

\begin{dmath}[label={eq:J1-J2-values}]
    {\mathcal{J}_1 = \frac{1}{4} \ \ \ \ \ \mathcal{J}_2 = 0}
\end{dmath}
Consequently, the complete form of the non-commutative prepotential is:

\begin{dmath}[label={eq:F2-nc-localization}]
    {\mathcal{F}^{\text{nc}}_2 = \widehat{\mathcal{Z}}_2 = \sum^N_{\substack{u,v =1 \\ u \neq v}} \frac{S_u (\phi^0_u) S_v (\phi^0_v)}{(\phi^0_u - \phi^0_v)^2} + \frac{1}{4} \sum^N_{u=1} S_u (\phi^0_u) \frac{\partial^2 S_u (\phi^0_u)}{\partial (\phi^0_u)^2} \  \text{for}  \ N_F < 2N}
\end{dmath}

While in the case of $N_F= 2N$, the necessary redefinition of the prepotential is as follows:

\begin{dmath}
    {\mathcal{F}_2 = \mathcal{F}^{\text{nc}}_2 + 2 \pi i c_1 \mathcal{F}^{\text{nc}}_1 + i \pi c_2 \sum^N_{u=1} (\phi^0_u)^2} 
    \\
    (\text{modulo constants})
\end{dmath}
where $c_1$ is as determined in \eqref{eq:c1} and $c_2$ in the case of $SU(2)$(so $N_F = 4$) can be computed with the help of \eqref{eq:F2-SU2-commutative}, \eqref{eq:F1-nc-localization}, and \eqref{eq:F2-nc-localization} can be determined:

\begin{dmath}
    c_2 |_{\text{SU(2)}} = -\frac{i}{2^3 \pi} \left( 1 + \frac{7}{2^4 3^5} - \frac{13}{2^4} \right)
\end{dmath}
   \end{itemize}

The functionality of the localization technique at higher orders is already proven in the softly broken $\mathcal{N}=4$ to $\mathcal{N}=2$, and they precisely agree with that of the Seiberg-Witten theory \cite{All orders}.

\end{itemize}

\chapter{Recurrence in SW Partition Function and Zamolodchikov's recursive relation, AGT correspondence}
\label{ch:recursion}

In this chapter, a general algorithmic and more mathematical formulation
for instanton counting through localization is presented. It turns out that this
formulation reveals a much deeper and fascinating link between $2d$ CFTs and $4d$
gauge theories. Indeed, this is the content of the AGT correspondence (duality).
The chapter opens with a self-contained account of the $\Omega$-background (Section \ref{sec:Omega}), on which everything that follows is built.
After a short review of Nekrasov's instanton counting method, an application of the
AGT correspondence is introduced, where the recurrence in Nekrasov's partition
function is related to Zamolodchikov's recursive relation for conformal blocks of the
4-point function of $2d$ CFT. The presentation of Sections \ref{sec:beyond2}--\ref{sec:agt}
follows \cite{Recursion}; in Section \ref{sec:pure} the recursion is specialised to the pure
$SU(2)$ theory and run explicitly, which prepares the comparison with topological
recursion carried out in Chapter \ref{ch:TR}.

\section{The $\Omega$-Background}
\label{sec:Omega}
Every formula of this chapter depends on two parameters, $\epsilon_1$ and $\epsilon_2$, which have not appeared in
the thesis so far. They describe a deformation of flat Euclidean space-time introduced by Nekrasov
\cite{Nekrasov} and called the \textbf{$\Omega$-background}. Both the counting of instantons and the dictionary of the AGT
correspondence rest on it. This section therefore explains, before any computation is done, where the
$\Omega$-background comes from, how it is defined, what the partition function on it means and why it is
the right tool for the problem at hand. The presentation follows Chapter 4 of \cite{Tachikawa}; what concerns the
general case $\epsilon_1+\epsilon_2\neq0$ is taken from \cite{Nekrasov}, \cite{NO} and \cite{NYlectures}.

\subsection{Motivation: what Chapter \ref{Localisation} left unsolved}
In Chapter \ref{Localisation} the instanton partition function \eqref{eq:instanton-partition-function} was computed for $k=1,2$ by localization.
The supercharge $\mathcal{Q}$ of \eqref{eq:BRST-charge} is nilpotent only up to an infinitesimal transformation of the auxiliary
group $U(k)$ and of the gauge group, the latter generated by the vacuum expectation values $\phi^0$, so that the integral over the resolved moduli space
localizes on the fixed points of the maximal torus $U(1)^N\subset U(N)$ of the gauge group. Two features made that
computation laborious, and both become worse as $k$ grows.
\begin{enumerate}
\item \emph{The fixed points are not isolated.} The gauge torus does not act on space-time, so it cannot fix the
positions of the instantons. Its fixed locus \eqref{eq:topicon-moduli-space} is a union of products of moduli spaces of $U(1)$
instantons, the topicons, which are free to move. An integral over each of these non-compact spaces is
left over: the integral \eqref{eq:relative-position-integral} over the relative position of two topicons, or the integrals over the
Eguchi-Hanson space whose values are quoted in \eqref{eq:J1-J2-values}.
\item \emph{The volume of space-time.} Translations of $\mathbb{R}^4$ move the instanton as a whole, so the integral over
its centre is the infinite volume of space-time. It had to be removed by hand, by passing to the centred
moduli space and by extracting the prepotential from a correlation function, \eqref{eq:four-fermion-correlator-prepotential} and \eqref{eq:four-fermion-correlator-instanton}.
\end{enumerate}
Both difficulties have the same root: the group used for the localization does not act on space-time.
The symmetries of $\mathbb{R}^4$ are translations and rotations. Translations have no fixed points at all, but a
generic rotation has exactly one, the origin. Nekrasov's idea is to let the \emph{rotations} of space-time take
part in the localization. The instantons are then forced to sit at the origin, the fixed points on the moduli
space become isolated, and the volume of space-time is replaced by a finite quantity. As a result one
needs only the geometry of the moduli space, and the detailed form of the instanton effective action can
be bypassed \cite{Tachikawa}. The $\Omega$-background is the physical setting which realises this idea.

\subsection{Five-dimensional origin and definition}
The construction is most transparent if the four-dimensional $\mathcal{N}=2$ theory is regarded as the limit of
the five-dimensional $\mathcal{N}=1$ super Yang-Mills theory compactified on a circle. The following facts are needed
\cite{Tachikawa}. The smallest supersymmetry algebra in five dimensions has eight supercharges, and its reduction
on a circle is the $\mathcal{N}=2$ algebra in four dimensions. The five-dimensional vector multiplet contains a
\emph{real} scalar; together with the Wilson line of the gauge field around the circle it forms the complex scalar
of the four-dimensional vector multiplet. Finally, the momentum along the circle appears in four
dimensions as the imaginary part of the central charge.

Let $y$ be the coordinate along the circle, of circumference $\beta$, and let $\Omega=(\Omega_{\mu\nu})$ be a constant
antisymmetric matrix, that is, a generator of the rotations $SO(4)$ of $\mathbb{R}^4$. The $\Omega$-background is the
five-dimensional space obtained from the strip $\mathbb{R}^4\times[0,\beta]$ by gluing its two ends after a rotation:
\begin{equation}
(\vec{x},y)\sim\left(e^{\beta\Omega}\,\vec{x},\,y+\beta\right)
\label{Omegaglue}
\end{equation}
The space is flat, and it is a fibration of $\mathbb{R}^4$ over the circle: a field transported once around the
circle comes back rotated. In the coordinates $\tilde{x}=e^{-y\Omega}\vec{x}$, which are periodic, the flat metric
$\mathrm{d}\vec{x}^{\,2}+\mathrm{d}y^2$ becomes
\begin{equation}
\mathrm{d}s^2=\left(\mathrm{d}\tilde{x}^\mu+V^\mu\,\mathrm{d}y\right)^2+\mathrm{d}y^2\quad,\quad V^\mu=\Omega^{\mu}{}_{\nu}\,\tilde{x}^\nu
\label{Omegametric}
\end{equation}
where $V$ is the vector field which generates the rotation. This is the form in which the background is
written in \cite{Nekrasov} and \cite{Tachikawa}. By a change of orthonormal basis $\Omega$ can always be brought to a block-diagonal form, a
rotation by independent angles in two orthogonal planes. In the complex coordinates $z_1=x^1+ix^2$,
$z_2=x^3+ix^4$ the identification \eqref{Omegaglue} then reads
\begin{equation}
(z_1,z_2,y)\sim\left(e^{i\beta\epsilon_1}z_1,\,e^{i\beta\epsilon_2}z_2,\,y+\beta\right)
\label{Omegaeps}
\end{equation}
which defines $\epsilon_1$ and $\epsilon_2$. They have the dimension of a mass, and the vector field $V$ vanishes only at the
origin $z_1=z_2=0$.

A rotation acts on spinors, hence on the supercharges, and a generic identification \eqref{Omegaglue} would break
supersymmetry completely. When $\epsilon_1+\epsilon_2=0$ the matrix $\Omega$ is (anti-)self-dual: the rotation lies in one
of the two factors of $SO(4)\simeq SU(2)\times SU(2)$, the half of the supercharges which are singlets under that
factor commute with it, and they are preserved \cite{Tachikawa}. When $\epsilon_1+\epsilon_2\neq0$ the rotation has a component in the
other factor as well, and supersymmetry survives only if this component is accompanied by an equal
rotation of the $SU(2)$ $R$-symmetry \cite{Nekrasov}, \cite{NO}. Identifying one $SU(2)$ factor of the rotation group with the
$R$-symmetry group is precisely the topological twist of Section \ref{subsec:Localization}, and the supercharge which is
invariant under the combined rotation is the scalar supercharge $\mathcal{Q}$ of \eqref{eq:BRST-charge}. In the general
$\Omega$-background, therefore, $\mathcal{Q}$ is preserved, and it is all one needs.

The vacuum expectation values are introduced in the same way. Since the four-dimensional complex
scalar is made of the five-dimensional scalar and of the Wilson line, and since the result is holomorphic
in it, it is enough to turn on the Wilson line \cite{Tachikawa}: around the circle the gauge field has the holonomy
$\exp\left(i\beta\,\mathrm{diag}(a_1,\dots,a_N)\right)$. Altogether, going once around the circle acts on the fields by the
element
\begin{equation}
g=\left(e^{i\beta\epsilon_1},e^{i\beta\epsilon_2}\,;\,e^{i\beta a_1},\dots,e^{i\beta a_N}\right)\in U(1)^2\times U(1)^N\subset SO(4)\times U(N)
\label{gelement}
\end{equation}
of the maximal torus of the rotation group and of the global gauge group. In units $\beta=1$ these are the
variables $(T_1,T_2)$ and $(e_1,\dots,e_N)$ of Section \ref{sec:beyond2}. The parameters $\epsilon_1,\epsilon_2$ and $a_1,\dots,a_N$ thus enter on
exactly the same footing.

\subsection{The partition function as an equivariant index}
Let the coordinate $y$ play the role of Euclidean time. In the Hamiltonian formalism the partition
function on the $\Omega$-background is a trace over the Hilbert space of the five-dimensional theory on
$\mathbb{R}^4$ \cite{Tachikawa}:
\begin{equation}
Z(a,\epsilon_1,\epsilon_2;\beta)=\mathrm{Tr}\,(-1)^F\,e^{-\beta H}\,e^{i\beta\left(\epsilon_1J_1+\epsilon_2J_2\right)}\,e^{i\beta\sum_\alpha a_\alpha J^\alpha}
\label{Zindex}
\end{equation}
where $H$ is the Hamiltonian, $J_1$ and $J_2$ generate the rotations of the planes $z_1$ and $z_2$ (accompanied by
the $R$-symmetry rotation discussed above) and $J^\alpha$ are the Cartan generators of the global gauge group. The
sign $(-1)^F$ is there because the fermions are periodic around the circle.

A trace of this kind is called an \emph{equivariant index}. Its basic property is the following \cite{Tachikawa}. Since $H$ is the
anticommutator of $\mathcal{Q}$ with its adjoint, every state of non-zero energy is paired by $\mathcal{Q}$ with a state of the
opposite $(-1)^F$; since $g$ commutes with $\mathcal{Q}$, the two partners have the same eigenvalue of $g$, and their
contributions cancel. Only the states annihilated by $\mathcal{Q}$, the BPS states, contribute. Two consequences
follow. First, $Z$ is protected: it does not change under deformations which preserve $\mathcal{Q}$, and it can be
computed exactly in the semi-classical approximation. Second, $Z$ is a counting function: it is the
character of the torus \eqref{gelement} on the space of BPS states, with $\epsilon_{1,2}$ and $a_\alpha$ playing the role of chemical
potentials for the angular momenta and for the gauge charges.

In each topological sector the configurations of lowest energy are the instantons, which in five
dimensions are particles. The field theory then reduces to a supersymmetric quantum mechanics on the
moduli space $\mathcal{M}_{k}$ of $k$ instantons, and
\begin{equation}
Z=\sum_{k\geq0}x^k\,\mathrm{Ind}_g\left(\mathcal{M}_{k}\right)
\label{Zsumind}
\end{equation}
where $x$ weighs the instanton number and $\mathrm{Ind}_g$ is the equivariant index of the moduli space \cite{Tachikawa}. Such an index
is given by a fixed-point formula. If the torus acts on a manifold $M$ with isolated fixed points $p$, then
\begin{equation}
\mathrm{Ind}_g(M)=\sum_{p}\,\prod_{i=1}^{\dim_{\mathbb{C}}M}\frac{1}{e^{i\beta w_i(p)/2}-e^{-i\beta w_i(p)/2}}
\label{fixedpoint}
\end{equation}
where $e^{i\beta w_i(p)}$ are the eigenvalues of $g$ on the tangent space at $p$; the numbers $w_i(p)$ are called the
\emph{weights}. The simplest example is space-time itself, $M=\mathbb{C}^2$, the moduli space of a single point
particle. There is one fixed point, the origin, with weights $\epsilon_1$ and $\epsilon_2$:
\begin{equation}
\mathrm{Ind}_g\left(\mathbb{C}^2\right)=\frac{1}{\left(e^{i\beta\epsilon_1/2}-e^{-i\beta\epsilon_1/2}\right)\left(e^{i\beta\epsilon_2/2}-e^{-i\beta\epsilon_2/2}\right)}\;\xrightarrow{\;\beta\to0\;}\;\frac{1}{(i\beta)^2\,\epsilon_1\epsilon_2}
\label{indC2}
\end{equation}
Without the rotation this number would be infinite: it would count the states of a particle which is free
to be anywhere in $\mathbb{R}^4$. The $\Omega$-background makes it finite, and $1/\epsilon_1\epsilon_2$ is what remains of the volume
of space-time.

On the moduli space of instantons the rotations move and rotate the instantons, and the gauge
torus rotates their orientation in the gauge group. A configuration is fixed by $g$ only if all the instantons
sit at the origin and each of them lies in one of the $N$ abelian factors. Such configurations are singular
(point-like instantons), and to make sense of them one passes to the resolved moduli space of Section
\ref{Non-Commutative Spacetime, Modified ADHM Constraints}; the result does not depend on the non-commutativity parameter. On the resolved space the
fixed points are isolated, and they are labelled by $N$-tuples of Young tableaux $\vec{Y}=(Y_1,\dots,Y_N)$ with
$k$ boxes in total: $Y_\alpha$ describes the $U(1)$ instantons of the $\alpha$-th factor piled up at the origin \cite{Tachikawa}. These are the
topicons of Chapter \ref{Localisation}, now pinned to the origin by the rotation. The weights at each fixed point are read off
from the ADHM data; they are collected in the character \eqref{character} of the next section, which is equation
(4.81) of \cite{Tachikawa} up to the sign convention for $a_\alpha$.

\subsection{The four-dimensional limit and the prepotential}
The four-dimensional theory is recovered in the limit $\beta\to0$ with $\epsilon_{1,2}$ and $a_\alpha$ fixed. As in \eqref{indC2}, each
factor of \eqref{fixedpoint} becomes $1/i\beta w_i$; the moduli space has complex dimension $2Nk$, and the resulting
powers of $\beta$ are absorbed in $x$, which becomes $\Lambda^{2N}$ for the theory without matter. The instanton
partition function of the four-dimensional theory is therefore
\begin{equation}
Z_{\rm inst}(a,\epsilon_1,\epsilon_2;x)=\sum_{k\geq0}x^k\sum_{\lvert\vec{Y}\rvert=k}\ \prod_{i=1}^{2Nk}\frac{1}{w_i(\vec{Y})}
\label{Zweights}
\end{equation}
a sum of rational functions of $a_\alpha$, $\epsilon_1$ and $\epsilon_2$: nothing is left to integrate. The $2Nk$ weights at the fixed point
$\vec{Y}$ are the numbers $E_{\alpha,\beta}(s)$ and $\epsilon-E_{\alpha,\beta}(s)$ of \eqref{Eab}, and \eqref{Zweights} is the formula \eqref{Zinst} with \eqref{Pgauge}.

The sum over fixed points has an equivalent integral form. By the theorem of Duistermaat and
Heckman \cite{Tachikawa},
\begin{equation}
\sum_{p}\prod_i\frac{1}{w_i(p)}=\int_{\mathcal{M}_k}e^{\,\omega+h}
\label{DH}
\end{equation}
where $\omega$ is the symplectic form of the moduli space and $h$ is the Hamiltonian function of the vector field
generated by $g$, linear in $\epsilon_{1,2}$ and $a_\alpha$. The reason is the following \cite{Tachikawa}. The form $\omega+h$ is closed with respect to the
equivariant differential $D=\mathrm{d}-\iota_V$, where $\iota_V$ is the contraction with the vector field, and the integral does not
change if a term $t\,D\alpha$ is added to the exponent, for any invariant one-form $\alpha$. Choosing $\alpha$ to be the vector field
itself, turned into a one-form by an invariant metric, $D\alpha$ contains $-\lvert V\rvert^2$, and for $t\to\infty$ the integrand concentrates
at the zeros of $V$, the fixed points, where a Gaussian integral gives the left-hand side. For $\epsilon_{1,2}=a_\alpha=0$ the right-hand side is the volume of the moduli
space, which is the divergent integral met in Chapter \ref{Localisation}. The factor $e^{h}$ is a Gaussian cut-off in all the
non-compact directions: on the centre of the instanton $h$ is proportional to $-(\epsilon_1\lvert z_1\rvert^2+\epsilon_2\lvert z_2\rvert^2)$, so that for positive $\epsilon_{1,2}$ the integral
over the centre gives $1/\epsilon_1\epsilon_2$ instead of the volume of $\mathbb{R}^4$. In this sense \eqref{Zweights} is the \emph{equivariant
volume} of the moduli space, and the $\Omega$-background is an infrared regulator which preserves supersymmetry.

The partition function $Z_{\rm inst}$ is not yet the prepotential. Its coefficient of $x^k$ has poles up to
$(\epsilon_1\epsilon_2)^{-k}$ when $\epsilon_{1,2}\to0$, one factor \eqref{indC2} for each of $k$ instantons far from each other. In the
logarithm only the connected part survives, in which the instantons move together and a single factor
of the volume is left:
\begin{equation}
Z_{\rm inst}(a,\epsilon_1,\epsilon_2;x)=\exp\left(-\frac{\mathcal{F}_{\rm inst}(a,\epsilon_1,\epsilon_2;x)}{\epsilon_1\epsilon_2}\right)
\label{Zexp}
\end{equation}
where $\mathcal{F}_{\rm inst}(a,\epsilon_1,\epsilon_2;x)$ is regular at $\epsilon_1=\epsilon_2=0$. This is the statement that the free energy is extensive: it is proportional to the volume, and the volume
is $1/\epsilon_1\epsilon_2$. At low energy the theory on its Coulomb branch is described by the effective action determined
by the prepotential, and on the $\Omega$-background this action reduces to the prepotential times the volume
\cite{Nekrasov}. This leads to Nekrasov's conjecture
\begin{equation}
\mathcal{F}_{\rm inst}(a,0,0;x)=\mathcal{F}^{SW}_{\rm inst}(a;x)
\end{equation}
that is, to the definition \eqref{Finstdef} used below. For $\epsilon_1=-\epsilon_2$ it is the relation between the prepotential and the
logarithm of the partition function on the $\Omega$-background of \cite{Tachikawa}. The conjecture has been proven in \cite{NO} and in \cite{NY}. In this
thesis the cancellation of the higher poles and the value of the limit are verified explicitly up to five
instantons for $SU(2)$ in Section \ref{sec:pure} and in Appendix \ref{app:checks}.

\subsection{The meaning of $\epsilon_1$ and $\epsilon_2$}
The parameters $\epsilon_{1,2}$ are not only a regulator. The $\Omega$-background with $\epsilon_1+\epsilon_2=0$ is closely related to
another five-dimensional geometry, the \emph{graviphoton background} \cite{Tachikawa}
\begin{equation}
\mathrm{d}s^2=\mathrm{d}\vec{x}^{\,2}+\left(\mathrm{d}y+A_\mu\,\mathrm{d}x^\mu\right)^2
\end{equation}
in which it is the circle that is fibred over $\mathbb{R}^4$, with a constant self-dual curvature $\partial_{[\mu}A_{\nu]}$. From the
four-dimensional point of view $A_\mu$ is the Kaluza-Klein gauge field, which in $\mathcal{N}=2$ supergravity is the
graviphoton: the background is a constant graviphoton field strength. The effective prepotential in such a
background depends on the field strength, and it is given by the logarithm of the partition function on the
$\Omega$-background, because both receive contributions from the same BPS states \cite{Tachikawa}. Writing
$\epsilon_1=-\epsilon_2=\hbar$, its expansion is
\begin{equation}
\log Z\Big\vert_{\epsilon_1=-\epsilon_2=\hbar}=\sum_{g\geq0}\hbar^{2g-2}F^{(g)}(a)
\label{genusOmega}
\end{equation}
The leading term $F^{(0)}$ is the prepotential in the absence of the graviphoton. The higher terms $F^{(g)}$ measure
the response of the gauge theory to a gravitational background, and for this reason they are called
gravitational corrections; when the gauge theory is obtained from string theory they are the topological
string amplitudes of genus $g$, the expansion parameter $\hbar^2$ being the square of the graviphoton field strength
(see \cite{Tachikawa} and the references given there). The term $F^{(1)}$ will be met again in Section \ref{sec:genus1}.

For independent $\epsilon_1$ and $\epsilon_2$ the function $\mathcal{F}_{\rm inst}(a,\epsilon_1,\epsilon_2;x)$ of \eqref{Zexp} is a two-parameter refinement of
\eqref{genusOmega}. It is symmetric under the exchange of $\epsilon_1$ and $\epsilon_2$, and the first terms of its expansion, which multiply
$\epsilon_1\epsilon_2$ and $(\epsilon_1+\epsilon_2)^2$, are displayed in \eqref{eps2}.

\subsection{Relevance for what follows}
The role of the $\Omega$-background in the rest of the thesis can be summarised in four points.
\begin{enumerate}
\item \emph{It turns integrals into sums.} The integrals over the moduli space of Chapter \ref{Localisation}, which could be carried out only
for $k\leq2$, are replaced for every $k$ by the finite sum \eqref{Zweights} over Young tableaux. This is the content
of Section \ref{sec:beyond2}.
\item \emph{It contains the Seiberg-Witten prepotential,} as the limit \eqref{Finstdef}, and more than that: the dependence on
$\epsilon_{1,2}$ is physical information about the same gauge theory.
\item \emph{It is what makes the AGT correspondence possible.} The dictionary of Sections \ref{sec:zam} and \ref{sec:agt} identifies
the $\Omega$-background with the parameters of the conformal field theory: by \eqref{Qeps} the central charge
depends only on the ratio $\epsilon_1/\epsilon_2$, while the vacuum expectation value and the masses, measured in units of
$\sqrt{\epsilon_1\epsilon_2}$, give the conformal dimensions. A conformal block at finite central charge and finite dimensions
therefore corresponds to the partition function at \emph{finite} $\epsilon_{1,2}$; the limit $\epsilon_{1,2}\to0$ at fixed ratio, which
gives the prepotential, is the limit of large dimensions at fixed central charge. Zamolodchikov's
recursion is a statement about the poles of the block at the degenerate dimensions \eqref{degdims}, that is, at
$2a=\pm(m\epsilon_1+n\epsilon_2)$: in the undeformed theory all these poles collapse onto the single point $a=0$.
\item \emph{It provides the genus expansion.} On the self-dual line $\epsilon_1+\epsilon_2=0$, which corresponds to the central charge
$c=1$, the expansion \eqref{genusOmega} has exactly the form of the expansion generated by the topological recursion.
This is the starting point of the comparison of Chapter \ref{ch:TR}.
\end{enumerate}

\section{Instanton Counting Beyond $k=2$}
\label{sec:beyond2}
As in the previous chapter, the localization technique was shown to significantly
simplify the computation of instanton coefficients. Nonetheless, computation above
$k=2$ is still challenging as the moduli space measure becomes increasingly complex.

One trick applied by Nekrasov \cite{Nekrasov}, was to include the maximal torus of the
$4d$ Euclidean rotational symmetry group $U(1)^2\subset SO(4)$\footnote{indeed, it turns out that there would be no chance for construction of non-trivial observables with
such a symmetry group, and a complex structure is assumed over the $4d$ Euclidean space which
reduces the group to that of $U(2)$ which after all has the same maximal torus!} apart from the maximal
torus of the non-commutative gauge group $U(1)^N\subset U(N)$ after application of a
shift generated by the rotational group in the localization process.

Let $(e_1,...,e_N)=(e^{ia_1},...,e^{ia_N})\in U(1)^N$ where $a_1,...,a_N$ are the VEVs of
the vector multiplets in $\mathcal{N}=2$ and $(T_1,T_2)=(e^{i\epsilon_1},e^{i\epsilon_2})$; $\epsilon_1,\epsilon_2$ characterize a
gravitational background named $\Omega$-background (Section \ref{sec:Omega}; here and below the units $\beta=1$ are used).

\begin{figure}[t]
\centering
\includegraphics[width=0.5\linewidth]{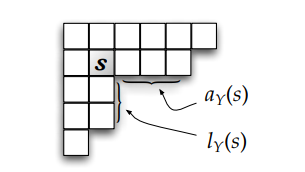}
\caption{The Young Tableaux with hook lengths of $Y=6\geq5\geq4\geq3\geq2\geq1$ (figure
from \cite{Tachikawa})}
\label{fig:young}
\end{figure}

The holomorphic tangent space of the moduli space decomposes into the sum of
[complex] one-dimensional irreducible representations of the Cartan subalgebra of
$U(N)\times U(2)$ \cite{Recursion} whose character is:
\begin{equation}
\chi=\sum_{\alpha,\beta=1}^{N}e_\alpha e_\beta^{-1}\left\{\sum_{s\in Y_\alpha}\left(T_1^{-l_{Y_\beta}(s)}T_2^{a_{Y_\alpha}(s)+1}\right)+\sum_{s\in Y_\beta}\left(T_1^{l_{Y_\alpha}(s)+1}T_2^{-a_{Y_\beta}(s)}\right)\right\}
\label{character}
\end{equation}
in which $a_{Y}(s)$, $l_{Y}(s)$ are the distances of the right edge of the box $s$ from the
limiting polygonal curve of the Young tableau $Y$ in the horizontal and vertical
direction, taken with the plus sign if the box $s\in Y$ and with the minus sign if $s\notin Y$ (Figure \ref{fig:young}).

Explicitly, if the tableau $Y$ has rows of lengths $Y_1\geq Y_2\geq\dots$ and columns of lengths $Y'_1\geq Y'_2\geq\dots$, then
for the box $s=(i,j)$ in the $i$-th row and the $j$-th column
\begin{equation}
a_Y(s)=Y_i-j\quad,\quad l_Y(s)=Y'_j-i
\label{armleg}
\end{equation}
and these numbers are negative when $s$ lies outside $Y$.

\subsubsection*{Fixed points and their weights}
The character \eqref{character} is obtained as follows \cite{Tachikawa}. In complex notation the ADHM data of $k$ instantons of
$U(N)$ are four matrices acting between the spaces $V=\mathbb{C}^k$ and $W=\mathbb{C}^N$,
\begin{equation}
B_1,B_2\in\mathrm{Hom}(V,V)\quad,\quad I\in\mathrm{Hom}(W,V)\quad,\quad J\in\mathrm{Hom}(V,W)
\end{equation}
In the notation of Chapter \ref{Localisation}, $B_1$ and $B_2$ are the two complex components of $a'_n$, and $I$, $J^\dagger$ are those of $w_{\dot\alpha}$. The data are
subject to the modified ADHM constraints \eqref{eq:modified-ADHM-constraints}, whose complex part reads $[B_1,B_2]+IJ=0$, and they are taken
modulo $U(k)$. The torus \eqref{gelement} acts on them: $B_1$ and $B_2$ rotate like the coordinates $z_1$ and $z_2$, the product $IJ$ rotates like
$z_1z_2$, and $I$, $J$ carry in addition the gauge weights of $W$. A point of the moduli space is fixed when the action of
the torus on the data can be undone by a transformation acting on $V$; this makes $V$ a representation of the torus. At
a fixed point one finds $J=0$, and $V$ is spanned by the vectors $B_1^{\,i-1}B_2^{\,j-1}I(w_\alpha)$, where $w_\alpha$ is the $\alpha$-th basis
vector of $W$. Since $B_1$ and $B_2$ commute when $J=0$, the pairs $(i,j)$ for which this vector does not vanish form, for
each $\alpha$, a Young tableau $Y_\alpha$: if the box $(i,j)$ is present, so are all the boxes above it and to its left. Therefore, as
representations of the torus,
\begin{equation}
V=\sum_{\alpha=1}^{N}e_\alpha\sum_{(i,j)\in Y_\alpha}T_1^{\,i-1}T_2^{\,j-1}\quad,\quad W=\sum_{\alpha=1}^{N}e_\alpha
\label{Vchar}
\end{equation}
The tangent space at the fixed point consists of the variations of the four matrices, minus the directions removed by
the complex constraint and by the transformations acting on $V$. Counting the weights of each piece,
\begin{equation}
\begin{split}
\chi&=(T_1+T_2)V^*V+T_1T_2\,W^*V+V^*W-T_1T_2\,V^*V-V^*V\\
&=-(1-T_1)(1-T_2)V^*V+T_1T_2\,W^*V+V^*W
\end{split}
\label{tangent}
\end{equation}
where the star denotes the dual representation, in which every $e_\alpha$ and $T_i$ is replaced by its inverse. Substituting
\eqref{Vchar} and reorganising the sum, the negative terms cancel and one obtains \eqref{character}. For instance, if $Y_\alpha$ is a single
box and the other tableaux are empty, $V=e_\alpha$ and
\begin{equation}
\chi=T_1+T_2+\sum_{\beta\neq\alpha}\left(T_1T_2\,e_\alpha e_\beta^{-1}+e_\beta e_\alpha^{-1}\right)
\label{onebox}
\end{equation}
which is what \eqref{character} gives: the first two terms describe the position of the instanton, the others its orientation in the
gauge group, and there are $2N$ of them, the complex dimension of the moduli space. In general the equality of
\eqref{tangent} and \eqref{character} is a combinatorial identity; it has been checked for all pairs of tableaux with up to five boxes
(Appendix \ref{app:checks}).

Each term of $\chi$ is a one-dimensional representation of the torus, on which the generating vector field acts with an
eigenvalue $w$, the weight of Section \ref{sec:Omega}. The weight is read off by replacing $e_\alpha\to a_\alpha$, $T_i\to\epsilon_i$ and products by sums. In
this way the first sum in \eqref{character} gives the weights $E_{\alpha,\beta}(s)$ defined in \eqref{Eab} below, and the second sum the weights
$\epsilon-E_{\beta,\alpha}(s)$ with $s\in Y_\beta$.

The contribution of the fixed points due to the gauge multiplets to Nekrasov's
partition function is the inverse determinant of the toric action vector field over the
fixed points:
\begin{equation}
P_{\rm gauge}(\vec{Y})=\prod_{\alpha,\beta=1}^{N}\prod_{s\in Y_\alpha}\frac{1}{E_{\alpha,\beta}(s)(\epsilon-E_{\alpha,\beta}(s))}
\label{Pgauge}
\end{equation}
in which
\begin{equation}
E_{\alpha,\beta}(s)=a_\alpha-a_\beta-\epsilon_1 l_{Y_\beta}(s)+\epsilon_2(a_{Y_\alpha}(s)+1)
\label{Eab}
\end{equation}
and the contribution of the $f\leq4$ matter fields in the antifundamental representation
of the gauge group is:
\begin{equation}
P_{\rm antifund}(\vec{Y})=\prod_{l=1}^{f}\prod_{\alpha=1}^{N}\prod_{s_\alpha\in Y_\alpha}(\chi_{\alpha,s_\alpha}+m_l)
\end{equation}
and a single hypermultiplet in the adjoint representation is:
\begin{equation}
P_{\rm adj}(\vec{Y})=\prod_{\alpha,\beta=1}^{N}\prod_{s\in Y_\alpha}(E_{\alpha,\beta}(s)-M)(\epsilon-E_{\alpha,\beta}(s)-M)
\end{equation}
where $m_l,M$ are the masses of the hypermultiplets and $\epsilon=\epsilon_1+\epsilon_2$ and
\begin{equation}
\chi_{\alpha,s_\alpha}=a_\alpha+(i_{s_\alpha}-1)\epsilon_1+(j_{s_\alpha}-1)\epsilon_2
\end{equation}
in which $i_{s_\alpha},j_{s_\alpha}$ are the coordinates of the box $s_\alpha$ in the Young tableau, the row and the column respectively.

The matter factors have the same origin. In the background of $k$ instantons a hypermultiplet in the
antifundamental representation has $k$ zero modes, which form over the moduli space a vector bundle with fibre $V$.
Integrating them out multiplies the measure by the determinant of the torus action on this fibre, with the mass
entering as the equivariant parameter of the flavour symmetry: this is the product over the boxes of the weights
$\chi_{\alpha,s}$ of \eqref{Vchar} shifted by $m_l$. For the adjoint representation the fibre is the tangent space itself, and the weights
$E_{\alpha,\beta}(s)$, $\epsilon-E_{\alpha,\beta}(s)$ are shifted by $M$.

The instanton partition function turns out to be:
\begin{equation}
Z_{\rm inst}=\sum_{\vec{Y}}x^{\lvert\vec{Y}\rvert}P_{\rm gauge}(\vec{Y})P_{\rm matter}(\vec{Y})
\label{Zinst}
\end{equation}
Here $P_{\rm matter}$ stands for $P_{\rm antifund}$ or $P_{\rm adj}$, and $P_{\rm matter}=1$ in the theory without matter. The sum runs over all
$N$-tuples of Young tableaux and $\lvert\vec{Y}\rvert$ is the total number of boxes, so that $x$ counts the instantons. For $f=2N$
antifundamental hypermultiplets, or for one adjoint hypermultiplet, the theory is conformal and $x=e^{2\pi i\tau_g}$ is
dimensionless, $\tau_g$ being the gauge coupling. For $f<2N$ the parameter $x$ has the dimension of a mass to the power
$2N-f$, and it is the corresponding power of the dynamical scale; in particular $x=\Lambda^4$ for pure $SU(2)$.

From now on $N=2$ and $a_1=-a_2=a$. As an illustration, for one instanton there are two fixed points, with a single
box in $Y_1$ or in $Y_2$. By \eqref{onebox} the weights of the first are $\epsilon_1$, $\epsilon_2$, $2a+\epsilon$ and $-2a$, those of the second are obtained by
$a\to-a$, and
\begin{equation}
Z_1=\frac{1}{\epsilon_1\epsilon_2}\left[\frac{\prod_{l=1}^{f}(m_l-a)}{2a\,(\epsilon-2a)}-\frac{\prod_{l=1}^{f}(m_l+a)}{2a\,(\epsilon+2a)}\right]
\label{Z1matter}
\end{equation}
is the coefficient of $x$ in \eqref{Zinst}. Each of the two terms has a pole at $a=0$, which cancels in the sum: this is the
simplest example of the cancellation of spurious poles mentioned in Section \ref{sec:pure}.

Generally speaking, as explained in Section \ref{sec:Omega}, the idea behind deriving the aforementioned results is to
define $\mathcal{N}=2$ SYM as the spatially twisted compactification of $5d$ $\mathcal{N}=1$ SYM with
\textbf{8} supercharges, where the compactification is along the fifth dimension to a circle of
circumference $\beta$:
\begin{equation}
(y,\vec{x})\sim(y+\beta,\exp(\beta\Omega).\vec{x})
\end{equation}
where $\Omega$ is the generator of $SO(4)$ rotations.

For a comprehensive review of the modern techniques used in localization, the
reader is referred to \cite{Nekrasov}, \cite{Tachikawa}, and \cite{Shadchin}.

\section{Zamolodchikov's $q$-recursion relation for CFT conformal block}
\label{sec:zam}
\subsubsection*{Conformal blocks}
The symmetry algebra of a two-dimensional conformal field theory is the Virasoro algebra
\begin{equation}
[L_m,L_n]=(m-n)L_{m+n}+\frac{c}{12}\,m(m^2-1)\,\delta_{m+n,0}
\label{Vir}
\end{equation}
where $c$ is the central charge. A primary field $V_\Delta(z)$ of dimension $\Delta$ corresponds to a highest-weight state $\lvert\Delta\rangle$,
\begin{equation}
L_0\lvert\Delta\rangle=\Delta\lvert\Delta\rangle\quad,\quad L_n\lvert\Delta\rangle=0\quad(n>0)
\end{equation}
The states $L_{-k_1}\cdots L_{-k_l}\lvert\Delta\rangle$ with $k_1\geq\dots\geq k_l\geq1$ are its descendants. They are labelled by the partitions
$Y=(k_1,\dots,k_l)$; the number $n=\lvert Y\rvert=k_1+\dots+k_l$ is called the level, and all together they span the Verma module of
$\lvert\Delta\rangle$. With $L_n^\dagger=L_{-n}$ the scalar products of the descendants are determined by \eqref{Vir} alone. They vanish between
different levels, and at level $n$ they form the Gram matrix
\begin{equation}
Q^{(n)}_\Delta(Y,Y')=\langle\Delta\rvert L_{k_l}\cdots L_{k_1}\,L_{-k'_1}\cdots L_{-k'_{l'}}\lvert\Delta\rangle
\label{Gram}
\end{equation}
For instance $Q^{(1)}_\Delta=2\Delta$, and at level two, in the basis $L_{-2}\lvert\Delta\rangle$, $L_{-1}^2\lvert\Delta\rangle$,
\begin{equation}
Q^{(2)}_\Delta=\begin{pmatrix}4\Delta+\frac{c}{2}&6\Delta\\[2pt]6\Delta&4\Delta(2\Delta+1)\end{pmatrix}
\end{equation}

Consider now the correlation function of four primary fields of dimensions $\Delta_1,\dots,\Delta_4$ placed at the points $x$, $0$, $1$ and
$\infty$ respectively. Inserting a complete set of states between the first two fields and the last two, the function
decomposes into a sum over the primary fields $V_\Delta$ of the theory which appear in the intermediate channel. Each term
is the product of two structure constants, which depend on the theory, and of a function which is fixed by the
Virasoro algebra alone, the \emph{conformal block} (times its complex conjugate):
\begin{equation}
\mathcal{F}(\Delta_i,\Delta,x)=x^{\Delta-\Delta_1-\Delta_2}\sum_{n=0}^{\infty}x^n\sum_{\lvert Y\rvert=\lvert Y'\rvert=n}\gamma_{12}(Y)\left[Q^{(n)}_\Delta\right]^{-1}\!(Y,Y')\ \gamma_{34}(Y')
\label{blockdef}
\end{equation}
\begin{equation}
\gamma_{12}(Y)=\prod_{i=1}^{l}\Big(\Delta+k_i\Delta_1-\Delta_2+\sum_{j>i}k_j\Big)
\label{gammadef}
\end{equation}
and $\gamma_{34}(Y)$ is the same expression with $(\Delta_1,\Delta_2)$ replaced by $(\Delta_3,\Delta_4)$. The factors $\gamma$ are the matrix elements of an
external field between a descendant and a highest-weight state; they follow from the commutation relation
$[L_n,V_\Delta(z)]=z^n\left(z\partial_z+(n+1)\Delta\right)V_\Delta(z)$. The first terms are
\begin{equation}
\mathcal{F}(\Delta_i,\Delta,x)=x^{\Delta-\Delta_1-\Delta_2}\left(1+\frac{(\Delta+\Delta_1-\Delta_2)(\Delta+\Delta_3-\Delta_4)}{2\Delta}\,x+\mathcal{O}(x^2)\right)
\label{block1}
\end{equation}
The conformal block thus depends on the central charge, on the four external dimensions, on the internal dimension
$\Delta$ and on $x$, and on nothing else. The definition \eqref{blockdef} is explicit, but the size of the Gram matrix grows quickly with the
level; Zamolodchikov's recursion is a much more efficient way to compute the same function.

As is known, the conformal blocks of the 4-point function can be reduced further to
a simpler block that depends only on the cross-ratio of the relative position of the
operator insertion points and the dimension of the external primary operators:
\begin{equation}
\begin{split}
\mathcal{F}(\Delta_i,\Delta,x)=(16q)^{-\alpha^2}&x^{Q^2/4-\Delta_1-\Delta_2}\\
&\times(1-x)^{Q^2/4-\Delta_1-\Delta_3}\theta_3(q)^{3Q^2-4(\Delta_1+\Delta_2+\Delta_3+\Delta_4)}H(\mu_i,\Delta,q)
\end{split}
\label{block}
\end{equation}
in which
\begin{equation}
\theta_3(q)=\sum_{n=-\infty}^{\infty}q^{n^2}
\end{equation}
where the dimensions $\Delta,\Delta_i$ can be reparameterized as:
\begin{equation}
\Delta_i=\frac{Q^2}{4}-\lambda_i^2\quad,\quad\Delta=\frac{Q^2}{4}-\alpha^2
\label{dims}
\end{equation}
where $Q$ is related to the central charge of the Virasoro algebra by:
\begin{equation}
c=1+6Q^2
\label{central}
\end{equation}
and
\begin{equation}
\begin{aligned}
\mu_1&=\lambda_1+\lambda_2+\frac{Q}{2}&\qquad\mu_2&=\lambda_1-\lambda_2+\frac{Q}{2}\\
\mu_3&=\lambda_3+\lambda_4+\frac{Q}{2}&\qquad\mu_4&=\lambda_3-\lambda_4+\frac{Q}{2}
\end{aligned}
\label{mus}
\end{equation}

Indeed, for the purpose of making the AGT correspondence more explicit, one can
write:
\begin{equation}
Q=\frac{\epsilon_1+\epsilon_2}{\sqrt{\epsilon_1\epsilon_2}}=\sqrt{\frac{\epsilon_1}{\epsilon_2}}+\sqrt{\frac{\epsilon_2}{\epsilon_1}}
\label{Qeps}
\end{equation}
a comparison with $Q=b+1/b$ results in $b=\sqrt{\epsilon_1/\epsilon_2}$. The parameter $q=e^{i\pi\tau}$
where:
\begin{equation}
\tau=i\frac{K(1-x)}{K(x)}\quad\text{where}\quad K(x)=\frac12\int_0^1\frac{dt}{\sqrt{t(1-t)(1-xt)}}
\label{nome}
\end{equation}
The converse would be as follows:
\begin{equation}
x=\frac{\theta_2^4(q)}{\theta_3^4(q)}
\label{xofq}
\end{equation}
where
\begin{equation}
\theta_2(q):=\sum_{n=-\infty}^{\infty}q^{(n+1/2)^2}
\end{equation}
An expansion up to the first few terms would look like:
\begin{equation}
16q=x+\frac{x^2}{2}+\frac{21x^3}{64}+\frac{31x^4}{128}+\mathcal{O}(x^5)
\label{qofx}
\end{equation}
As for the asymptotic behavior of the conformal block for $\Delta\to\infty$, Zamolodchikov
has shown \cite{ZM2}:
\begin{equation}
H=1+\mathcal{O}(1/\Delta)
\label{Hasym}
\end{equation}
The recursion relation is as follows:
\begin{equation}
H(\mu_i,\Delta,q)=1+\sum_{m,n=1}^{\infty}\frac{q^{mn}R^{(4)}_{m,n}}{\Delta-\Delta_{m,n}}H(\mu_i,\Delta_{m,n}+mn,q)
\label{Zrec}
\end{equation}
where:
\begin{equation}
\Delta_{m,n}=\frac{Q^2}{4}-\lambda_{m,n}^2\quad\text{where}\quad\lambda_{m,n}=\frac{m\epsilon_1+n\epsilon_2}{2\sqrt{\epsilon_1\epsilon_2}}
\label{degdims}
\end{equation}
which are exactly the internal degenerate dimensions. Also:
\begin{equation}
R^{(f)}_{m,n}=\frac{2\prod_{r,s}\prod_{i=1}^{f}(\mu_i-\frac{Q}{2}-\lambda_{r,s})}{\prod_{k,l}'\lambda_{k,l}}
\label{Rmn}
\end{equation}
where:
\[
r=-m+1,-m+3,...,m-1,\quad s=-n+1,-n+3,...,n-1
\]
\[
k=-m+1,-m+2,...,m-1,m,\quad l=-n+1,-n+2,...,n-1,n
\]
the prime over the product stands for excluding the pairs $(0,0)$ and $(m,n)$ in the product.

The recursive relation forces $H$ to be symmetric in $\mu_i$ and to be symmetric under
the following simultaneous transformations:
\begin{equation}
\mu_i\to Q-\mu_i\qquad q\to-q
\end{equation}
which is the result of $\lambda_{-r,-s}=-\lambda_{r,s}$.

The first terms of $H$ under the recursion relation turn out to be:
\begin{equation}
H=1+\frac{R^{(f)}_{1,1}\,q}{\Delta-\Delta_{1,1}}+\left(\frac{\big(R^{(f)}_{1,1}\big)^2}{\Delta-\Delta_{1,1}}+\frac{R^{(f)}_{1,2}}{\Delta-\Delta_{1,2}}+\frac{R^{(f)}_{2,1}}{\Delta-\Delta_{2,1}}\right)q^2+\mathcal{O}(q^3)
\end{equation}
It is worth mentioning that the poles at the order $q^l$ occur at $\Delta=\Delta_{m,n}$ with $mn\leq l$.

\subsubsection*{The origin of the recursion}
The relation \eqref{Zrec} rests on three properties of the conformal block.
\begin{enumerate}
\item \emph{Poles.} The coefficients of \eqref{blockdef} are rational functions of $\Delta$, and they can have poles only where the Gram matrix
is not invertible. Its determinant is given by the formula of Kac,
\begin{equation}
\det Q^{(n)}_\Delta=C_n\prod_{\substack{m,m'\geq1\\ mm'\leq n}}\left(\Delta-\Delta_{m,m'}\right)^{p(n-mm')}
\label{Kac}
\end{equation}
where $C_n$ is a constant, $p(n)$ is the number of partitions of $n$ and $\Delta_{m,m'}$ are the degenerate dimensions \eqref{degdims}.
Hence the block has poles at $\Delta=\Delta_{m,n}$, starting from the level $mn$. For $n=1$ this is the statement $Q^{(1)}_\Delta=2\Delta$, with
$\Delta_{1,1}=0$.
\item \emph{Residues.} When $\Delta=\Delta_{m,n}$ the Verma module contains at the level $mn$ a \emph{null vector}: a descendant which is
itself a highest-weight state, of dimension $\Delta_{m,n}+mn$, and which is orthogonal to every state. This vector and its
descendants, which form a Verma module of their own, are responsible for the pole. The residue of the block at
$\Delta=\Delta_{m,n}$ is therefore proportional to $x^{mn}$ times the block with internal dimension $\Delta_{m,n}+mn$. The coefficient of
proportionality is the product of the matrix elements of the null vector with the two pairs of external fields,
polynomials which vanish when the fusion with the degenerate field is allowed and which give the numerator of
\eqref{Rmn}, divided by the derivative of the norm of the null vector, which gives the denominator.
\item \emph{Behaviour at large $\Delta$.} For $\Delta\to\infty$ the block behaves like $(16q)^{\Delta}$, where $q$ is the elliptic nome \eqref{nome}; more
precisely, after the factors displayed in \eqref{block} have been extracted, what is left tends to $1$, which is \eqref{Hasym}.
\end{enumerate}
A function of $\Delta$ which has only simple poles with known residues and which tends to $1$ at infinity is the sum
over its poles, and this is the recursion \eqref{Zrec}. The first two properties are elementary consequences of the
representation theory of the Virasoro algebra. The explicit form \eqref{Rmn} of the coefficients and the asymptotic
behaviour \eqref{Hasym} are the results of Zamolodchikov \cite{ZM2}, \cite{ZM}, and they are not derived here. They can however be tested
against the definition \eqref{blockdef}. At first order this is done by hand. Since $\Delta_{1,1}=0$ and $R^{(4)}_{1,1}=8\prod_{i=1}^4(\mu_i-\frac{Q}{2})$, the
recursion gives
\begin{equation}
H=1+\frac{\prod_{i=1}^{4}\left(\mu_i-\frac{Q}{2}\right)}{2\Delta}\,16q+\mathcal{O}(q^2)
\end{equation}
By \eqref{dims} and \eqref{mus}, $(\mu_1-\frac{Q}{2})(\mu_2-\frac{Q}{2})=\Delta_2-\Delta_1$ and $(\mu_3-\frac{Q}{2})(\mu_4-\frac{Q}{2})=\Delta_4-\Delta_3$, so that the pole of $H$ at
$\Delta=0$ is $(\Delta_1-\Delta_2)(\Delta_3-\Delta_4)\,x/2\Delta$: this is exactly the pole of the first coefficient in \eqref{block1}. Its regular part,
$\frac12(\Delta+\Delta_1-\Delta_2+\Delta_3-\Delta_4)$, is reproduced by the expansion of the three prefactors of \eqref{block} with $16q=x+\dots$ and
$\theta_3(q)=1+2q+\dots$. At higher orders the comparison was made by computer: the series \eqref{blockdef}, computed from the
Gram matrices, and the series obtained from \eqref{block} and \eqref{Zrec} coincide up to the order $x^3$ (Appendix \ref{app:checks}).

\section{AGT correspondence, Recursion in 4-Point Conformal Blocks and Nekrasov's SW Prepotential}
\label{sec:agt}
By the AGT correspondence \cite{AGT} one can write for $f=4$ antifundamental hypermultiplets
and $N=2$:
\begin{equation}
\begin{split}
Z^{(4)}_{\rm inst}(a,m_i,x)&=x^{\Delta_1+\Delta_2-\Delta}(1-x)^{2(\lambda_1+\frac{Q}{2})(\lambda_3+\frac{Q}{2})}\mathcal{F}(\Delta,\Delta_i,x)\\
&=\left(\frac{x}{16q}\right)^{\alpha^2}(1-x)^{\frac14(Q-\sum_{i=1}^4\mu_i)^2}\,[\theta_3(q)]^{2\sum_{i=1}^4(\mu_i^2-Q\mu_i)+Q^2}H(\mu_i,\Delta,q)
\end{split}
\label{AGT4}
\end{equation}
where $Z^{(4)}_{\rm inst}$ is given by \eqref{Zinst}. The VEV $a=\alpha\sqrt{\epsilon_1\epsilon_2}$ and the masses $m_i=\mu_i\sqrt{\epsilon_1\epsilon_2}$
are related to the conformal dimensions $\Delta,\Delta_i$ through \eqref{dims} and \eqref{mus}.

The two factors in front of $\mathcal{F}$ in the first line of \eqref{AGT4} have a simple meaning. The power of $x$ removes the leading
power of \eqref{blockdef}, and the power of $1-x$ accounts for the difference between the gauge group $U(2)$, for which \eqref{Zinst} is
written, and $SU(2)$. The second line follows from the first by \eqref{block}, \eqref{dims} and \eqref{mus}. At the lowest order the relation
can be checked by hand: by \eqref{block1} it states that the one-instanton coefficient \eqref{Z1matter} with $f=4$ equals
\begin{equation}
Z_1=\frac{(\Delta+\Delta_1-\Delta_2)(\Delta+\Delta_3-\Delta_4)}{2\Delta}-\frac12(\mu_1+\mu_2)(\mu_3+\mu_4)
\label{AGT1}
\end{equation}
which is an identity between two rational functions of $a$, $m_i$ and $\epsilon_{1,2}$. The comparison was carried out with exact
arithmetic up to three instantons, the left-hand side being computed from the sum over Young tableaux \eqref{Zinst} and the
right-hand side from the Gram matrices \eqref{blockdef} (Appendix \ref{app:checks}). In \cite{AGT} the relation \eqref{AGT4} was a conjecture; the cases in
which it has become a theorem are recalled in Section \ref{sec:pure}.

Therefore, the large VEV behavior of the instanton partition function can be
determined:
\begin{equation}
Z^{(4)}_{\rm inst}\sim\left(\frac{x}{16q}\right)^{\frac{a^2}{\epsilon_1\epsilon_2}}(1-x)^{\frac{1}{4\epsilon_1\epsilon_2}(\epsilon-\sum_{i=1}^4m_i)^2}\times[\theta_3(q)]^{\frac{1}{\epsilon_1\epsilon_2}\sum_{i=1}^4\left(m_i^2+(\epsilon-m_i)^2-3\epsilon^2/4\right)}
\end{equation}

The instanton part of the Seiberg-Witten prepotential is:
\begin{equation}
\mathcal{F}^{SW}_{\rm inst}=-\lim_{\substack{\epsilon_1\to0\\ \epsilon_2\to0}}\epsilon_1\epsilon_2\log Z_{\rm inst}
\label{Finstdef}
\end{equation}
resulting in:
\begin{equation}
\mathcal{F}^{SW}_{\rm inst}\sim a^2\log\frac{16q}{x}-\frac14\left(\sum_{i=1}^4m_i\right)^2\log(1-x)-2\sum_{i=1}^4m_i^2\log\theta_3(q)
\label{F4asym}
\end{equation}
which, by application of \eqref{xofq} and \eqref{qofx}, the leading term becomes:
\begin{equation}
\mathcal{F}^{SW}_{\rm inst}\sim a^2\log\frac{16q}{x}=a^2\left(\frac{x}{2}+\frac{13x^2}{64}+\frac{23x^3}{192}+\frac{2701x^4}{32768}+\frac{5057x^5}{81920}+\cdots\right)
\end{equation}
which is in complete agreement with B3 of \cite{AGT}.

The analysis can be carried out more or less the same way for $f=3$ and
$f=2,1,0$ \cite{Recursion}; the case $f=0$ is worked out in the next section.
For a more recent and comprehensive work comparing recursive relations in $2d$
CFT and the SW partition function, the reader is referred to \cite{Recursion2}.

\section{The Pure $SU(2)$ Theory: Running the Recursion}
\label{sec:pure}
The theory without matter is reached from \eqref{AGT4} by decoupling the hypermultiplets: $m_i\to\infty$ and $x\to0$ with
the product $x\,m_1m_2m_3m_4$ kept fixed. This product has the dimension of a mass to the fourth power; it is the instanton
parameter of the pure theory and it is denoted again by $x$. On the gauge theory side the limit is obvious from
\eqref{Zinst}: each term of $P_{\rm antifund}$ tends to $(m_1m_2m_3m_4)^{k}$, and the partition function with four flavours becomes the one
without matter. On the CFT side one takes the same limit in the second line of \eqref{AGT4}. By \eqref{qofx}, $q\simeq x/16\to0$, and by
\eqref{Rmn}, $R^{(4)}_{m,n}\simeq(\mu_1\mu_2\mu_3\mu_4)^{mn}R^{(0)}_{m,n}$, so that $q^{mn}R^{(4)}_{m,n}$ has a finite limit, in which $q$ is replaced by $q\,\mu_1\mu_2\mu_3\mu_4$.
The three prefactors of $H$ tend to $1$, because their exponents grow like $\mu_i^2$ while $x$ and $q$ vanish like $\mu_i^{-4}$.
One is left with \cite{Recursion}:
\begin{equation}
Z_{\rm inst}(a,x)=H^{(0)}(\Delta,q)\quad,\quad q=\frac{x}{16(\epsilon_1\epsilon_2)^2}
\label{pureH}
\end{equation}
where $H^{(0)}$ satisfies \eqref{Zrec} with $R^{(4)}_{m,n}$ replaced by $R^{(0)}_{m,n}$ of \eqref{Rmn} (the product over $i$ being empty).
Since $\Delta-\Delta_{m,n}=\lambda_{m,n}^2-\alpha^2$ and since the shifted dimension $\Delta_{m,n}+mn$ corresponds
to $\alpha^2=\lambda_{m,n}^2-mn=(m\epsilon_1-n\epsilon_2)^2/4\epsilon_1\epsilon_2$, the relation can be rewritten purely in
gauge theory terms:
\begin{equation}
Z_{\rm inst}(a,x)=1-\sum_{m,n=1}^{\infty}\frac{x^{mn}R_{m,n}}{4a^2-(m\epsilon_1+n\epsilon_2)^2}\,Z_{\rm inst}\!\left(\frac{m\epsilon_1-n\epsilon_2}{2},x\right)
\label{purerec}
\end{equation}
\begin{equation}
R_{m,n}=2\,{\prod_{k,l}}'\,\frac{1}{k\epsilon_1+l\epsilon_2}
\label{pureR}
\end{equation}
with the range of $k,l$ and the meaning of the prime as in \eqref{Rmn}.

The function $H^{(0)}$ is again an object of the representation theory of the Virasoro algebra, an \emph{irregular} conformal
block \cite{Gaiotto}. In the Verma module of $\lvert\Delta\rangle$ consider the state
\begin{equation}
\lvert G\rangle=\sum_{n=0}^{\infty}\mathfrak{x}^{n/2}\sum_{\lvert Y\rvert=n}\left[Q^{(n)}_\Delta\right]^{-1}\!(1^n,Y)\ L_{-Y}\lvert\Delta\rangle\quad,\quad\mathfrak{x}=\frac{x}{(\epsilon_1\epsilon_2)^2}
\end{equation}
where $1^n$ denotes the partition $(1,\dots,1)$ and $L_{-Y}=L_{-k_1}\cdots L_{-k_l}$. It satisfies $L_1\lvert G\rangle=\mathfrak{x}^{1/2}\lvert G\rangle$ and $L_2\lvert G\rangle=0$, and it is
called a Whittaker, or Gaiotto, state. Its norm is the limit of the four-point block described above, and
\begin{equation}
Z_{\rm inst}(a,x)=\langle G\vert G\rangle=\sum_{n=0}^{\infty}\mathfrak{x}^n\left[Q^{(n)}_\Delta\right]^{-1}\!(1^n,1^n)\quad,\quad\Delta=\frac{\epsilon^2-4a^2}{4\epsilon_1\epsilon_2}
\label{Gnorm}
\end{equation}
At first order $[Q^{(1)}_\Delta]^{-1}=1/2\Delta$, and \eqref{Gnorm} gives $Z_1=2/\epsilon_1\epsilon_2(\epsilon^2-4a^2)$, in agreement with \eqref{Z1matter} for $f=0$. The
equality \eqref{Gnorm} was checked with exact arithmetic up to $n=3$ (Appendix \ref{app:checks}).

Two remarks are in order. First, in \cite{Recursion} the relation \eqref{purerec} was a consequence of the
AGT conjecture. It has since been proven directly from the combinatorial formula
\eqref{Zinst}, together with the corresponding statement for the irregular blocks, for $f=0,1,2$ in \cite{HJS}
(see also \cite{FL} for the theory with an adjoint hypermultiplet). For the pure theory, therefore,
\eqref{purerec} is a theorem, and so is the AGT relation \eqref{pureH}. Second, \eqref{purerec} is far more economical
than \eqref{Zinst}: at order $x^k$ only the finitely many pairs with $mn\leq k$ contribute, the only poles in $a$
are the simple poles at $2a=\pm(m\epsilon_1+n\epsilon_2)$, and their residues are the lower orders evaluated
at the degenerate values $2a=m\epsilon_1-n\epsilon_2$. In the sum \eqref{Zinst} the number of terms
grows rapidly and most of the poles of the individual terms cancel.

Writing $Z_{\rm inst}=1+\sum_{k\geq1}Z_kx^k$, the first two orders of \eqref{purerec} are:
\begin{equation}
Z_1=-\frac{2}{\epsilon_1\epsilon_2\left(4a^2-\epsilon^2\right)}
\label{Z1}
\end{equation}
\begin{equation}
Z_2=\frac{8a^2-8\epsilon_1^2-17\epsilon_1\epsilon_2-8\epsilon_2^2}{\epsilon_1^2\epsilon_2^2\left(4a^2-\epsilon^2\right)\left(4a^2-(2\epsilon_1+\epsilon_2)^2\right)\left(4a^2-(\epsilon_1+2\epsilon_2)^2\right)}
\label{Z2}
\end{equation}
Both coincide with the sum over pairs of Young tableaux \eqref{Zinst} with $\lvert\vec Y\rvert=1,2$. More generally,
\eqref{purerec} and \eqref{Zinst} were compared with exact rational arithmetic at generic rational values
of $(a,\epsilon_1,\epsilon_2)$ and agree identically up to $k=5$ (Appendix \ref{app:checks}).

Finally, the prepotential \eqref{Finstdef} follows by taking the logarithm:
\begin{equation}
\mathcal{F}^{SW}_{\rm inst}=\sum_{k=1}^{\infty}F_k\frac{x^k}{a^{4k-2}}
\label{Fkdef}
\end{equation}
with the coefficients collected in Table \ref{tab:Fk}.
\begin{table}[h]
\caption{The instanton coefficients $F_k$ of the pure $SU(2)$ prepotential \eqref{Fkdef} generated by the recursion \eqref{purerec}.}
\label{tab:Fk}
\centering
\renewcommand{\arraystretch}{1.5}
\begin{tabular}{c|ccccc}
$k$&1&2&3&4&5\\\hline
\rule[-2.4ex]{0pt}{6ex}$F_k$&$\dfrac12$&$\dfrac{5}{64}$&$\dfrac{3}{64}$&$\dfrac{1469}{32768}$&$\dfrac{4471}{81920}$
\end{tabular}
\end{table}

The first two entries close a circle inside this thesis. Specializing the one- and
two-instanton results of the localization computation of Chapter \ref{Localisation}, \eqref{eq:F1-nc-localization} and \eqref{eq:F2-nc-localization}, to
$N=2$, $N_F=0$ and $\phi^0_1=-\phi^0_2=a$, one has $S_1(a)=S_2(-a)=1/4a^2$ in \eqref{eq:S-u-definition}, so that
\begin{equation}
\widehat{Z}_1=\frac{2}{(2a)^2}=\frac{1}{2a^2}\quad,\quad
\widehat{Z}_2=\frac{2}{(2a)^2}\frac{1}{(4a^2)^2}+\frac14\cdot2\cdot\frac{1}{4a^2}\cdot\frac{6}{(2a)^4}=\frac{5}{64a^6}
\end{equation}
which are precisely $F_1/a^2$ and $F_2/a^6$. The remaining entries of Table \ref{tab:Fk} extend the
comparison with the Seiberg-Witten curve to five instantons; this is done in
Section \ref{sec:meeting}, where the same numbers are obtained from the periods of the curve.

What the recursion \eqref{purerec} cannot see is equally important for what follows. It is
exact in $a$, $\epsilon_1$ and $\epsilon_2$ at each order in $x$, but it is an expansion around $a=\infty$: it
knows nothing about the region of the moduli space where $x/a^4$ is not small, which
is where the monopole and the dyon of Chapter \ref{ch:Seiberg-Witten-theory} live. Moreover, in the limit
$\epsilon_{1,2}\to0$ all of its poles $2a=\pm(m\epsilon_1+n\epsilon_2)$ collapse onto the single point $a=0$, so that
the recursion has no term-by-term classical limit: the prepotential emerges only after
the logarithm is taken. The next chapter describes a second recursion which is, in
both respects, complementary.

\chapter{Topological Recursion on the Seiberg-Witten Curve: a Synthesis}
\label{ch:TR}

The prepotential has by now been obtained in three ways: from the periods of the
Seiberg-Witten curve (Chapter \ref{ch:Seiberg-Witten-theory}), from the localization of the instanton measure
(Chapter \ref{Localisation}), and from Zamolodchikov's recursion through the AGT correspondence
(Chapter \ref{ch:recursion}). All three reach the instanton coefficients one by one, as an expansion
around weak coupling. This final chapter adds a fourth construction which works
the other way around. The topological recursion of Eynard and Orantin \cite{EO} is run on one
fixed, smooth Seiberg-Witten curve and produces the Taylor expansion of the
prepotential around the corresponding point of the moduli space, each coefficient
being exact in $\Lambda$ \cite{Chaimanowong}. The two recursions are therefore two expansions of one
and the same function in transverse directions, and the purpose of the chapter is to make them meet
explicitly. For pure $SU(2)$ it is shown that:
\begin{itemize}
\item the first two non-trivial steps of the topological recursion can be carried out in closed
form (Proposition \ref{prop:genus0}) and reproduce, order by order, the instanton series generated
by Zamolodchikov's recursion (Section \ref{sec:meeting});
\item unlike the instanton series, the result remains valid at strong coupling, where it
detects the massless monopole (Section \ref{sec:monopole});
\item the first corrections in $\epsilon_1,\epsilon_2$ produced by Zamolodchikov's recursion are again
functions on the same curve, and on the self-dual background they are reproduced by the
genus-one step of the topological recursion (Section \ref{sec:genus1}).
\end{itemize}
The last point suggests a conjecture and a few concrete open problems, which close
the thesis (Section \ref{sec:outlook}). Every formula of this chapter that is not quoted from the
literature has been verified by computer algebra and numerically; the checks are
listed in Appendix \ref{app:checks}.

\section{Conventions}
\label{sec:conv}
Throughout this chapter the gauge group is $SU(2)$ without matter and the Seiberg-Witten
curve is taken in the form
\begin{equation}
\Sigma_u:\qquad\Lambda^2\left(w+\frac1w\right)=z^2-u
\label{SWcurve}
\end{equation}
In terms of $y:=\Lambda^2(w-1/w)$ this is the hyperelliptic curve
\begin{equation}
y^2=(z^2-u)^2-4\Lambda^4\quad,\quad\Delta:=u^2-4\Lambda^4
\label{hyper}
\end{equation}
that is, the curve of Section \ref{subsec:Singularity-Structure} for $N=2$, $N_f=0$ (with $\Lambda^4$ there equal to $4\Lambda^4$ here). The curve is
smooth for $\Delta\neq0$; the two singular points $u=\pm2\Lambda^2$ are the monopole and the dyon
point. The Seiberg-Witten differential and its periods are
\begin{equation}
\mathrm{d} S=z\,\frac{\mathrm{d} w}{w}=\frac{2z^2\,\mathrm{d} z}{y}\quad,\quad a=\frac{1}{2\pi i}\oint_A\mathrm{d} S\quad,\quad a_D=\frac{1}{2\pi i}\oint_B\mathrm{d} S
\label{periods}
\end{equation}
where, for $u>2\Lambda^2$, $A$ is the lift of the circle $\lvert w\rvert=1$ to the sheet $z>0$ and $B$ is the
cycle with $A\circ B=+1$; other values of $u$ are reached by analytic continuation.
The prepotential is normalised by
\begin{equation}
\frac{\partial\mathcal{F}}{\partial a}=2\pi i\,a_D\quad,\quad\frac{\partial^2\mathcal{F}}{\partial a^2}=2\pi i\,\tau
\label{Fnorm}
\end{equation}
where $\tau$ is the period of the curve \eqref{SWcurve}, $\mathrm{Im}\,\tau>0$. It is shown in Section \ref{sec:su2curve} that at weak coupling the
periods of the curve give
\begin{equation}
\mathcal{F}(a)=-4a^2\ln\frac{2a}{\Lambda}+6a^2+\sum_{k=1}^{\infty}F_k\frac{\Lambda^{4k}}{a^{4k-2}}
\label{Ffull}
\end{equation}
with coefficients $F_k$ determined by the period integral. Nekrasov's conjecture, proven in \cite{NO} and \cite{NY}, states that these
coefficients are those of the instanton prepotential \eqref{Finstdef}, \eqref{Fkdef} with $x=\Lambda^4$. The conjecture is not assumed in this
chapter: the two sets of coefficients are computed independently and compared in Section \ref{sec:meeting}. In other words, $\mathcal{F}$ of this
chapter is $2\pi i$ times the prepotential of Section \ref{sec:SW-curves-SQCD}, whose one-loop part \eqref{eq:one-loop-prepotential}
is indeed $(2i/\pi)\,a^2\ln(2a/\Lambda)$ for $a_1=-a_2=a$.

\begin{remark}
\label{rem:dictionary}
The curve \eqref{SWcurve} and the curve \eqref{eq:SU2-elliptic-curve} of Section \ref{Witten---Seiberg-Theory} are related by an isogeny. In the units $\Lambda=1$
used there, the two descriptions are matched by $u=2\tilde u$, where $\tilde u$ denotes the modulus of \eqref{eq:SU2-elliptic-curve}; the period $a$ is the same
function, given by \eqref{eq:a-hypergeometric}, while $a_D$ and $\tau$ of this chapter are twice those of \eqref{eq:aD-hypergeometric} and \eqref{eq:tau-elliptic}. For
instance, at $u=6\Lambda^2$ the modulus in \eqref{eq:tau-elliptic} is $k^2=1/2$, so that $\tau=2i$ exactly.
\end{remark}

\section{The Eynard-Orantin Topological Recursion}
\label{sec:EO}
A \emph{spectral curve} $(\Sigma,X,Y,\omega_{0,2})$ consists of a compact Riemann surface $\Sigma$ with a chosen
symplectic basis of cycles, two functions $X$, $Y$ on it (possibly multivalued, with $\mathrm{d} X$ and $\mathrm{d} Y$
meromorphic) and the \emph{Bergman kernel} $\omega_{0,2}$: the unique symmetric bidifferential on $\Sigma\times\Sigma$
whose only singularity is a double pole on the diagonal,
\begin{equation}
\omega_{0,2}(p_1,p_2)=\frac{\mathrm{d} t_1\,\mathrm{d} t_2}{(t_1-t_2)^2}+\text{regular}
\label{Bergdef}
\end{equation}
in any local coordinate $t$, and whose $A$-periods vanish \cite{Fay}. Its $B$-period is the
normalised holomorphic differential $\omega$:
\begin{equation}
\oint_{p_1\in B}\omega_{0,2}(p_1,p_2)=2\pi i\,\omega(p_2)\quad,\quad\oint_A\omega=1\quad,\quad\oint_B\omega=\tau
\label{Bergper}
\end{equation}
The first relation in \eqref{Bergper} follows from the definition. The $B$-period of $\omega_{0,2}$ in its first argument is a holomorphic
differential in the second, because the double pole has no residue; and its $A$-period equals $2\pi i$, because the two
integrations over $A$ and over $B$ fail to commute precisely by the residue of the double pole at the point where the
two cycles intersect. A symplectic basis is a pair of cycles with intersection number $A\circ B=1$; on a curve of genus
one, such as \eqref{SWcurve}, there is a single pair and a single holomorphic differential.
The zeros of $\mathrm{d} X$ are assumed simple and are called \emph{ramification points}; near each
of them, $r$, there is a local involution $\sigma_r$ with $X\circ\sigma_r=X$ and $\sigma_r(r)=r$. Starting from
\begin{equation}
\omega_{0,1}=Y\,\mathrm{d} X
\end{equation}
and $\omega_{0,2}$, the recursion defines symmetric multi-differentials $\omega_{g,n}$ for $2g-2+n>0$ by \cite{EO}
\begin{equation}
\begin{split}
\omega_{g,n}(p_1,\dots,p_n)=\sum_r\mathop{\mathrm{Res}}_{p\to r}K(p_1,p)\Big[&\omega_{g-1,n+1}(p,\sigma_r(p),p_2,\dots,p_n)\\
&+\sum{}^{'}\,\omega_{g_1,1+\lvert I_1\rvert}(p,p_{I_1})\,\omega_{g_2,1+\lvert I_2\rvert}(\sigma_r(p),p_{I_2})\Big]
\end{split}
\label{TR}
\end{equation}
where the primed sum runs over $g_1+g_2=g$ and $I_1\sqcup I_2=\{2,\dots,n\}$, the terms containing
$\omega_{0,1}$ being excluded, and where the recursion kernel is
\begin{equation}
K(p_1,p)=-\frac12\,\frac{\int_{\sigma_r(p)}^{p}\omega_{0,2}(p_1,\cdot\,)}{\omega_{0,1}(p)-\omega_{0,1}(\sigma_r(p))}
\label{kernel}
\end{equation}
The recursion is on the number $2g-2+n$, minus the Euler characteristic of a surface of
genus $g$ with $n$ boundaries, whence the name. Each $\omega_{g,n}$ is a meromorphic differential in every
variable, with poles only at the ramification points and with vanishing $A$-periods.

It is instructive to compare \eqref{TR} with the recursion \eqref{purerec} at this structural level.
Both express an object through residues of simpler objects of the same kind. But in
\eqref{purerec} the residues are taken at the degenerate values of the modulus $a$ and the relation is linear,
whereas in \eqref{TR} they are taken at points of the curve, at fixed modulus, and the relation is
quadratic.

\section{The Prepotential Theorem}
\label{sec:theorem}
The Seiberg-Witten curve \eqref{SWcurve} becomes a spectral curve by taking
\begin{equation}
X=\ln w\quad,\quad Y=z\quad,\quad\omega_{0,1}=\mathrm{d} S
\label{spectral}
\end{equation}
The choice of $X$ is not arbitrary: the differential $\mathrm{d} S$ has the defining property that its
derivative with respect to $u$ \emph{at fixed} $w$ is holomorphic (Section \ref{subsec:SW-curves-general-remarks}), and it is precisely the
fibration $w=\text{const}$ that is singled out in this way \cite{Chaimanowong}. The following theorem is the
bridge between the topological recursion and Seiberg-Witten theory.

\begin{theorem}[\cite{Chaimanowong}, see also \cite{CNST,BH}]
\label{thm:prepotential}
Let $\Sigma_{u_0}$ be a smooth Seiberg-Witten curve, $a_0=a(u_0)$, and let $\omega_{0,n}$ be the genus-zero
multi-differentials produced by the topological recursion on the spectral curve \eqref{spectral} at $u=u_0$. Then, in a neighbourhood of $a_0$,
\begin{equation}
\mathcal{F}(a)=\mathcal{F}(a_0)+2\pi i\,a_D(a_0)\,(a-a_0)+\sum_{n=2}^{\infty}\frac{(-1)^n}{n!}\left[\oint_{B}\!\!\cdots\!\oint_{B}\omega_{0,n}\right](a-a_0)^n
\label{Taylor}
\end{equation}
that is,
\begin{equation}
\frac{\partial^n\mathcal{F}}{\partial a^n}=(-1)^n\oint_{B}\!\!\cdots\!\oint_{B}\omega_{0,n}\qquad(n\geq2)
\label{derivs}
\end{equation}
\end{theorem}

Some comments are necessary. In \cite{Chaimanowong} the theorem is stated for $SU(N+1)$, with the periods
$\oint_A\mathrm{d} S$ as coordinates and with the prepotential defined by $\partial\mathcal{F}/\partial a=\oint_B\mathrm{d} S$; equation \eqref{Taylor} is
equation (117) there, rewritten in the normalisation \eqref{periods}, \eqref{Fnorm}. The theorem holds more generally
for families of curves in a foliated symplectic surface; this is proven in \cite{CNST} by means of
the Airy structures of Kontsevich and Soibelman \cite{KS}, and for spectral curves of
Hitchin systems in \cite{BH}. For $n=2$, equation \eqref{derivs} is the second relation in \eqref{Fnorm}, by \eqref{Bergper}. For
$n=3$ it is the classical variational formula of Rauch \cite{Rauch} for the period matrix. The
sign $(-1)^n$ is a matter of convention: it disappears if $Y$ is replaced by $-Y$.

The mechanism behind the theorem is a general property of the topological recursion \cite{EO}. Suppose that the
spectral curve depends on a parameter and that the variation of $\omega_{0,1}$ with respect to it, taken at fixed $X$, is a
$B$-period of $\omega_{0,2}$. Then the variation of every $\omega_{g,n}$ at fixed $X$ is, up to a sign fixed by the conventions, the
$B$-period of $\omega_{g,n+1}$. The reason is that the recursion \eqref{TR} builds $\omega_{g,n+1}$ out of the kernel \eqref{kernel} and of the lower
differentials in the same way in which the variation acts on them; the first step, the variation of $\omega_{0,2}$ itself, is
Rauch's formula. For the Seiberg-Witten family the hypothesis holds with the period $a$ as parameter: differentiating
\eqref{SWcurve} at fixed $w$ gives $\partial_a\mathrm{d} S=2\pi i\,\omega$, which by \eqref{Bergper} is the $B$-period of $\omega_{0,2}$. Starting from $n=2$ and integrating
all the arguments over $B$ one obtains \eqref{derivs} by induction. This argument is given here only to explain why the
theorem is true. Nothing below depends on it: for $n=3$ and $n=4$, and for the genus-one analogue, the statement is
proven directly in Sections \ref{sec:genus0} and \ref{sec:genus1}.

The theorem makes the contrast with Chapter \ref{ch:recursion} precise. The topological recursion needs only \emph{one} curve
of the family, at which it is exact in $\Lambda$, and returns the germ of $\mathcal{F}$ at that point;
Zamolodchikov's recursion needs no curve at all and returns $\mathcal{F}$ as a series in $\Lambda^4$, exact in $a$. Neither
\cite{Chaimanowong} nor \cite{Recursion} computes the same quantity in both ways. This is done in the next three sections.

\section{The $SU(2)$ Spectral Curve in Closed Form}
\label{sec:su2curve}
\subsubsection*{Periods}
On the cycle $A$ one can put $w=e^{i\theta}$, which gives
\begin{equation}
a(u)=\frac1\pi\int_0^\pi\sqrt{u+2\Lambda^2\cos\theta}\;\mathrm{d}\theta
\label{aint}
\end{equation}
Differentiating \eqref{SWcurve} at fixed $w$ gives $\partial_u\mathrm{d} S=\mathrm{d} z/y$, hence, with $a':=\mathrm{d} a/\mathrm{d} u$,
\begin{equation}
\oint_A\frac{\mathrm{d} z}{y}=2\pi i\,a'\quad,\quad\omega=\frac{1}{2\pi i\,a'}\,\frac{\mathrm{d} z}{y}\quad,\quad\tau=\frac{a_D'}{a'}
\label{holo}
\end{equation}
Both periods satisfy the Picard-Fuchs equation \cite{KLT}
\begin{equation}
\left(u^2-4\Lambda^4\right)\frac{\mathrm{d}^2\Pi}{\mathrm{d} u^2}+\frac14\,\Pi=0\quad,\quad\Pi\in\{a,a_D\}
\label{PF}
\end{equation}
The proof is one line. At fixed $w$ one has $\mathrm{d} S=z\,\mathrm{d} w/w$ with $z^2=u+\Lambda^2(w+1/w)$, hence $\partial_u^2z=-1/4z^3$, and a direct
computation gives
\begin{equation}
\left[\left(u^2-4\Lambda^4\right)\partial_u^2+\frac14\right]\mathrm{d} S=\mathrm{d}\!\left(\frac{y}{2z}\right)
\label{PFexact}
\end{equation}
The right-hand side is the differential of a function on the curve, and its integral over any closed cycle vanishes.

Since \eqref{PF} has no first-derivative term, the combination $a\,a_D'-a_D\,a'$ of two solutions is constant. Its value is found
at large $u$. There \eqref{aint} gives $a\simeq\sqrt{u}$, while the $B$-period of $\mathrm{d} z/y$ is twice the integral between the two inner branch
points $z=\pm\sqrt{u-2\Lambda^2}$, a complete elliptic integral which diverges logarithmically when the branch points
$\pm\sqrt{u-2\Lambda^2}$ and $\pm\sqrt{u+2\Lambda^2}$ approach each other:
\begin{equation}
\frac{\mathrm{d} a_D}{\mathrm{d} u}\simeq\frac{i}{\pi\sqrt{u}}\,\ln\frac{4u}{\Lambda^2}\quad,\quad a_D\simeq\frac{4i}{\pi}\,a\left(\ln\frac{2a}{\Lambda}-1\right)\qquad(u\to\infty)
\label{aDasym}
\end{equation}
the sign being fixed by $\mathrm{Im}\,\tau>0$. Inserting these expressions,
\begin{equation}
a\,\frac{\mathrm{d} a_D}{\mathrm{d} u}-a_D\,\frac{\mathrm{d} a}{\mathrm{d} u}=\frac{2i}{\pi}
\label{Legendre}
\end{equation}
The weak-coupling form \eqref{Ffull} of the prepotential now follows. By \eqref{Fnorm} and \eqref{Legendre},
\begin{equation}
\frac{\mathrm{d}}{\mathrm{d} u}\left(2\mathcal{F}-a\frac{\partial\mathcal{F}}{\partial a}\right)=\left(\frac{\partial\mathcal{F}}{\partial a}-a\frac{\partial^2\mathcal{F}}{\partial a^2}\right)\frac{\mathrm{d} a}{\mathrm{d} u}=2\pi i\left(a_D\,a'-a\,a_D'\right)=4
\end{equation}
so that, with a suitable choice of the additive constant in $\mathcal{F}$, one obtains the relation of Matone \cite{Matone}:
\begin{equation}
u=\frac14\left(2\mathcal{F}-a\frac{\partial\mathcal{F}}{\partial a}\right)=a^2+\sum_{k=1}^\infty kF_k\frac{\Lambda^{4k}}{a^{4k-2}}
\label{Matone}
\end{equation}
The first equality is a differential equation which determines $\mathcal{F}(a)$ once $u(a)$ is known. Inverting the expansion of
\eqref{aint} in powers of $\Lambda^4/u^2$ gives $u(a)$ as a series in $\Lambda^4/a^4$, whose coefficients are written $kF_k$ in the second equality. A term
$F_k\Lambda^{4k}/a^{4k-2}$ of $\mathcal{F}$ contributes $4kF_k\Lambda^{4k}/a^{4k-2}$ to $2\mathcal{F}-a\,\partial_a\mathcal{F}$, a term $-4a^2\ln a$ contributes $4a^2$, and a term proportional to
$a^2$ contributes nothing; the coefficient of the latter is fixed by \eqref{aDasym} and $\partial_a\mathcal{F}=2\pi i\,a_D$. This is \eqref{Ffull}.

\subsubsection*{Ramification points}
The differential $\mathrm{d} X=\mathrm{d} w/w=2z\,\mathrm{d} z/y$ vanishes at the two points over $z=0$:
\begin{equation}
r_\pm:\qquad z=0\quad,\quad y(r_\pm)=\pm\sqrt{\Delta}
\label{ram}
\end{equation}
These are \emph{not} the four branch points $y=0$ of the hyperelliptic description \eqref{hyper}, at which
$\mathrm{d} X\neq0$. Near $r_\pm$, $z$ is a good local coordinate and the local involution is the restriction of the
global automorphism
\begin{equation}
\sigma:\ (w,z)\mapsto(w,-z)\quad,\quad\sigma^*\mathrm{d} z=-\mathrm{d} z\quad,\quad\sigma^*y=y\quad,\quad\sigma^*\omega=-\omega
\end{equation}
so that
\begin{equation}
\omega_{0,1}-\sigma^*\omega_{0,1}=\frac{4z^2\,\mathrm{d} z}{y}\quad,\quad y=y(r_\pm)\left(1-\frac{u}{\Delta}z^2+\mathcal{O}(z^4)\right)
\label{localdata}
\end{equation}

\subsubsection*{Bergman kernel}
For two points $p_i=(z_i,y_i)$ of the curve, consider
\begin{equation}
\omega_{0,2}(p_1,p_2)=\left[\frac{y_1y_2+(z_1z_2-u)^2-4\Lambda^4}{2\,y_1y_2\,(z_1-z_2)^2}-\frac{a}{4a'}\,\frac{1}{y_1y_2}\right]\mathrm{d} z_1\,\mathrm{d} z_2
\label{Bergman}
\end{equation}
The first term has the double pole \eqref{Bergdef} when $p_1\to p_2$, because $(z^2-u)^2-4\Lambda^4=y^2$; it is
regular when $p_1$ approaches the conjugate point $(z_2,-y_2)$, because the numerator then vanishes
to second order in $z_1-z_2$; and it is regular at the two points at infinity. The second term is a multiple
of $\omega(p_1)\omega(p_2)$, and its coefficient is fixed by the vanishing of the $A$-period. Indeed, at $p_2=r_\pm$ the bracket
in \eqref{Bergman} equals $1/2z_1^2+y(r_\pm)/2y_1z_1^2-a/4a'y_1y(r_\pm)$, and, using
\begin{equation}
\mathrm{d}\!\left(\frac{y}{z}\right)=\left(\frac{z^2}{y}-\frac{\Delta}{z^2y}\right)\mathrm{d} z\quad\Longrightarrow\quad\oint_A\frac{\mathrm{d} z}{z^2y}=\frac1\Delta\oint_A\frac{z^2\mathrm{d} z}{y}=\frac{\pi i\,a}{\Delta}
\end{equation}
together with \eqref{holo}, the $A$-period in $p_1$ vanishes. Being a holomorphic differential in $p_2$ which vanishes
at a point where $\omega\neq0$, it vanishes identically. Thus \eqref{Bergman} is the Bergman
kernel of $\Sigma_u$. The entire dependence on the choice of cycles sits in the single
number $a/a'$.

\section{The Three- and Four-Point Functions}
\label{sec:genus0}
\begin{proposition}
\label{prop:genus0}
For the spectral curve \eqref{spectral} of pure $SU(2)$:
\begin{align}
\oint_{B}\!\oint_{B}\omega_{0,3}(p,\cdot,\cdot)&=\frac{1}{2{a'}^2}\left(\frac{1}{z^2}-\frac{a}{2a'\Delta}\right)\frac{\mathrm{d} z}{y}\label{om03p}\\[4pt]
\oint_{B}\!\oint_{B}\!\oint_{B}\omega_{0,3}&=\frac{1}{\Delta\,{a'}^3}\label{om03}\\[4pt]
\oint_{B}\!\oint_{B}\!\oint_{B}\!\oint_{B}\omega_{0,4}&=\frac{1}{\Delta^2{a'}^4}\left(2u-\frac{3a}{4a'}\right)\label{om04}
\end{align}
where $a'=\mathrm{d} a/\mathrm{d} u$ and $\Delta=u^2-4\Lambda^4$.
\end{proposition}
\begin{proof}
Let $\mathcal{K}(p):=\oint_{p_1\in B}K(p_1,p)$. By \eqref{Bergper}, \eqref{holo} and \eqref{localdata}, in the coordinate $z$ near $r_\pm$,
\begin{equation}
\mathcal{K}=-\pi i\,\frac{\int_{\sigma(p)}^{p}\omega}{\omega_{0,1}-\sigma^*\omega_{0,1}}=-\frac{I(z)\,y(z)}{8a'z^2\,\mathrm{d} z}\quad,\quad I(z):=\int_{-z}^{z}\frac{\mathrm{d}\zeta}{y(\zeta)}
\label{Kint}
\end{equation}
with $I(z)\,y(z)=2z\,(1-\frac{2u}{3\Delta}z^2+\dots)$ and $I(z)/y(z)=\frac{2z}{\Delta}(1+\frac{4u}{3\Delta}z^2+\dots)$.

For $n=3$ the bracket in \eqref{TR} is $\omega_{0,2}(p,p_2)\omega_{0,2}(\sigma(p),p_3)+\omega_{0,2}(p,p_3)\omega_{0,2}(\sigma(p),p_2)$. Integrating $p_2$
and $p_3$ over $B$ turns it into $2(2\pi i)^2\omega\,\sigma^*\omega=-2(2\pi i)^2\omega^2$. If $p_1$ is kept free, the kernel \eqref{kernel} is needed with
\eqref{Bergman}; since $\int_{-z'}^{z'}\omega_{0,2}(p,\cdot)=2z'\,\omega_{0,2}(p,r_\pm)/\mathrm{d} z'+\mathcal{O}({z'}^3)$, each residue is elementary and
\begin{equation}
\begin{split}
\oint_{B}\!\oint_{B}\omega_{0,3}(p,\cdot,\cdot)&=\frac{1}{2{a'}^2}\sum_{r_\pm}\frac{1}{y(r_\pm)}\frac{\omega_{0,2}(p,r_\pm)}{\mathrm{d} z'}\\
&=\frac{1}{2{a'}^2}\sum_{\pm}\left[\pm\frac{1}{2\sqrt\Delta\,z^2}+\frac{1}{2yz^2}-\frac{a}{4a'\Delta\,y}\right]\mathrm{d} z
\end{split}
\end{equation}
which is \eqref{om03p}. If $p_1$ is integrated as well, $\mathcal{K}$ replaces $K$ and
\begin{equation}
\oint_{B}\!\oint_{B}\!\oint_{B}\omega_{0,3}=-2(2\pi i)^2\sum_{r_\pm}\mathop{\mathrm{Res}}_{z=0}\mathcal{K}\,\omega^2=\sum_{r_\pm}\mathop{\mathrm{Res}}_{z=0}\frac{I(z)}{4{a'}^3z^2y(z)}\,\mathrm{d} z=\frac{2}{2{a'}^3\Delta}
\end{equation}

For $n=4$ the bracket in \eqref{TR} contains, for each of the three ways of singling out one of the points
$p_2$, $p_3$, $p_4$, the two terms $\omega_{0,3}(p,\cdot,\cdot)\omega_{0,2}(\sigma(p),\cdot)+\omega_{0,3}(\sigma(p),\cdot,\cdot)\omega_{0,2}(p,\cdot)$. Let $\Phi$ denote the
differential \eqref{om03p}. It is odd under $\sigma$, like $\omega$, so that after the three integrations the bracket becomes
$3\cdot2\pi i\,(\Phi\,\sigma^*\omega+\sigma^*\Phi\,\omega)=-12\pi i\,\Phi\,\omega$, and
\begin{equation}
\oint_{B}\!\oint_{B}\!\oint_{B}\!\oint_{B}\omega_{0,4}=-12\pi i\sum_{r_\pm}\mathop{\mathrm{Res}}_{z=0}\mathcal{K}\,\Phi\,\omega=\sum_{r_\pm}\mathop{\mathrm{Res}}_{z=0}\frac{3}{8{a'}^4}\,\frac{I(z)}{z^2y(z)}\left(\frac1{z^2}-\frac{a}{2a'\Delta}\right)\mathrm{d} z
\end{equation}
The residue is $\frac{3}{8{a'}^4}\cdot\frac{2}{\Delta^2}\left(\frac{4u}{3}-\frac{a}{2a'}\right)$ at each of the two points, which gives \eqref{om04}.
\end{proof}

Proposition \ref{prop:genus0} can now be confronted with Theorem \ref{thm:prepotential}, independently of the general
proof of the latter. By \eqref{Fnorm} and \eqref{holo}, and then by \eqref{PF} and \eqref{Legendre},
\begin{equation}
\frac{\partial^3\mathcal{F}}{\partial a^3}=\frac{2\pi i}{a'}\frac{\mathrm{d}}{\mathrm{d} u}\frac{a_D'}{a'}=\frac{2\pi i}{{a'}^3}\left(a'a_D''-a''a_D'\right)=-\frac{2\pi i}{4\Delta {a'}^3}\left(a'a_D-a\,a_D'\right)=-\frac{1}{\Delta\,{a'}^3}
\label{d3F}
\end{equation}
in agreement with \eqref{om03} and \eqref{derivs}. Differentiating once more, and using \eqref{PF} in the form $a''=-a/4\Delta$,
\begin{equation}
\frac{\partial^4\mathcal{F}}{\partial a^4}=-\frac1{a'}\frac{\mathrm{d}}{\mathrm{d} u}\frac{1}{\Delta {a'}^3}=\frac{1}{\Delta^2{a'}^4}\left(2u-\frac{3a}{4a'}\right)
\label{d4F}
\end{equation}
which is \eqref{om04}. The agreement at $n=4$ is the first genuine test of the recursion: the
four-point function is built out of the three-point function \eqref{om03p} with one leg left
open, and the two terms in \eqref{om04} have different origins, the first coming from the
expansion of the curve around the ramification points and the second from the
normalisation of the Bergman kernel. Note also that integrating \eqref{om03p} over $B$ must give
back \eqref{om03}; with $\oint_B z^2\mathrm{d} z/y=\pi i\,a_D$ and $\oint_B\mathrm{d} z/y=2\pi i\,a_D'$ this is seen to be
equivalent to the Legendre relation \eqref{Legendre}, that is, to Matone's relation \eqref{Matone}.

Equation \eqref{om03} can also be read as
\begin{equation}
\frac{\mathrm{d}\tau}{\mathrm{d} u}=\frac{i}{2\pi}\,\frac{1}{\left(u^2-4\Lambda^4\right)\left(\mathrm{d} a/\mathrm{d} u\right)^2}
\end{equation}
which shows directly that $\tau$ is a local coordinate on the moduli space away from the singular points.

\section{Meeting Zamolodchikov's Recursion}
\label{sec:meeting}
The two recursions can now be compared on their common ground, the weak-coupling
region. On the one hand, inverting the expansion of the period \eqref{aint} in powers of $\Lambda^4/u^2$
gives $u(a)$ and hence, through \eqref{om03} and \eqref{om04}, the expansion of the $B$-periods of $\omega_{0,3}$ and $\omega_{0,4}$
in powers of $\Lambda^4/a^4$. On the other hand, the recursion \eqref{purerec} gives the coefficients $F_k$ of Table \ref{tab:Fk}, hence the
derivatives of \eqref{Ffull}. The outcome is
\begin{align}
\oint_{B}\!\oint_{B}\!\oint_{B}\omega_{0,3}&=\frac{1}{a}\sum_{k=0}^\infty t_k\frac{\Lambda^{4k}}{a^{4k}}=-\frac{\partial^3\mathcal{F}}{\partial a^3}\quad,\quad t_k=(4k-2)(4k-1)(4k)\,F_k\quad(k\geq1)\\
\oint_{B}\!\oint_{B}\!\oint_{B}\!\oint_{B}\omega_{0,4}&=\frac{1}{a^2}\sum_{k=0}^\infty(4k+1)\,t_k\frac{\Lambda^{4k}}{a^{4k}}=\frac{\partial^4\mathcal{F}}{\partial a^4}
\end{align}
with $t_0=8$ the one-loop term, and the coefficients computed in the two ways coincide up to the order
considered, $k=5$ (Table \ref{tab:meeting}). In the same way the coefficients of $u(a)$ obtained from the period
integral coincide with $kF_k$, as required by \eqref{Matone}.

\begin{table}[h]
\caption{Coefficients of the weak-coupling expansion obtained independently from the Seiberg-Witten curve
(period integral \eqref{aint} and topological recursion, equations \eqref{om03}, \eqref{om04}) and from Zamolodchikov's recursion \eqref{purerec}. The two computations give identical numbers.}
\label{tab:meeting}
\centering
\renewcommand{\arraystretch}{1.6}
\begin{tabular}{c|cccccc}
$k$&0&1&2&3&4&5\\\hline
\rule[-2.4ex]{0pt}{6ex}$u(a)$: $\;kF_k$&$1$&$\dfrac12$&$\dfrac{5}{32}$&$\dfrac{9}{64}$&$\dfrac{1469}{8192}$&$\dfrac{4471}{16384}$\\
\rule[-2.4ex]{0pt}{6ex}$\oint^3\omega_{0,3}$: $\;t_k$&$8$&$12$&$\dfrac{105}{4}$&$\dfrac{495}{8}$&$\dfrac{154245}{1024}$&$\dfrac{764541}{2048}$\\
\rule[-2.4ex]{0pt}{6ex}$\oint^4\omega_{0,4}$: $\;(4k+1)t_k$&$8$&$60$&$\dfrac{945}{4}$&$\dfrac{6435}{8}$&$\dfrac{2622165}{1024}$&$\dfrac{16055361}{2048}$
\end{tabular}
\end{table}

This is the precise sense in which the two recursions are transverse. Each number $t_k$
of Table \ref{tab:meeting} is one coefficient of one $B$-period: Zamolodchikov's recursion produces the table
column by column, all derivatives at a given instanton number, while the topological
recursion produces it row by row, all instanton numbers at a given order of the Taylor expansion in $a$. A
single step of either recursion is an infinite resummation from the point of view of the other.

The difference becomes visible as soon as one leaves weak coupling. The functions \eqref{om03}, \eqref{om04}
are exact, whereas the instanton series converges at best for $\Lambda^4/\lvert a\rvert^4<(\pi/4)^4\simeq0.38$, the value
reached at the monopole and dyon points, where $a=4\Lambda/\pi$ up to a phase by \eqref{aint}. Table \ref{tab:numerics} shows three
representative points: $u=6\Lambda^2$, where $\tau=2i$; $u=2.5\Lambda^2$, close to the monopole point; and the
$\mathbb{Z}_2$-symmetric point $u=0$ in the middle of the strong-coupling region, where the curve is perfectly
smooth and the topological recursion applies, but the instanton series diverges.

\begin{table}[h]
\caption{The third derivative of the prepotential in units $\Lambda=1$: exact value from the topological recursion, equation \eqref{om03},
against the instanton series truncated at five instantons. At $u=0$ the period is continued from $\mathrm{Im}\,u>0$.}
\label{tab:numerics}
\centering
\renewcommand{\arraystretch}{1.3}
\begin{tabular}{c|c|c|c|c}
$u$&$a$&$\Lambda^4/\lvert a\rvert^4$&$\oint^3\omega_{0,3}=-\partial_a^3\mathcal{F}$&five instantons\\\hline
$6$&$2.432013$&$0.029$&$3.4399565$&$3.4399563$\\
$2.5$&$1.504079$&$0.195$&$8.1364$&$8.0677$\\
$0$&$0.539353\,(1+i)$&$2.95$&$2.432417\,(1-i)$&divergent
\end{tabular}
\end{table}

\section{Strong Coupling: the Monopole Point}
\label{sec:monopole}
Theorem \ref{thm:prepotential} holds for any choice of symplectic basis. Near $u=2\Lambda^2$ the natural basis is the dual one,
$(\tilde A,\tilde B)=(B,-A)$, in which the local coordinate is $a_D$ and the prepotential is the dual
prepotential $\mathcal{F}_D(a_D)$, $\partial\mathcal{F}_D/\partial a_D=-2\pi i\,a$. Nothing has to be recomputed: in the Bergman kernel
\eqref{Bergman} and in Proposition \ref{prop:genus0} the cycles enter only through $a$ and $a'$, so that
\begin{equation}
-\frac{\partial^3\mathcal{F}_D}{\partial a_D^3}=\oint_{\tilde B}\!\oint_{\tilde B}\!\oint_{\tilde B}\tilde\omega_{0,3}=\frac{1}{\Delta\,(\mathrm{d} a_D/\mathrm{d} u)^3}
\label{dual3}
\end{equation}
Near the monopole point the $B$-cycle shrinks and \eqref{periods} gives $a_D=i\,(u-2\Lambda^2)/2\Lambda+\dots$, hence
\begin{equation}
\frac{\partial^3\mathcal{F}_D}{\partial a_D^3}=+\frac{1}{a_D}+\dots\qquad\text{to be compared with}\qquad\frac{\partial^3\mathcal{F}}{\partial a^3}=-\frac{8}{a}+\dots\quad(a\to\infty)
\label{poles}
\end{equation}
These two numbers have a clear physical meaning. In the normalisation \eqref{Fnorm} a hypermultiplet of
charge $n$ under the low-energy $U(1)$ contributes $n^2/a$ to the third derivative of the
prepotential at one loop, and a vector multiplet $-2n^2/a$. Indeed, a multiplet which becomes massless at $a=0$
contributes to $\mathcal{F}$ a term proportional to $n^2a^2\ln(na)$, and the coefficient is $-1$ for a vector multiplet, as one reads off from
\eqref{Ffull} for the $W$-boson of charge $2$, and $+\frac12$ for a hypermultiplet, which contributes to the $\beta$-function
$-\frac12$ of what a vector multiplet does, as in the factor $2N-N_f$ of \eqref{eq:one-loop-prepotential}. At infinity one finds the $W$-boson,
a vector multiplet of charge $2$: $-2\cdot2^2=-8$. At $u=2\Lambda^2$ one finds $+1$: a single hypermultiplet of
unit magnetic charge, the monopole of Seiberg and Witten \cite{SW2}, whose condensation was
discussed in Section \ref{sec:Less-Supersymmetric-SYM}. The $B$-period of $\omega_{0,3}$ thus has a simple pole at each cusp of the
moduli space, and its residue counts the BPS states which become massless there. The instanton
expansion, and hence Zamolodchikov's recursion in the form \eqref{purerec}, has no access to the second of the two numbers in \eqref{poles}.

\section{Beyond Genus Zero: the $\Omega$-Background}
\label{sec:genus1}
Zamolodchikov's recursion contains much more than the prepotential, namely the
complete dependence on $\epsilon_1,\epsilon_2$. Expanding the logarithm of the solution of \eqref{purerec} to
the next order one finds
\begin{equation}
-\epsilon_1\epsilon_2\log Z_{\rm inst}=\mathcal{F}_{\rm inst}+(\epsilon_1+\epsilon_2)^2\sum_{k\geq1}h_k\frac{\Lambda^{4k}}{a^{4k}}+\epsilon_1\epsilon_2\sum_{k\geq1}g_k\frac{\Lambda^{4k}}{a^{4k}}+\mathcal{O}(\epsilon^4)
\label{eps2}
\end{equation}
with the coefficients of Table \ref{tab:eps2}.
\begin{table}[h]
\caption{The coefficients of the first $\epsilon$-corrections \eqref{eps2} generated by the recursion \eqref{purerec}.}
\label{tab:eps2}
\centering
\renewcommand{\arraystretch}{1.6}
\begin{tabular}{c|ccccc}
$k$&1&2&3&4&5\\\hline
\rule[-2.4ex]{0pt}{6ex}$h_k$&$\dfrac18$&$\dfrac{21}{128}$&$\dfrac{55}{192}$&$\dfrac{18445}{32768}$&$\dfrac{193401}{163840}$\\
\rule[-2.4ex]{0pt}{6ex}$g_k$&$0$&$-\dfrac{1}{32}$&$-\dfrac{1}{12}$&$-\dfrac{1647}{8192}$&$-\dfrac{981}{2048}$
\end{tabular}
\end{table}

These numbers are not new data: they are again functions on the Seiberg-Witten
curve. Up to the order considered,
\begin{equation}
-\epsilon_1\epsilon_2\log Z_{\rm inst}=\mathcal{F}_{\rm inst}-\frac{\epsilon_1\epsilon_2}{2}\ln\!\left(\frac{1}{2a}\frac{\mathrm{d} u}{\mathrm{d} a}\right)-\frac{\epsilon_1^2+\epsilon_2^2}{24}\ln\frac{\Delta}{a^4}+\mathcal{O}(\epsilon^4)
\label{genusone}
\end{equation}
where $u(a)$ is given by \eqref{Matone}. This is the theorem of Nakajima and Yoshioka \cite{NYlectures}: the
coefficients of $\epsilon_1\epsilon_2$ and of $(\epsilon_1^2+\epsilon_2^2)/3$ are the logarithms of the two measure factors
$(\mathrm{d} u/\mathrm{d} a)^{1/2}$ and $\Delta^{1/8}$ of the $u$-plane integral of Moore and Witten \cite{MW}. Here it is recovered, with its
normalisation, from Zamolodchikov's recursion alone.

In the language of the conformal block \eqref{pureH} the statement \eqref{genusone} reads as follows. Let $c$ be fixed and
let $\alpha^2=Q^2/4-\Delta_{\rm int}\to\infty$ with $\xi:=16q/\alpha^4=\Lambda^4/a^4$ fixed, where $\Delta_{\rm int}$ denotes the internal dimension of \eqref{dims}. Then
\begin{equation}
H^{(0)}(\Delta_{\rm int},q)\simeq\exp\!\Big[-\alpha^2\sum_{k\geq1}F_k\xi^k\Big]\left(\mathsf{f}-2\xi\mathsf{f}'\right)^{\frac12}\left(\mathsf{f}^2-4\xi\right)^{\frac{c-13}{144}}
\label{blocklimit}
\end{equation}
where $\mathsf{f}(\xi)=1+\sum_{k\geq1}kF_k\xi^k$, up to corrections of order $1/\alpha^2$, since $(\epsilon_1^2+\epsilon_2^2)/\epsilon_1\epsilon_2=Q^2-2=(c-13)/6$. The irregular
conformal block exponentiates in this limit, the exponent being the prepotential, and the first
correction is a product of two factors which carry the whole dependence on the
central charge; the second factor is a power of the discriminant of the Seiberg-Witten curve and disappears at $c=13$.

On the self-dual background $\epsilon_1=-\epsilon_2=\hbar$, i.e. $c=1$, the expansion organises itself in even
powers of $\hbar$,
\begin{equation}
\log Z=\sum_{g\geq0}\hbar^{2g-2}F^{(g)}(a)\quad,\quad F^{(0)}=\mathcal{F}\quad,\quad F^{(1)}=-\frac12\ln\frac{\mathrm{d} a}{\mathrm{d} u}-\frac1{12}\ln\Delta
\label{genusexp}
\end{equation}
where the instanton part of $F^{(1)}$ is what \eqref{genusone} gives at $\epsilon_1+\epsilon_2=0$, and an additive constant has been dropped. This is exactly the
shape of the expansion produced by the topological recursion, which defines $\omega_{g,n}$ for all
genera. The first of them is the one-point function $\omega_{1,1}(p_1)=\sum_r\mathop{\mathrm{Res}}_{p\to r}K(p_1,p)\,\omega_{0,2}(p,\sigma_r(p))$.

\begin{proposition}
\label{prop:genus1}
For the spectral curve \eqref{spectral} of pure $SU(2)$,
\begin{equation}
\oint_{B}\omega_{1,1}=\frac{1}{\Delta\,a'}\left(\frac{u}{6}-\frac{a}{8a'}\right)=-\frac{\partial F^{(1)}}{\partial a}
\label{om11}
\end{equation}
with $F^{(1)}$ as in \eqref{genusexp}.
\end{proposition}
\begin{proof}
Evaluating \eqref{Bergman} at $p_2=\sigma(p_1)$, i.e. $z_2=-z_1=-z$, $y_2=y_1=y$, and using
$(z^2+u)^2-4\Lambda^4=y^2+4uz^2$,
\begin{equation}
\omega_{0,2}(p,\sigma(p))=-\left[\frac{1}{4z^2}+\frac{1}{y^2}\left(\frac u2-\frac{a}{4a'}\right)\right]\mathrm{d} z^2
\end{equation}
With $\mathcal{K}$ of \eqref{Kint},
\begin{equation}
\begin{split}
\oint_{B}\omega_{1,1}&=\sum_{r_\pm}\mathop{\mathrm{Res}}_{z=0}\frac{I(z)y(z)}{8a'z^2}\left[\frac{1}{4z^2}+\frac{1}{y^2}\left(\frac u2-\frac{a}{4a'}\right)\right]\mathrm{d} z\\
&=\frac{2}{4a'}\left[-\frac{u}{6\Delta}+\frac1\Delta\left(\frac u2-\frac{a}{4a'}\right)\right]
\end{split}
\end{equation}
which is the first equality. For the second, $\partial_aF^{(1)}=\frac{1}{a'}\left(-\frac{a''}{2a'}-\frac{u}{6\Delta}\right)$ and $a''=-a/4\Delta$ by \eqref{PF}.
\end{proof}

Thus the genus-one step of the topological recursion on the Seiberg-Witten curve
reproduces the first $\hbar$-correction of the Nekrasov partition function on the self-dual
background, with the same sign rule as in \eqref{derivs}. At weak coupling, \eqref{om11} expands as
\begin{equation}
\oint_{B}\omega_{1,1}=\frac1a\left(-\frac16+\frac14\frac{\Lambda^8}{a^8}+\frac{\Lambda^{12}}{a^{12}}+\frac{1647}{512}\frac{\Lambda^{16}}{a^{16}}+\frac{4905}{512}\frac{\Lambda^{20}}{a^{20}}+\dots\right)
\end{equation}
and the coefficients are $-4kg_k$ with $g_k$ of Table \ref{tab:eps2}, as they should: the instanton part
of $F^{(1)}$ is $-\sum_kg_k\Lambda^{4k}/a^{4k}$.

\section{A Conjecture, Open Problems and Conclusions}
\label{sec:outlook}
The results of Sections \ref{sec:genus0}--\ref{sec:genus1} are instances of one statement.

\begin{conjecture}
\label{conj}
Let $F^{(g)}(a)$ be the coefficients of the expansion \eqref{genusexp} of the pure $SU(2)$ Nekrasov partition
function on the self-dual $\Omega$-background, and let $\omega_{g,n}$ be the multi-differentials produced by the
topological recursion on the Seiberg-Witten spectral curve \eqref{spectral}. Then, for all $g\geq0$ and
$n\geq1$ with $2g-2+n>0$,
\begin{equation}
\oint_{B}\!\!\cdots\!\oint_{B}\omega_{g,n}=(-1)^n\frac{\partial^nF^{(g)}}{\partial a^n}
\label{conjeq}
\end{equation}
\end{conjecture}

For $g=0$ this is Theorem \ref{thm:prepotential}, verified here explicitly for $n=3,4$; for $(g,n)=(1,1)$ it is
Proposition \ref{prop:genus1}. Since a $B$-period of $\omega_{g,n+1}$ is the derivative of $\omega_{g,n}$ with respect to the
modulus $a$ at fixed $X$ \cite{EO}, the conjecture amounts to one statement for each genus. It is certainly
not unexpected: the self-dual Nekrasov partition function can be written as a matrix
model whose spectral curve is the Seiberg-Witten curve \cite{KlemmSulkowski}, and the Whittaker vectors
whose norm is the pure-gauge partition function \cite{Gaiotto} have recently been shown to be
computed by a topological recursion \cite{BBCC}. What the present chapter adds is an elementary and
completely explicit verification, in closed form, at the first three non-trivial steps
$(g,n)=(0,3),(0,4),(1,1)$, together with the corresponding output of Zamolodchikov's recursion.

Several questions remain open, and they are listed here in increasing order of
difficulty.

\begin{enumerate}
\item \emph{Higher steps.} The next cases, $(g,n)=(1,2)$ and $(2,1)$, can be attacked with the same closed
formulas \eqref{Bergman}, \eqref{om03p}. The target for genus two is the coefficient of $\epsilon^4$ in \eqref{eps2}, which the
recursion \eqref{purerec} generates without difficulty.

\item \emph{Generic $\epsilon_1,\epsilon_2$.} Away from the self-dual line the coefficients $h_k$ of Table \ref{tab:eps2} appear, and by
\eqref{genusone} they are governed by the discriminant alone. The natural candidate on the side of the
curve is the $\beta$-deformed version of the topological recursion \cite{ChekhovEynard}. The numbers $h_k$ and
the closed form \eqref{genusone} are the first test that any such identification must pass.

\item \emph{The self-dual line is singular for Zamolodchikov's recursion.} The two recursions
cannot be mapped into each other term by term. The coefficients \eqref{pureR} have poles at every
rational value of $\epsilon_1/\epsilon_2$; for instance the two contributions to $Z_2$ with $mn=2$ are
\begin{equation}
\frac{1}{\epsilon_1\epsilon_2^3\left(\epsilon_1^2-\epsilon_2^2\right)\left(4a^2-(\epsilon_1+2\epsilon_2)^2\right)}\quad\text{and}\quad-\frac{1}{\epsilon_1^3\epsilon_2\left(\epsilon_1^2-\epsilon_2^2\right)\left(4a^2-(2\epsilon_1+\epsilon_2)^2\right)}
\end{equation}
each of which diverges on the line $\epsilon_1+\epsilon_2=0$, where the topological recursion lives,
while their sum \eqref{Z2} is finite. On that line the simple poles of \eqref{purerec} in $a$ merge into double
poles and into a pole at $a=0$. Understanding how the residues at the ramification
points of \eqref{TR} organise this cancellation would amount to a derivation of one recursion from the other.

\item \emph{Matter.} For $f>0$ the Seiberg-Witten differential acquires poles with residues
given by the masses (Section \ref{subsec:SW-curves-general-remarks}), and Theorem \ref{thm:prepotential} does not apply as it stands. For $f=4$ the
elliptic nome $q$ of \eqref{nome} is the modulus of the massless curve, so that the leading term of
\eqref{F4asym} is the $n=2$ term of \eqref{Taylor}; the mass-dependent terms of \eqref{F4asym} should correspond to
the additional periods around the poles.
\end{enumerate}

\subsubsection*{Conclusions}
The path followed in this thesis started from the ADHM construction and the
measure on the moduli space of instantons, went through the Seiberg-Witten solution
and the localization formulae which reproduce it instanton by instanton, and arrived
at the AGT correspondence, where the same partition function is a conformal block
and obeys Zamolodchikov's recursion. In this last chapter that recursion was run for
the pure $SU(2)$ theory and confronted with the topological recursion on the Seiberg-Witten
curve. The two turned out to be complementary descriptions of the same object. One is linear, exact in
the vacuum expectation value and in the $\Omega$-background, perturbative in the instanton
parameter. The other is quadratic, exact in the instanton parameter at a fixed vacuum,
perturbative in the displacement from it and in the genus. They agree wherever they
were compared: on the prepotential to five instantons through its third and fourth
derivatives, and on the first gravitational correction through the genus-one one-point
function. The curve that Seiberg and Witten introduced to encode the monodromies
of a low-energy effective action is, at the same time, the initial datum of a recursion
which reconstructs the partition function around any vacuum, including those in which
the light degrees of freedom are magnetic monopoles.

\section*{Acknowledgement}

This work would be impossible without the resources provided by INFN and, in particular, Professor Francesco Fucito and Fabio Riccioni. The reviews by N. Dorey \textit{et. al.} \cite{Instanton Calculus} and S. Vandoren \textit{et. al.} \cite{Instantons} had a great impact on this work.

\appendix
\chapter{Independent Checks of the Formulae of the Last Two Chapters}
\label{app:checks}

The statements of Section \ref{sec:pure} and of Chapter \ref{ch:TR} which are not quoted from the literature
were checked independently of the derivations given in the text. Two kinds of checks were used:
exact ones, carried out in rational arithmetic, and numerical ones, carried out with 30 significant
digits. They are listed here so that they can be repeated.

\section{The Instanton Sum and the Recursion}
\begin{enumerate}
\item \emph{Recursion against the sum over Young tableaux.} The coefficients $Z_k$, $k\leq5$, of the sum \eqref{Zinst}, with
\eqref{Pgauge} and \eqref{Eab} for $N=2$, $a_1=-a_2=a$ and no matter, and the coefficients generated by the recursion
\eqref{purerec}, \eqref{pureR}, were evaluated as exact rational numbers at the three points
\[
(a,\epsilon_1,\epsilon_2)=\left(\tfrac73,\tfrac25,\tfrac3{11}\right),\quad\left(-\tfrac{13}7,\tfrac59,-\tfrac2{13}\right),\quad\left(\tfrac{31}{10},\tfrac17,\tfrac43\right)
\]
They coincide for all $k\leq5$.
\item \emph{Closed forms.} At the same points \eqref{Z1} and \eqref{Z2} coincide with $Z_1$ and $Z_2$, and the two contributions
with $mn=2$ displayed in Section \ref{sec:outlook}, added to the contribution of $(m,n)=(1,1)$, give $Z_2$.
\item \emph{Expansion in the $\Omega$-background.} With $\epsilon_1=t\,p$ and $\epsilon_2=t\,q$, every term of \eqref{Zinst} was expanded as a
Laurent series in $t$ with exact coefficients, at
\[
(a,p,q)=\left(\tfrac32,1,\tfrac{17}{13}\right),\quad\left(\tfrac53,1,-\tfrac{19}7\right),\quad\left(\tfrac74,1,-1\right),\quad\left(\tfrac95,\tfrac{23}9,1\right)
\]
In $-\epsilon_1\epsilon_2\log Z_{\rm inst}$ all negative powers of $t$ cancel up to $k=5$. The term of order $t^0$ gives the
coefficients $F_k$ of Table \ref{tab:Fk} at all four points, and the term of order $t$ vanishes. The term of order $t^2$ has the form
$\left[(p+q)^2h_k+pq\,g_k\right]a^{-4k}$ required by \eqref{eps2}: the first two points determine $h_k$ and $g_k$ of Table \ref{tab:eps2}, and the
other two, one of which is the self-dual background, confirm them.
\item \emph{Four flavours.} The expansion \eqref{qofx} and the series for $\log(16q/x)$ below \eqref{F4asym} were re-derived from \eqref{xofq}.
\item \emph{Weights at the fixed points.} The equality of the two expressions \eqref{tangent} and \eqref{character} for the character of the tangent
space was verified, as an identity between Laurent polynomials, for all the $74$ pairs of Young tableaux with at most
five boxes ($N=2$).
\end{enumerate}

\section{Conformal Blocks and the AGT Relation}
In the following checks $\epsilon_1=b$, $\epsilon_2=1/b$ and the two sets of values
\[
(b,a;\mu_1,\mu_2,\mu_3,\mu_4)=\left(\tfrac32,\tfrac75;\tfrac13,\tfrac27,\tfrac54,-\tfrac3{11}\right),\quad\left(\tfrac57,-\tfrac94;2,\tfrac15,-\tfrac43,\tfrac79\right)
\]
were used; all the quantities are then rational numbers.
\begin{enumerate}
\item \emph{Zamolodchikov's recursion.} The coefficients of $x$, $x^2$ and $x^3$ of the conformal block were computed from the definition
\eqref{blockdef}, with the Gram matrices \eqref{Gram} obtained from the Virasoro algebra \eqref{Vir}, and from the representation \eqref{block} with
$H$ generated by the recursion \eqref{Zrec}, \eqref{Rmn}. They coincide.
\item \emph{AGT relation with four flavours.} The coefficients of $x$, $x^2$ and $x^3$ of the sum over Young tableaux \eqref{Zinst} with $f=4$
and those of the right-hand side of the first line of \eqref{AGT4}, computed from \eqref{blockdef}, coincide; in particular \eqref{AGT1} holds,
with $Z_1$ given by \eqref{Z1matter}. The agreement requires the sign conventions of \eqref{Eab} and \eqref{Vchar} as they are written in
this thesis.
\item \emph{Pure theory.} The norm \eqref{Gnorm} of the Whittaker state coincides with the instanton sum up to $n=3$, and the
conditions $L_1\lvert G\rangle=\mathfrak{x}^{1/2}\lvert G\rangle$, $L_2\lvert G\rangle=0$ were verified up to the level five.
\end{enumerate}

\section{The Seiberg-Witten Curve}
\begin{enumerate}
\item \emph{Weak-coupling series.} Expanding the integrand of \eqref{aint} gives
\[
\frac{a}{\sqrt u}=\sum_{n=0}^{\infty}\binom{1/2}{2n}\binom{2n}{n}\left(\frac{\Lambda^4}{u^2}\right)^{n}
\]
This series was inverted exactly up to the order $\Lambda^{20}$. The coefficients of $u(a)$ are $kF_k$, which is \eqref{Matone};
inserted in \eqref{om03}, \eqref{om04} and \eqref{om11} the series gives the numbers $t_k$ and $(4k+1)t_k$ of Table \ref{tab:meeting}
and the coefficients $-4kg_k$; inserted in the right-hand side of \eqref{genusone} it gives $h_k$ as the coefficient
of $\epsilon_1^2+\epsilon_2^2$ and $2h_k+g_k$ as the coefficient of $\epsilon_1\epsilon_2$.
\item \emph{Picard-Fuchs equation.} The identity \eqref{PFexact} was verified by computer algebra, and the asymptotic form \eqref{aDasym} of $\mathrm{d}a_D/\mathrm{d}u$
was compared numerically with the exact elliptic integral, the ratio being $1.0002$ at $u=50\Lambda^2$ and $1.000002$ at $u=500\Lambda^2$.
\item \emph{Residues.} The three residues in the proofs of Propositions \ref{prop:genus0} and \ref{prop:genus1}, the values of the Bergman
kernel \eqref{Bergman} at $p_2=r_\pm$ and at $p_2=\sigma(p_1)$, and the identities \eqref{d3F}, \eqref{d4F} and the second equality in \eqref{om11}
were recomputed by computer algebra, the last three from \eqref{PF} and \eqref{Legendre} alone.
\item \emph{Periods.} At $u/\Lambda^2=6$, $2.5$ and $3+0.7\,i$ the integral \eqref{aint} equals the hypergeometric expression \eqref{eq:a-hypergeometric}
evaluated at $u/2$; this function and twice the function \eqref{eq:aD-hypergeometric} evaluated at $u/2$ satisfy the Picard-Fuchs
equation \eqref{PF} and the relation \eqref{Legendre}, and their ratio of derivatives is $\tau=2i$ at $u=6\Lambda^2$.
\item \emph{Derivatives of the prepotential.} At the same three points $\partial_a^3\mathcal{F}$ and $\partial_a^4\mathcal{F}$, obtained by numerical
differentiation of $\tau=a_D'/a'$, agree to ten digits with \eqref{om03} and \eqref{om04}, with the signs of \eqref{derivs}; in the same way
$-\partial_aF^{(1)}$ agrees with \eqref{om11}.
\item \emph{Bergman kernel.} The $A$-period of \eqref{Bergman} in $p_1$, at a generic point $p_2$ and for $u=6\Lambda^2$, vanishes within
the numerical precision ($10^{-29}$), and $\oint_A\mathrm{d}z/y=2\pi i\,a'$, $\oint_A\mathrm{d}z/z^2y=\pi i\,a/\Delta$.
\item \emph{Strong coupling.} The entries of Table \ref{tab:numerics} were recomputed; at $u=2\Lambda^2$ one finds $a=4\Lambda/\pi$, and for
$u\to2\Lambda^2$ one finds $a_D/(u-2\Lambda^2)\to i/2\Lambda$ and $a_D\,\partial^3\mathcal{F}_D/\partial a_D^3\to1$, as in \eqref{poles}.
\end{enumerate}

\backmatter
\cleardoublepage
\phantomsection 
\addcontentsline{toc}{chapter}{\bibname}

\end{document}